%% file: sample62.tex
\documentclass[twocolumn]{aastex63}

\graphicspath{{./}{figures/}}
\usepackage{makecell, mathtools}
\usepackage{hyperref}
\usepackage{multirow}

\shorttitle{30 Tidal Disruption Events from the ZTF-I Survey}
\shortauthors{Hammerstein et al.}

\begin{document}

\title{The Final Season Reimagined: 30 Tidal Disruption Events from the ZTF-I Survey}

\author[0000-0002-5698-8703]{Erica Hammerstein}
\affil{Department of Astronomy, University of Maryland, College Park, MD 20742, USA}
\affil{Astrophysics Science Division, NASA Goddard Space Flight Center, 8800 Greenbelt Road, Greenbelt, MD 20771, USA}
\affil{Center for Research and Exploration in Space Science and Technology, NASA/GSFC, Greenbelt, MD 20771, USA}

\author[0000-0002-3859-8074]{Sjoert van Velzen}
\affil{Leiden Observatory, Leiden University, PO Box 9513, 2300 RA Leiden, The Netherlands}

\author[0000-0003-3703-5154]{Suvi Gezari}
\affil{Space Telescope Science Institute, 3700 San Martin Drive, Baltimore, MD 21218, USA}

\author[0000-0003-1673-970X]{S.~Bradley Cenko}
\affil{Astrophysics Science Division, NASA Goddard Space Flight Center, 8800 Greenbelt Road, Greenbelt, MD 20771, USA}
\affil{Joint Space-Science Institute, University of Maryland, College Park, MD 20742, USA}

\author[0000-0001-6747-8509]{Yuhan Yao}
\affil{Division of Physics, Mathematics, and Astronomy, California Institute of Technology, Pasadena, CA 91125, USA}

\author[0000-0002-4557-6682]{Charlotte Ward}
\affil{Department of Astronomy, University of Maryland, College Park, MD 20742, USA}

\author[0000-0001-9676-730X]{Sara Frederick}
\affil{Department of Physics and Astronomy, Vanderbilt University, VU Station 1807, Nashville, TN 37235, USA}

\author[0000-0001-6917-4656]{Natalia Villanueva}
\affil{Department of Astronomy, Harvard University, Cambridge, MA 02138, USA}

\author[0000-0001-8426-5732]{Jean J.~Somalwar}
\affil{Division of Physics, Mathematics, and Astronomy, California Institute of Technology, Pasadena, CA 91125, USA}

\author[0000-0002-3168-0139]{Matthew J.~Graham}
\affil{Division of Physics, Mathematics, and Astronomy, California Institute of Technology, Pasadena, CA 91125, USA}

\author[0000-0001-5390-8563]{Shrinivas R.~Kulkarni}
\affil{Division of Physics, Mathematics, and Astronomy, California Institute of Technology, Pasadena, CA 91125, USA}

\author[0000-0003-2686-9241]{Daniel Stern}
\affil{Jet Propulsion Laboratory, California Institute of Technology, 4800 Oak Grove Drive, Pasadena, CA 91109, USA}

\author[0000-0002-8977-1498]{Igor Andreoni}
\altaffiliation{Neil Gehrels Fellow}
\affil{Joint Space-Science Institute, University of Maryland, College Park, MD 20742, USA}
\affil{Department of Astronomy, University of Maryland, College Park, MD 20742, USA}
\affil{Astrophysics Science Division, NASA Goddard Space Flight Center, 8800 Greenbelt Road, Greenbelt, MD 20771, USA}

\author[0000-0001-8018-5348]{Eric C. Bellm}
\affiliation{DIRAC Institute, Department of Astronomy, University of Washington, 3910 15th Avenue NE, Seattle, WA 98195, USA}

\author[0000-0002-5884-7867]{Richard Dekany}
\affil{Caltech Optical Observatories, California Institute of Technology, Pasadena, CA 91125, USA}

\author{Suhail Dhawan}
\affil{Institute of Astronomy and Kavli Institute for Cosmology, University of Cambridge, Madingley Road, Cambridge CB3 0HA, UK}

\author{Andrew J.~Drake}
\affil{Division of Physics, Mathematics, and Astronomy, California Institute of Technology, Pasadena, CA 91125, USA}

\author{Christoffer Fremling}
\affil{Division of Physics, Mathematics, and Astronomy, California Institute of Technology, Pasadena, CA 91125, USA}

\author{Pradip Gatkine}
\affil{Division of Physics, Mathematics, and Astronomy, California Institute of Technology, Pasadena, CA 91125, USA}

\author[0000-0001-5668-3507]{Steven L. Groom}
\affiliation{IPAC, California Institute of Technology, 1200 E. California Blvd, Pasadena, CA 91125, USA}

\author[0000-0002-9017-3567]{Anna Y. Q.~Ho}
\affiliation{Department of Astronomy, Cornell University, Ithaca, NY 14853, USA}

\author[0000-0002-5619-4938]{Mansi M. Kasliwal}
\affil{Division of Physics, Mathematics, and Astronomy, California Institute of Technology, Pasadena, CA 91125, USA}

\author{Viraj Karambelkar}
\affil{Division of Physics, Mathematics, and Astronomy, California Institute of Technology, Pasadena, CA 91125, USA}

\author[0000-0002-7252-3877]{Erik C.~Kool}
\affil{The Oskar Klein Centre, Department of Astronomy, Stockholm University, AlbaNova, SE-10691, Stockholm, Sweden}

\author[0000-0002-8532-9395]{Frank J. Masci}
\affiliation{IPAC, California Institute of Technology, 1200 E. California Blvd, Pasadena, CA 91125, USA}

\author[0000-0002-7226-0659]{Michael S. Medford}
\affil{University of California, Berkeley, Department of Astronomy, Berkeley, CA 94720, USA}
\affil{Lawrence Berkeley National Laboratory, 1 Cyclotron Rd., Berkeley, CA 94720, USA}

\author[0000-0001-8472-1996]{Daniel A. Perley}
\affil{Astrophysics Research Institute, Liverpool John Moores University, IC2, Liverpool Science Park, 146 Brownlow Hill, Liverpool L3 5RF, UK}

\author[0000-0003-1227-3738]{Josiah Purdum}
\affiliation{Caltech Optical Observatories, California Institute of Technology, Pasadena, CA 91125, USA}

\author[0000-0002-2626-2872]{Jan van~Roestel}
\affiliation{Division of Physics, Mathematics, and Astronomy, California
Institute of Technology, Pasadena, CA 91125, USA}

\author[0000-0003-4531-1745]{Yashvi Sharma}
\affiliation{Division of Physics, Mathematics, and Astronomy, California
Institute of Technology, Pasadena, CA 91125, USA}

\author[0000-0003-1546-6615]{Jesper Sollerman}
\affiliation{The Oskar Klein Centre, Department of Astronomy, Stockholm University, AlbaNova, SE-10691, Stockholm, Sweden}

\author[0000-0002-5748-4558]{Kirsty Taggart}
\affil{Department of Astronomy and Astrophysics, University of California, Santa Cruz, CA 95064, USA}

\author{Lin Yan}
\affiliation{Caltech Optical Observatories, California Institute of Technology, Pasadena, CA 91125, USA}

\correspondingauthor{Erica Hammerstein}
\email{ekhammer@astro.umd.edu}

\begin{abstract}
Tidal disruption events (TDEs) offer a unique way to study dormant black holes. While the number of observed TDEs has grown thanks to the emergence of wide-field surveys in the past few decades, questions regarding the nature of the observed optical, UV, and X-ray emission remain. We present a uniformly selected sample of 30 spectroscopically classified TDEs from the Zwicky Transient Facility Phase I survey operations with follow-up \textit{Swift} UV and X-ray observations. Through our investigation into correlations between light curve properties, we recover a shallow positive correlation between the peak bolometric luminosity and decay timescales. We introduce a new spectroscopic class of TDE, TDE-featureless, which are characterized by featureless optical spectra. The new TDE-featureless class shows larger peak bolometric luminosities, peak blackbody temperatures, and peak blackbody radii. We examine the differences between the X-ray bright and X-ray faint populations of TDEs in this sample, finding that X-ray bright TDEs show higher peak blackbody luminosities than the X-ray faint sub-sample. This sample of optically selected TDEs is the largest sample of TDEs from a single survey yet, and the systematic discovery, classification, and follow-up of this sample allows for robust characterization of TDE properties, an important stepping stone looking forward toward the Rubin era.
\end{abstract}

\section{Introduction}
A tidal disruption event (TDE) occurs when a star passes close enough to a massive black hole (MBH), such that the tidal forces are stronger than the star's self-gravity and the star is ripped apart, causing a luminous flare of radiation from $\sim$half of the stellar debris that circularizes into an accretion disk and is accreted \citep{Rees88, Evans1989, Ulmer1999}. While these events were first predicted theoretically almost 50 years ago \citep[e.g.,][]{Hills75, Lidskii1979}, the advent of all-sky surveys across the electromagnetic spectrum in the past several decades has been a catalyst for the discovery of these transients.

TDEs have now been observed from the radio to the X-rays, with wide-field optical surveys at the forefront of these discoveries, including iPTF \citep{Blagorodnova2017, Hung2017, Blagorodnova2019}, ASAS-SN \citep{Holoien2014, Holoien2016, Holoien2016b, Wevers2019, Holoein2019, Hinkle2021}, Pan-STARRS \citep{Gezari2012, Chornock2014, Holoien2019b, Nicholl2019}, SDSS \citep{vanVelzen2011}, and ZTF \citep{vanVelzen2019, vanVelzen21}, and now X-ray surveys, such as SRG/eROSITA \citep{Sazonov2021}. The growing number of TDEs discovered through these surveys is making their use as probes of MBH demographics, accretion, jet formation, and shock physics a reality. However, the origin of the strong optical and UV emission seen in these transients is still under debate and a resolution is required before these transients can be used to robustly study the properties of the MBHs behind these events \citep[i.e.,][]{Mockler2019}.

While the soft X-ray emission seen in some optically selected TDEs can be explained by thermal emission from the inner portions of the accretion disk \citep[][for a review]{Ulmer1999, Saxton2021}, the origin of the UV and optical emission is more puzzling. The blackbody radii measured from the UV/optical light curves are much larger than expected for the newly formed accretion disk \citep[for a review, see][]{Gezari2021}, which has spurred several theories as to the nature of this larger structure. Outflows and winds have been proposed as the origin of this emission \citep{Metzger2016, Dai2018}, as well as shocks from the intersecting debris streams \citep{Piran2015, Jiang2016}. To further complicate the picture of TDE emission, the lack of an X-ray component in most optically selected TDEs is also not well understood. The most common explanations for this lack of X-ray emission are the absorption of the X-ray photons from the disk and subsequent reprocessing into optical/UV wavelengths \citep{Guillochon2014, Auchettl2017}, and the delayed onset of accretion and therefore X-ray emission due to the time it takes for the tidal debris to circularize and form an accretion disk \citep{Piran2015, Krolik2016, Gezari2017}. The model of \citet{Dai2018} proposes instead that viewing angle is responsible for the lack of X-rays in some optical TDEs and the detection of X-rays in others. The characterization of both the optical/UV and X-ray light curves is thus crucial to determining which of these models is at play.

The features observed in the optical spectra of TDEs are varied, with some having shown only He II emission \citep{Gezari2012} and others showing evidence for Bowen fluorescence lines \citep{Leloudas2019, Blagorodnova2019b}. Building on the classification scheme of \citet{Arcavi14}, \citet{vanVelzen21} presented a scheme for classifying the optical spectra into three categories with varying strengths of hydrogen and helium emission lines. Explanations for this observed diversity in spectroscopic features include the composition of the disrupted star due to stellar evolution \citep{Kochanek2016}, details in the physics of photoionization \citep{Guillochon2014, Gaskell2014, Roth2016}, and viewing angle effects \citep{Holoien2019b, Hung2020}. Here, we investigate whether the spectroscopic classes of TDEs show differences in their light curve and host galaxy properties.

In this paper, we present a sample of 30 spectroscopically classified TDEs from the ZTF survey, the largest systematically selected sample of TDEs from a single survey yet. We present our method for candidate selection and details on the sample in Section \ref{sec:tdesearch}. We briefly discuss the host galaxy properties in Section \ref{sec:host} and describe the follow-up observations for each TDE in Section \ref{sec:followup}. We describe our methods for the analysis of the optical/UV light curves in Section \ref{sec:analysis} and we present our results in Section \ref{sec:results}, an estimation of the MBH mass in Section \ref{sec:bhmass}, and a discussion in Section \ref{sec:discussion}. We conclude with a summary in Section \ref{sec:conclusion}. Throughout this paper, we adopt a flat cosmology with $\Omega_\Lambda = 0.7$ and $H_0 = 70 ~\mathrm{km~s^{-1}~Mpc^{-1}}$. All magnitudes are reported in the AB system.

\section{The Search for TDEs in ZTF-I} \label{sec:tdesearch}
\subsection{TDE Candidate Selection}
The first phase of the ZTF survey \citep[hereafter ZTF-I;][]{Graham19, Bellm2019} completed operations in October 2020. Over the course of the 2.6 year survey (March 2018 -- October 2020), we conducted a systematic search for TDEs almost entirely within the public MSIP data \citep{Bellm19}, which observed the entire visible Northern sky every 3 nights in both $g$- and $r$-bands. The multi-band observations were key to the efficient filtering of the ZTF alert stream \citep{Patterson2019}, as they allowed us to narrow our search for TDEs to a specific subset of photometric properties that aid in the discrimination between TDEs and other nuclear transients, such as active galactic nuclei (AGN) and nuclear supernovae (SNe). 

We will summarize the key aspects of our ZTF-I TDE search here, but we point the reader to \citet{vanVelzen21}, where our filtering of the ZTF-I alert stream is described in more detail. Our filtering techniques included rejecting galaxies classified as broad-line AGN, but otherwise was not restricted to host galaxy type. We filtered known AGN using the Million quasar catalog \citep[][v.~5.2]{Flesch2015} and constructed neoWISE \citep{Mainzer2011} light curves to reject any galaxy with significant variability or a mean W1$-$W2 color consistent with the AGN threshold of \citet{Stern2012}. We used the ZTF observations to filter on photometric properties which can discriminate TDEs from AGN and nuclear supernovae. These properties included $g-r$ color and rate of color change, in addition to rise and fade timescales. Specifically, our filters included rejecting transients that are significantly offset from the known galaxy host (mean offset $> 0\farcs4$), have significant $g-r$ color evolution ($d(g-r)/dt > 0.015$ day$^{-1}$), or show only a modest flux increase in the difference flux compared to the PSF flux in the ZTF reference image ($m_{\rm diff}-m_{\rm ref} > -1.5$ mag). This filtering allowed for a more focused spectroscopic follow-up effort, which allowed for further filtering of AGN and nuclear SNe based on features present in follow-up spectra. To manage data for the candidates, including photometry and spectra, we made use of the GROWTH Marshal \citep{Kasliwal2019} and Fritz \citep{vanderWalt2019, Duev2019}.

\subsection{The ZTF-I TDE Sample} \label{sec:sample}
We present the entire sample of 30 TDEs classified during ZTF-I in Table \ref{tab:sample}, along with the IAU name, ZTF name, our internal nickname, names given by other surveys, and reference to the first classification as a TDE. The bolded names credit the first detection of the transient reported to the Transient Name Server (TNS). ZTF was the first to report 22/30 of the TDEs in this sample, with ATLAS providing 4 discoveries, ASAS-SN providing 2 discoveries, and PS1 and Gaia each providing 1 discovery.

Sixteen of these TDEs were originally presented as part of a ZTF-I sample in \citet{vanVelzen21}. We note the exclusion of AT2019eve, which was included in \citet{vanVelzen21}, but is not included here as the properties and evolution of the light curve and spectra of the source give rise to uncertainty in this classification.\footnote{AT2019eve was a sole outlier in light curve properties as compared to the rest of the sample in \citet{vanVelzen21}, which led to the reconsideration of its classification. In addition to a fast rise and some reddening in the post-peak light curve, the source has only faint UV detections, all of which make the TDE classification less favorable. The H$\alpha$ emission in the spectra that was originally used to classify the transient as a TDE persists over one year post peak, making association with the transient less likely. \citet{vanVelzen21} do not list a TDE spectral classification for this object.} We note that this issue is unlikely to affect other objects in our sample, which have much better spectral coverage post-peak.

\subsection{Spectroscopic Classifications} \label{sec:spectra}
We classify the TDEs into four spectroscopic classes, largely following the spectroscopic classification scheme given in \citet{vanVelzen21}, which divides TDEs into three spectroscopic classes:
\begin{itemize}
    \item[\it i.]  {\it TDE-H}: broad H$\alpha$ and H$\beta$ emission lines. 
    
    \item[\it ii.] {\it TDE-H+He}: broad H$\alpha$ and H$\beta$ emission  lines and a broad complex of emission lines around He II $\lambda$4686 \AA. The majority of the sources in this class also show \ion{N}{3} $\lambda$4640 \AA  and emission at $\lambda$4100 \AA (identified as \ion{N}{3} $\lambda$4100 \AA instead of H$\delta$), and in some cases also \ion{O}{3} $\lambda$3760 \AA.
    
    \item[\it iii.] {\it TDE-He}: no broad Balmer emission lines, a broad emission line near He~II~$\lambda4686$ \AA only.
\end{itemize}
In addition to these three classes, we present a fourth spectroscopic class for TDEs: 
\begin{itemize}
    \item[\it iv.] {\it TDE-featureless}: no discernible emission lines or spectroscopic features present in the three classes above, although host galaxy absorption lines can be observed.
\end{itemize}
Despite the lack of observed features in the optical spectra of these transients, they are nonetheless classified as TDEs due to their coincidence with galaxy nuclei, persistent blue optical colors, and other light curve properties consistent with the TDEs of other spectroscopic classes. We discuss the properties of this class of TDEs further in Section \ref{sec:discussion}.

\begin{figure*}
    \centering
    \includegraphics[width=0.9\textwidth]{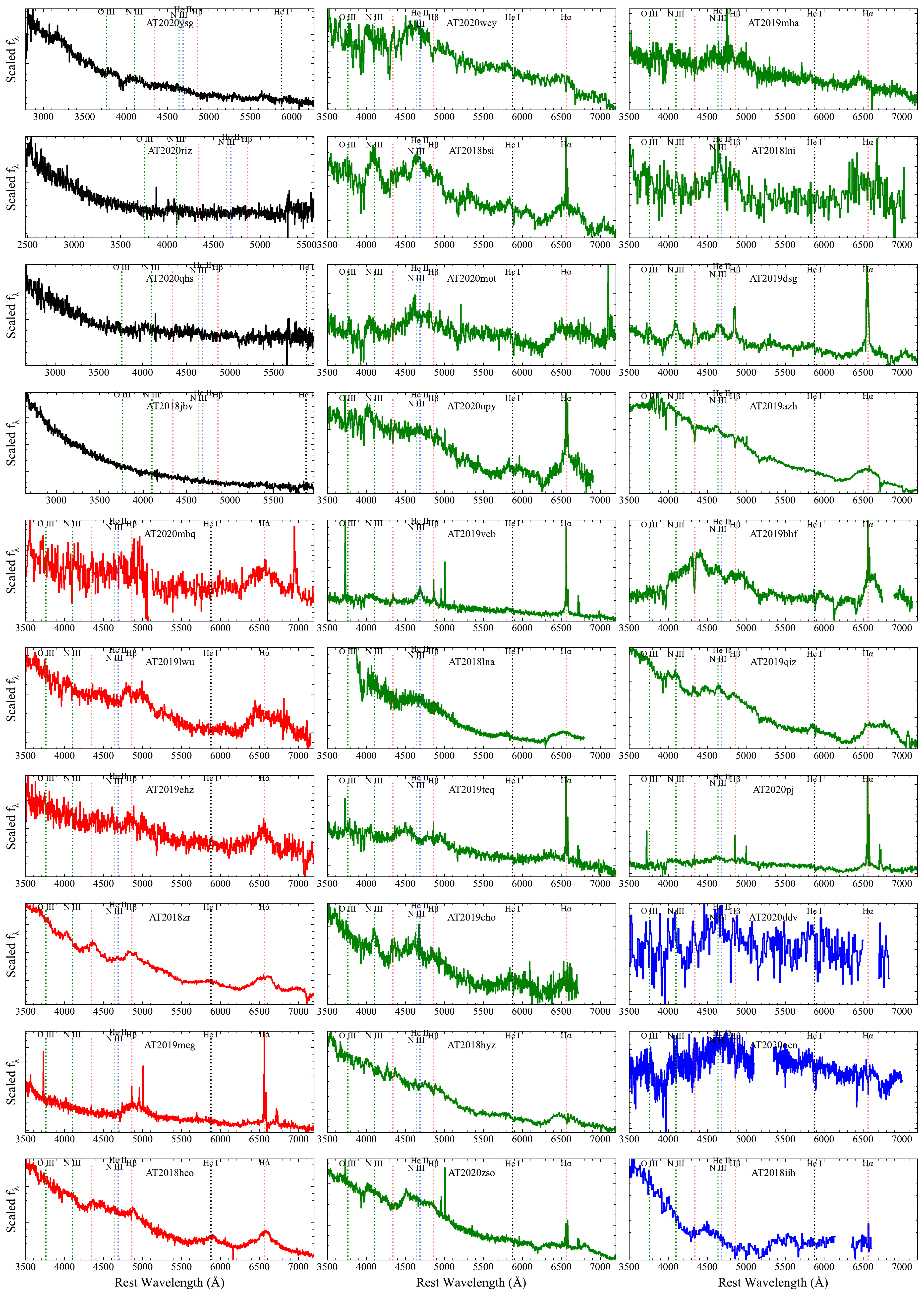}
    \caption{Spectral classifications for the ZTF-I TDE sample, which can also be found in Table \ref{tab:sample}, with black being TDE-featureless, red being TDE-H, green is TDE-H+He, and blue is TDE-He. Spectra have not been host galaxy subtracted. Details regarding the spectral classifications and more spectra are in the Appendix.}
    \label{fig:spectra}
\end{figure*}

Our sample of TDEs contains 6 TDE-H, 3 TDE-He, 17 TDE-H+He, and 4 TDE-featureless, which we show in Figure \ref{fig:spectra}. We note that the spectra used to classify these events have not been host galaxy subtracted, as host galaxy spectra are not yet available for all objects. We discuss the individual spectroscopic classifications and provide early- and late-time spectra for each object, when available, in the Appendix. While the four spectroscopic classes illustrate a clean division among spectroscopic features, there are still subtle differences among the spectra even within a particular class. TDEs in the TDE-H class all show strong, broad H$\alpha$ and H$\beta$ emission and lack He II, \ion{N}{3}, and \ion{O}{3} emission lines, but some also show evidence for H$\gamma$ emission. Furthermore, there is evidence for He I $\lambda$5876 in several TDE-H TDEs, such as AT2018zr and AT2018hco. The TDE-H+He shows similar variety in the lines that appear, with some showing hydrogen lines bluer than H$\beta$, some showing \ion{O}{3} and \ion{N}{3}, and He I $\lambda$5876 \AA. A more detailed analysis of the spectral features, including temporal evolution, present in this sample of TDEs will be presented in a forthcoming publication. For the purposes of this work we will only consider the spectroscopic class assigned to each TDE according to Table \ref{tab:sample}.

\movetabledown=1.5in
\tabletypesize{\footnotesize}
\begin{rotatetable*}
\begin{deluxetable*}{l l l l l  ll }
\tablecaption{ZTF-I TDEs}
\tablehead{ \colhead{IAU Name} & \colhead{ZTF Name} & \colhead{GoT Name} & \colhead{Other/Discovery Name} & \colhead{First TDE Classification} & \colhead{Spectroscopic Class} & \colhead{Redshift}}
\startdata
AT2018zr & ZTF18aabtxvd	& Ned & \textbf{PS18kh} & ATel \#11444 &  TDE-H & 0.075\\
AT2018bsi & \textbf{ZTF18aahqkbt} & Jon & & ATel \#12035 &  TDE-H+He & 0.051\\
AT2018hco & ZTF18abxftqm & Sansa & \textbf{ATLAS18way} & ATel \#12263 &  TDE-H & 0.088 \\
AT2018iih & ZTF18acaqdaa & Jorah & \textbf{ATLAS18yzs}, Gaia18dpo & \citet{vanVelzen21} &   TDE-He & 0.212 \\
AT2018hyz & ZTF18acpdvos & Gendry & \textbf{ASASSN-18zj}, ATLAS18bafs & ATel \#12198 & TDE-H+He & 0.046 \\
AT2018lni & \textbf{ZTF18actaqdw} & Arya & & \citet{vanVelzen21} &  TDE-H+He & 0.138 \\
AT2018lna & \textbf{ZTF19aabbnzo} & Cersei & & ATel \#12509 &  TDE-H+He & 0.091 \\
AT2018jbv & \textbf{ZTF18acnbpmd} & Samwell & ATLAS19acl, PS19aoz & This paper &  TDE-featureless & 0.340\\
AT2019cho & \textbf{ZTF19aakiwze} & Petyr & & \citet{vanVelzen21} &  TDE-H+He & 0.193 \\
AT2019bhf & \textbf{ZTF19aakswrb} & Varys & & \citet{vanVelzen21} &  TDE-H+He & 0.121 \\
AT2019azh & ZTF17aaazdba & Jaime & \textbf{ASASSN-19dj}, Gaia19bvo & ATel \#12568\tablenotemark{*} &  TDE-H+He & 0.022 \\
AT2019dsg & \textbf{ZTF19aapreis} & Bran & ATLAS19kl & ATel \#12752 &   TDE-H+He & 0.051 \\
AT2019ehz & ZTF19aarioci & Brienne & \textbf{Gaia19bpt} & ATel \#12789 &  TDE-H & 0.074\\
AT2019mha & ZTF19abhejal & Bronn & \textbf{ATLAS19qqu} & \citet{vanVelzen21} &   TDE-H+He  & 0.148 \\
AT2019meg & \textbf{ZTF19abhhjcc} & Margaery & Gaia19dhd & AN-2019-88  &  TDE-H & 0.152\\
AT2019lwu & \textbf{ZTF19abidbya} & Robb & ATLAS19rnz, PS19ega & \citet{vanVelzen21} & TDE-H & 0.117 \\
AT2019qiz & \textbf{ZTF19abzrhgq} & Melisandre & ATLAS19vfr, Gaia19eks, PS19gdd & ATel \#13131 &  TDE-H+He & 0.015 \\
AT2019teq & \textbf{ZTF19accmaxo} & Missandei & & TNSCR \#7482 & TDE-H+He & 0.087 \\
AT2020pj  & \textbf{ZTF20aabqihu} & Gilly & ATLAS20cab & TNSCR \#7481 & TDE-H+He & 0.068 \\
AT2019vcb & \textbf{ZTF19acspeuw} & Tormund & Gaia19feb, ATLAS19bcyz & TNSCR \#7078  & TDE-H+He & 0.088 \\
AT2020ddv & ZTF20aamqmfk & Shae & \textbf{ATLAS20gee} & ATel \#13655 & TDE-He & 0.160 \\
AT2020ocn & \textbf{ZTF18aakelin} & Podrick & & ATel \#13859 &  TDE-He & 0.070 \\
AT2020opy & \textbf{ZTF20abjwvae} & High Sparrow & PS20fxm & ATel \#13944 &  TDE-H+He & 0.159 \\
AT2020mot & \textbf{ZTF20abfcszi} & Pycelle & Gaia20ead & ATel \#13944 & TDE-H+He & 0.070 \\
AT2020mbq & \textbf{ZTF20abefeab} & Yara & ATLAS20pfz, PS20grv & This paper &   TDE-H & 0.093 \\
AT2020qhs & \textbf{ZTF20abowque} & Loras & ATLAS20upw, PS20krl & This paper &  TDE-featureless & 0.345 \\
AT2020riz & \textbf{ZTF20abrnwfc} & Talisa & PS20jop & This paper &  TDE-featureless & 0.435\\
AT2020wey & \textbf{ZTF20acitpfz} &	Roose & ATLAS20belb, Gaia20fck & TNSCR \#7769 &  TDE-H+He & 0.027 \\
AT2020zso & \textbf{ZTF20acqoiyt} &	Hodor & ATLAS20bfok & TNSCR \#8025 &  TDE-H+He & 0.057\\
AT2020ysg & \textbf{ZTF20abnorit} &	Osha & ATLAS20bjqp, PS21cru & This paper &  TDE-featureless & 0.277
\enddata
\label{tab:sample}
\tablecomments{The names of each the 30 TDEs detected in ZTF-I, with boldface indicating the discovery name, i.e.~the first survey to report photometry of the transient detection to the TNS, and the GoT name is the ZTF TDE Working Group nickname which references characters from the popular television show \textit{Game of Thrones}. We also include the first TDE classification report, with abbreviations ATel corresponding to the Astronomer's Telegram\tablenotemark{a}, AN corresponding to AstroNotes \tablenotemark{b}, and TNSCR corresponding to TNS classification reports. The last two columns contain the TDE spectral class, as described in Section \ref{sec:spectra}, and the redshift. Redshifts were determined using host galaxy stellar absorption lines or narrow emission lines associated with star formation, namely Ca II H and K or narrow H$\alpha$ emission.}
\tablenotetext{*}{See also \citet{Hinkle2021}.}
\tablenotetext{a}{\url{https://astronomerstelegram.org/}}
\tablenotetext{b}{\url{https://www.wis-tns.org/astronotes}}
\end{deluxetable*}
\end{rotatetable*}

\section{Host Galaxy Properties} \label{sec:host}
\begin{figure*}
    \centering
    \includegraphics[width=\textwidth]{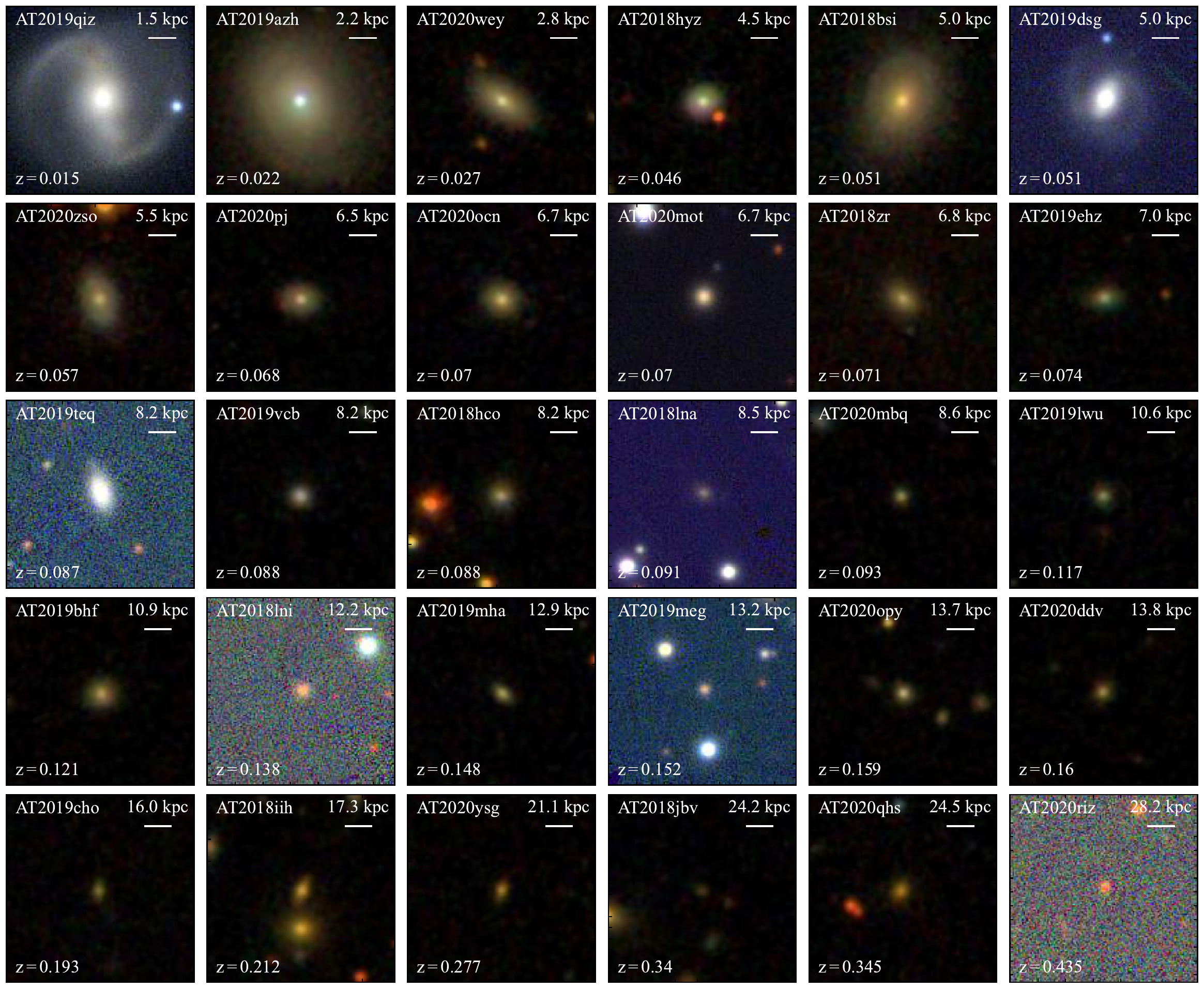}
    \caption{SDSS and Pan-STARRS $gri$ images of the TDE host galaxies in order of increasing redshift. All images are $34\arcsec \times 34 \arcsec$. The morphology of the TDE hosts appears to be dominated by elliptical components, with only the lowest redshift TDEs showing discernible disk components.}
    \label{fig:hosts}
\end{figure*}

\begin{figure}
    \centering
    \includegraphics[width=\columnwidth]{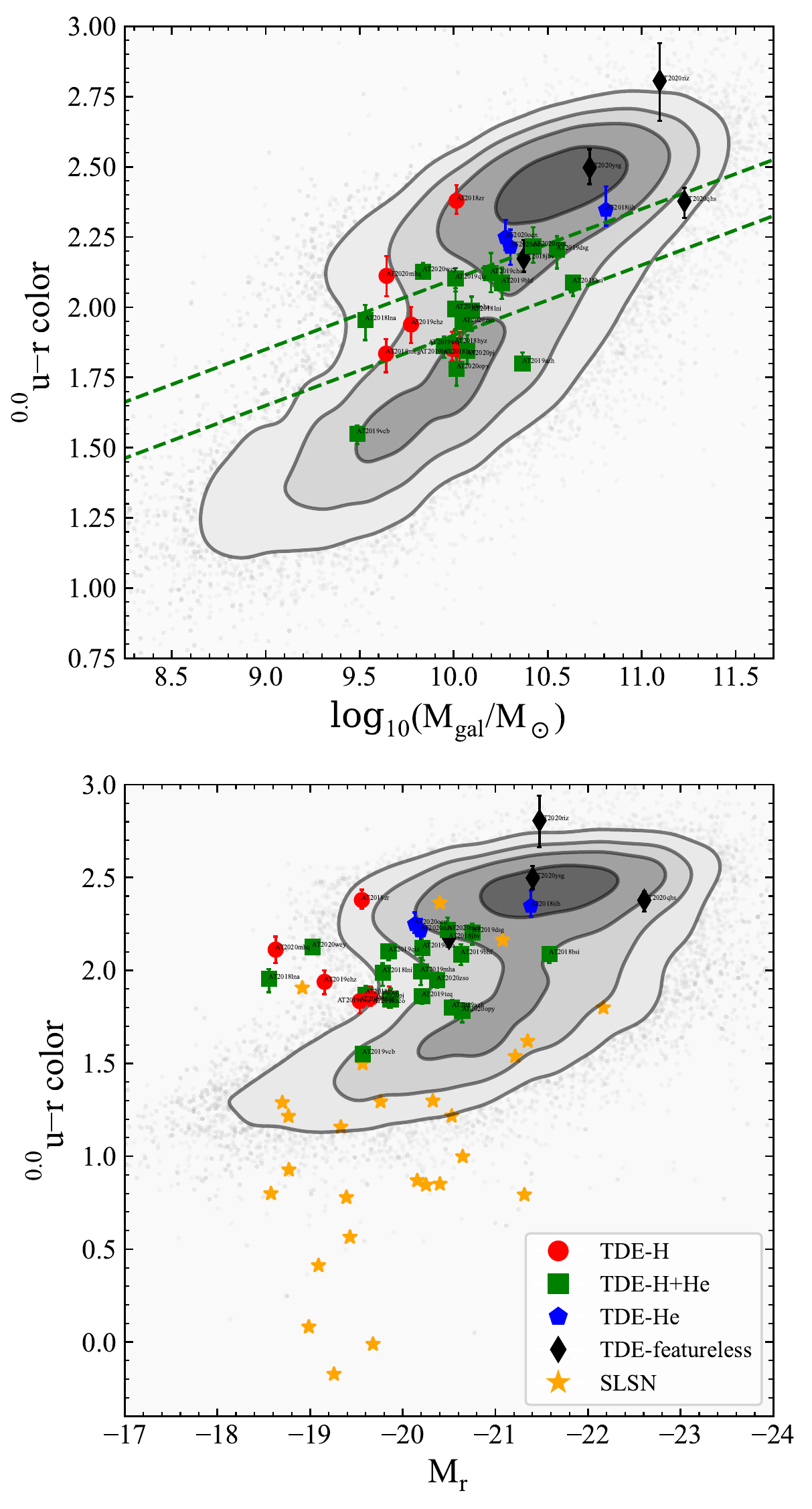}
    \caption{\textit{Top}: The extinction-corrected, rest-frame $u-r$ color vs.~total stellar mass of the TDE hosts, estimated from the stellar population synthesis fits to the pre-flare photometry. 47\% of the TDE hosts are within the limits of the green valley, and 69\% of the hosts outside of the green valley are within 0.12 mag of the boundary. Red circles are TDE-H; green squares are TDE-H+He; blue pentagons are TDE-He; and black diamonds are TDE-featureless. \textit{Bottom}: The extinction-corrected, rest-frame $u-r$ color vs.~absolute $r$-band magnitude of the TDE hosts, plus a selection of SLSN hosts from TNS with SDSS observations. The SLSN hosts are largely blue, star-forming galaxies, while the TDE hosts are dominated by green and red galaxies.}
    \label{fig:urslsn}
\end{figure}

In Figure \ref{fig:hosts} we show false-color $gri$ cut-outs of the 30 TDE host galaxies from SDSS and Pan-STARRS, in order of increasing redshift. The majority of the hosts appear to be dominated by an elliptical component, with only the lowest redshift host galaxies showing a disk component accompanying a compact core. This may be an artifact of distance; in Figure \ref{fig:urslsn} however, we show that very few of the TDE host galaxies fall in the blue cloud, a region where blue, disk-like galaxies are expected to reside. Additionally, \citet{Hammerstein2021} found that many of the TDE hosts in this sample show morphological structure closer to that of red, elliptical galaxies despite falling in the green valley. Galaxies within the green valley, where a large number of TDE hosts fall, may still maintain a disk component, and better imaging is required to determine whether a disk component is present in these galaxies.

Using the pipeline of \citet{vanVelzen21}, we fit SEDs of the TDE host galaxies constructed from pre-flare photometry in order to estimate the total stellar masses. This includes either SDSS model magnitudes or Pan-STARRS Kron magnitudes (if a source is outside the SDSS footprint), as well as GALEX NUV and FUV photometry. We use the Prospector software \citep{Johnson2021} to run a Markov Chain Monte Carlo (MCMC) sampler \citep{Foreman-Mackey2013}, with 100 walkers and 1000 steps, to obtain the posterior distributions of the Flexible Stellar Population Synthesis models \citep{Conroy2009}. We discard the first 500 steps to ensure proper sampling of the posterior distribution. We follow the procedure of \citet{Mendel14}, adopting the same parameter choices for the 5 free parameters: stellar mass, \citet{Calzetti00} dust model optical depth, stellar population age, metallicity, and the e-folding time of the star formation history. The results of this fitting are given in Table \ref{tab:host}.

Figure \ref{fig:urslsn} shows the extinction corrected, rest-frame $u-r$ color vs.~total stellar mass of the TDE hosts estimated from the stellar population synthesis fits to the pre-flare photometry. Both panels in this figure show the same background sample of 30,000 SDSS galaxies taken from the \citet{Mendel14} catalog of total stellar mass estimates, constructed in the same manner as in \citet{Hammerstein2021}, which corrects for the flux-limited nature of SDSS and produces a sample representative of the galaxies our search for TDEs is sensitive to. The top panel of Figure \ref{fig:urslsn} also shows the limits of the green valley, the transition region between blue, star-forming galaxies and red, quiescent galaxies, originally defined by \citet{Schawinski14}.

Previous studies of TDE host galaxies have found that a majority of TDE hosts are green \citep{LawSmith17, Hammerstein2021}. Most recently, \citet{Sazonov2021} found that a sample of X-ray bright TDE hosts discovered within the \textit{SRG}/eROSITA survey were predominantly green. \citet{Hammerstein2021} found that of the first 19 TDEs in this sample, 63\% of them fell within the limits of the green valley. With an additional 11 TDE hosts, we find that 47\% of the hosts fall within the green valley limits as defined in \citet{Hammerstein2021} compared to only 13\% of the background sample, with 9/30 TDE hosts in the red sequence and 7/30 in the blue cloud. However, 11/17 of the blue and red galaxies fall within 0.12 mag of the green valley limit, which can be difficult to define due to differences in sample selection and redshift cuts. We perform a binomial test to determine whether the number of TDE hosts within the green valley differs significantly from what is expected given the background sample of SDSS galaxies. We find that we can reject the null hypothesis that the TDE hosts are drawn uniformly from the sample of SDSS galaxies with a $p$-value $=6.5\times10^{-6}$.

It is important to compare the properties of the TDE-featureless class to those of possible impostor transients and look-alikes. One such class of impostor are superluminous supernovae (SLSN). The early-time light curves of TDEs and SLSN can be difficult to differentiate, and the optical spectra of SLSN can show features that can be mistaken for features characteristic of the 4 TDE spectroscopic classes described in Section \ref{sec:sample} \citep{Gal-Yam2012, Zabludoff2021}. The early-time spectra of SLSN-II can even be featureless, making the classification of a transient as TDE-featureless more complicated. Figure \ref{fig:urslsn} also shows the extinction corrected, rest-frame $u-r$ color vs.~absolute $r$-band magnitude of the TDE hosts, along with a selection of SLSN host galaxies from TNS. SLSN hosts were chosen from those classified as SLSN-I and SLSN-II and were required to have SDSS observations for ease of data access. The distribution of SLSN hosts is not surprising, given previous studies of SLSN hosts \citep[e.g.][]{Lunnan2014, Leloudas2015, Perley2016, Schulze2018, Hatsukade2018, Orum2020, Taggart2021, Schulze2021}. The majority of SLSN hosts shown in Figure \ref{fig:urslsn} are blue, star-forming hosts, while all 4 TDE-featureless hosts are near or above the red edge of the green valley. This type of host color distinction, which has previously been discussed in \citet{French2018}, will be important for distinguishing TDEs from impostors in the age of the Rubin Observatory. A more careful examination of the 30 TDE hosts in this sample, including spectroscopic $M_{\rm BH}-\sigma$ black hole mass estimates, will be presented in a forthcoming publication.

\begin{deluxetable*}{l r r r r r r}
\tablecaption{Host Properties}
\tablehead{ \colhead{IAU Name} & \colhead{$\log M / M_\odot$} & \colhead{$^{0.0}u-r$} & \thead{dust \\ $E(B-V)$} & \thead{age \\ $\log$ Gyr} & \thead{$\tau_{\rm sfh}$ \\ $\log$ Gyr} & \colhead{$\log Z/Z_\odot$}}
\startdata
\input{tables/host_tab}
\enddata
\label{tab:host}
\tablecomments{The properties of the ZTF-I TDE host galaxies, as estimated from the SED fitting described in Section \ref{sec:host}. We include the total stellar mass, the $u-r$ color, the color excess, age of the galaxy, the star formation e-folding timescale, and the metallicity. A rough star formation rate (SFR) can be calculated using the relation SFR $\propto e^{-t/\tau_{\rm sfh}}$.}
\end{deluxetable*}

\section{Observations} \label{sec:followup}
\subsection{ZTF Forced Photometry}
We performed forced point spread function (PSF) photometry to extract precise flux measurements of each source through the ZTF forced-photometry service \citep{Masci19}. The position of each source fed to the pipeline was taken as the median of the coordinates of all epochs in which the source was detected. The typical RMS scatter in R.A.~and Dec.~was 0\farcs19 and 0\farcs14 respectively. Photometry was gathered using ZTF DR12, including partnership data. 
Following the recommendations listed in \citet{Masci19}, we cleaned the resulting light curves by filtering out epochs that may have been impacted by bad pixels, and requiring thresholds for the signal-to-noise of the observations, seeing, the sigma-per-pixel in the input science image, and the 1-$\sigma$ uncertainty on the difference image photometry measurement.

To correct for any systematic offsets in the forced photometric flux measurements, we define a temporal baseline for each ZTF field and filter combination. The baseline is defined using all observations up to 100 days prior to the peak of the flare (via visual inspection we confirm no pre-peak emission is included in the baseline window). For each field/filter combination, the median flux of the baseline is subtracted from all forced photometry flux measurements. We typically find small ($\sim 10$ $\mu$Jy), but significant offsets. In addition, we account for any systematic uncertainty by increasing the reported uncertainty with $\sqrt{\chi^2}/{\rm dof}$ of the observations in the baseline period. 

We only accept photometry from ZTF field/filter combinations that have at least 5 observations in the baseline period. We also exclude fields for which the reference images were obtained after the baseline period (i.e., fields for which the transient is ``caught in the reference frame"). An exception to the latter requirement is made for AT2018zr, AT2018bsi, and AT2018hyz. For these sources we allow a post-peak baseline using the last 180 days of observations in the light curve. The baseline corrections and the resulting ZTF forced photometry light curves will be available in machine-readable format at the journal website.


The forced-photometry light curves allow for detections over 800 days post-peak for some TDEs. The resulting forced-photometry light curves, which can be found in the Appendix, along with the follow-up observations described in the remainder of this section, are used in the analysis described below.  

\subsection{Swift (UVOT \& XRT)} \label{sec:Swift}
All 30 TDEs were followed up with observations from the \textit{Neil Gehrels Swift Observatory} \citep{Gehrels2004} in the UV with UVOT \citep{Roming2005} and the X-ray with XRT \citep{Burrows2005}. We used the \texttt{uvotsource} package to analyze the \textit{Swift} UVOT photometry, using an aperture of 5$\arcsec$ for all sources except AT2019azh, AT2019bsi, AT2019qiz, and AT2019dsg, which required a larger aperture to capture the host galaxy light. We subtracted the host galaxy flux estimated from the population synthesis described in Section \ref{sec:host}.

The 0.3--10 keV X-ray light curves for the 9/30 TDEs with XRT detections were produced using the UK Swift Data center online XRT data products tool, which uses the HEASOFT v6.22 software \citep{Arnaud96}. We used a fixed aperture at the ZTF coordinate of the transient, and converted to flux using the best fit blackbody model to the stacked XRT spectrum.  The XRT stacked spectra were processed by the XRT Products Page \citep{Evans2009}, with Galactic extinction fixed to values from the HI4PI survey \citep{2016A&A...594A.116H} shown in Table \ref{tab:xrt}. The blackbody temperatures used to convert from counts/sec to flux using the online PIMMS tool\footnote{\url{https://cxc.harvard.edu/toolkit/pimms.jsp}} are also shown in Table \ref{tab:xrt}.

\begin{deluxetable}{l c c}
\tablecaption{XRT reduction parameters}
\tablehead{ \colhead{Name} & \colhead{$N_{\rm H}$/$10^{20}$ cm$^{-2}$} & \colhead{$kT$/keV}}
\startdata
AT2018zr & 4.4 & 0.100 \\
AT2018hyz & 2.59 & 0.132 \\
AT2019azh & 4.16 & 0.053 \\
AT2019dsg & 6.46 & 0.071 \\
AT2019ehz & 1.42 & 0.101 \\
AT2019teq & 4.54 & 0.200 \\
AT2019vcb & 1.45 & 0.100 \\
AT2020ddv & 1.35 & 0.081 \\
AT2020ocn & 0.93 & 0.120
\enddata
\label{tab:xrt}
\tablecomments{Galactic extinction values and blackbody temperatures (for converting counts/sec to flux) used in the XRT reduction.}
\end{deluxetable}

While all 30 TDEs have at least one epoch of simultaneous UVOT and XRT observations, it is difficult to define ``X-ray bright'' and ``X-ray faint'' classifications of the 30 TDEs, as there may be higher redshift TDEs which have X-ray emission that is below the flux limit for XRT and will thus go undetected. To account for this, we set a luminosity limit of $\log L_{\rm X} = 42$ ergs/s, and define a redshift cut-off, $z=0.075$, beyond which that luminosity would no longer be detected by the typical XRT observation of 2.0 ks. We define ``X-ray bright'' to be any TDE with an XRT detection above the luminosity cutoff \textit{and} a redshift below the redshift cutoff. We have therefore excluded one X-ray detected TDE from the ``X-ray bright'' group, AT2018zr, which has no detections above $\log L_{\rm X} = 42$ ergs/s, and moved it into the ``X-ray faint'' sample.  We define ``X-ray faint'' (or dim) as any TDE below the redshift cut-off which has no XRT detections above the luminosity cutoff. This X-ray faint sample includes AT2018zr, AT2018bsi, AT2019qiz, AT2020pj, AT2020mot, AT2020wey, and AT2020zso.

\subsection{ATLAS}
We obtained additional forced photometry of all 30 TDEs from the Asteroid Terrestrial-impact Last Alert System (ATLAS) survey using the ATLAS forced photometry service\footnote{\url{https://fallingstar-data.com/forcedphot/}} \citep{Tonry2018, Smith2020}. The ATLAS difference image forced photometry is less straightforward to clean in a similar manner to the ZTF forced photometry, as the metadata for each observation is not as comprehensive. We removed epochs with significantly negative flux measurements and large errors, as well as significant outliers.

The ATLAS forced photometry is included in the light curve fitting for the majority of the TDEs in this sample. For some TDEs, however, the reference image used for the difference image photometry changed partway through the event to a reference image that included the flare itself. This led to incorrect baselines for the difference image photometry, and without knowledge of which observations belong to which reference image, there is no straightforward way to perform robust baseline corrections as for the ZTF forced photometry. Therefore, we do not use the ATLAS forced photometry when fitting the light curves of the following 210 TDEs: AT2018bsi, AT2018iih, AT2018jbv, AT2019cho, AT2019dsg, AT2019ehz, AT2019mha, AT2019meg, AT2019lwu, and AT2020wey.


\section{Light Curve Analysis} \label{sec:analysis}

\subsection{Model Fitting} \label{sec:modelfitting}
Similar to \citet{vanVelzen21}, we consider two models to describe the TDE light curve and fit the K-corrected multi-band data: an exponential decay and a power-law decay, both combined with a Gaussian rise. The Gaussian rise is chosen to be consistent with \citet{vanVelzen21} and avoids the addition of the power-law index as a free parameter in a rise characterized by a power-law. The first of these models, which is fit to only the first 100 days post-peak, is described by the following equation:
\begin{equation}
     \label{eq:exp}
     \begin{aligned}
     L_\nu(t) &= L_{\nu_0~\rm{peak}} \frac{B_\nu (T_0)}{B_{\nu_0}(T_0)} \\
     & \times
     \begin{cases}
     e^{-(t-t_{\rm{peak}})^2/2\sigma^2} & t \leq t_{\rm{peak}}\\
     e^{-(t-t_{\rm peak})/\tau} & t > t_{\rm peak}
     \end{cases}
     \end{aligned}
\end{equation}
In this equation, $\nu_0$ refers to the reference frequency, which we have chosen to be the $g$-band ($6.3\times10^{14}$ Hz), and thus $L_{\nu_0~\rm{peak}}$ is the luminosity at peak in this band. The $g$-band is chosen as the reference frequency to minimize the K-correction applied to the ZTF data. This model fits for only one temperature, $T_0$, which is used to predict the luminosity in the other bands at all times by assuming the spectrum follows a blackbody, $B_\nu(T_0)$.

We fit the long-term light curve ($\leq$350 days post-peak) with a Gaussian rise and power-law decay, to more accurately capture the deviation from exponential decay that most TDEs show \citep[e.g.][]{vanVelzen21}. Fits to the photometry at times much longer than 400 days post-peak would require an additional constant component in the model to capture the plateaus that are seen in late-time TDE light curves \citep{vanVelzen2019b}. This model is described by the following equation:
\begin{equation}
     \label{eq:PL}
     \begin{aligned}
     L(t, \nu) &= L_{\rm{peak}} \frac{\pi B_\nu (T(t))}{\sigma_{\rm SB} T^4(t)} \\
     & \times
     \begin{cases}
     e^{-(t-t_{\rm{peak}})^2/2\sigma^2} & t \leq t_{\rm{peak}}\\
     [(t-t_{\rm peak}+t_0)/t_0]^p & t > t_{\rm peak}
     \end{cases}
     \end{aligned}
\end{equation}
We consider two types of temperature evolution with this model: linear and non-parametric evolution which allows for much more freedom in the way the temperature can evolve. In this more flexible, non-parametric temperature model, we fit the temperature at grid points spaced $\pm$30 days apart beginning at peak and use a log-normal Gaussian prior at each grid point centered on the mean temperature obtained from Equation \ref{eq:exp}. The resolution of the temperature grid is chosen so that this method of fitting is applicable to all objects in our sample. While UV coverage at a resolution finer than 30 days is available for some objects, this is not the case for all objects in the sample.

\begin{figure*}
    \centering
    \includegraphics[width=0.7\textwidth]{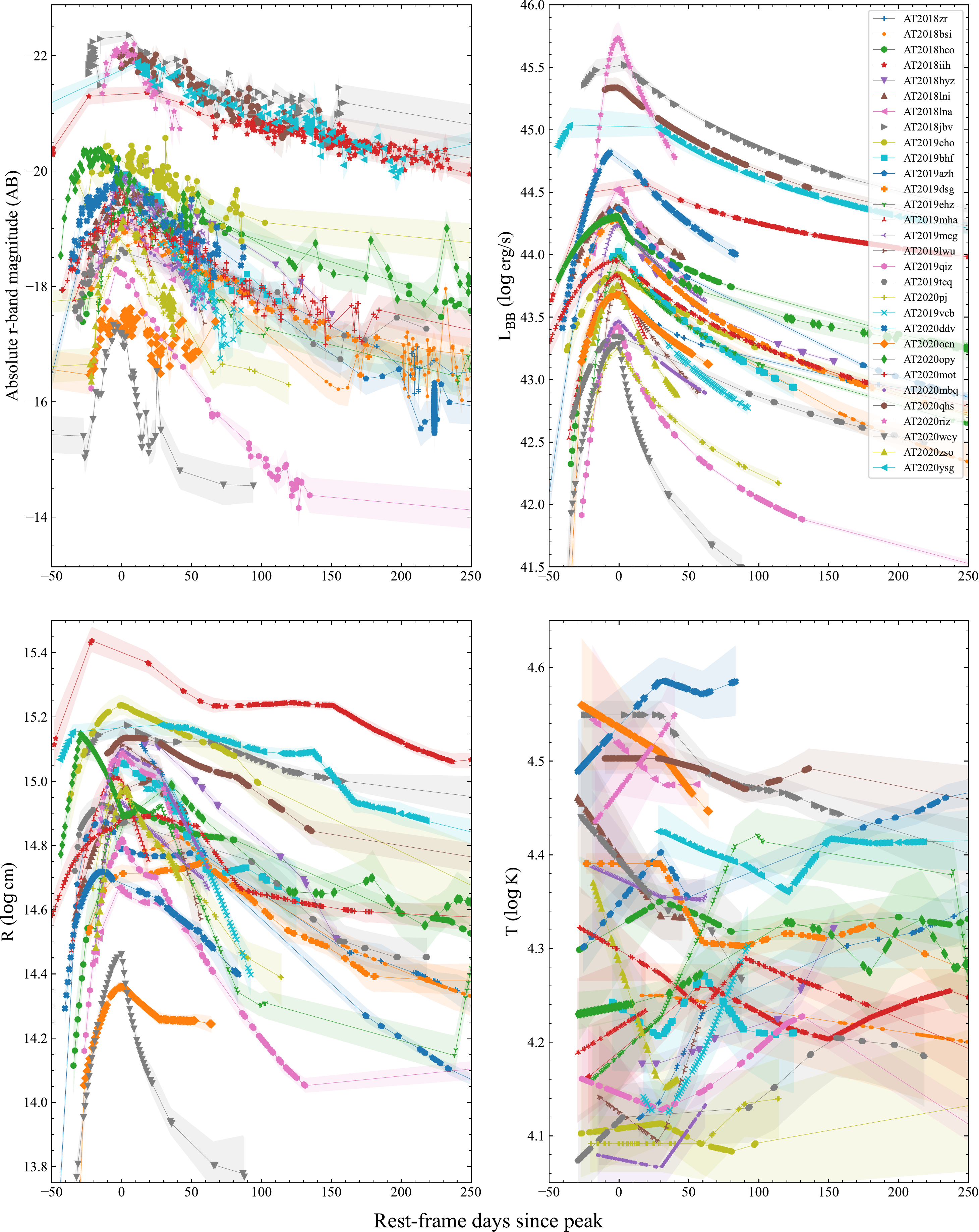}
    \caption{The $r$-band absolute magnitude, blackbody luminosity, blackbody radius, and blackbody temperature for the TDEs in our sample. The TDE-featureless class shows a distinct separation from the other classes in absolute magnitude and blackbody luminosity. All TDEs show a decrease in radius after peak and there is a spread in the change in temperature, with some events showing a modest decrease in temperature and some showing a modest increase in temperature.}
    \label{fig:MLRT}
\end{figure*}

To estimate the parameters of the models above we use the \texttt{emcee} sampler \citep{Foreman-Mackey2013} using a Gaussian likelihood function that includes a ``white noise'' term, ln($f$), that accounts for any variance in the data not captured by the reported uncertainties and flat priors for all parameters (except when employing the flexible temperature evolution as described above). We use 100 walkers and 2000 steps, discarding the first 1500 steps to ensure convergence. The free parameters of the models are listed in Table \ref{tab:priors}. We show the rest-frame absolute $r$-band magnitude, and derived blackbody luminosity, radius, and temperature with time in Figure \ref{fig:MLRT}.

\begin{deluxetable}{l l l}
\tablecaption{Free Parameters and Priors}
\tablehead{\colhead{Parameter} & \colhead{Description} & \colhead{Prior}}
\startdata
$\log L_{\rm peak}$ & Peak luminosity & [$L_{\rm max}/2$, $2L_{\rm max}$] \\
$t_{\rm peak}$ & Time of peak & [$-20$, $20$] days \\
$\log T_0$ & Mean temperature & [4, 5] Kelvin \\
$\log \sigma$ & Gaussian rise time & [0, 1.5] days \\
$\log \tau$ & Exponential decay time & [0, 3] days \\
$p$ & Power-law index & [$-$5, 0] \\
$\log t_0$ & Power-law normalization & [0, 3] days \\
$dT/dt$ & Temperature change & [$-$200, 200] K day$^{-1}$ \\
$\ln{f}$ & White noise factor & [$-$5, $-$1.8] 
\enddata
\label{tab:priors}
\tablecomments{The free parameters and corresponding priors for the light curve analysis described in Section \ref{sec:modelfitting}. $L_{\rm max}$ is the observed maximum luminosity.}
\end{deluxetable}

\subsection{Empirical Timescale Estimates}
To ensure that any correlations found between light curve properties, particularly the timescales, are not simply a product of the chosen model, we also measured the rise and peak timescales empirically. We calculate the time between the peak magnitude, $m_{\rm peak}$, and one magnitude fainter than peak, $m_{\rm peak}+1$ mag, on both sides of the estimated peak of the light curve to measure the rise and decay timescales. The value $m_{\rm peak}+1$ mag often fell between two observed points on the light curve. We fit for $t_{m_{\rm peak}+1}$ on both sides of the peak in order to obtain the most likely value and uncertainties to accurately estimate the empirical rise and fade timescales, accounting for the uncertainties on the adjacent points and the uncertainty on the slope between them. These empirical rise and decay timescales are positively correlated with rise and decay timescales measured in Section \ref{sec:analysis}, which implies that the light curve properties and resulting correlations found from our fits are not merely a product of our chosen model.

\section{Results} \label{sec:results}
We present the results of the analysis described in Section \ref{sec:analysis}. In the following sections, we will discuss our search for correlations between the light curve parameters and the host galaxy properties. We also investigate differences between the spectroscopic classes of TDEs and the light curve classes of TDEs, as well as the differences between the X-ray bright and X-ray faint events. We note the caveat that the results presented in this section, particularly the $p$-values, do not include a correction for the ``look-elsewhere'' effect. We discuss this in Section \ref{sec:lookelse}.

\subsection{Light Curve Property Correlations}
We searched for correlations between all of the parameters in the light curve fitting described in Section \ref{sec:modelfitting} using a Kendall's tau test \citep{Kendall1938}, the results of which are shown in Table \ref{tab:kendalltau}. We consider a correlation to be significant if we can reject the null hypothesis that the variables are uncorrelated at a significance level of $p < 0.05$.

\begin{figure}
    \centering
    \includegraphics[width=\columnwidth]{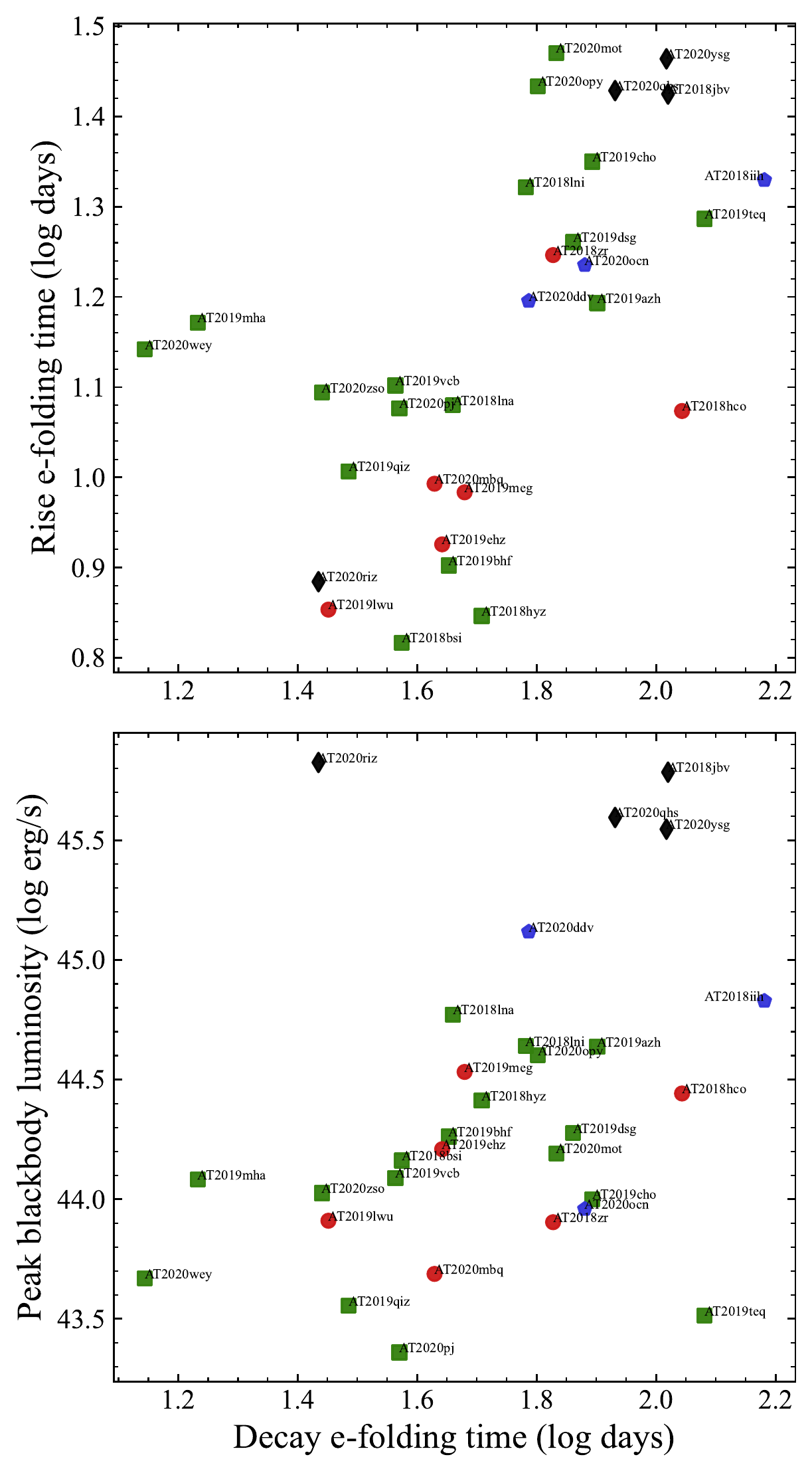}
    \caption{We show the peak blackbody luminosity and the rise time compared with the decay timescale. We find that both the blackbody luminosity and the rise timescale are positively correlated with the decay timescale. The colors and symbols are the same as in Figure \ref{fig:urslsn}.}
    \label{fig:lcfitresults}
\end{figure}

We find significant correlations between the peak luminosity and the radius, as is expected from $L_{\rm BB} \propto R^2 T^4$. In Figure \ref{fig:lcfitresults}, we show the peak blackbody luminosity and the rise timescale compared to the decay timescale. We find a significant, although shallow, positive correlation between the peak luminosity and the decay e-folding timescale ($p$-value $=0.031$). We find that the rise timescale and the decay e-folding timescale are weakly positively correlated ($p$-value $<0.001$), however we find no significant correlation between the rise timescale and the luminosity.

\begin{figure*}
    \centering
    \includegraphics[width=\textwidth]{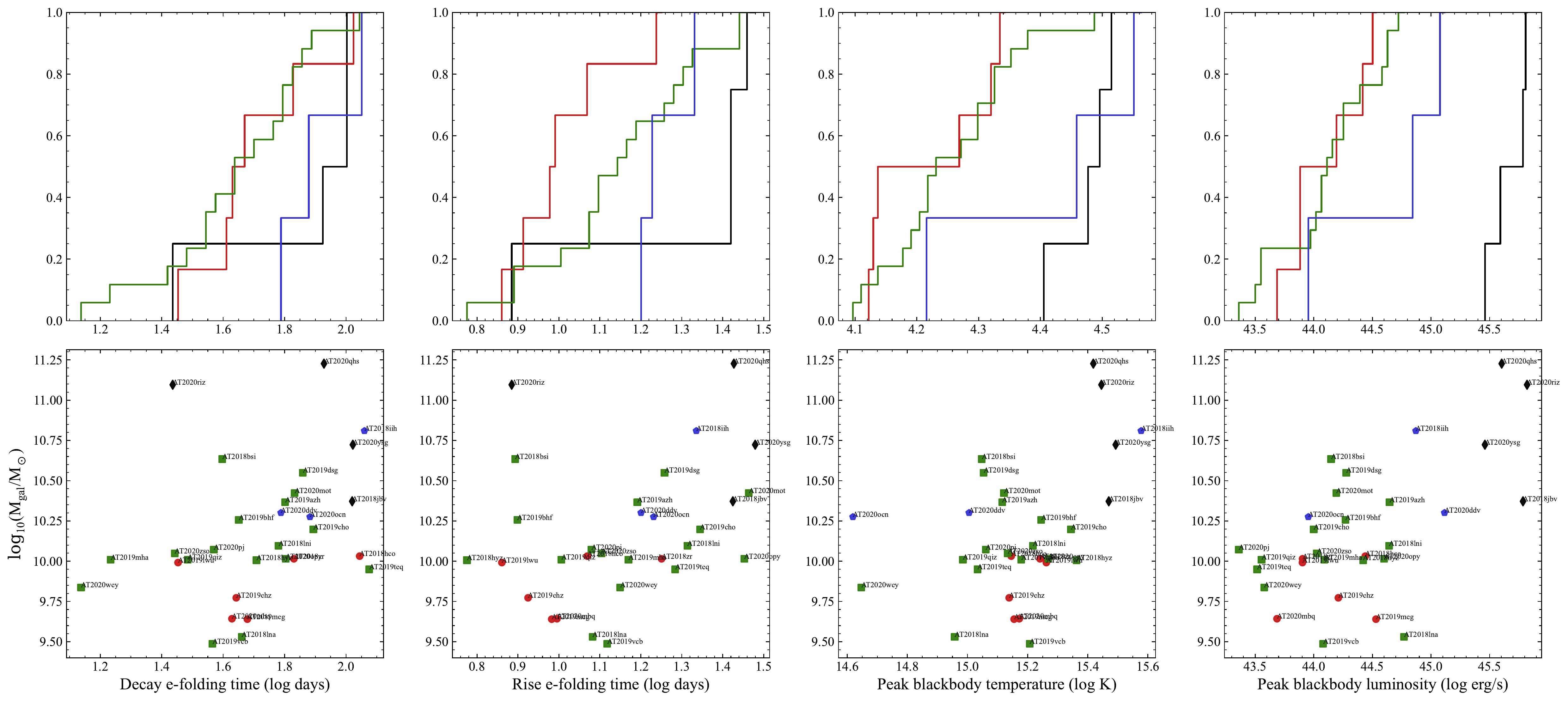}
    \caption{Selected properties measured from the fits to the multi-band light curves compared to the host galaxy stellar mass, with the cumulative distributions of spectroscopic classes. We find significant correlations between the host galaxy stellar mass and the properties shown, which include the decay e-folding timescale, the rise timescale, and the peak blackbody temperature and luminosity. We do not find that the spectroscopic classes show significant differences in their light curve decay or rise timescales, but the TDE-featureless class shows higher luminosities, temperatures, and radii than the other three classes. Both the TDE-He and TDE-featureless classes show significantly more massive host galaxies. The colors and symbols are the same as in Figure \ref{fig:urslsn}.}
    \label{fig:lcfithost}
\end{figure*}

We now turn to the correlations between the light curve properties and the host galaxy properties, particularly $M_{\rm gal}$. The properties of the light curve can be expected to be correlated with host galaxy mass, as the properties of the MBH should be imprinted on the TDE light curve and the host galaxy mass is correlated with the MBH mass. We show a selection of light curve properties vs.~the host stellar mass in Figure \ref{fig:lcfithost}. We find that the peak blackbody luminosity as well as the peak blackbody temperature are positively correlated with the mass of the host galaxy ($p$-value $=0.005$ and $=0.031$, respectively). We also find that the rise timescale and decay e-folding timescale is positively correlated with the mass of the host galaxy ($p$-value $=0.019$ and $=0.016$, respectively). We find no significant correlation with the fallback time-scale, defined as $t_0$ when $p=-5/3$. This may be due to late-time plateaus in the post-peak light curve.

\subsection{Spectral Class Correlations}
We used an Anderson-Darling test \citep{AD1954} to assess whether the four spectroscopic classes of TDEs show differences in their light curve or host galaxy properties. The results of this test are shown in Table \ref{tab:spectest}. We consider a result to be significant if we can reject the null hypothesis that the two samples are drawn from the same parent population at a significance level of $p < 0.05$. We also show the cumulative distributions of the light curve properties and the cumulative distributions of the host galaxy mass in Figures \ref{fig:lcfitresults} and \ref{fig:lcfithost}.

We first examine the properties measured from the light curves. We do not find any significant ($p$-value $<0.05$) differences in the rise timescales of the light curves for the four classes. We note that the spectral classifications in \citet{vanVelzen21} contained many more TDE-H objects, including events prior to ZTF-I, while three have been reclassified here as another class following more spectroscopic observations, which may explain why we no longer find a difference between the rise times of these two classes. We find that the TDE-featureless class has significantly hotter temperatures and larger radii when compared to the TDE-H and TDE-H+He classes, and higher peak blackbody and $g$-band luminosities when compared to all other classes.

Both TDE-He and TDE-featureless show significant differences in their host galaxy properties when compared to TDE-H and TDE-H+He. The TDE-featureless class shows a distribution favoring more massive and redder galaxies when compared to both TDE-H and TDE-H+He. The TDE-He possesses more massive galaxies as compared to the TDE-H class, with redder galaxies compared to the TDE-H+He class.

\subsection{X-ray Correlations}
We also employed an Anderson-Darling test to evaluate the differences in the X-ray bright and X-ray faint populations in this sample and test the null hypothesis that these two samples are drawn from the same parent population. As described in Section \ref{sec:Swift}, we define X-ray bright to be a TDE with at least one detection of $\log L_{\rm X} \geq 42$ ergs/s and below a redshift of $z=0.075$. We define X-ray faint to be any TDE below a redshift of $z=0.075$ without an XRT detection. This gives an X-ray faint sample of 6 TDEs, compared to 8 X-ray bright TDEs. One TDE detected with XRT, AT2018zr, has no detections with $\log L_{\rm X} \geq 42$ ergs/s but is within the redshift cutoff, and so we include this object in the X-ray faint sample.

\begin{figure}
    \centering
    \includegraphics[width=\columnwidth]{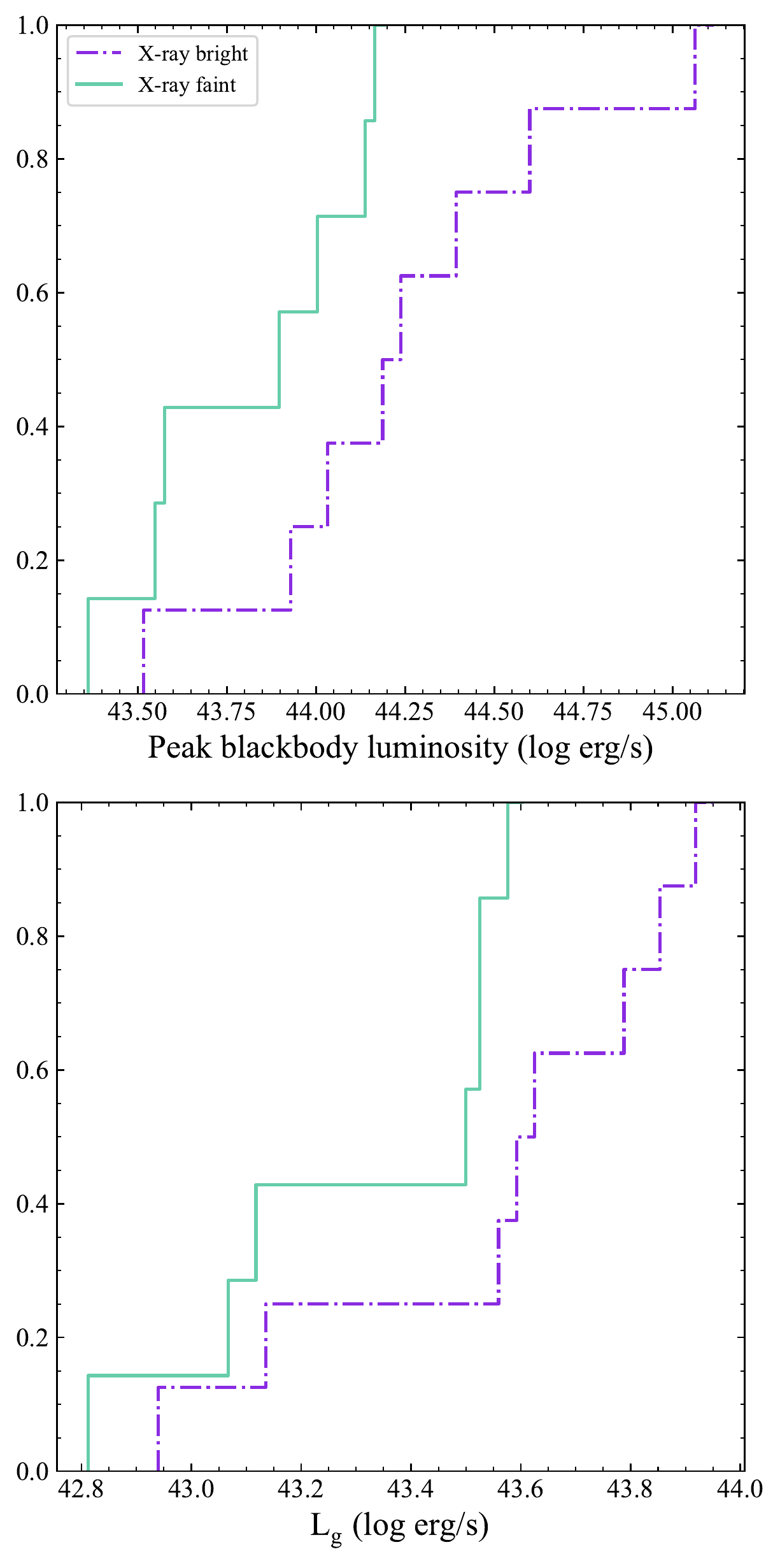}
    \caption{Cumulative distributions of selected properties of the TDE light curves for the X-ray bright (purple dot-dashed line) and X-ray faint (aqua, solid line) populations of TDEs in the ZTF-I sample. We find that the X-ray bright TDEs have significantly higher blackbody and $g$-band luminosities.}
    \label{fig:Xrayvnot}
\end{figure}

We find that the X-ray bright and X-ray faint TDEs differ only in their peak luminosities, with both the peak blackbody luminosity and peak $g$-band luminosity of the X-ray bright TDEs being more luminous ($p$-value $=0.049$ and $=0.045$, respectively). We show the results of the Anderson-Darling tests in Table \ref{tab:spectest}. We also show the cumulative distributions of the selected properties in Figure \ref{fig:Xrayvnot}.

\subsection{The Look-Elsewhere Effect} \label{sec:lookelse}
We search for correlations among the light curve properties and perform a total of 36 different Kendall's tau tests. Because of the size of this parameter space, it is important to address the ``look-elsewhere'' effect, which is a phenomenon in which statistically significant observations result by chance due to the large size of the parameter space being searched. If the parameter that are investigated are independent, for 36 Kendall's tau tests for correlations, we would expect a $p$-value of 0.05 to occur by chance once every 20 tests, or $\approx$2/36. The probability from a binomial distribution of having $\geq$1 significant ($p < 0.05$) outcome by chance is 84\%. However, we have 12 significant outcomes. The probability of this happening by chance is $\approx10^{-7}$. This low probability demonstrates that most of the significant correlations between parameters are not due to the look-elsewhere effect. We anticipate this happens because a large fraction of the parameters we investigate are not independent. However tracing the direction of causality (i.e., the fundamental relation which underpins the multiple correlations we observe here) is beyond the scope of this work.

We perform 70 different Anderson-Darling tests to assess whether there are differences in the properties of the spectral classes and the X-ray bright and X-ray faint samples. If all of the parameters that are tested are independent of each other, we expect a significant outcome to occur by chance every 20 tests, or $\approx$4/70. The probability from a binomial distribution of having $\geq$1 significant ($p < 0.05$) outcome by chance is 93\%. We found 19 significant outcomes. The probability of this happening by chance is $\approx 10^{-9}$.

For both of these tests, we can account for the look-elsewhere effect by dividing our significance threshold by the number of degrees of freedom in the tests. If we take this to be the number of tests, this would reduce the threshold to $p < 0.001$ for the Kendall's tau tests and $p < 0.0007$ for the Anderson-Darling tests. However, our tests are not completely independent as we expect there to be some correlation between the parameters, such as between $L_{\rm peak}$, $T_{\rm peak}$, and $R_{\rm peak}$. We conclude that it is unlikely that the correlations we have found here are due to chance (i.e. the look-elsewhere effect), given the low probabilities for the number of significant outcomes we find occurring due to chance.

\subsection{Optical to X-ray Ratio}
\begin{figure}
    \centering
\includegraphics[width=\columnwidth]{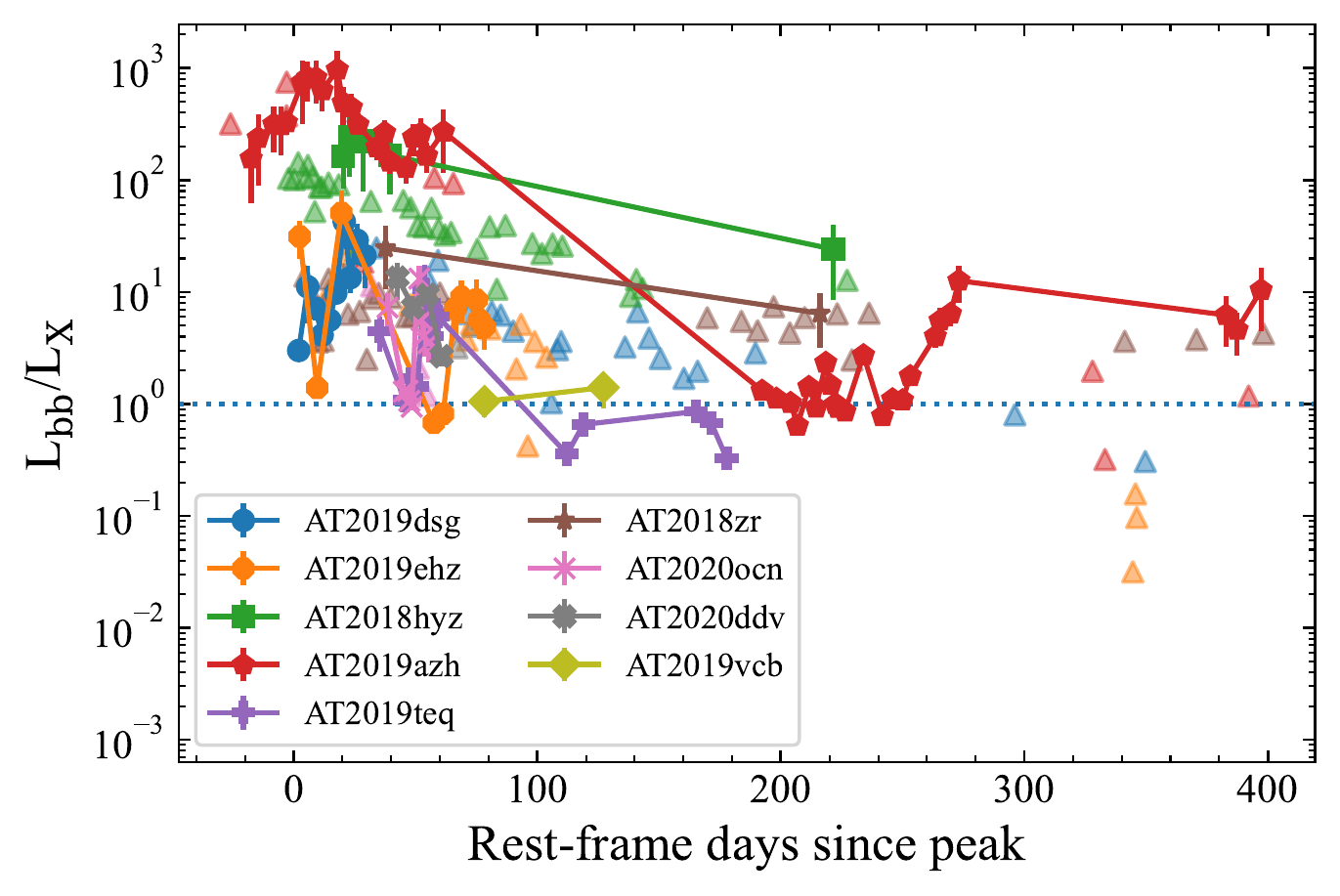}
    \caption{The ratio of the blackbody luminosity, derived from the optical and UV light curves, to the 0.3--10 keV X-ray luminosity from \textit{Swift}/XRT. Triangles are 3$\sigma$ lower limits.}
    \label{fig:XrayOpt}
\end{figure}

In Figure \ref{fig:XrayOpt}, we show the ratio of blackbody luminosity derived from the fits to the UV/optical light curves to the 0.3--10 keV luminosity from the \textit{Swift}/XRT observations, for 9 TDEs with \textit{Swift} XRT detections. We also show the 0.3--10 keV light curves compared to the optical/UV blackbody light curves in the figures in the Appendix. Four of these TDEs were presented in \citet{vanVelzen21}, including AT2018hyz, AT2019dsg, AT2019ehz, and AT2019azh. We present additional observations for each of these, in addition to 5 more TDEs not presented in that paper.

\citet{vanVelzen21} noted the large amplitude flaring of AT2019ehz, and the increase in luminosity over timescales of several months for other TDEs like AT2019azh. The 9 TDEs in Figure \ref{fig:XrayOpt} show a similar long term increase in luminosity, and we note the general trend of $L_{\rm bb}/L_{\rm X}$ towards 1 at later times.

\section{Black Hole Mass Estimates} \label{sec:bhmass}
\subsection{\rm \texttt{MOSFiT}}
\begin{figure}
    \centering
    \includegraphics[width=\columnwidth]{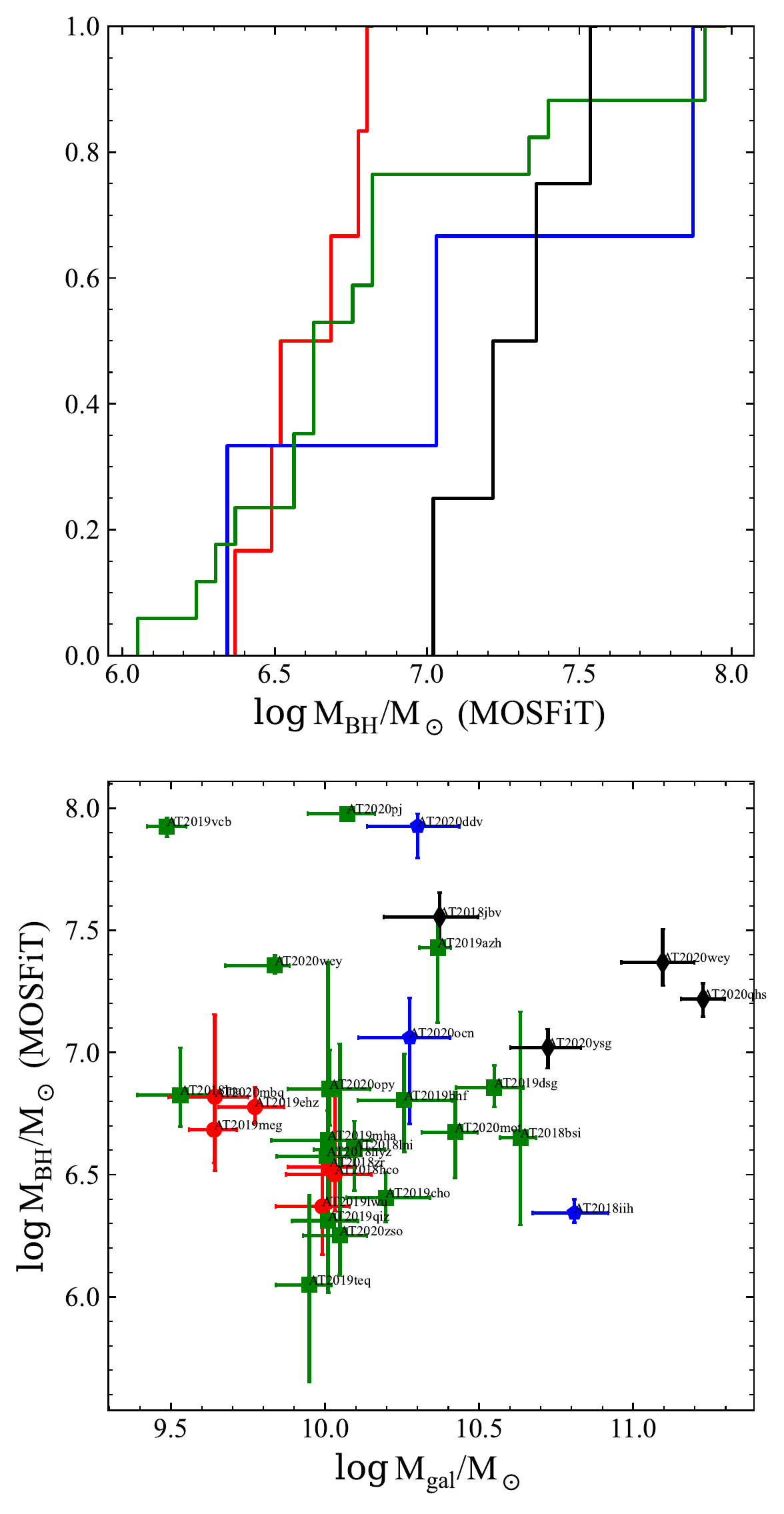}
    \caption{The black hole mass estimated from the \texttt{MOSFiT} fits to the optical/UV light curves vs.~the total stellar mass of the host galaxies measured from the SED fits to the pre-flare photometry. We find no significant correlation between the black hole mass and the galaxy stellar mass. The TDE-featureless events are shown to have more massive black holes as compared to the remainder of the sample. Colors and symbols are the same as in Figure \ref{fig:urslsn}.}
    \label{fig:mosfit}
\end{figure}

\begin{figure}
    \centering
    \includegraphics[width=\columnwidth]{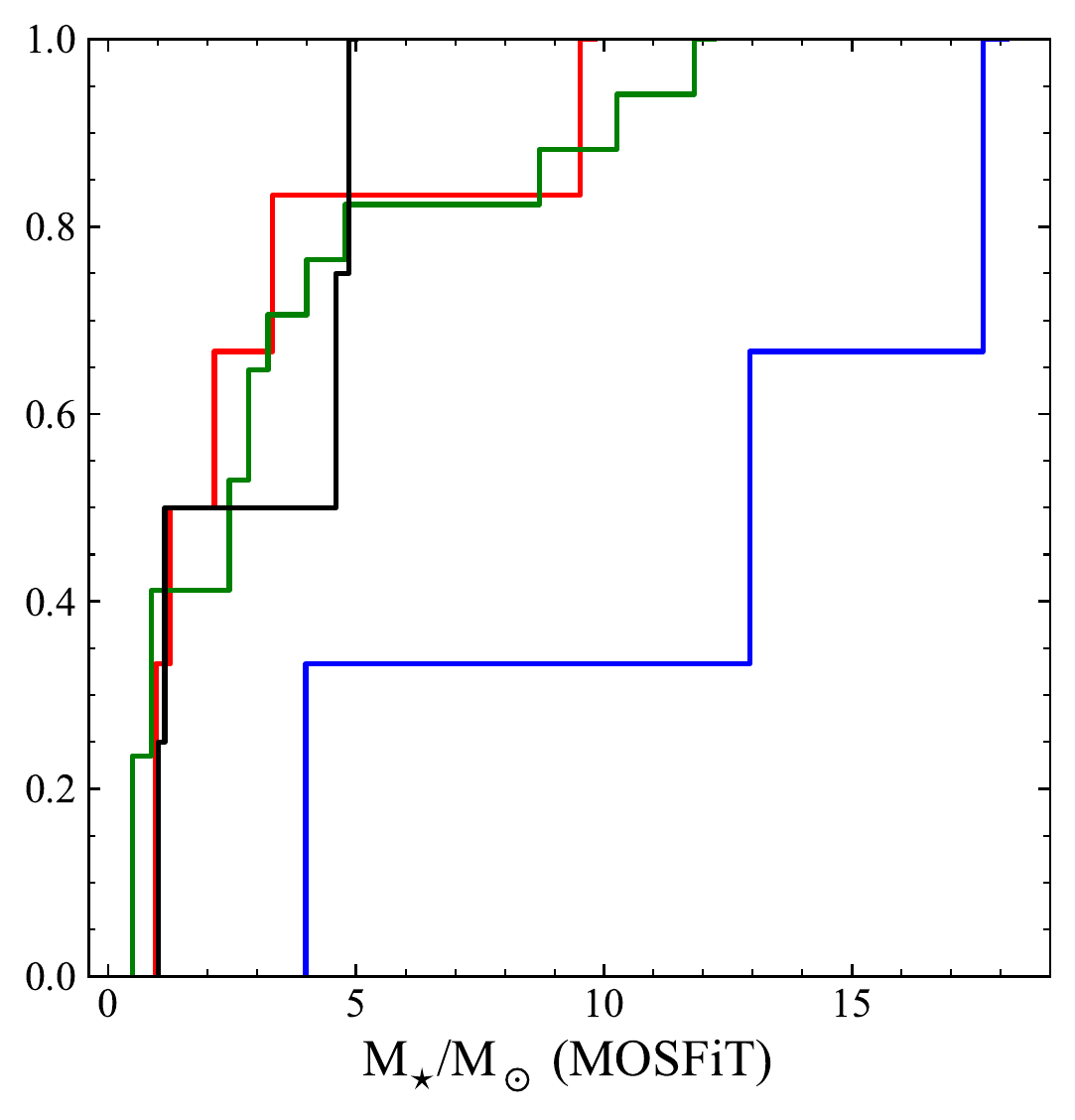}
    \caption{The disrupted star mass estimated from the \texttt{MOSFiT} fits to the optical/UV light curves for each of the TDE spectral types. We find that TDE-He events show significantly larger disrupted star masses as compared to the remainder of the sample. Colors are the same as in Figure \ref{fig:urslsn}.}
    \label{fig:mosfitstar}
\end{figure}

In addition to the light curve fitting described in Section \ref{sec:modelfitting}, we use the Modular Open-Source Fitter for Transients \citep[\texttt{MOSFiT};][]{Guillochon2018, Mockler2019} to fit the light curves of the 30 TDEs in the ZTF-I sample. The TDE module in \texttt{MOSFiT} generates bolometric light curves via hydrodynamical simulations and passes them through viscosity and reprocessing transformation functions to create the single-band light curves. These single-band light curves are then used to fit the multi-band data to obtain the light curve properties and information on the physical parameters of the disrupted star, the tidal encounter, and the MBH. In this analysis, we are most interested in the properties of \texttt{MOSFiT}'s ability to estimate the parameters of the MBH and the disrupted star from the TDE light curve. \texttt{MOSFiT} is particularly sensitive to plateaus in the late-time data as well as the slope of the pre-peak rise. We therefore only fit our forced photometry data between $-10 \mathrm{~days} \leq t_{\rm peak} \leq +300 \mathrm{~days}$. We show the black hole mass estimated from this fitting compared to the host galaxy stellar mass in Figure \ref{fig:mosfit} and the mass of the disrupted star in Figure \ref{fig:mosfitstar}.

We find that the MBH masses range from $\approx 6.0 \leq \log(M_{\rm BH}/M_\odot) \leq 7.9$. We evaluate the black hole masses vs. the galaxy stellar mass for correlation with a Kendall's tau test and find no significant correlation between the two parameters. This is surprising, given that one expects the mass of the galaxy to scale positively with the mass of its central MBH. Furthermore, this is in conflict with \citet{Mockler2019}, who found that their estimates of the black hole mass are consistent with the estimates from the bulk galaxy properties. We point out that two joint papers which were released shortly before the submission of this manuscript, \citet{Nicholl2022} and \citet{Ramsden2022}, find a positive correlation between black hole mass measured from \texttt{MOSFiT} and host galaxy bulge mass measured from stellar population synthesis fitting. Our use of total galaxy mass instead of bulge mass may be the source of the discrepancy. While \citet{Ramsden2022} derive the host galaxy masses in a similar manner to the one presented here and are generally consistent with those in Table \ref{tab:host}, they perform bulge-disk decompositions on SDSS and PanSTARRS imaging of the TDE hosts. \citet{Hammerstein2021} note that imaging from ground-based observatories may not provide the resolution required to study galaxy morphology at the redshifts of the TDE hosts. We therefore maintain our use of the total stellar mass instead of the bulge mass. Using an Anderson-Darling test, we find that the TDE-featureless events have significantly larger black holes ($p$-value $=0.04$) as compared to the remainder of the sample. We also find that TDE-He events show larger disrupted star masses when compared to the rest of the sample ($p$-value $=0.008$).

\subsection{\rm \texttt{TDEmass}}
\begin{figure}
    \centering
    \includegraphics[width=\columnwidth]{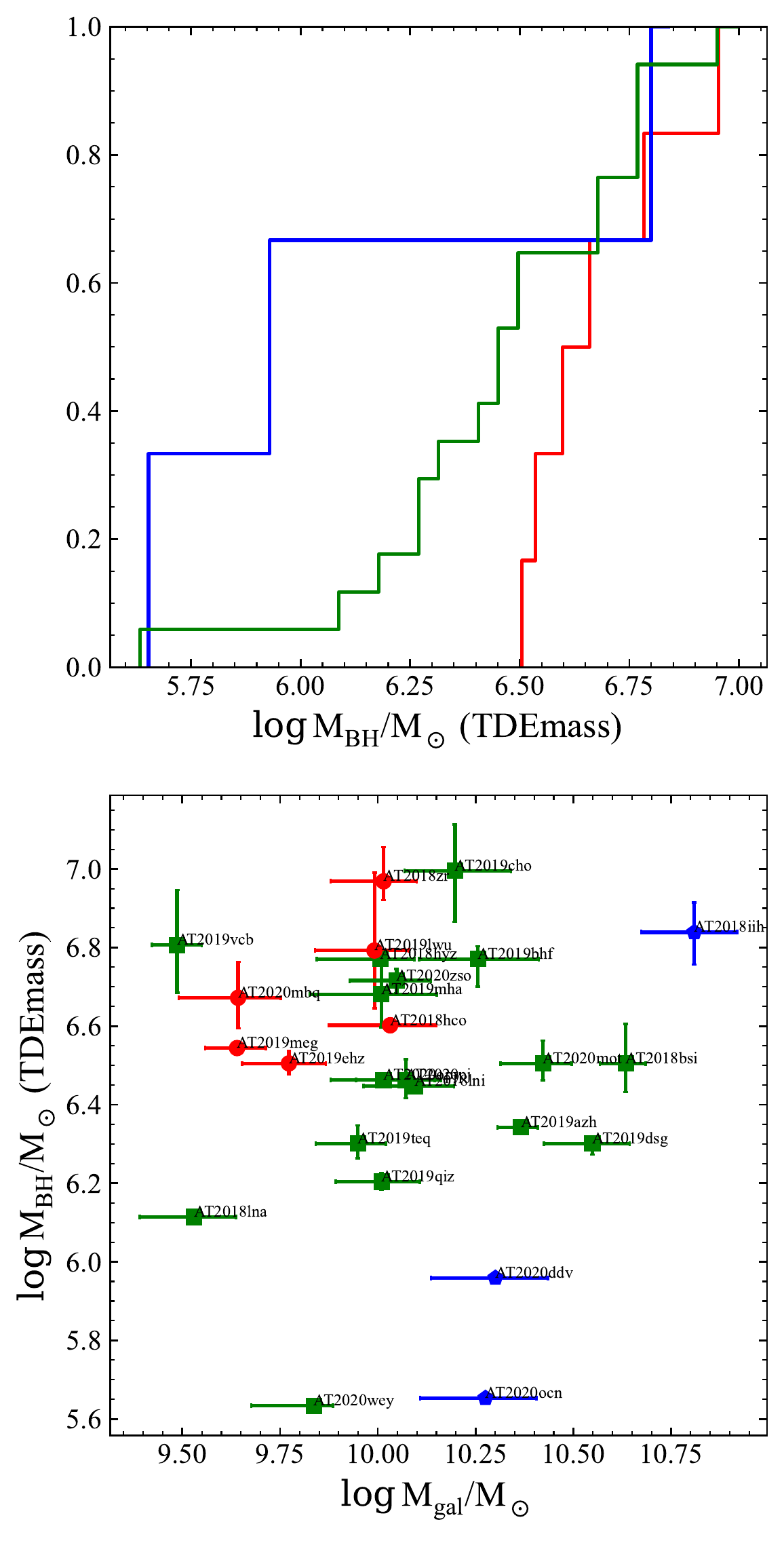}
    \caption{The black hole mass estimated from \texttt{TDEmass} vs.~the total stellar mass of the host galaxies measured from the SED fits to the pre-flare photometry. We find no significant correlation between the black hole mass and the galaxy stellar mass. Colors and symbols are the same as in Figure \ref{fig:urslsn}.}
    \label{fig:tdemass}
\end{figure}

\begin{figure}
    \centering
    \includegraphics[width=\columnwidth]{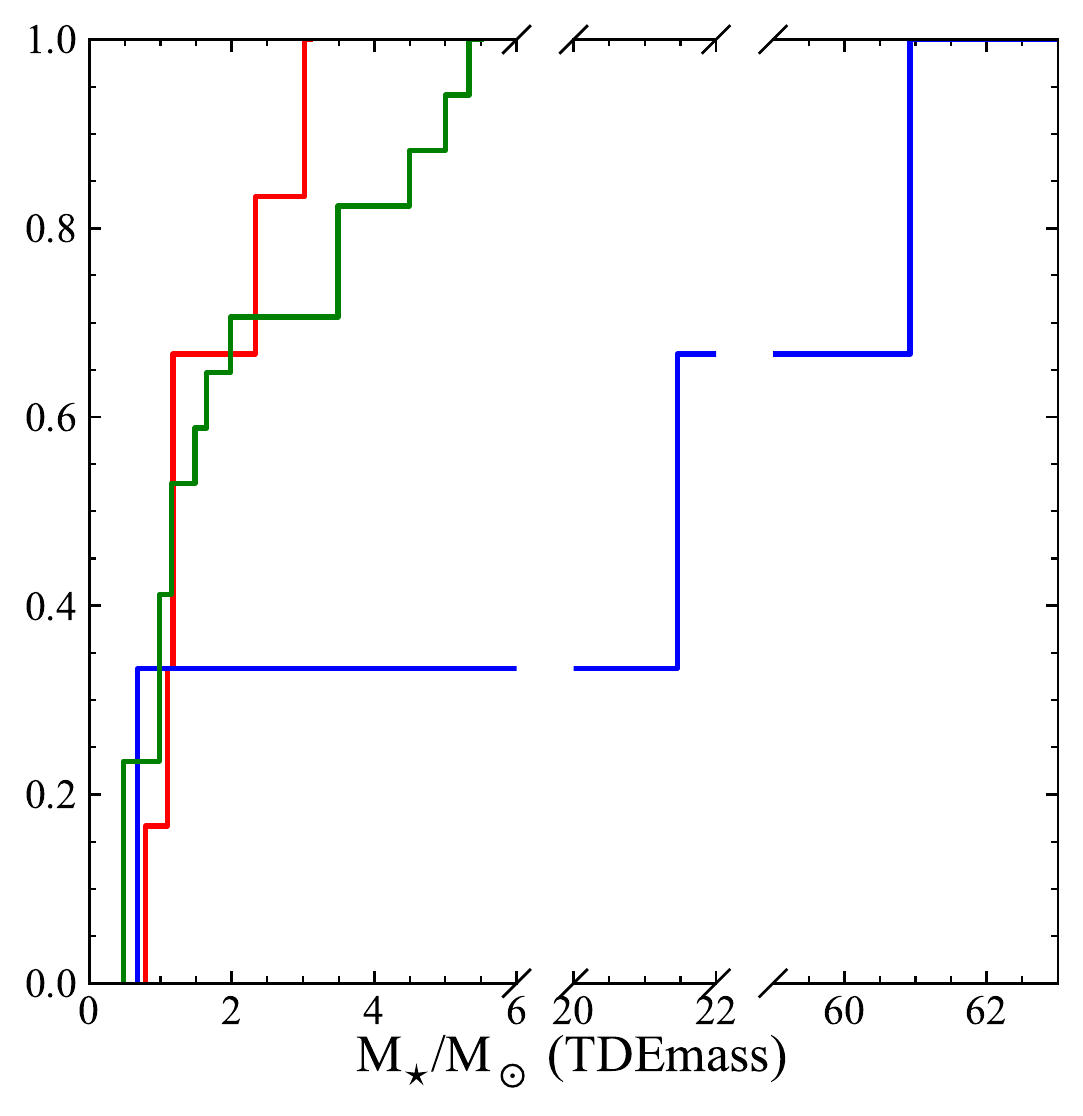}
    \caption{The disrupted star mass estimated from \texttt{TDEmass} split by the TDE spectral types. We find that TDE-He events have significantly larger disrupted star masses as compared to TDE-H and TDE-H+He events. We note the broken axis to accommodate the large star masses in the TDE-He class. Colors are the same as in Figure \ref{fig:urslsn}.}
    \label{fig:tdemassstar}
\end{figure}

We also estimate the MBH mass from \texttt{TDEmass} \citep{Ryu2020}, which takes the peak luminosity and color temperature of the flare as input to calculate the masses of the MBH and the disrupted star. This method of estimating the MBH mass assumes that circularization happens slowly, and that the UV/optical emission arises from shocks in the intersecting debris streams instead of in an outflow or wind. We show the MBH mass estimated from \texttt{TDEmass} compared to the host galaxy stellar mass in Figure \ref{fig:tdemass} and the mass of the disrupted star in Figure \ref{fig:tdemassstar}.

Using this method, we find MBH masses in the range $5.6 \leq \log M_{\rm BH} \leq 7.0$, which is less massive than found with \texttt{MOSFiT}. We point out that we were not able to obtain masses for the four featureless events with \texttt{TDEmass}, as the peak luminosities and temperatures are outside of the limits explored by the model. Again, we find no significant correlation with host galaxy stellar mass. \citet{Ryu2020} did, however, found that their estimates for the MBH mass were roughly consistent with the masses estimated from bulge properties. Again, we use the total stellar mass which may be the source for this discrepancy. Additionally, we find a negative correlation between the MBH mass estimated from \texttt{MOSFiT} and that estimated from \texttt{TDEmass}, with the \texttt{MOSFiT} estimates larger by factor of at least an order of magnitude in most cases. This large difference is perhaps not surprising, as the two methods for estimating the black hole mass employ completely different models for the origin of the UV/optical emission. Estimates of the black hole mass from other, light curve independent methods, such as via the $M_{\rm BH}-\sigma$ relation, will help to narrow down which of these mass estimates is more favorable. We find again that the TDE-He events show significantly larger disrupted star masses as compared to TDE-H and TDE-H+He events ($p$-value $=$0.04).

\section{Discussion} \label{sec:discussion}
We have investigated several correlations among the properties of the light curves presented in this paper, as well as the differences between sub-populations based on spectroscopic class, light curve shape, and X-ray detection. \citet{vanVelzen21}, who analyzed the first 16 TDEs in this paper (plus an additional 22 from the literature) and whose light curve fitting methods we have reproduced here, found a correlation between the decay timescale and the host galaxy stellar mass. With an additional 15 events in our analysis, we find a similar correlation here, consistent with \citet{vanVelzen21} and other previous studies \citep{Blagorodnova2017, Wevers2017}. Our results imply that the decay timescale of the optical/UV light curve follows the fall-back rate, which is crucial for light curve fitting methods that produce an estimate of the black hole mass, such as \texttt{MOSFiT}. This has already been tested against a small sample of post-peak light curves \citep{Mockler2019}. They find evidence that the light curves fitted there are also consistent with tracing the fallback rate. We also recover a weak positive correlation between the peak blackbody luminosity and the decay timescale, which is consistent with a correlation found in \citet{Hinkle20}.

We do find a correlation, although shallow, between the rise timescale and the host galaxy stellar mass, which was not present in \citet{vanVelzen21}. \citet{vanVelzen21} attributed this lack of correlation between rise timescale and host galaxy mass to two possible models, photon advection \citep{Metzger2016} and diffusion \citep{Piran2015}. In the advection model, the optical radiation is advected through a wind until it reaches the trapping radius, which is the radius at which the radiative diffusion time through the debris is shorter than the outflow expansion time. One feature of this model is that for low mass black holes ($\lesssim 7 \times 10^6 M_\odot$), the correlation between the peak luminosity and the black hole mass is weak. However, we do find a correlation between the peak luminosity and the host galaxy stellar mass, which may weaken the plausibility of this model being at play here. 

We find no differences between the TDE spectroscopic classes in terms of rise and decay timescales. \citet{vanVelzen21} found that the TDE-H+He class shows longer rise times and smaller blackbody radii than other spectroscopic classes. They attributed this to the idea that the Bowen fluorescence lines which are sometimes seen in the TDE-H+He class require high densities, which lead to longer diffusion timescales and can be reached at the smaller blackbody radii they found in the class.

A significant difference between blackbody radius and rise times for TDE-H and TDE-H+He was discovered by \citet{vanVelzen21} and confirmed by \citet{Nicholl2022}. These works are based on a larger sample of TDEs compared to the ZTF-only collection presented in this work. \citet{vanVelzen21} contains 13 TDE-H, while our sample contains only 6 events in this spectral class. As such, our ZTF-only sample has less statistical power to uncover differences between the TDE-H and TDE-H+He populations. However, we can use the newly discovered TDEs in our sample to confirm the earlier conclusion that below a radius of $10^{15.1}$~cm, all TDEs between the two classes are classified as TDE-H+He. The same is true for the rise time, where above a rise time of $\sim$16 days, all TDEs between the two classes are TDE-H+He. Our work thus supports the idea that the TDE-H+He events require high density environments, and that the rise times of the light curves are governed not by the fallback timescale, but by the diffusion of photons through the tidal debris.

The TDE-featureless class is characterized by high luminosities, large blackbody radii, and high blackbody temperatures at peak, particularly when compared to the TDE-H and TDE-H+He classes. The spectra of TDE-featureless events are just that, lacking any discernible emission features present in the other three spectroscopic classes. While the four TDE-featureless events we present here are among the highest redshift events in this sample, this, supported by the high luminosities of this class, can be attributed to the rarity of these events, i.e., a larger volume is required to observe them. Additionally, the lack of spectral features is unlikely to be an artifact of their higher redshift, given that the observation of spectral features associated with the host galaxy stellar population, seen most prominently in the spectrum of AT2020ysg, is not uncommon. The host galaxies for the TDE-featureless class are generally more massive than TDE-H and TDE-H+He classes, in addition to being redder in color. We also point out the peculiar event AT2020riz, which shows a particularly fast rise and decay as compared to the other TDE-featureless events. A larger sample of TDE-featureless events is needed in order to understand the diversity of this class of TDEs.

We find that the X-ray bright and X-ray faint events differ in their peak blackbody and $g$-band luminosities. The lack of differences in other properties is surprising. In the reprocessing scenario for explaining the lack of X-rays in some optically selected TDEs, one might expect larger blackbody radii for the X-ray faint sample, as the blackbody radius is that of the larger reprocessing medium and not that of the smaller accretion disk. While it is not possible to entirely rule out the delayed onset of accretion due to circularization of the tidal debris to explain the lack of X-rays, the correlation we have found between the decay timescale and the host galaxy stellar mass makes this less likely as it appears the decay timescale closely follows the fallback rate. In the viewing angle model of \citet{Dai2018}, the X-ray bright and X-ray faint TDEs differ only in whether or not X-rays are visible along the observer's line of sight. Thus, it is less likely that differences among other properties, such as the blackbody radius, will be as important. The lack of difference in host galaxy mass also favors the viewing angle model. One might expect a difference between the two populations in host galaxy mass (as a proxy for black hole mass) for several reasons, whether it be accretion disk temperature \citep[e.g.][]{Dai2015}, rapid circularization \citep[e.g.][]{Guillochon2015}, or the result of the Eddington ratio of the newly formed accretion disk \citep[e.g.][]{Mummery2021}. While other studies, such as \citet{French2020}, have found a difference between the X-ray bright and X-ray faint populations in terms of host galaxy mass, we find no such difference in the sample presented here. However, a measurement of the black hole mass, as opposed to using the host galaxy mass as a proxy, will help to truly discern whether or not there are differences between the two populations.

\begin{figure*}
    \centering
    \includegraphics[width=0.85\textwidth]{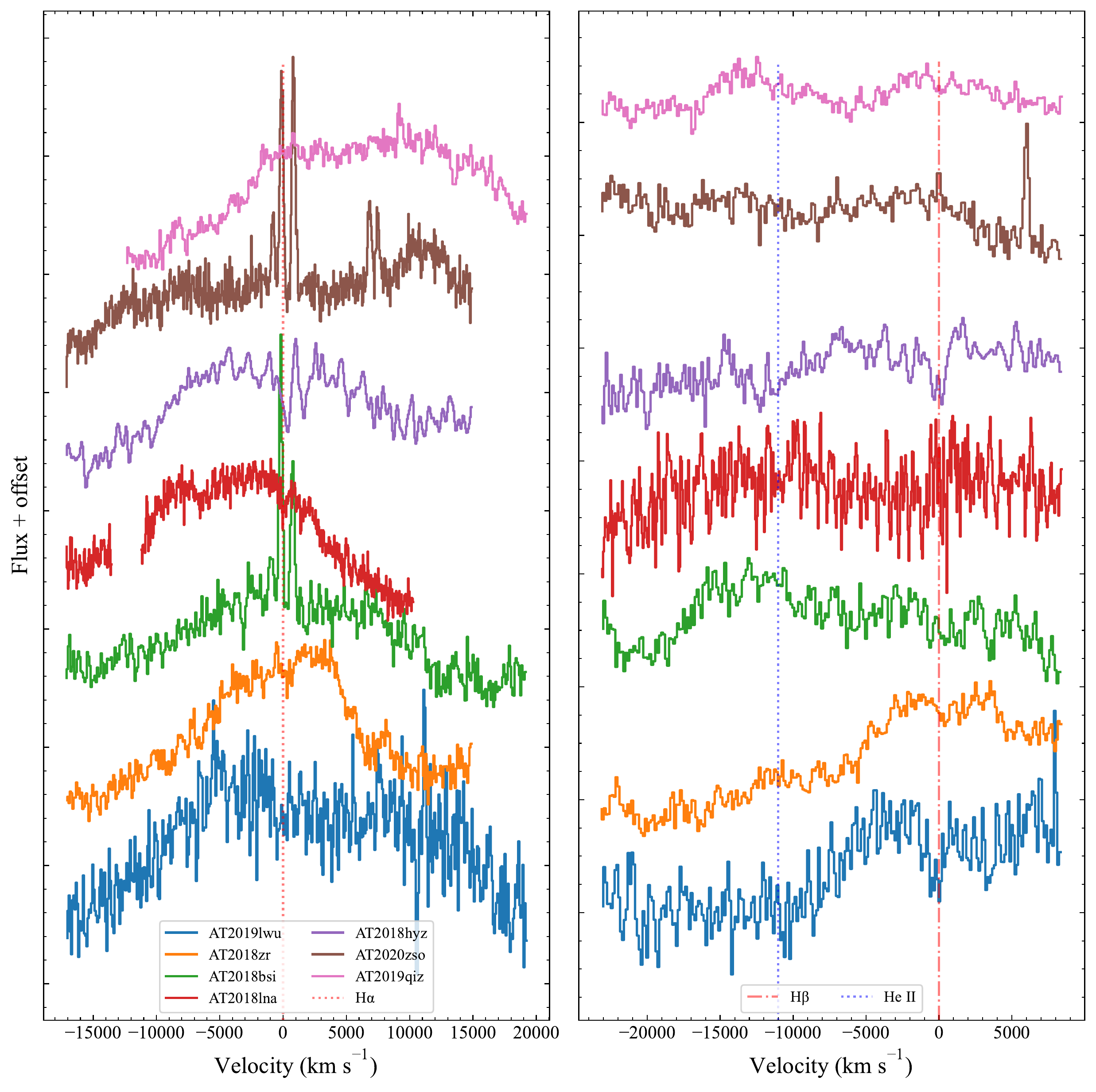}
    \caption{Left: The H$\alpha$ regions of the most extreme broad/flat-topped or double-peaked TDEs in our sample. H$\alpha$ is marked with a dotted red line. Right: The H$\beta$ and He II region of the same objects shown in the left panel, with H$\beta$ marked with a red dot-dashed line and He II marked with a blue dotted line. }
    \label{fig:zoomspec}
\end{figure*}

While this work focuses largely on the light curve properties of these TDEs, the spectra play an important role in the follow-up and classification of candidates as TDEs. The classification of a candidate as a TDE and subsequent sub-classification as one of the spectral types presented in Section \ref{sec:spectra} and in \citet{vanVelzen21} is dependent on the appearance of broad hydrogen and helium emission lines in spectra. The profiles of these broad lines are varied, as seen in Figures \ref{fig:spec1}, \ref{fig:spec2}, and \ref{fig:spec3}, and the differences can give information on potential outflows and the geometry of the system. In particular, double-peaked emission lines, which are seen in some AGN, are thought to originate from the outer regions of an inclined accretion disk. \citet{Wevers2022} examined the line profiles of AT2020zso, a TDE we have included in our sample, and found that the emission lines after peak can be reproduced with a highly inclined, highly elliptical, and relatively compact accretion disk, further supporting the unification picture where viewing angle determines the observed properties of a TDE. In Figure \ref{fig:zoomspec}, we show our spectrum of AT2020zso along with several other extreme broad and flat-topped/double-peaked TDEs in our sample. Of those shown, 2 are of the TDE-H class while the remaining 5 are of the TDE-H+He class. Two of these, AT2018zr and AT2018hyz, are also X-ray detected. The large fraction of X-ray dim TDE-H+He with extreme broad, flat-topped lines in this sample lends further support to the unification picture, but more work is needed to understand why these line profiles are not exclusive to X-ray brightness or spectral class.

\citet{Charalampopoulos22} studied a larger sample of TDE spectra and quantified the evolution of prominent TDE lines with time, such as H$\alpha$, He II, and Bowen lines. They present a scheme for sub-classification of the spectral types of TDEs, with TDE-H and TDE-H+He having X-ray bright and X-ray dim sub-categories which show different spectroscopic features such as double-peaked lines, Fe lines, and \ion{N}{3} lines. They conclude that the large spectroscopic diversity of TDEs, for which they have determined subcategories, can be attributed to viewing angle effects. Although a detailed study of the spectroscopic features of the TDEs is beyond the scope of this work, a cursory examination of the spectra reveals some agreement with these sub-classes. Specifically, AT2018zr and AT2018hyz show evidence for double-peaked Balmer lines accompanied with detected X-ray emission, which is in line with the sub-category of the TDE-H class presented by \citet{Charalampopoulos22}. A more thorough analysis of the spectra and investigation of emission lines will be necessary to understand these sub-categories further.

\section{Conclusions} \label{sec:conclusion}
We have presented a sample of 30 systematically gathered TDEs with light curves from ZTF and \textit{Swift} UVOT and XRT observations, the largest sample of TDEs from a single survey yet. We estimated the parameters of the UV/optical light curves by fitting the multi-band data with two models and examined correlations between the light curve parameters and host galaxy properties, as well as differences among the different sub-classes of TDEs. We summarize our main conclusions below.
\begin{itemize}
    \item Our sample can be split into four spectroscopic classes, with 6 TDE-H, 3 TDE-He, 17 TDE-H+He, and 4 TDEs of the new TDE-featureless class, which we present here for the first time.
    \item Only 47\% of the TDE host galaxies within this sample are in the green valley, although 11/17 of those outside the green valley are within 0.12 mag of its upper or lower bounds.
    \item Using \texttt{MOSFiT}, we find that the TDE-featureless events have significantly larger black hole masses as compared to the rest of the classes. We also find that both \texttt{MOSFiT} and \texttt{TDEmass} yield significantly higher disrupted star masses for the TDE-He class as compared to the rest of the spectral classes. This may hint at the reason for the different spectral classes of TDEs.
    \item We find a correlation between the decay timescale and the host galaxy stellar mass, which is consistent with previous findings from \citet{vanVelzen21}, and is consistent with the picture where the post-peak TDE light curve follows the fallback rate.
    \item We recover a weak correlation between the peak luminosity and the decline rate, where more luminous TDEs decay more slowly, consistent with a correlation found in \citet{Hinkle2020}.
    \item We find that the X-ray bright TDEs show significantly higher peak blackbody and $g$-band luminosities. The lack of differences among other properties such as blackbody radius and host galaxy mass makes the viewing angle model of \citet{Dai2018} for explaining the lack of X-rays in some TDEs more favorable.
\end{itemize}

\acknowledgements
We thank the anonymous referee for their helpful comments towards improving this manuscript. We also thank B.~Mockler and M.~Nicholl for their help with understanding and running the \texttt{MOSFiT} tool.

Based on observations obtained with the Samuel Oschin Telescope 48-inch and the 60-inch Telescope at the Palomar Observatory as part of the Zwicky Transient Facility project. ZTF is supported by the National Science Foundation under Grant No. AST-1440341 and a collaboration including Caltech, IPAC, the Weizmann Institute for Science, the Oskar Klein Center at Stockholm University, the University of Maryland, the University of Washington, Deutsches Elektronen-Synchrotron and Humboldt University, Los Alamos National Laboratories, the TANGO Consortium of Taiwan, the University of Wisconsin at Milwaukee, and Lawrence Berkeley National Laboratories. Operations are conducted by COO, IPAC, and UW. SED Machine is based upon work supported by the National Science Foundation under Grant No. 1106171. The ZTF forced-photometry service was funded under the Heising-Simons Foundation grant \#12540303 (PI: Graham). This work was supported by the GROWTH project funded by the National Science Foundation under Grant No 1545949.

The material is based upon work supported by NASA under award number 80GSFC21M0002. ECK acknowledges support from the G.R.E.A.T research environment funded by {\em Vetenskapsr\aa det}, the Swedish Research Council, under project number 2016-06012, and support from The Wenner-Gren Foundations.

\bibstyle{aasjournals}
\bibliography{sample62}

\appendix
\section{Detailed Spectra}
We describe the spectra for each event presented in this sample and justify our TDE spectral type classification. For each event, we provide an early-time spectrum and a late-time spectrum when available. We detail any evolution which may appear from the early to late time spectra provided. We note that some events do not have pre-peak or even near peak spectra, with the first medium-to-high resolution spectra available over 2 months post-peak. However, this is likely not a problem when investigating spectral class evolution, as most evolution from one class to another for a single object occurs from pre- or near peak to post-peak. All spectra presented here will be made publicly available upon publication.

For the objects that show evolution in their spectra or are unclear in their classification, namely AT2018hyz, AT2019bhf, and AT2019mha, we re-investigate the significance of the spectral class differences after changing their spectral type from what is presented in Table \ref{tab:sample}. 

\subsection{AT2018zr}
We classify AT2018zr as a TDE-H. This is consistent with the original classification given by \citet{Tucker2018}, which reports broad Balmer emission lines 18 days after first detection. We provide an early-time spectrum of this source from the Lowell Discovery Telescope (LDT) DeVeny spectrograph on 2018 Apr 4, which shows broad H$\alpha$, H$\beta$, and H$\gamma$ emission lines and evidence for He I $\lambda$5876. We provide a late-time spectrum from LDT/DeVeny on 2018 May 19, which again shows broad H$\alpha$, H$\beta$, and H$\gamma$ emission lines and evidence for He I $\lambda$5876.

\subsection{AT2018bsi}
We classify AT2018bsi as a TDE-H+He. This classification is consistent with \citet{Gezari2018}, which reports broad hydrogen and helium lines 8 days after first detection. We present an early-time low resolution spectrum from the Palomar P60 SED machine \citep[SEDM;][]{Blagorodnova2018, Rigault2019} on 2018 Apr 18 which shows broad Balmer lines and broad He II $\lambda$4686. We provide a late-time spectrum from LDT/DeVeny on 2018 May 19, which additionally shows \ion{N}{3} $\lambda$4100. We do not interpret this as evolution given that the SEDM spectrum is very low resolution.

\subsection{AT2018hco}
We classify AT2018hco as TDE-H. This is consistent with the classification given in \citet{vanVelzen2018hco}, which classifies AT2018hco as a TDE-H object with broad H$\alpha$ emission and evidence for He I emission. We present an early-time low resolution spectrum from SEDM on 2018 Oct 26 which shows a blue continuum. We also provide a spectrum from the Keck Low Resolution Imaging Spectrograph (LRIS) 2018 Dec 1, which shows broad H$\alpha$ and He I emission. \citet{Reynolds2018hco} reported a weak He II $\lambda$4686 emission line on 2018 Dec 5 in a spectrum from the Nordic Optical Telescope (NOT) Alhambra Faint Object Spectrograph and Camera (ALFOSC). The LRIS spectrum from 4 days prior does indeed show weak emission closer to \ion{N}{3} $\lambda$4640 than He II $\lambda$4686. However, when comparing this host+transient spectra to the host spectrum in \citet{Hammerstein2021}, we find there is a persistent feature near \ion{N}{3} $\lambda$4640. We therefore keep the original classification of TDE-H.

\subsection{AT2018iih}
We classify AT2018iih as a TDE-He, consistent with the classification presented in \citet{vanVelzen21}. We provide a spectrum from LDT/DeVeny on 2019 Mar 10, which shows a steep blue continuum and emission near $\lambda$4500 that we interpret as broad, blueshifted He II. While the redshift of the source places H$\alpha$ nearly out of the wavelength range of the spectrograph, we do not observe broad H$\beta$, which typically accompanies broad H$\alpha$ emission in TDEs. We provide a late-time spectrum from the Palomar P200 Double Spectrograph (DBSP) on 2019 Oct 3, which shows a flattening in the continuum, although still blue, and does indeed cover the wavelength range of H$\alpha$. The telluric-corrected DBSP spectrum shows the He II emission detected at early times, but no evidence for broad H$\alpha$ emission.

\subsection{AT2018hyz}
We classify AT2018hyz as a TDE-H+He. AT2018hyz is one event where evolution of the spectral features has been noted. \citet{Dong2018hyz} found broad H$\alpha$ and weaker broad H$\beta$ emission, but no He II emission in a Lick/Kast spectrum from 2018 Nov 9. \citet{Arcavi2018hyz} noted similar features in a spectrum from the Faulkes Telescope North (FTN) Floyds on 2018 Nov 9, which we provide here as an early-time spectrum. \citet{vanVelzen21} classified AT2018hyz as a TDE-H and performed their analysis with this classification. However, \citet{Hung2020} and \citet{Short2020} presented a suite of spectra which showed evolution in He II and \ion{N}{3}. We show a spectrum from \citet{Short2020} from the Magellan-Baade Inamori Magellan Areal Camera and Spectrograph (IMACS) from 2019 Jun 6 as an example of a late-time spectrum of AT2018hyz. Because of this evolution, we perform our investigation into differences among the spectral class properties again, with AT2018hyz classified as TDE-H but keeping all other classifications as shown in Table \ref{tab:sample}.

If we change the classification of AT2018hyz to TDE-H, as it was in \citet{vanVelzen21}, the difference in rise time between TDE-H and TDE-H+He events is now significant with $p$-value $= 0.012$, which is consistent with the result from \citet{vanVelzen21}. We also find that the difference in rise time between TDE-H and TDE-He events are significant with $p = 0.044$. The difference in $t_0$ between the TDE-H+He and TDE-He class is no longer significant. There are no changes to the other comparisons between light curve classes which would make an insignificant correlation now significant or vice versa.

\subsection{AT2018lni}
We classify AT2018lni as a TDE-H+He. This is consistent with the classification given by \citet{Frederick2019lni} which details the detection of broad H$\alpha$ and He II emission. We provide a spectrum from Palomar/DBSP on 2019 Jan 7, which is detailed in \citet{Frederick2019lni} and shows broad H$\alpha$ and He II emission. We provide a spectrum from LDT/DeVeny on 2019 Mar 1, which also shows evidence for broad H$\alpha$ and He II emission.

\subsection{AT2018lna}
We classify AT2018lna as a TDE-H+He event. \citet{vanVelzen2019lna} did not note any He II in the spectrum from DBSP on 2019 Jan 26 that was used to classify AT2018lna as a TDE, although we provide this observation as an example of an early-time spectrum and now note that there is evidence for He II emission. We present a late-time spectrum from LDT/DeVeny on 2019 Mar 28, which shows further evidence for strong Balmer, He II, and \ion{N}{3} emission.

\subsection{AT2018jbv}
We classify AT2018jbv as a TDE-featureless event. We provide a spectrum from LDT/DeVeny on 2019 Mar 28 as an early-time spectrum. The early-time spectroscopic follow-up of AT2018jbv with medium-to-high resolution spectrographs is limited. This is likely because there were no ZTF $g$-band observations pre-peak, which resulted in AT2018jbv not being flagged in our TDE search until $g$-band observations were performed post-peak. While this spectrum does not cover H$\alpha$, there is no evidence for broad emission near H$\beta$.

\subsection{AT2019cho}
We classify AT2019cho as a TDE-H+He, consistent with the classification in \citet{vanVelzen21}. We provide an early-time spectrum from SEDM on 2019 Mar 4, which shows a blue continuum and evidence for broad H$\alpha$ emission. Due to the low resolution obtained by SEDM, it is difficult to determine whether there is broad He II present in this spectrum. The late-time spectrum we provide was obtained on 2019 May 2 with LDT/DeVeny. This spectrum shows broad Balmer emission accompanied by broad He II and \ion{N}{3} emission.

\subsection{AT2019bhf}
We classify AT2019bhf as a TDE-H+He. This object was originally classified as TDE-H in \citet{vanVelzen21}, however, further examination of the available spectra revealed broad bumps near He II and \ion{N}{3} $\lambda$4640. This has led to the reclassification of this object as TDE-H+He. We provide one early-time spectrum from SEDM on 2019 Mar 30, which shows broad H$\alpha$ emission, and a broad bump in the H$\beta$, He II, \ion{N}{3} region. The late-time spectrum we provide is from LDT/DeVeny on 2019 Jun 29, which again shows broad H$\alpha$ and a broad bump near H$\beta$, He II, and \ion{N}{3}. We perform our search for correlations among light curve and host properties again, with AT2019bhf classified as TDE-H.

After performing our investigation into the spectral class differences with AT2019bhf classified as TDE-H, we find several differences. The difference between TDE-H and TDE-H+He rise times ($\sigma$) is now significant with a $p$-value $=0.021$. The difference between the TDE-H and TDE-He rise times is also significant with $p$-value $=0.044$. The difference in $t_0$ between the TDE-H+He and TDE-He classes is no longer significant. The remaining comparisons are unchanged.

\subsection{AT2019azh}
We classify AT2019azh as a TDE-H+He. \citet{vanVelzen21} classified this object as TDE-H+He based on follow-up spectra, which evolved from featureless to show broad Balmer emission and evidence for He II and \ion{N}{3} emission. We provide a spectrum near peak from LDT/DeVeny on 2019 Mar 10, which shows evidence for broad Balmer emission and a steep blue continuum, although there is Balmer absorption from the host galaxy. Our late-time spectrum from LDT/DeVeny on 2019 May 2 shows strong broad H$\alpha$, a broad bump near H$\beta$, and emission near He II and \ion{N}{3}. \citet{Hinkle20} also examined spectra of AT2019azh and found that there are Bowen fluorescence lines that appear post-peak in addition to the broad Balmer emission, although the spectra are dominated by Balmer emission at early times.

\subsection{AT2019dsg}
We classify AT2019dsg as TDE-H+He, consistent with the classification in \citet{vanVelzen21}. We include an early-time spectrum from New Technology Telescope (NTT) ESO Faint Object Spectrograph and Camera v.2 (EFOSC2) on 2019 May 13, which shows broad Balmer emission, broad He II, and broad \ion{N}{3} emission \citep{Short2019dsg}. We provide a late-time spectrum from LDT/DeVeny on 2019 Jun 29, which shows a flattening in the continuum, but persistent broad Balmer, He II, and \ion{N}{3} emission.

\subsection{AT2019ehz}
We classify AT2019ehz as a TDE-H object. The early-time spectrum we present is from the Liverpool Telescope (LT) SPectrograph for the Rapid Acquisition of Transients (SPRAT) on 2019 May 10. This spectrum is blue and mostly featureless. Our late-time spectrum from LDT/DeVeny on 2019 Jun 29 shows broad H$\alpha$ emission and possible broad H$\beta$ emission.

\subsection{AT2019mha}
We classify AT2019mha as TDE-H+He. We have only one early-time spectrum for this source from DBSP on 2019 Aug 27, which shows host galaxy lines at $z=0.148$ but broad Balmer emission, and He II and \ion{N}{3} emission blueshifted by $\sim$5000 km s$^{-1}$ with respect to the host galaxy lines. Because this source was reclassified from \citet{vanVelzen21}, we have performed the investigation into spectral class differences again, with AT2019mha classified as TDE-H while keeping all other classifications in Table \ref{tab:sample} the same. We find that changing the classification of AT2019mha to TDE-H does not affect any comparisons between spectral classes.

\subsection{AT2019meg}
We classify AT2019meg as TDE-H. This is consistent with the classification given by \citet{vanVelzen2019meg}. We provide one early time spectrum from SEDM on 2019 Jul 31, which shows a blue continuum and broad H$\alpha$ and H$\beta$ emission lines. The late-time spectra of this object are limited, but we provide a later time spectrum from DBSP on 2019 Aug 10, which also shows a blue continuum, but the broad H$\beta$ emission is now more prominent.

\subsection{AT2019lwu}
We classify AT2019lwu as TDE-H, consistent with the classification given in \citet{vanVelzen21}. We provide an early-time spectrum from SEDM on 2019 Aug 8, which shows a blue continuum, however no discernible broad emission features are seen in the low-resolution spectrum. The late-time spectra of AT2019lwu are limited, but we provide another spectrum from LDT/DeVeny on 2019 Aug 27 which shows a blue continuum and now broad H$\alpha$ and H$\beta$ emission lines.

\subsection{AT2019qiz}
We classify AT2019qiz as a TDE-H+He object. We provide an early-time spectrum from SEDM on 2019 Sept 24, which shows a blue continuum and potential for broad emission lines near H$\alpha$, H$\beta$, He II, and \ion{N}{3}. A late-time spectrum from LDT/DeVeny on 2019 Nov 5 confirms that the existence of broad Balmer emission, as well as broad He II and \ion{N}{3} emission.

\subsection{AT2019teq}
We classify AT2019teq as a TDE-H+He object. We provide one early-time spectrum of this object from LDT/DeVeny on 2019 Oct 23, which shows broad Balmer emission that is potentially blueshifted by $\sim$8000 km s$^{-1}$. The He II and \ion{N}{3} emission is also blueshifted by this same amount. The classification report for this object \citep{Hammerstein2020teq} notes the possibility for the presence of Fe II narrow line complex near He II. We provide a later-time spectrum from  LDT/DeVeny on 2019 Nov 5, which shows stronger evidence for blueshifted TDE-like lines.

\subsection{AT2020pj}
We classify AT2020pj as a TDE-H+He object. We provide an early-time spectrum from LT/SPRAT on 2020 Jan 15, which shows a blue continuum and a broad bump near H$\beta$ and He II. We note a peculiar absorption line near H$\alpha$ which is due to an error in the telluric absorption correction. We also note that this galaxy is a star-forming galaxy and possesses narrow H$\alpha$ emission. The late-time spectrum that we provide is from LDT/DeVeny on 2020 Feb 26. This spectrum shows a blue continuum and a broad base to the narrow H$\alpha$ emission. It also shows a broad base to the H$\beta$ emission and broad He II and \ion{N}{3}.

\subsection{AT2019vcb}
We classify AT2019vcb as TDE-H+He. We provide an early-time low-resolution spectrum from LT/SPRAT on 2019 Dec 28, which shows a strong blue continuum and a broad base to the narrow H$\alpha$ from the host galaxy. There are also potential broad bumps near H$\beta$ and He II. We provide a late-time spectrum from Keck/LRIS on 2020 Feb 18, which additionally shows broad He II and \ion{N}{3} emission.

\subsection{AT2020ddv}
We classify AT2020ddv as a TDE-He object. The follow-up spectra of this object are unfortunately limited, but we provide an early-time spectrum from DBSP on 2020 Feb 27, which shows a blue continuum and lack of obvious broad Balmer emission. There is, however, a broad bump near He II, which points towards the classification of this object as TDE-He. We provide a late-time spectrum of this object from LDT/DeVeny on 2020 Jun 9, which shows a flattening in the continuum and broad emission near He II, but again no obvious broad Balmer emission lines.

\subsection{AT2020ocn}
We classify AT2020ocn as a TDE-He object. We provide an early-time spectrum of AT2020ocn from DBSP on 2020 Jun 17, which shows a blue continuum a broad emission near He II and potentially \ion{N}{3}. There is no obvious broad Balmer emission. We provide a later time spectrum from DBSP on 2020 Jul 16, which shows flattening in the continuum but the broad emission near He II remains. Again, there is no obvious broad Balmer emission. 

\subsection{AT2020opy}
We classify AT2020opy as a TDE-H+He object. We provide an early-time spectrum from LDT/DeVeny on 2020 Aug 19, which shows a blue continuum, a broad base to the narrow H$\alpha$ from the host galaxy, and broad emission near He II and H$\beta$. We provide a late-time spectrum from LDT/DeVeny on 2020 Oct 11. The continuum has now flattened, but the broad emission near He II and \ion{N}{3} is now more apparent, accompanied by the broad Balmer emission.

\subsection{AT2020mot}
We classify AT2020mot as a TDE-H+He object. The spectra of this object are unfortunately limited. We provide a low-resolution spectrum from LT/SPRAT on 2020 Jul 29, which shows a broad emission feature near He II and H$\beta$. There is also a potential broad emission feature near H$\alpha$. We provide a spectrum from LDT/DeVeny on 2020 Aug 19 as a late-time spectrum. This spectrum shows a broad emission feature near H$\alpha$ and H$\beta$, as well as broad emission from He II and \ion{N}{3}.

\subsection{AT2020mbq}
We classify AT2020mbq as a TDE-H object. The available spectra for this source are unfortunately very limited. We provide one spectrum from DBSP on 2020 Aug 14, which shows a blue continuum and broad H$\alpha$ and H$\beta$ emission.

\subsection{AT2020qhs}
We classify AT2020qhs as a TDE-featureless object. Similar to AT2018jbv, we were unable to classify this object close to peak as the ZTF survey did not observe this object until it had already started to decline. We did not obtain a first spectrum of this object until roughly 77 days post-peak. We provide this spectrum from LDT/DeVeny on 2020 Oct 11 as the earliest-time spectrum available. The spectrum shows a steep blue continuum with no obvious emission lines. Although H$\alpha$ is not within the wavelength range observed by DeVeny, there is no broad H$\beta$ emission, which typically accompanies any broad H$\alpha$. We provide a late-time spectrum from Keck/LRIS on 2020 Nov 20, which also shows a steep blue continuum and no obvious broad emission lines. This spectrum does cover H$\alpha$, and no obvious broad emission is present.

\subsection{AT2020riz}
We classify AT2020riz as a TDE-featureless object. The follow-up spectra for this object are unfortunately very limited. We show one spectrum from LDT/DeVeny on 2020 Oct 15, which shows a steep blue continuum and no obvious broad emission features. While some TDEs do evolve from featureless to having broad emission features, this typically occurs pre-peak to post-peak, as we have discussed above. The spectrum we provide here is sufficiently post-peak that this is likely not what is occurring in this spectrum.

\subsection{AT2020wey}
We classify AT2020wey as a TDE-H+He object. This object was originally classified by \citet{Arcavi2020wey} as a TDE-H+He object. We provide the spectrum used in this classification as an example of an early-time spectrum. This spectrum is from FTN/Floyds on 2020 Oct 22. We provide one additional spectrum from DBSP on 2020 Nov 12, which shows a similar blue continuum, and more prominent H$\alpha$ emission. The broad He II emission is still present.

\subsection{AT2020zso}
We classify AT2020zso as a TDE-H+He object. The available spectra for this event are not spread over a large span of time, but we provide one earlier-time spectrum from SEDM on 2020 Nov 25, which shows a blue continuum and evidence for broad Balmer, He II, and \ion{N}{3} emission. The later-time spectrum we provide is from Keck/LRIS on 2020 Dec 12, which now shows the broad H$\alpha$ and H$\beta$ emission more prominently, and confirms the presence of broad He II and \ion{N}{3}.

\subsection{AT2020ysg}
We clasify AT2020ysg as a TDE-featureless object. We provide one early-time spectrum from LDT/DeVeny on 2020 Dec 6, which shows a steep blue continuum and no apparent broad emission features. We provide another spectrum from LDT/DeVeny on 2021 Jan 11, which still shows the steep blue continuum and lack of broad emission features. We note that these spectra are over 50 days post-peak. AT2020ysg suffers from a similar predicament as AT2018jbv, where the peak was missed by the ZTF survey and no color information was available pre-peak. This delayed the classification of this object and subsequent follow-up efforts until sufficiently post-peak that the classification was secure. We note that the first spectrum was taken approximately 50 days after post-peak color information became available. Additionally, any evolution from featureless to the emergence of broad lines that we have noted in the spectra presented in this Appendix typically occurs from pre-peak to post-peak. These spectra are sufficiently post-peak that evolution would likely have already taken place.

\begin{deluxetable}{l l r l | l r l}
\tablecaption{Spectroscopic Observations}
\tablehead{ \colhead{IAU Name} & \colhead{Date} & \colhead{Phase} & \colhead{Telescope/Inst.} & \colhead{Date} & \colhead{Phase} & \colhead{Telescope/Inst.}}
\startdata
AT2018zr & 2018 Apr 04 & 7 & LDT/DeVeny & 2018 May 19 & 52 & LDT/DeVeny \\
AT2018bsi & 2018 Apr 18 & 1 & P60/SEDM & 2018 May 19 & 32 & LDT/DeVeny \\
AT2018hco & 2018 Oct 26 & 12 & P60/SEDM & 2018 Dec 01 & 48 & Keck/LRIS \\
AT2018iih & 2019 Mar 10 & 90 & LDT/DeVeny & 2019 Oct 03 & 297 & P200/DBSP \\
AT2018hyz & 2018 Nov 9 & 3 & FTN/Floyds\tablenotemark{a} & 2019 Jun 06 & 213 & Magellan-Baade/IMACS\tablenotemark{b} \\
AT2018lni & 2019 Jan 07 & 23 & P200/DBSP & 2019 Mar 01 & 76 & LDT/DeVeny \\
AT2018lna & 2019 Jan 26 & 0 & P200/DBSP & 2019 Mar 28 & 61 & LDT/DeVeny \\
AT2018jbv & 2019 Mar 28 & 101 & LDT/DeVeny &  \\
AT2019cho & 2019 Mar 04 & 0 & P60/SEDM & 2019 May 02 & 58 & LDT/DeVeny \\
AT2019bhf & 2019 Mar 30 & 28 & P60/SEDM & 2019 Jun 29 & 119 & LDT/DeVeny \\
AT2019azh & 2019 Mar 10 & $-$6 & LDT/DeVeny & 2019 May 02 & 46 & LDT/DeVeny \\
AT2019dsg & 2019 May 13 & 12 & NTT/EFOSC2\tablenotemark{c} & 2019 Jun 29 & 59 & LDT/DeVeny \\
AT2019ehz & 2019 May 10 & 0 & LT/SPRAT & 2019 Jun 29 & 50 & LDT/DeVeny \\
AT2019mha & 2019 Aug 27 & 18 & P200/DBSP \\
AT2019meg & 2019 Jul 31 & $-$1 & P60/SEDM & 2019 Aug 10 & 8 & P200/DBSP \\
AT2019lwu & 2019 Aug 08 & 11 & P60/SEDM & 2019 Aug 27 & 30 & LDT/DeVeny \\
AT2019qiz & 2019 Sep 24 & $-$13 & P60/SEDM & 2019 Nov 05 & 28 & LDT/DeVeny \\
AT2019teq & 2019 Oct 23 & $-$15 & LDT/DeVeny & 2019 Nov 05 & $-$2 & LDT/DeVeny \\
AT2020pj & 2020 Jan 15 & 1 & LT/SPRAT & 2020 Feb 26 & 43 & LDT/DeVeny \\
AT2019vcb & 2019 Dec 28 & 16 & LT/SPRAT & 2020 Feb 18 & 68 & Keck/LRIS \\
AT2020ddv & 2020 Feb 27 & $-$9 & P200/DBSP & 2020 Jun 09 & 93 & LDT/DeVeny \\
AT2020ocn & 2020 Jun 17 & 30 & P200/DBSP & 2020 Jul 16 & 59 & P200/DBSP \\
AT2020opy & 2020 Aug 19 & $-$9 & LDT/DeVeny & 2020 Oct 11 & 43 & LDT/DeVeny \\
AT2020mot & 2020 Jul 29 & 7 & LT/SPRAT & 2020 Aug 19 & 13 & LDT/DeVeny \\
AT2020mbq & 2020 Aug 14 & 55 & P200/DBSP \\
AT2020qhs & 2020 Oct 11 & 77 & LDT/DeVeny & 2020 Nov 20 & 117 & Keck/LRIS \\
AT2020riz & 2020 Oct 15 & 57 & LDT/DeVeny \\
AT2020wey & 2020 Oct 22 & -5 & FTN/Floyds\tablenotemark{d} & 2020 Nov 12 & 15 & P200/DBSP \\
AT2020zso & 2020 Nov 25 & $-$14 & P60/SEDM & 2020 Dec 12 & 2 & Keck/LRIS\\
AT2020ysg & 2020 Dec 06 & 50 & LDT/Deveny & 2021 Jan 11 & 86 & LDT/DeVeny
\enddata
\label{tab:spec}
\tablecomments{Information for all spectra shown in Figures \ref{fig:spec1}, \ref{fig:spec2}, and \ref{fig:spec3}. We include the date the spectrum was observed, the approximate phase from estimated peak the spectrum was observed in days, and the telescope and instrument. The phase is approximate to within one day of when the spectrum was observed.}
\tablenotetext{a}{\citet{Arcavi2018}}
\tablenotetext{b}{\citet{Short2020}}
\tablenotetext{c}{\citet{Short2019dsg}}
\tablenotetext{d}{\citet{Arcavi2020}}
\end{deluxetable}

\begin{figure*}
\gridline{	\fig{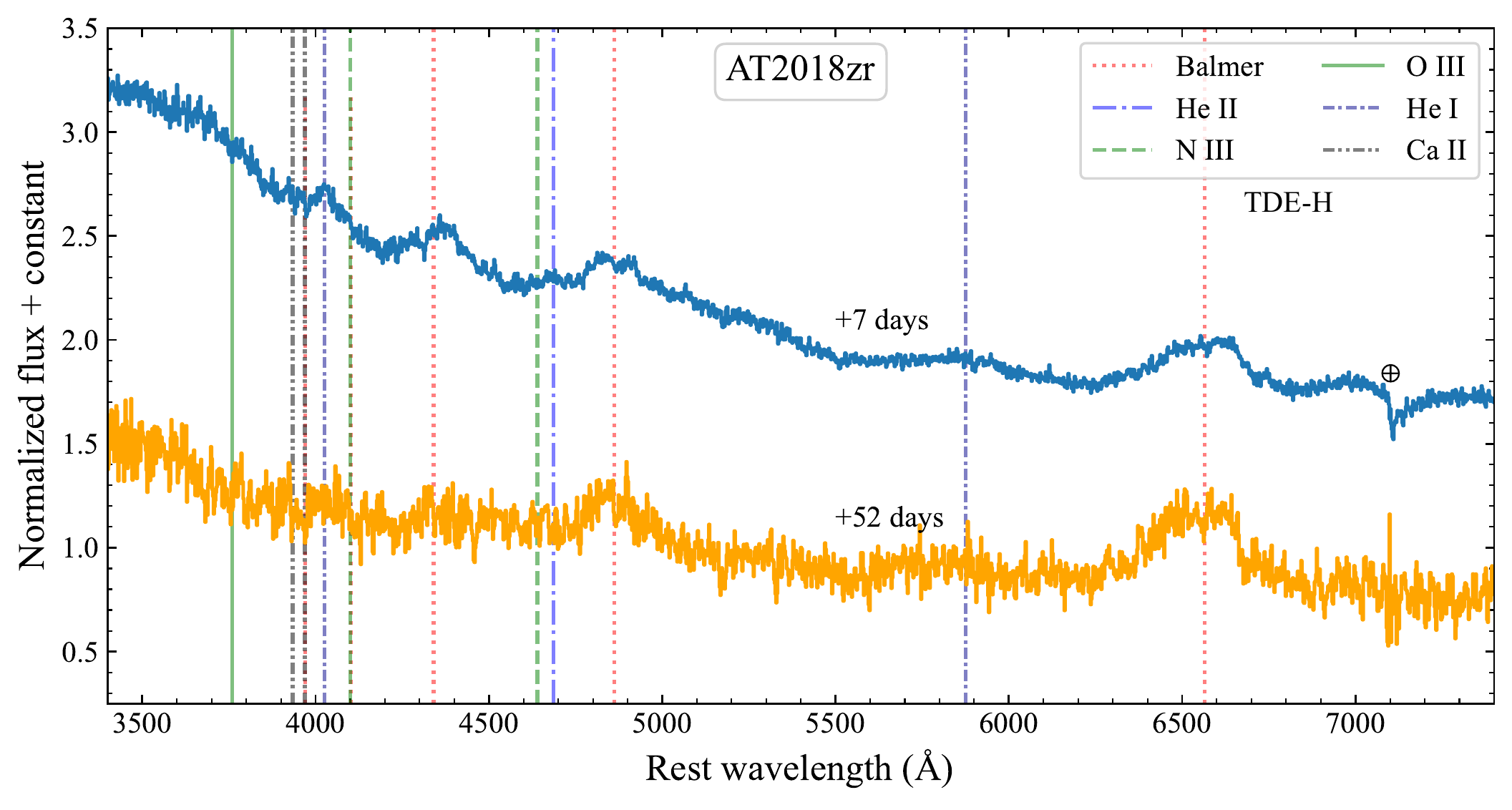}{0.45 \textwidth}{} 
			\fig{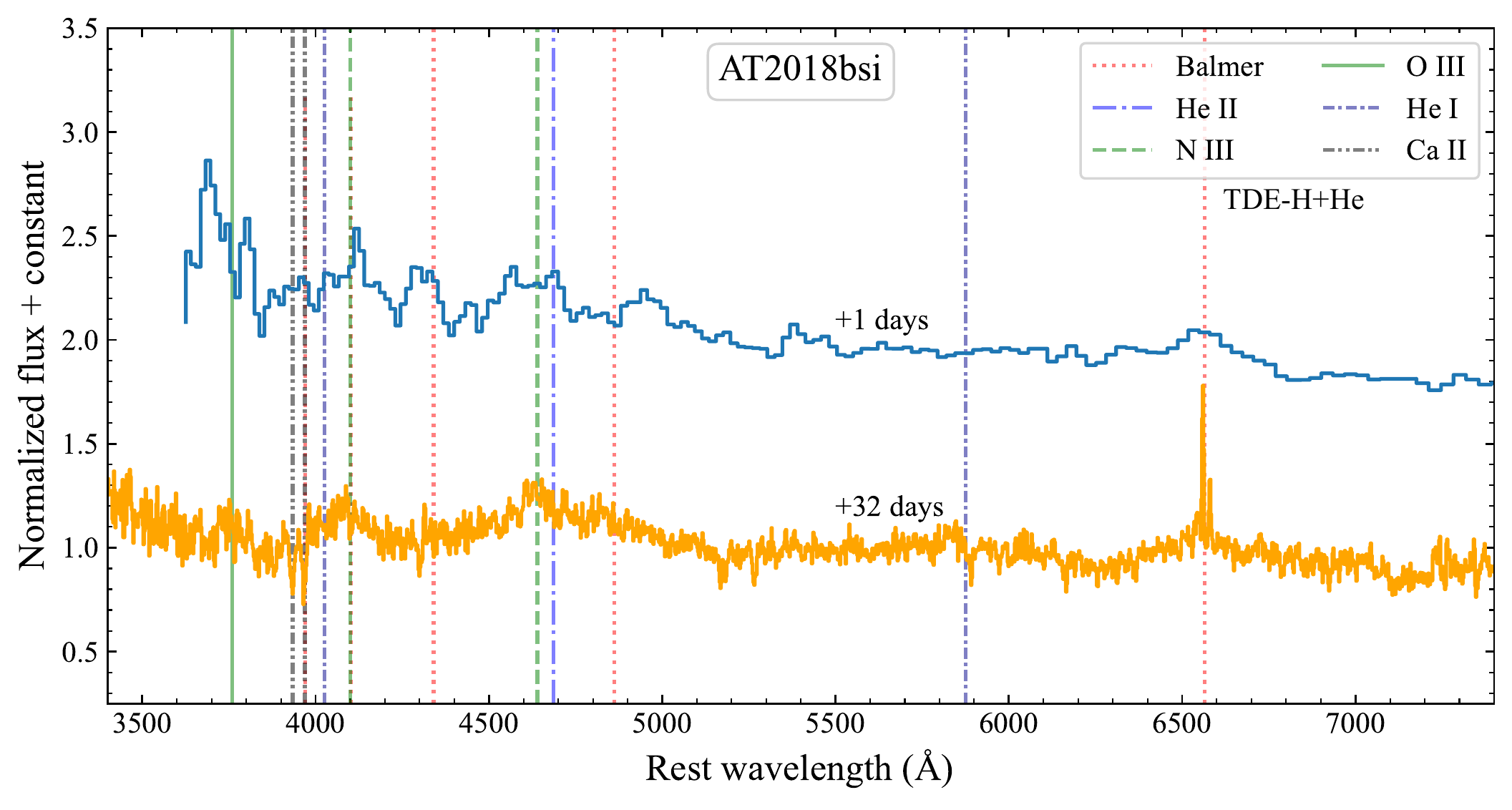}{0.45 \textwidth}{} 
			\\[-30pt]}
			
\gridline{  \fig{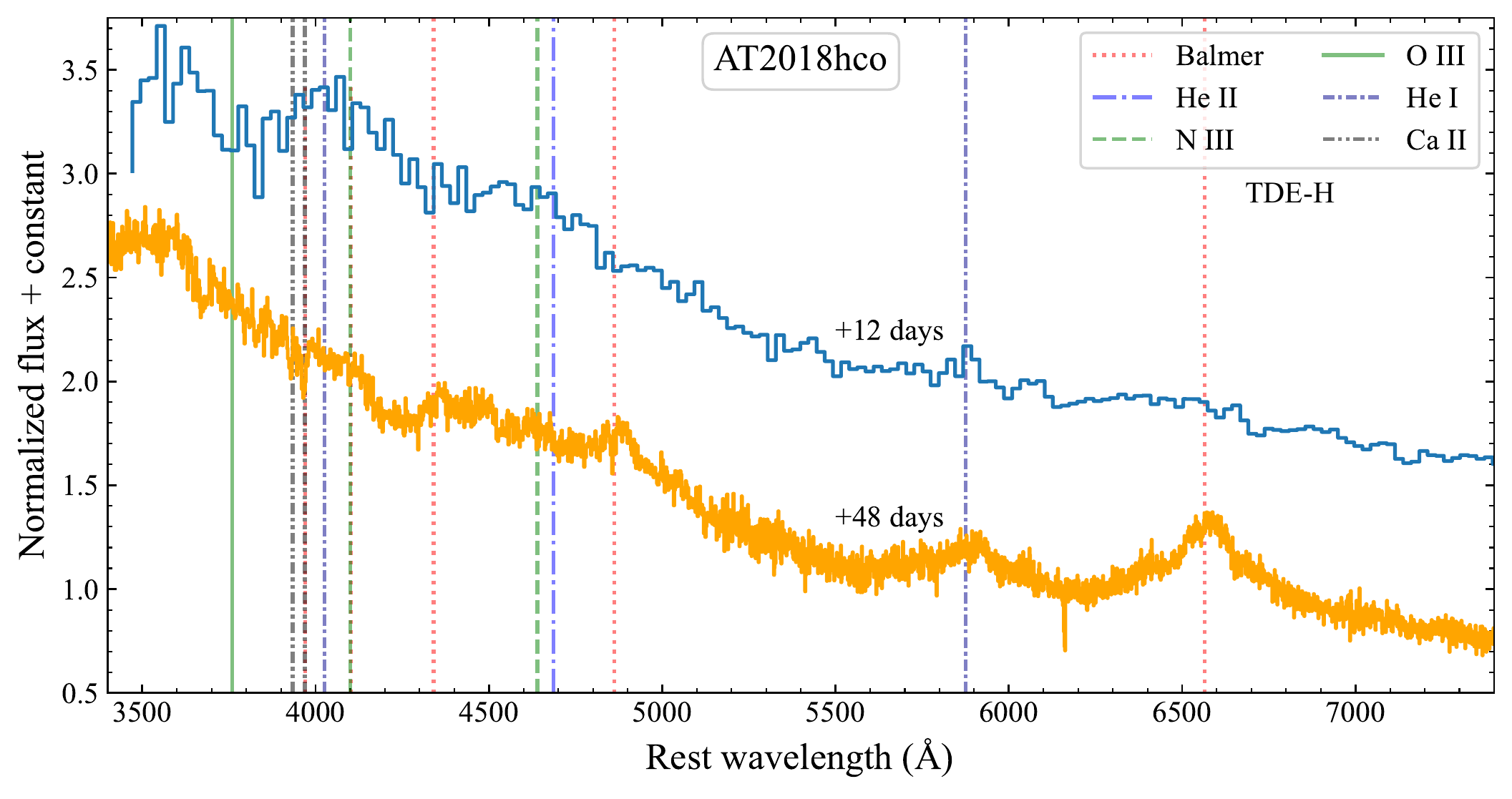}{0.45 \textwidth}{}
            \fig{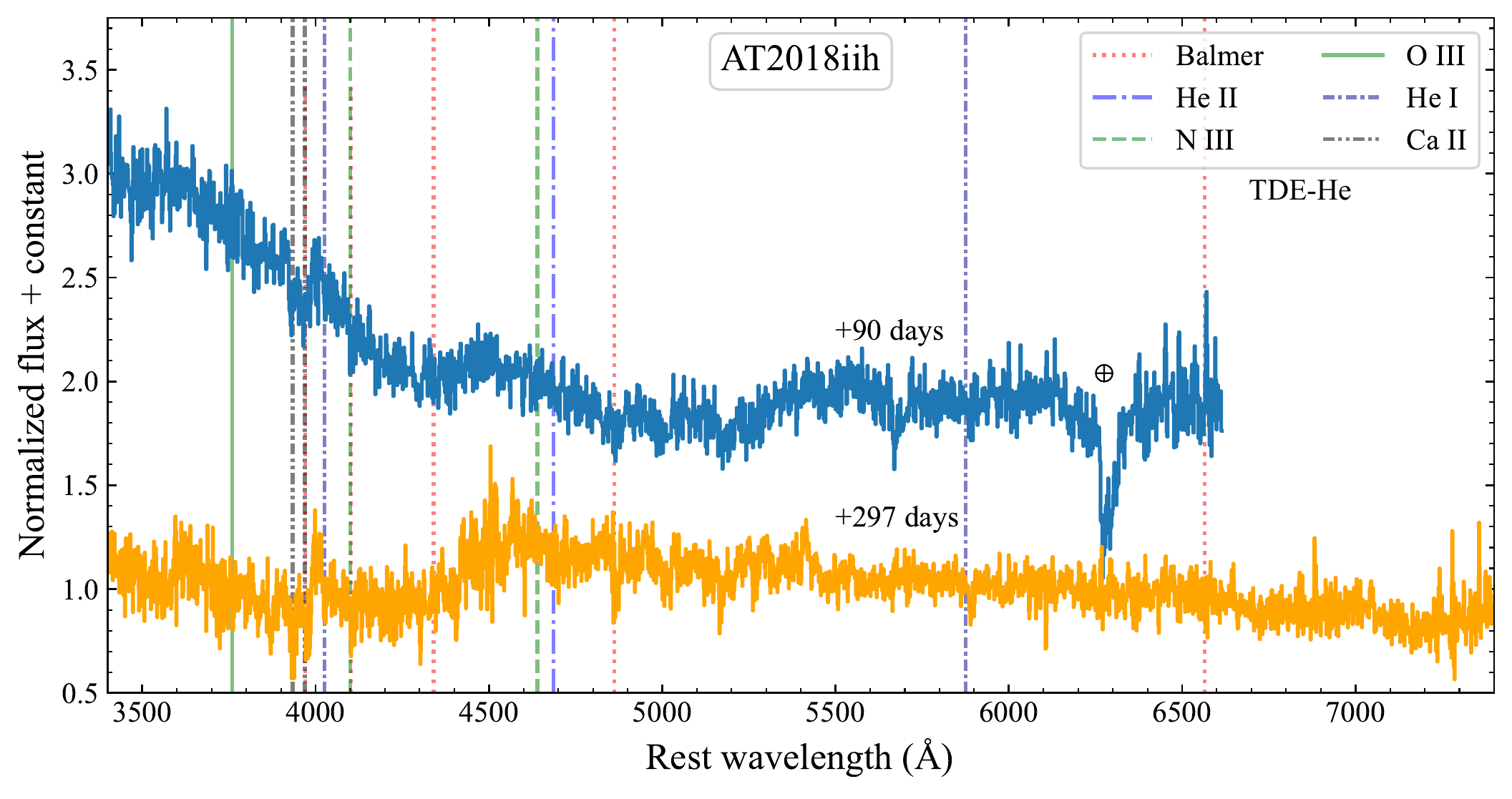}{0.45 \textwidth}{} 
            \\[-30pt]}
            
\gridline{  \fig{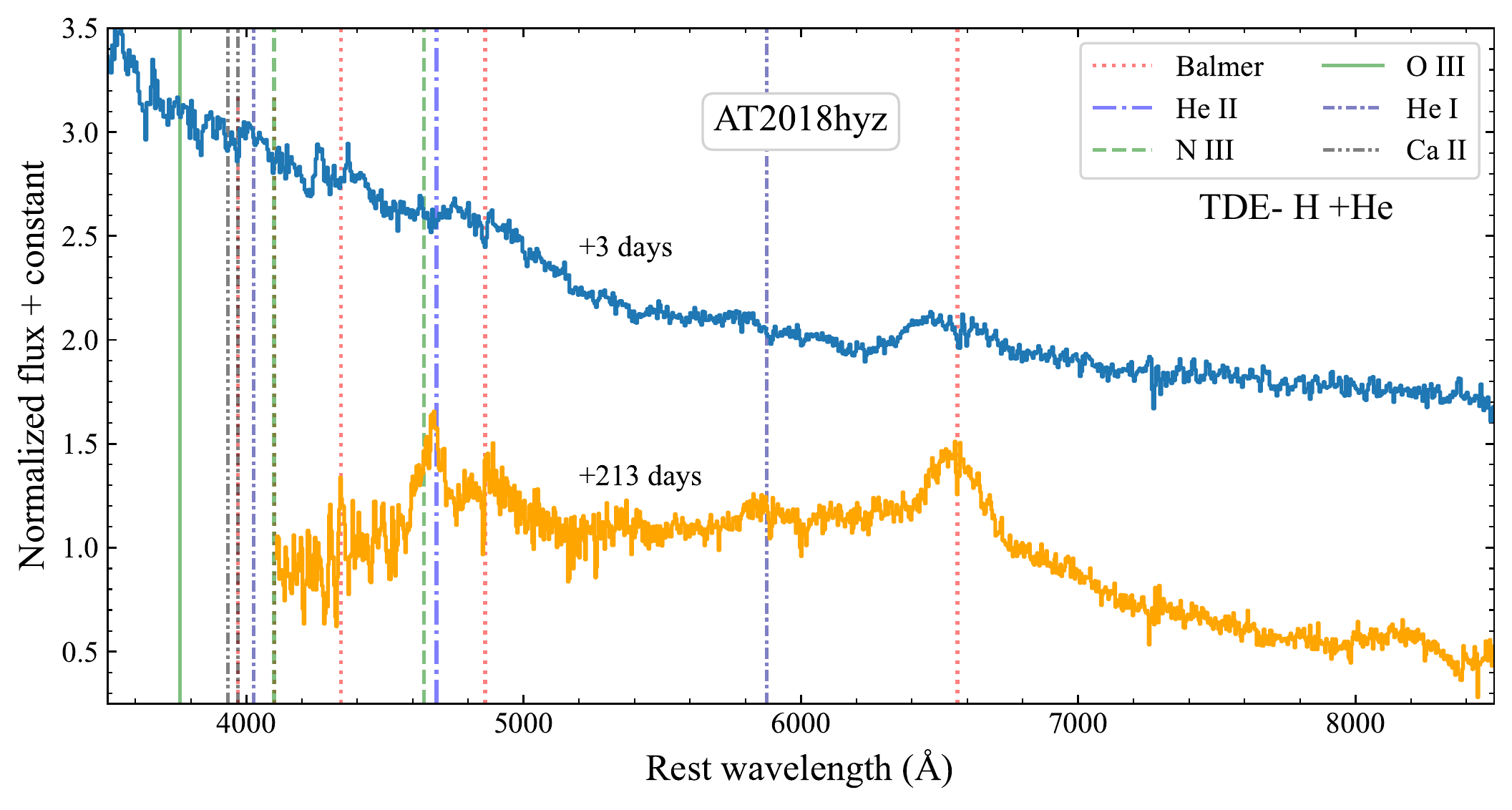}{0.45 \textwidth}{} 
            \fig{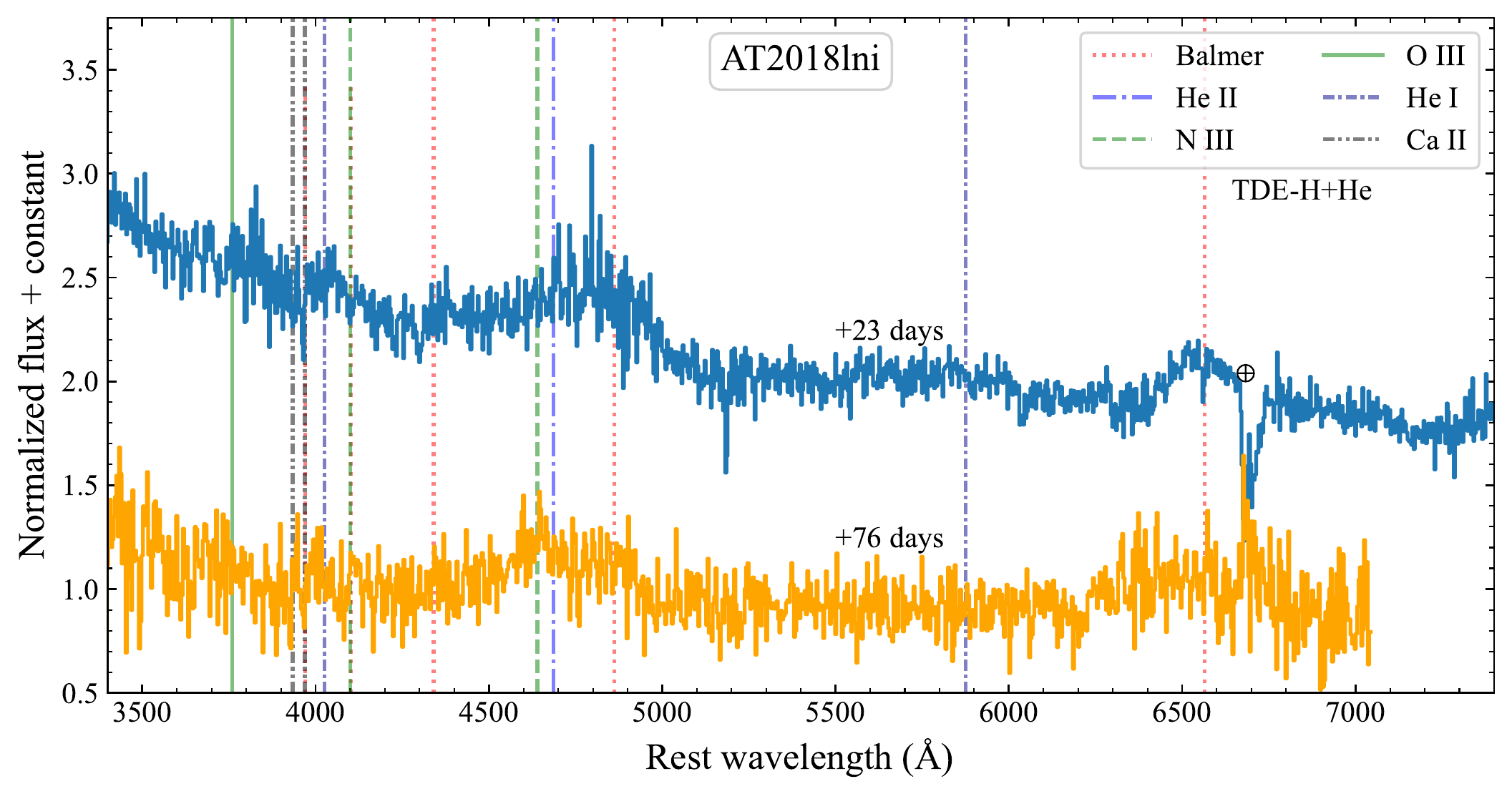}{0.45 \textwidth}{}
            \\[-30pt]}          

\gridline{  \fig{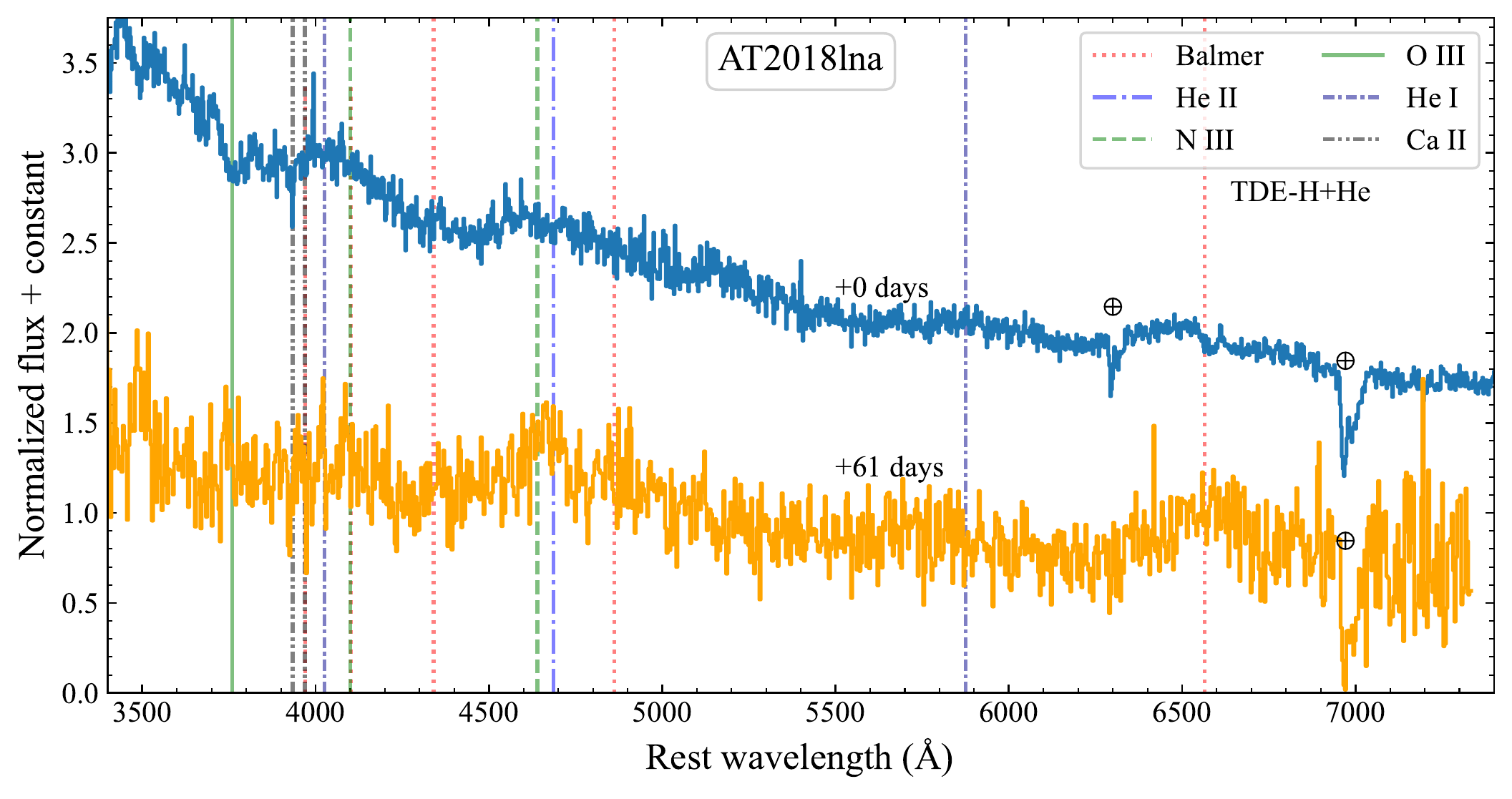}{0.45 \textwidth}{}
            \fig{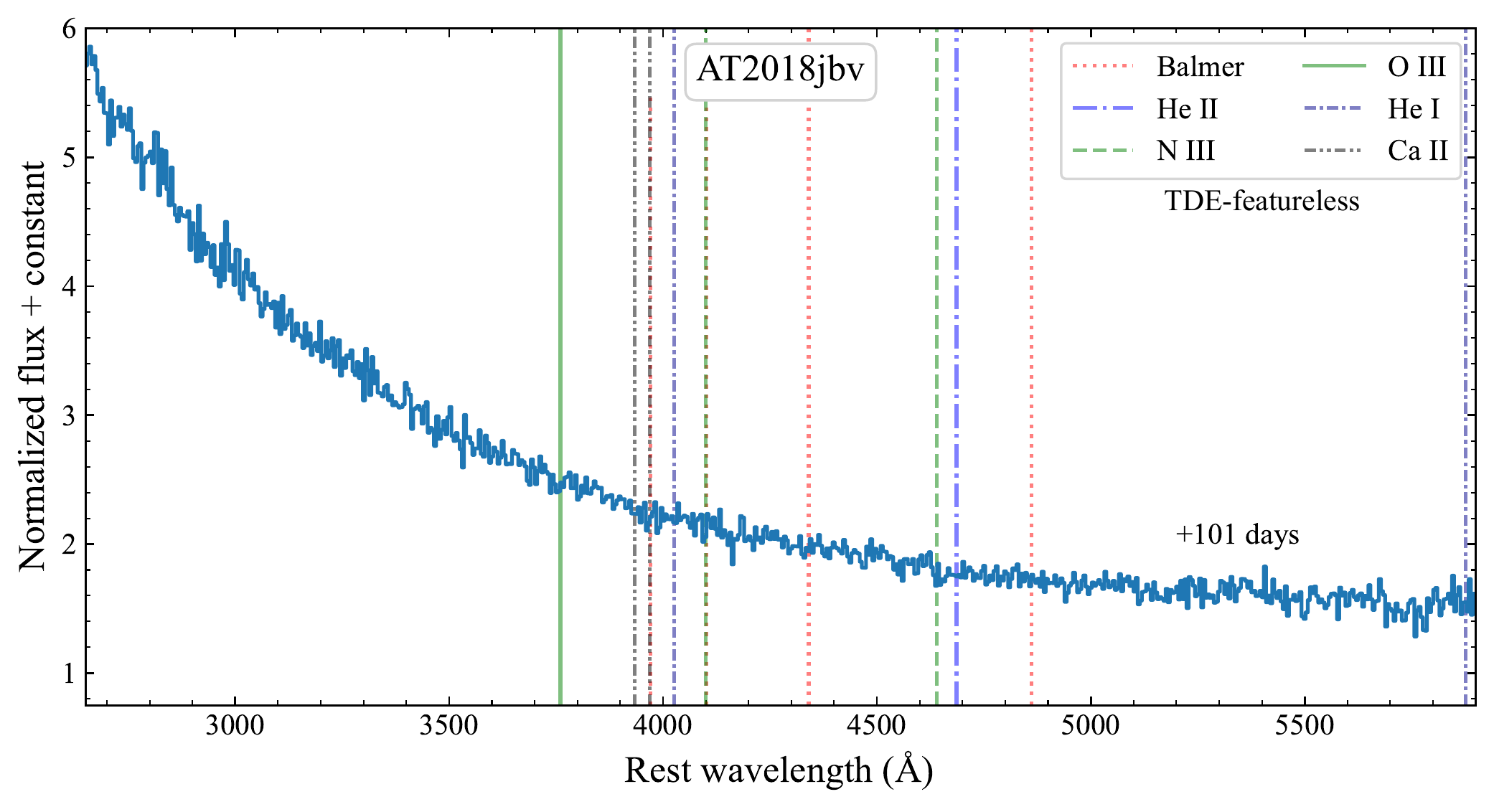}{0.45 \textwidth}{}
             \\[-30pt]}

\gridline{  \fig{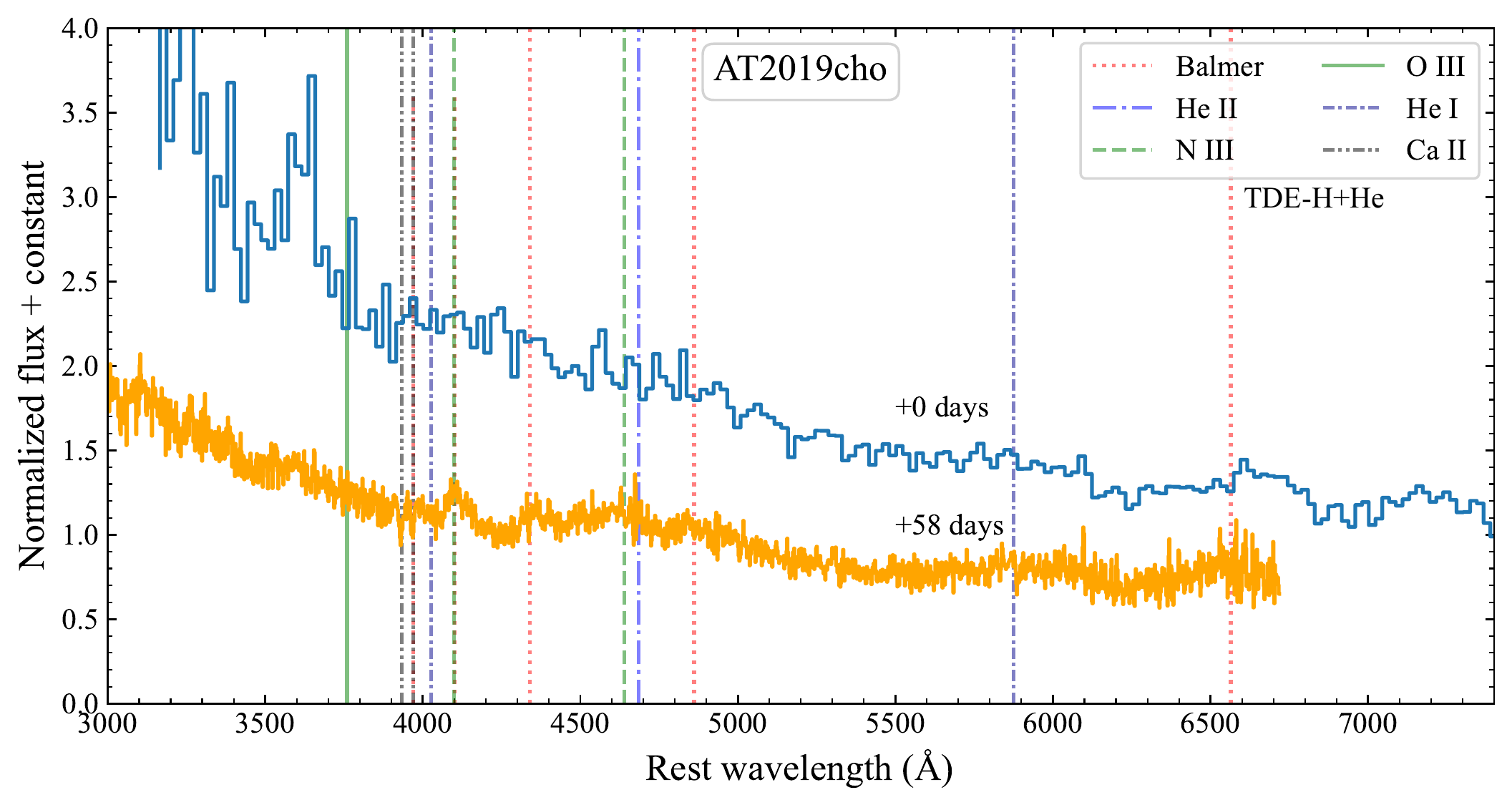}{0.45 \textwidth}{} 
            \fig{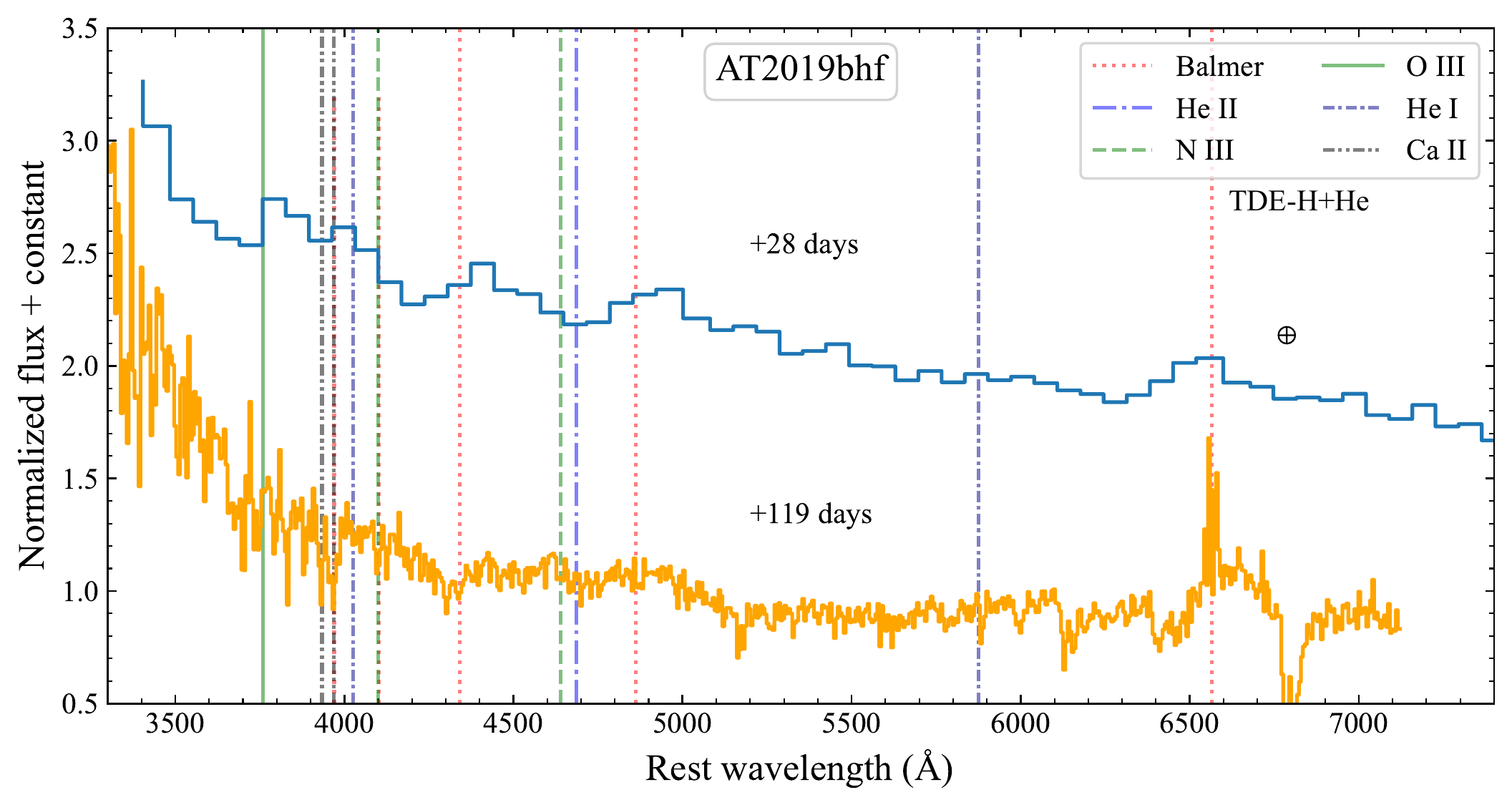}{0.45 \textwidth}{}
            \\[-30pt]}   

\caption{Optical spectra for the events in this sample. We provide an early and late time spectrum for each event when available and provide the approximate phase relative to peak that the spectrum was taken. We label common TDE emission lines and galaxy absorption lines. Spectra have not been host subtracted. Some spectra still contain telluric absorption lines, which have been labeled. Data for Figures 14--16 are available in the related files associated with Figure 14.}\label{fig:spec1}
\end{figure*}

\begin{figure*}
\gridline{	\fig{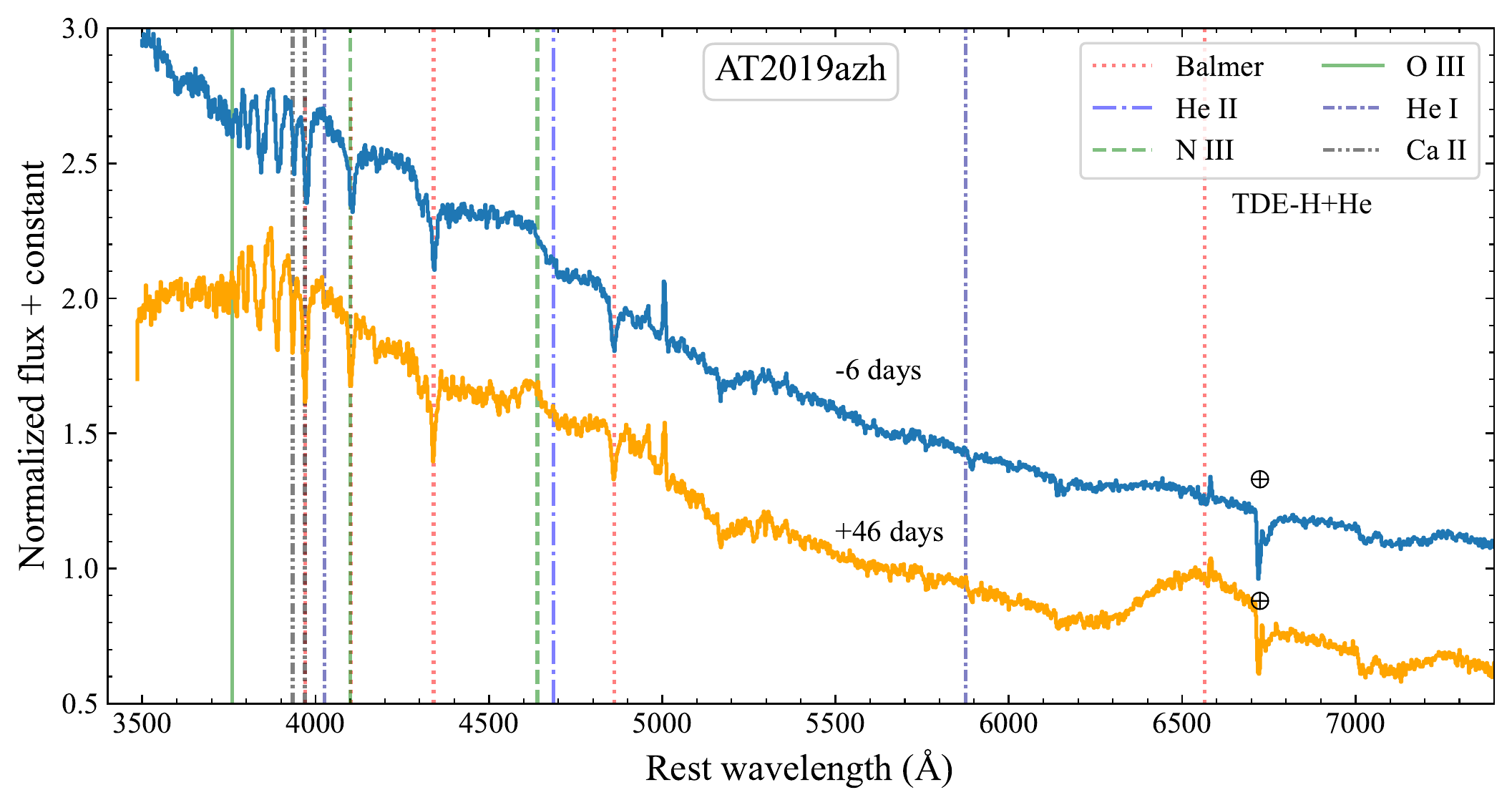}{0.45 \textwidth}{} 
			\fig{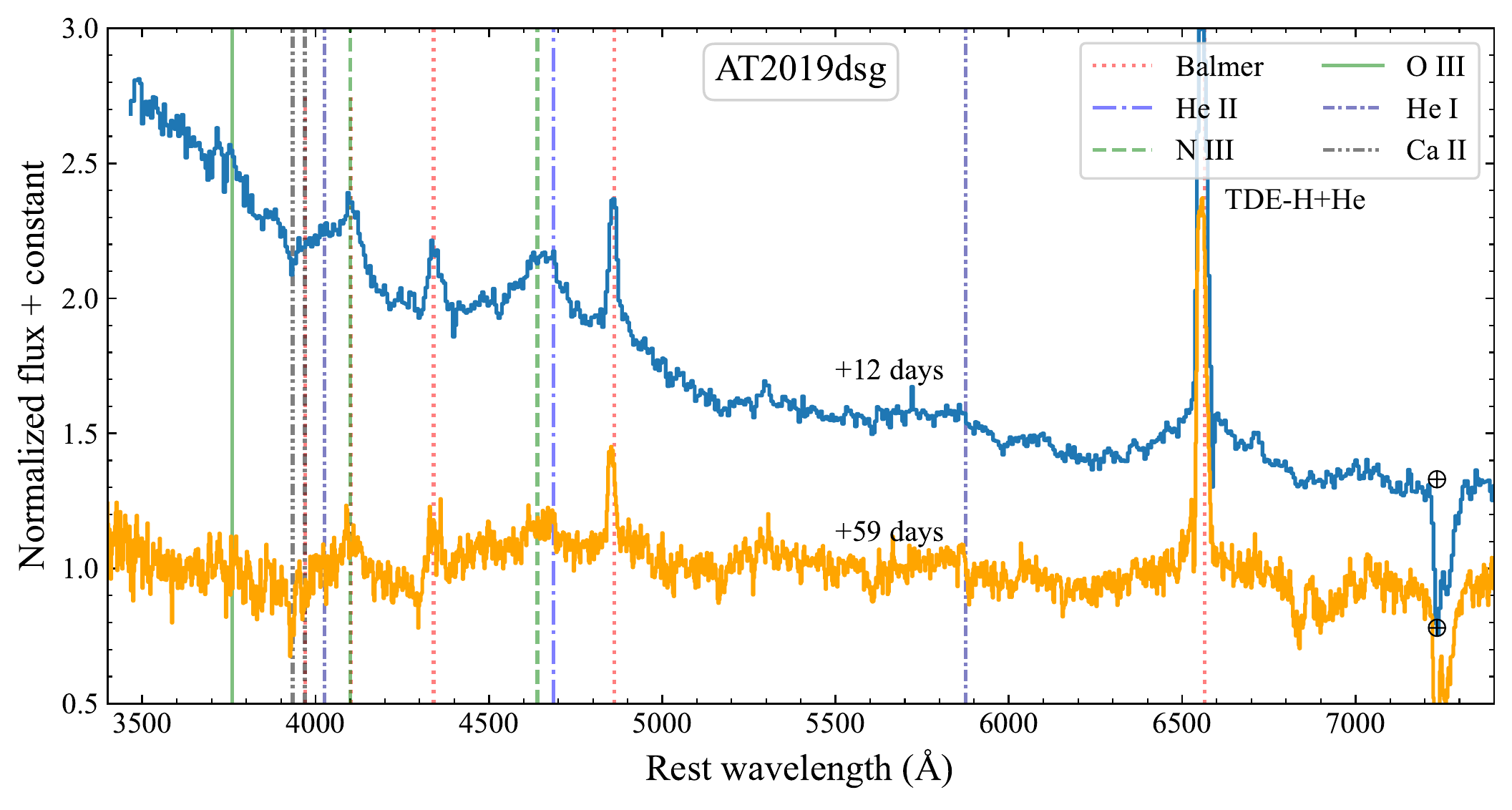}{0.45 \textwidth}{} 
			\\[-30pt]}
			
\gridline{  \fig{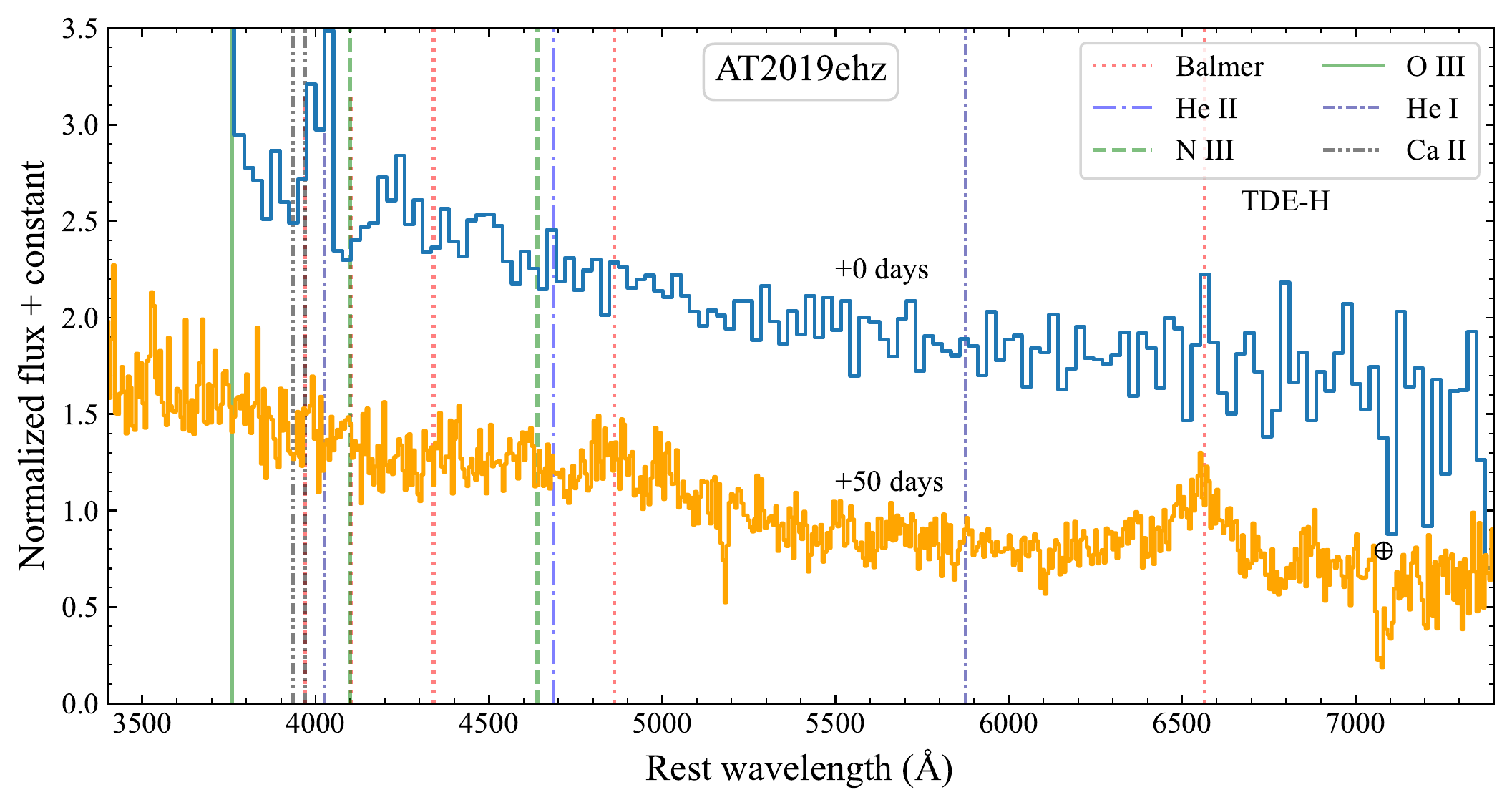}{0.45 \textwidth}{}
            \fig{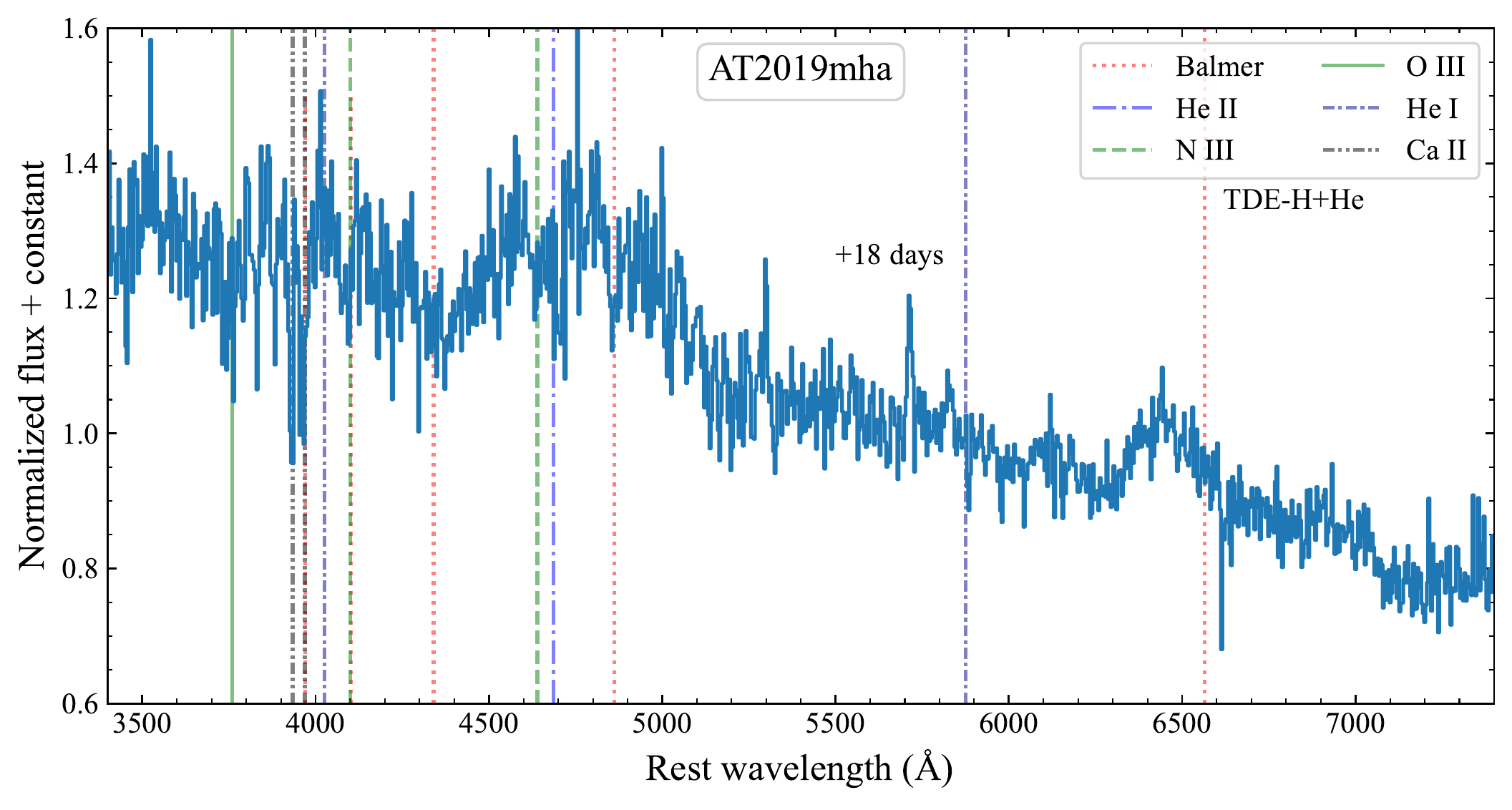}{0.45 \textwidth}{} 
            \\[-30pt]}
            
\gridline{  \fig{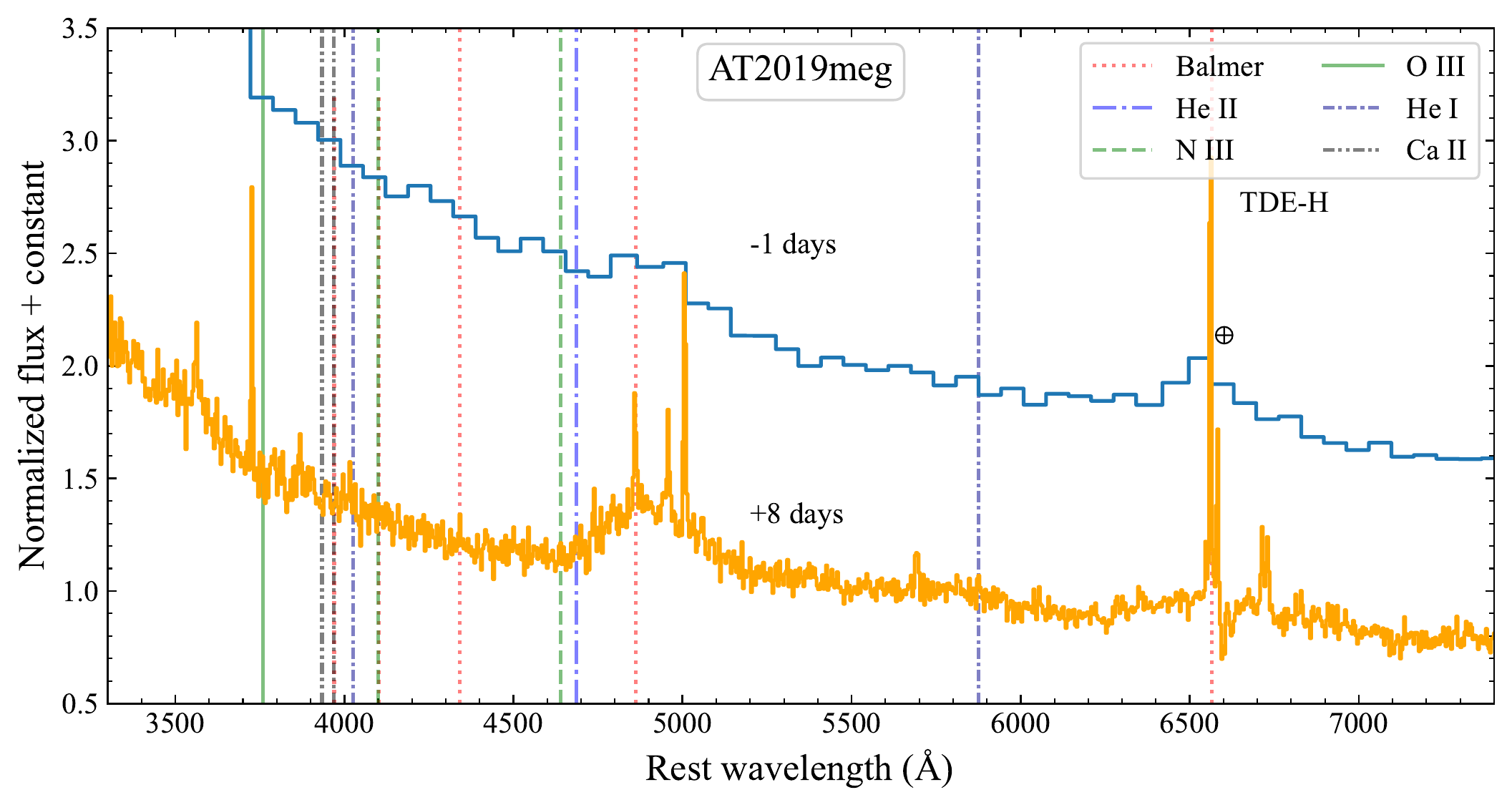}{0.45 \textwidth}{} 
            \fig{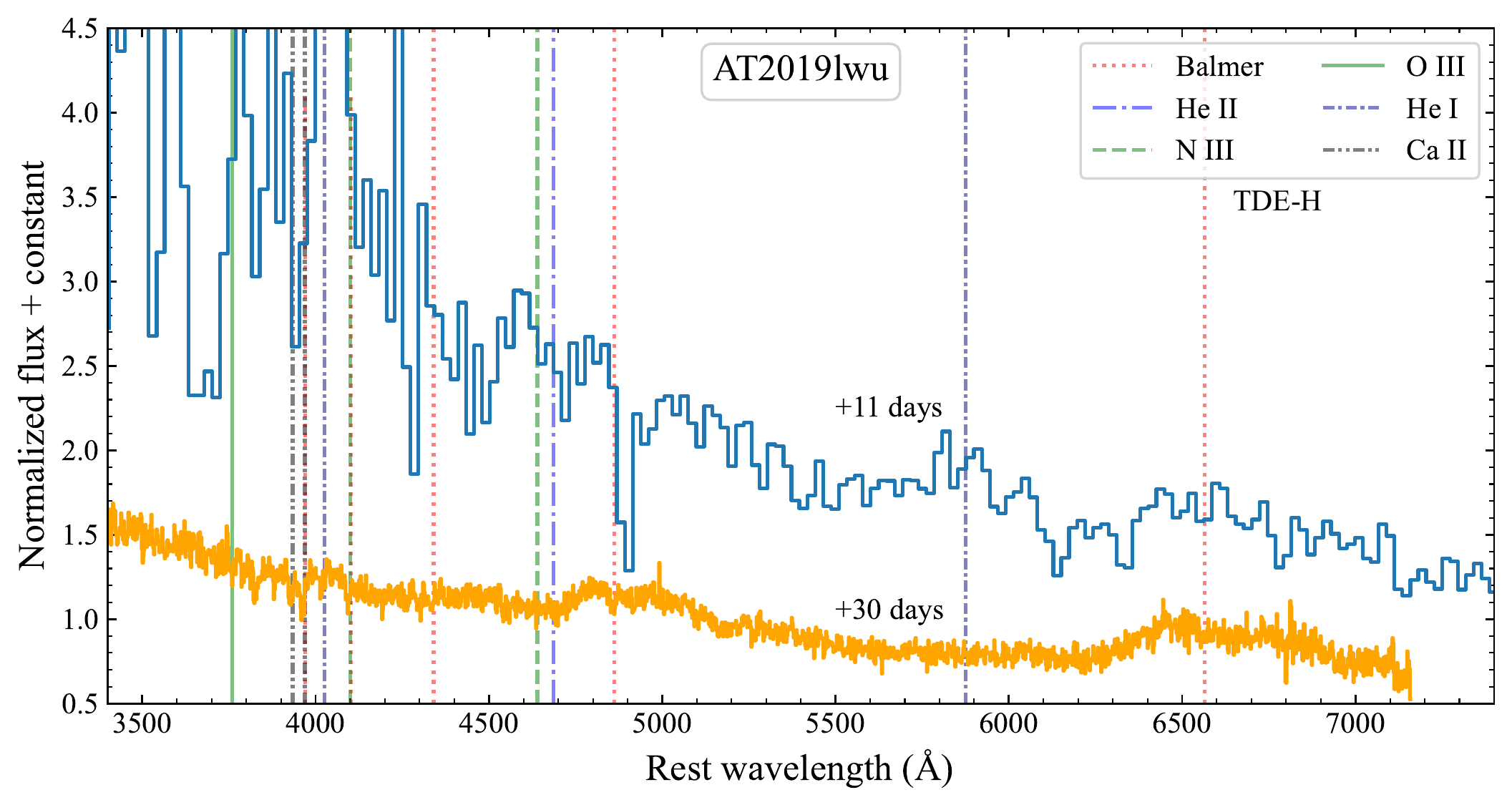}{0.45 \textwidth}{}
            \\[-30pt]}          

\gridline{  \fig{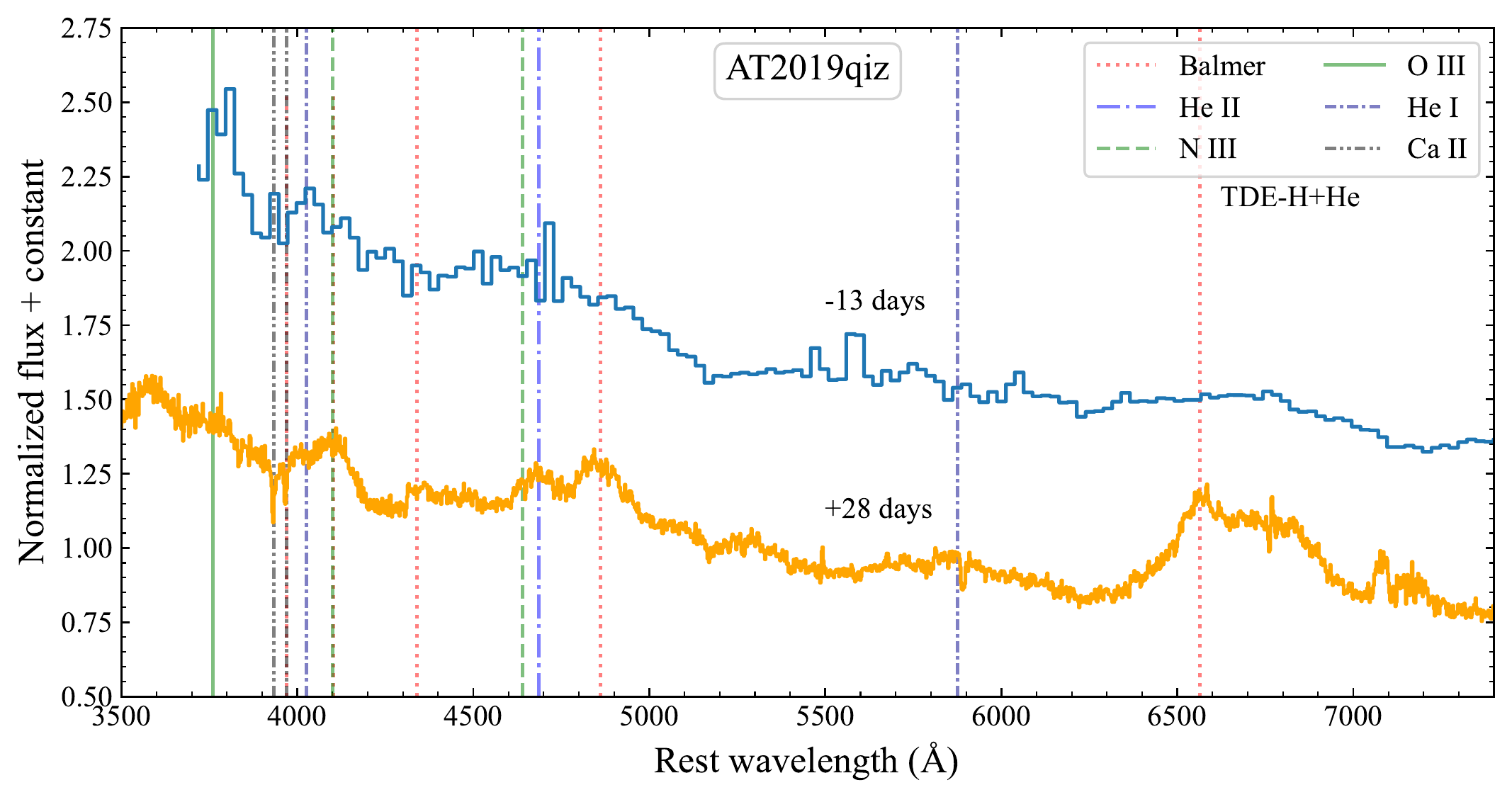}{0.45 \textwidth}{}
            \fig{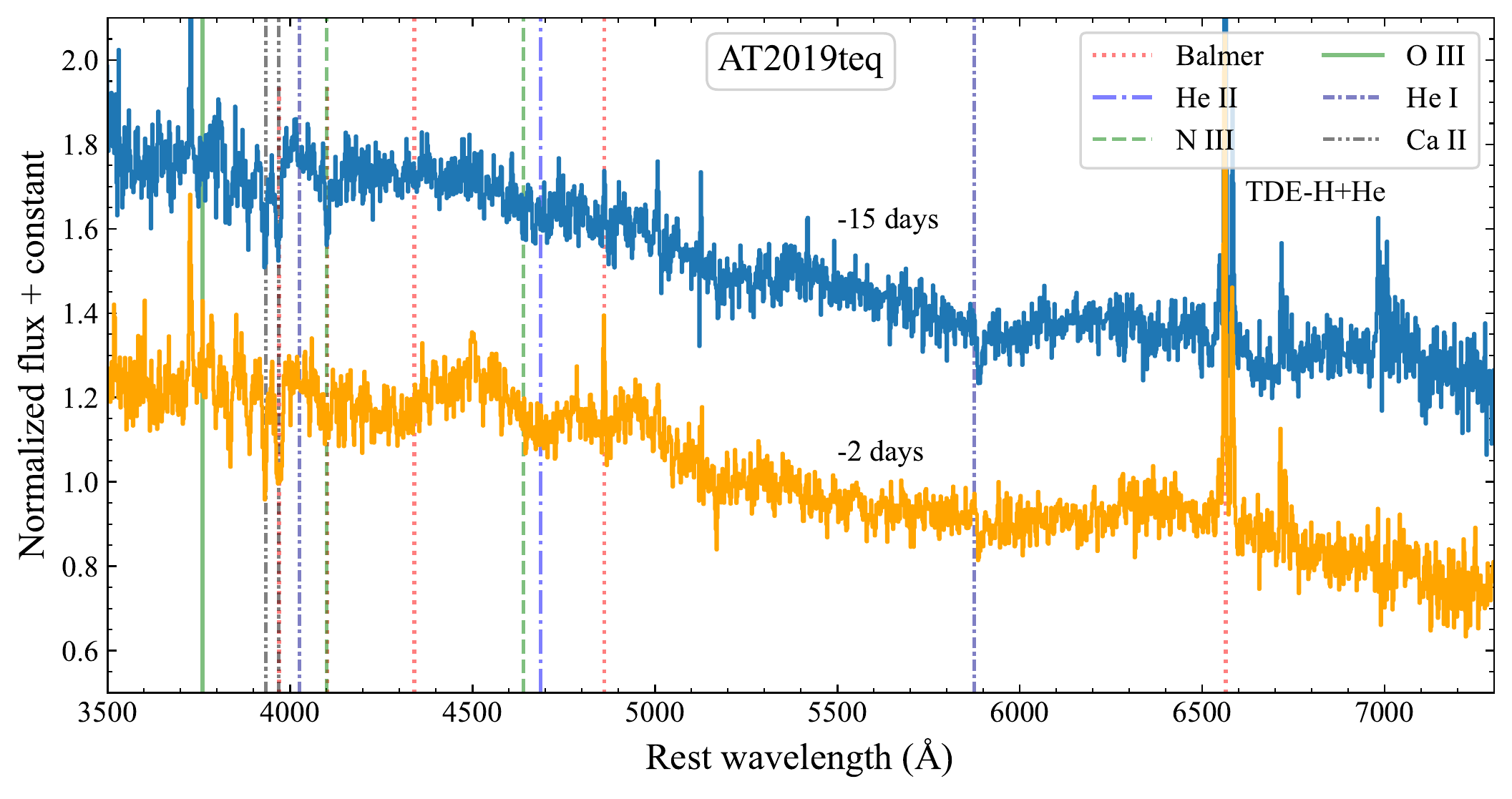}{0.45 \textwidth}{}
             \\[-30pt]}

\gridline{  \fig{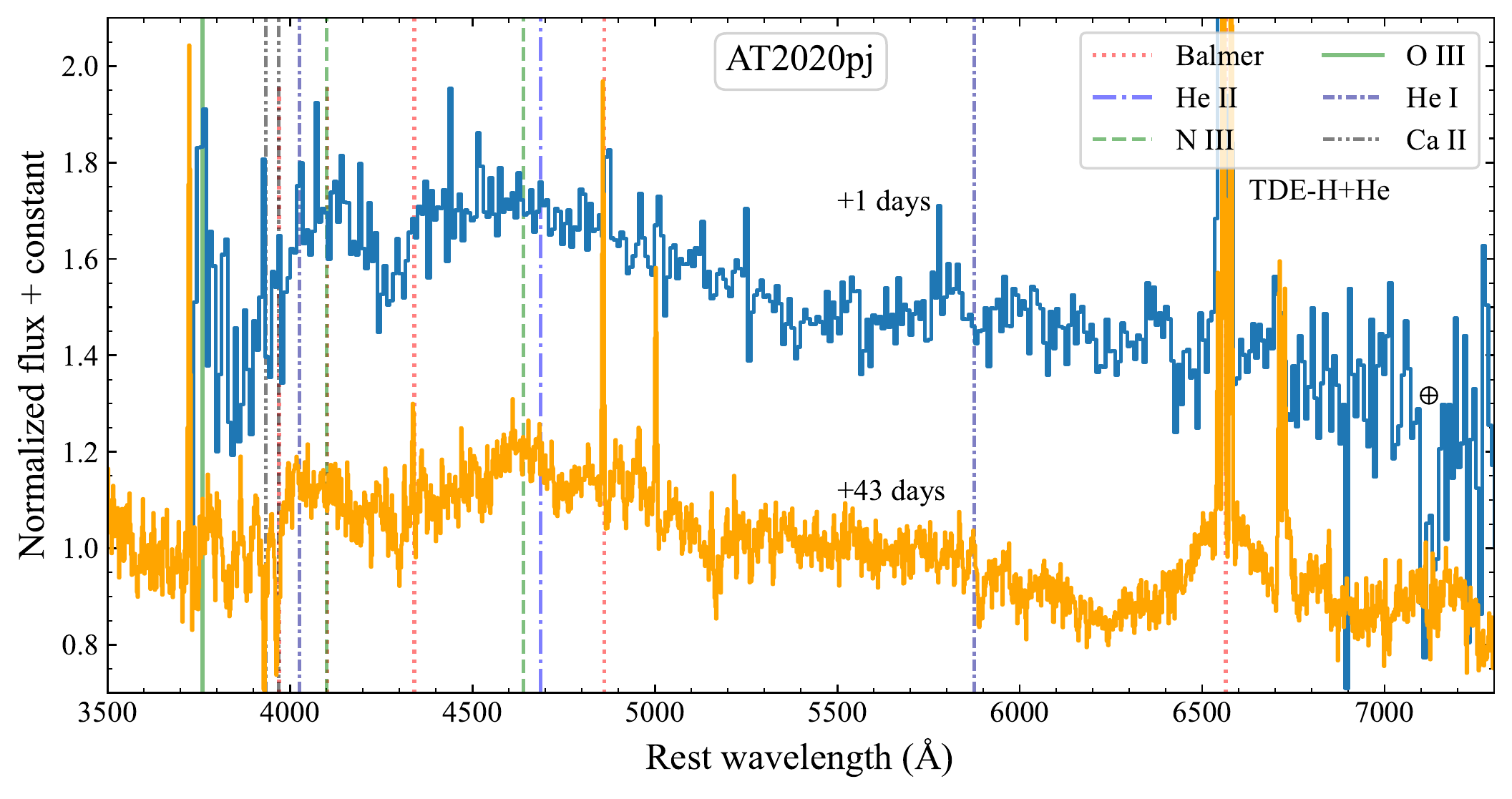}{0.45 \textwidth}{} 
            \fig{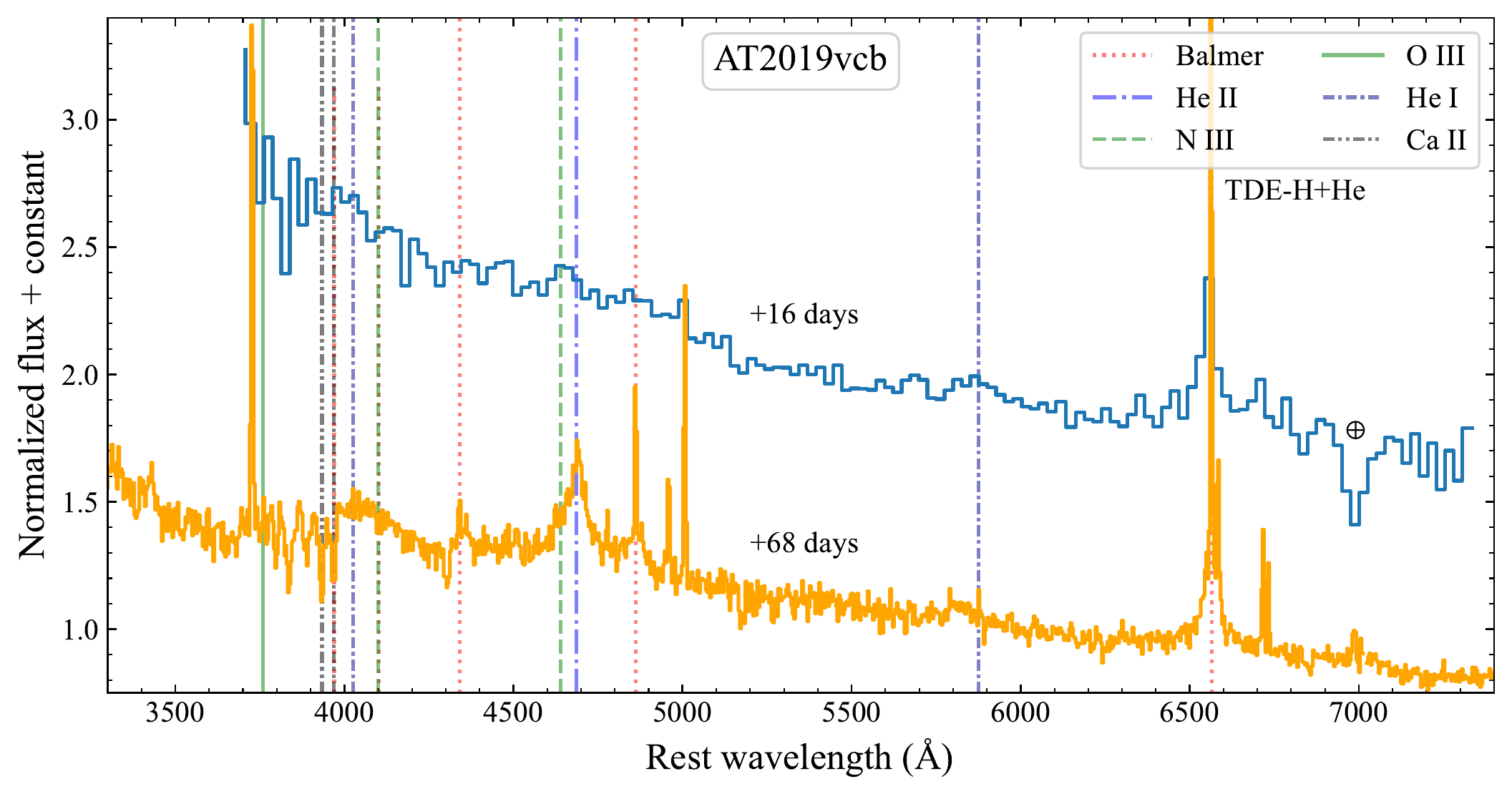}{0.45 \textwidth}{}
            \\[-30pt]}   

\caption{Optical spectra for the events in this sample. We provide an early and late time spectrum for each event when available and provide the approximate phase relative to peak that the spectrum was taken. We label common TDE emission lines and galaxy absorption lines. Spectra have not been host subtracted. Some spectra still contain telluric absorption lines, which have been labeled.}\label{fig:spec2}
\end{figure*}

\begin{figure*}
\gridline{	\fig{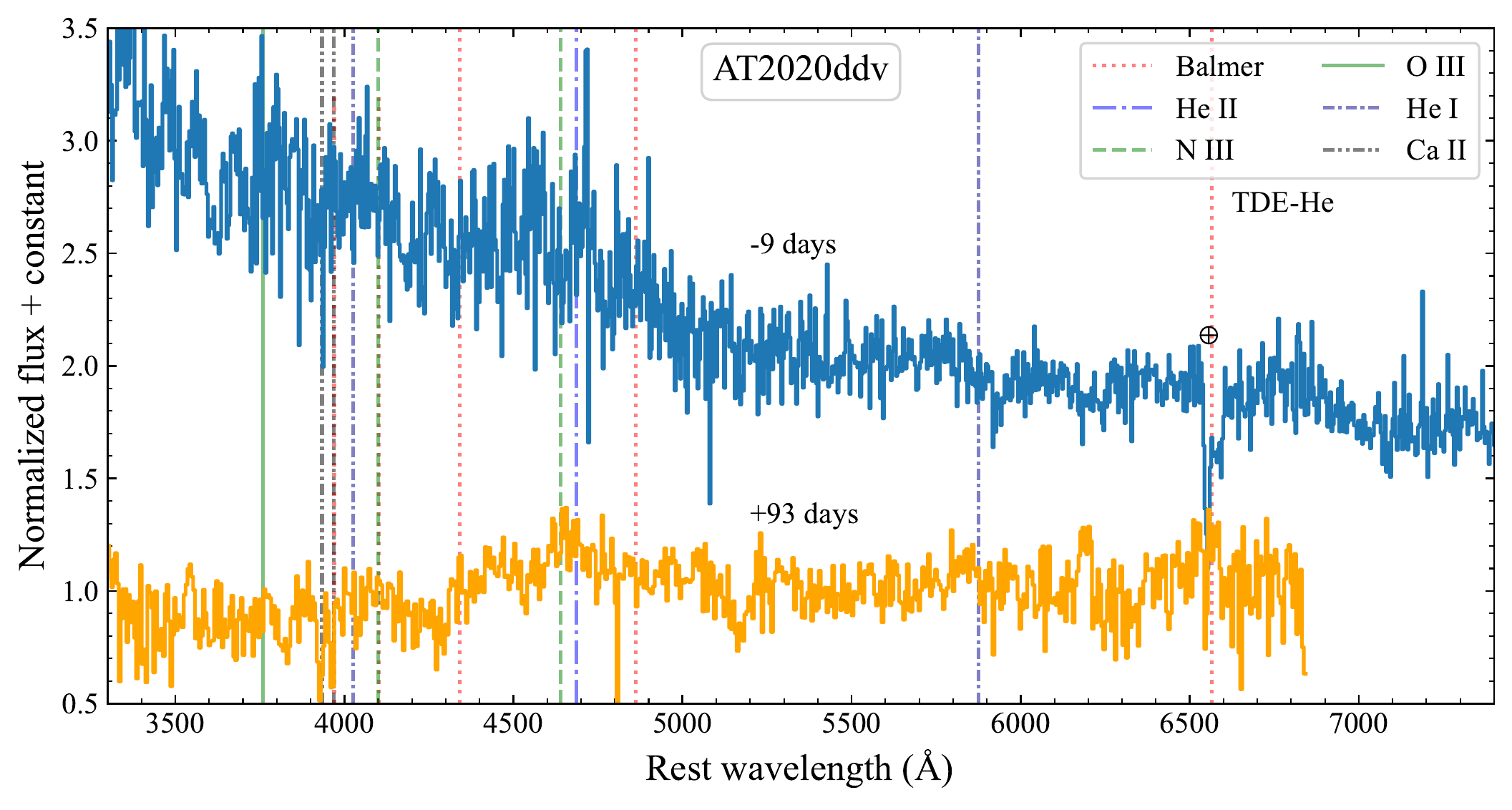}{0.45 \textwidth}{} 
			\fig{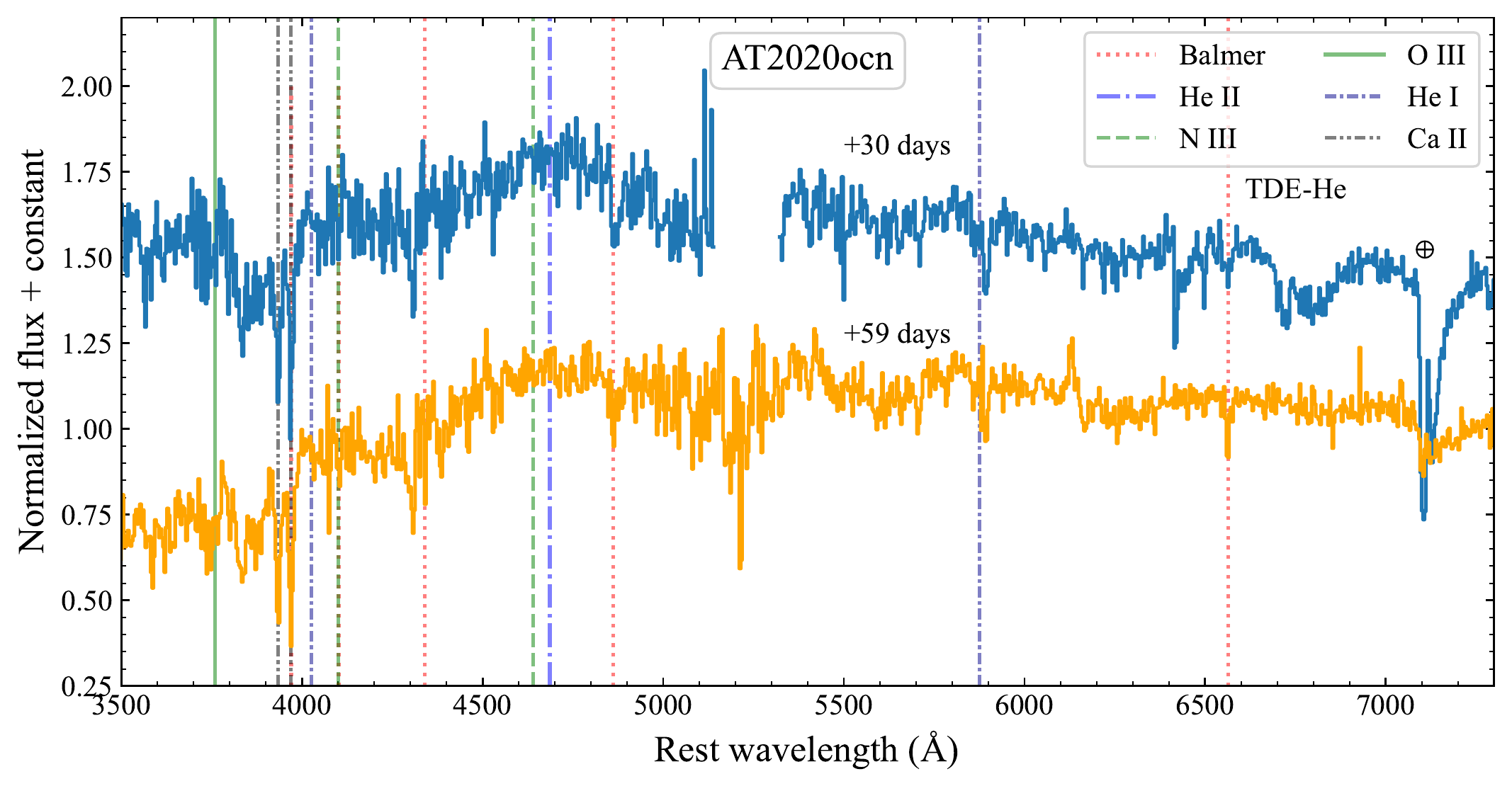}{0.45 \textwidth}{} 
			\\[-30pt]}
			
\gridline{  \fig{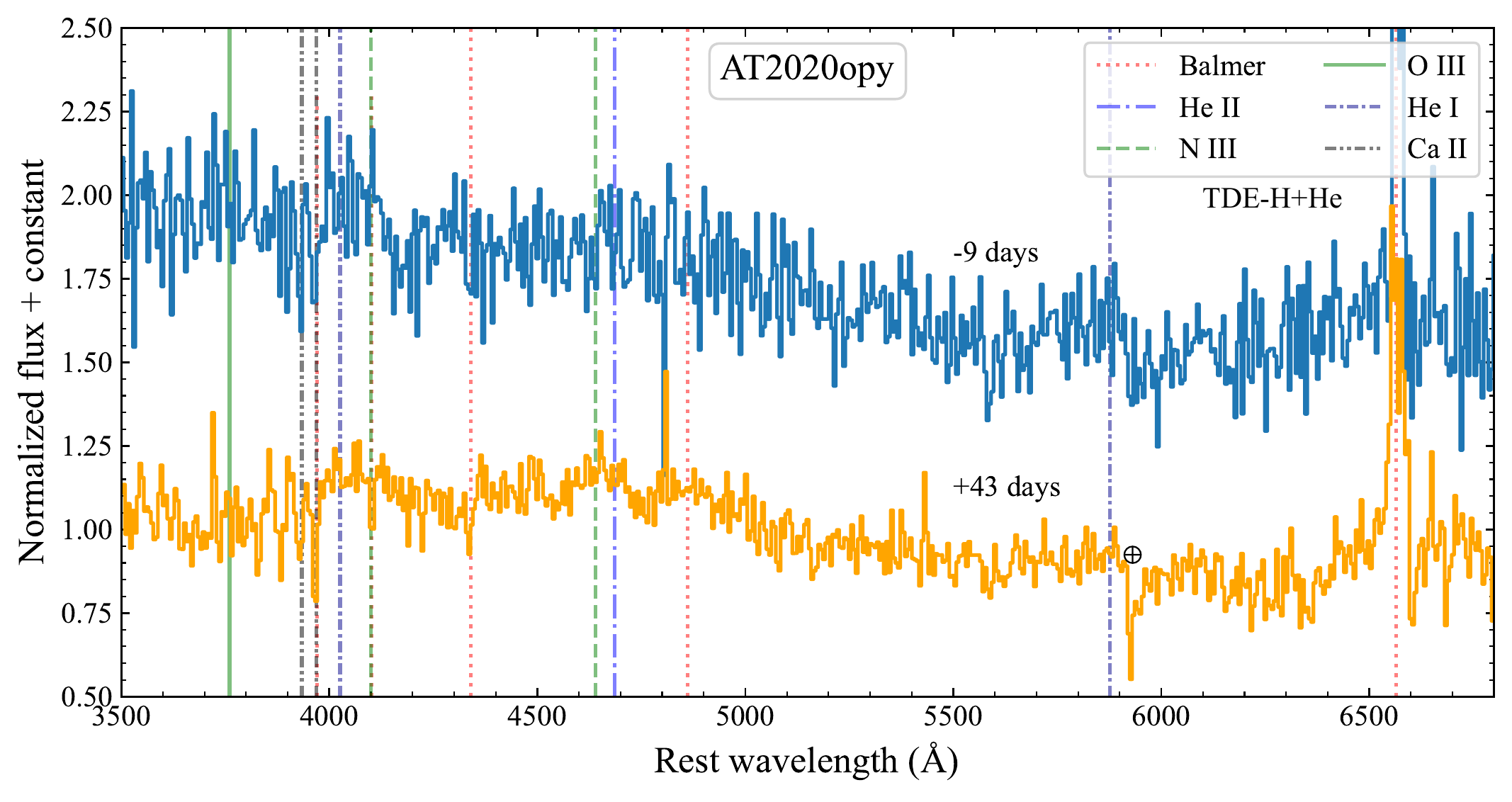}{0.45 \textwidth}{}
            \fig{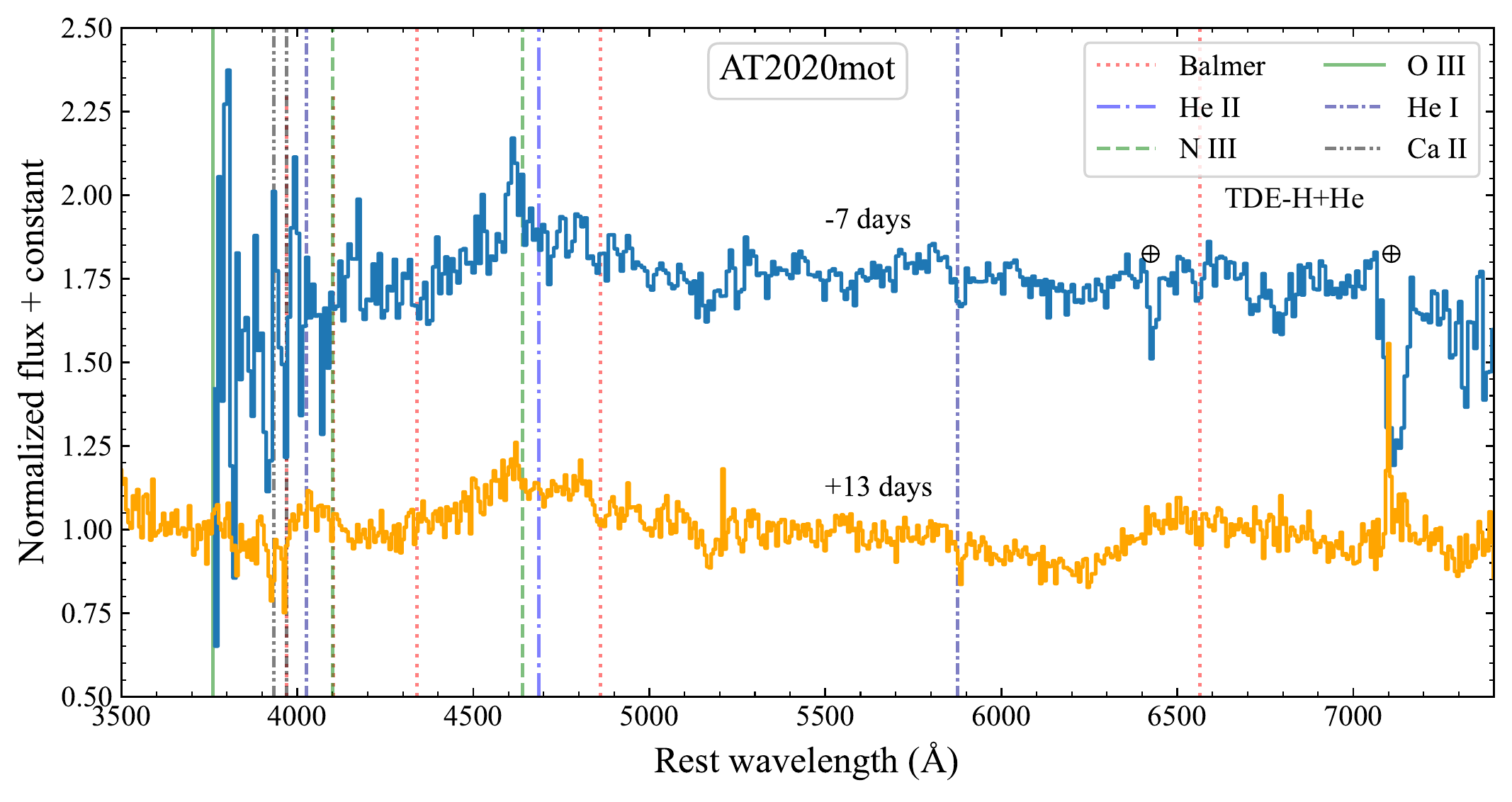}{0.45 \textwidth}{} 
            \\[-30pt]}
            
\gridline{  \fig{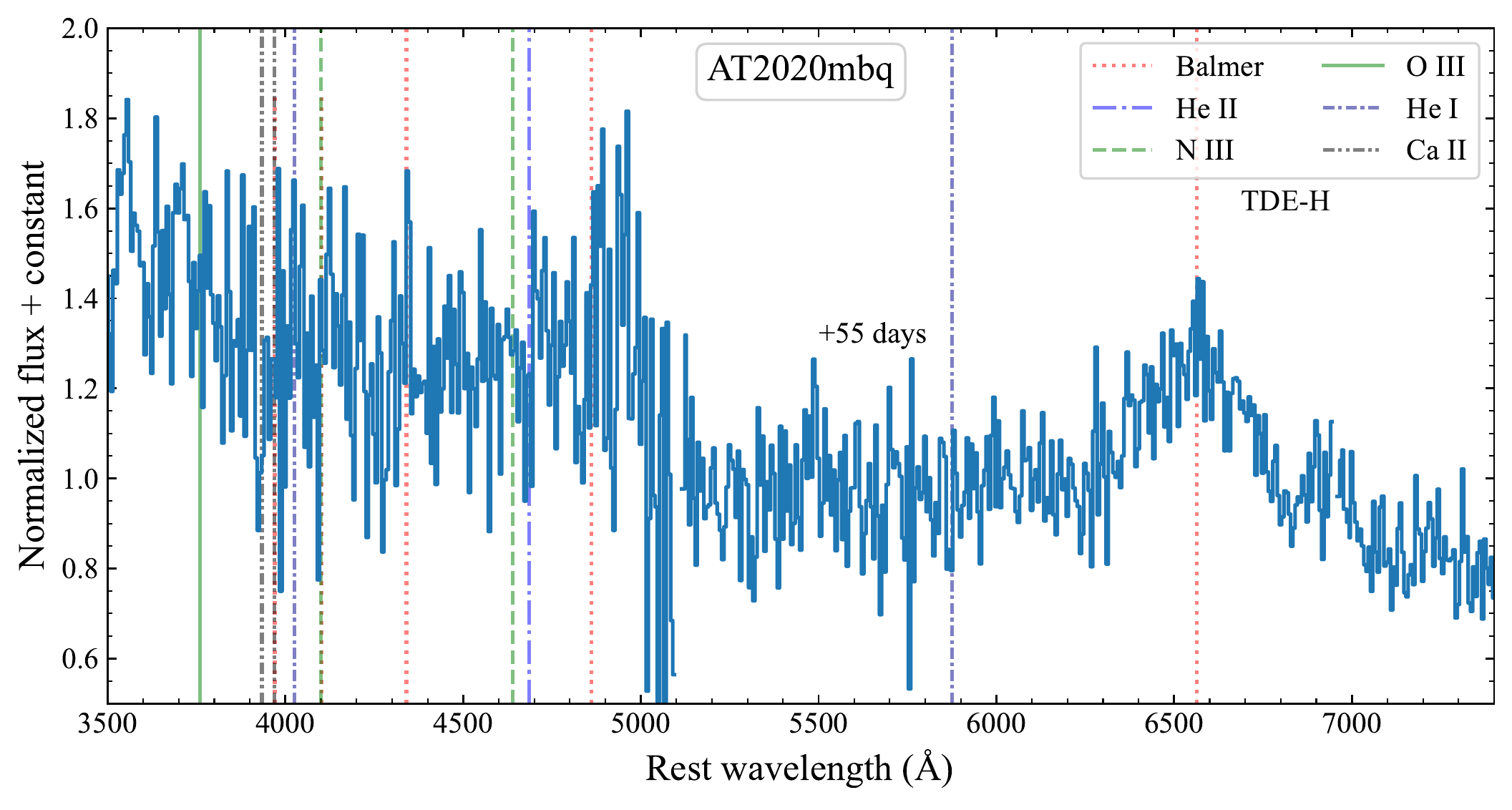}{0.45 \textwidth}{} 
            \fig{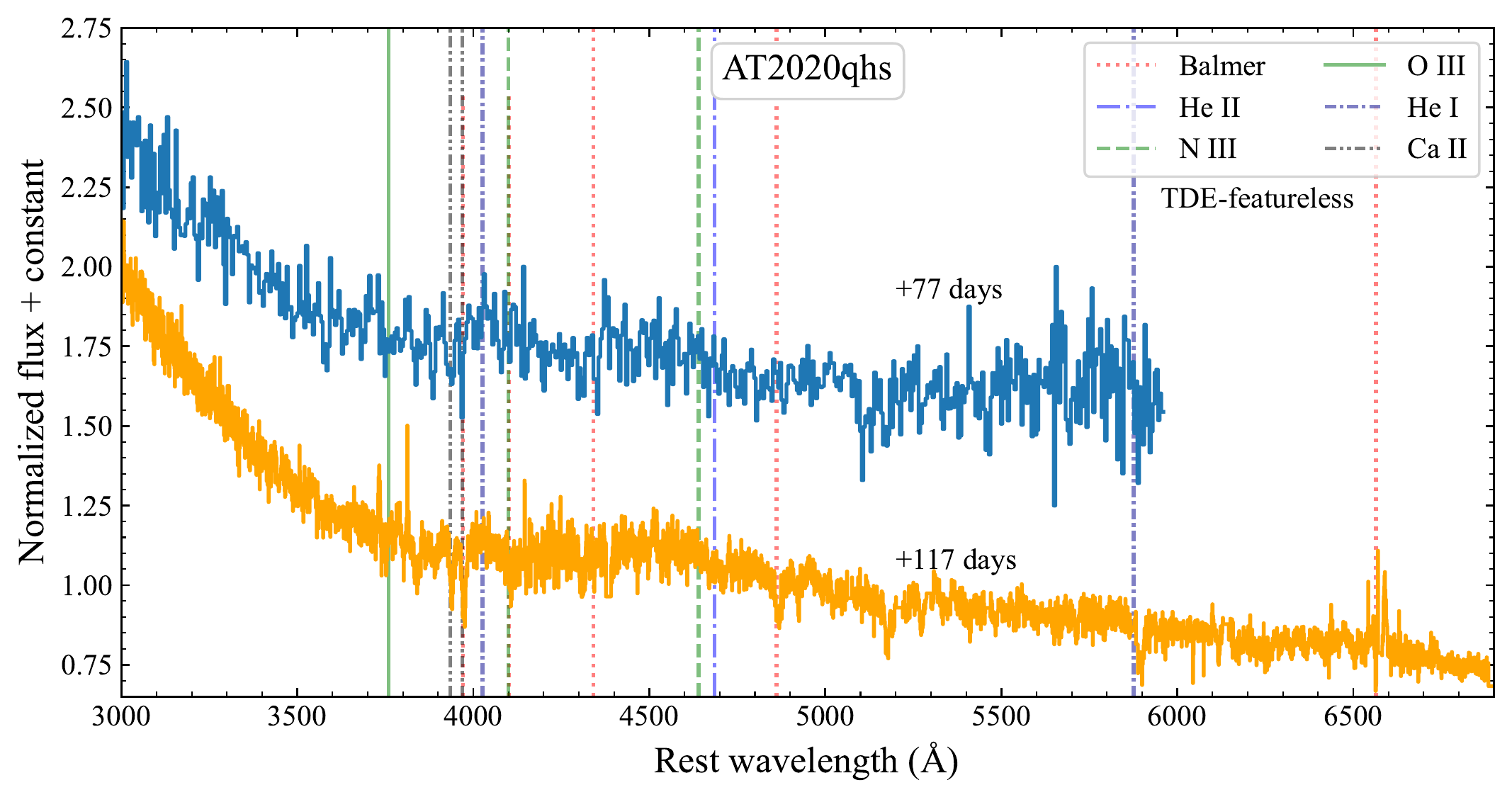}{0.45 \textwidth}{}
            \\[-30pt]}          

\gridline{  \fig{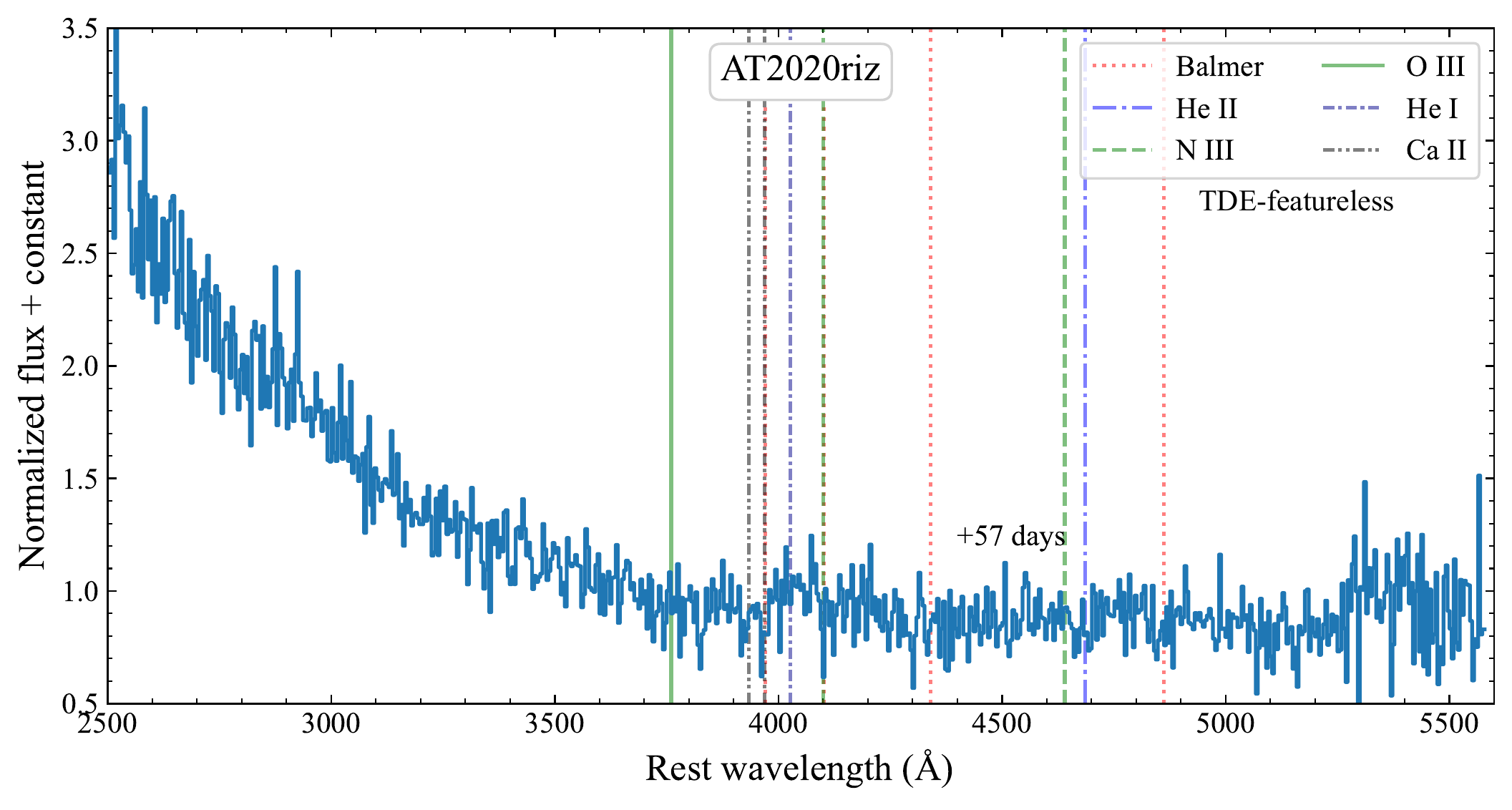}{0.45 \textwidth}{}
            \fig{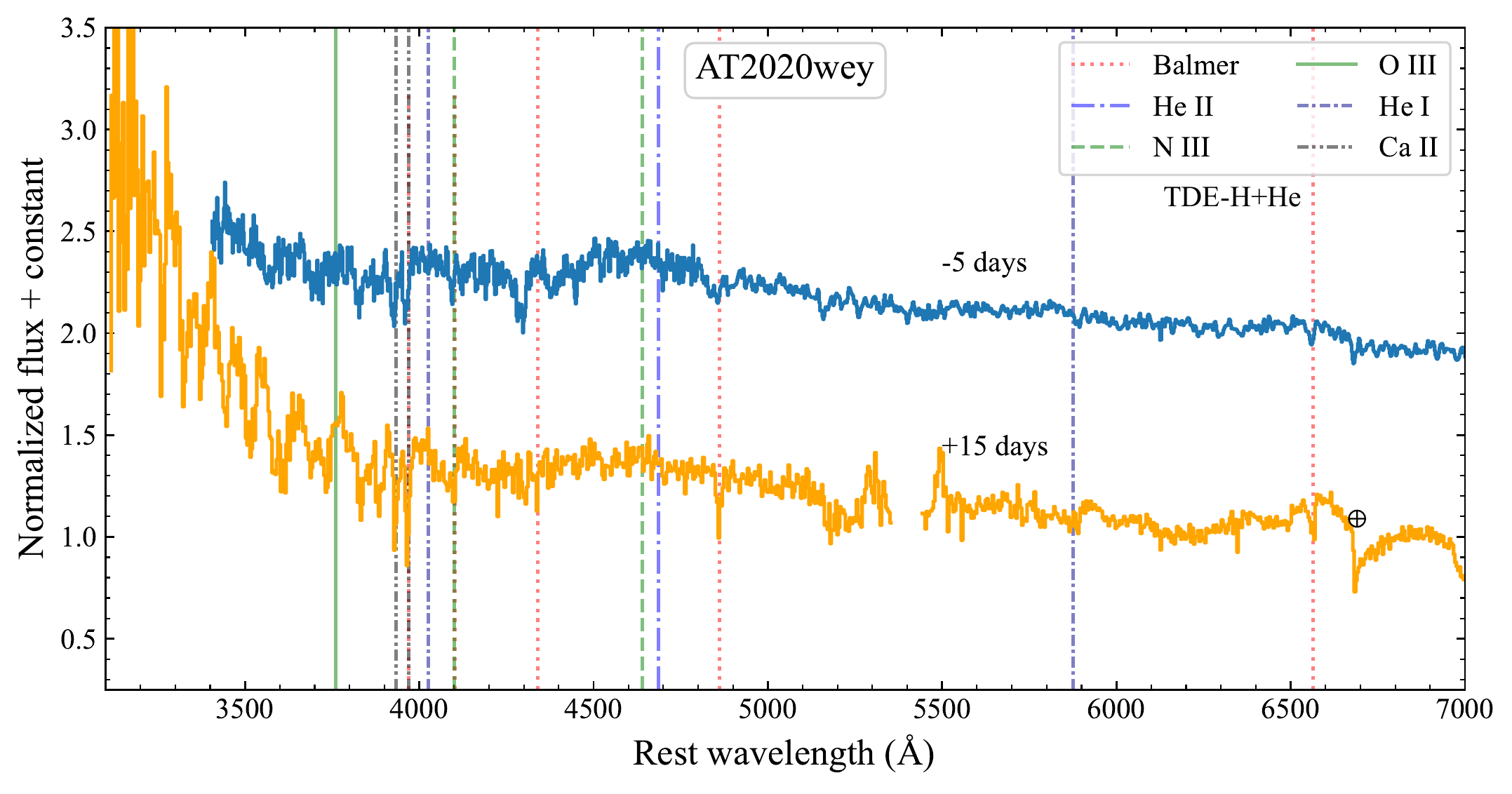}{0.45 \textwidth}{}
             \\[-30pt]}

\gridline{  \fig{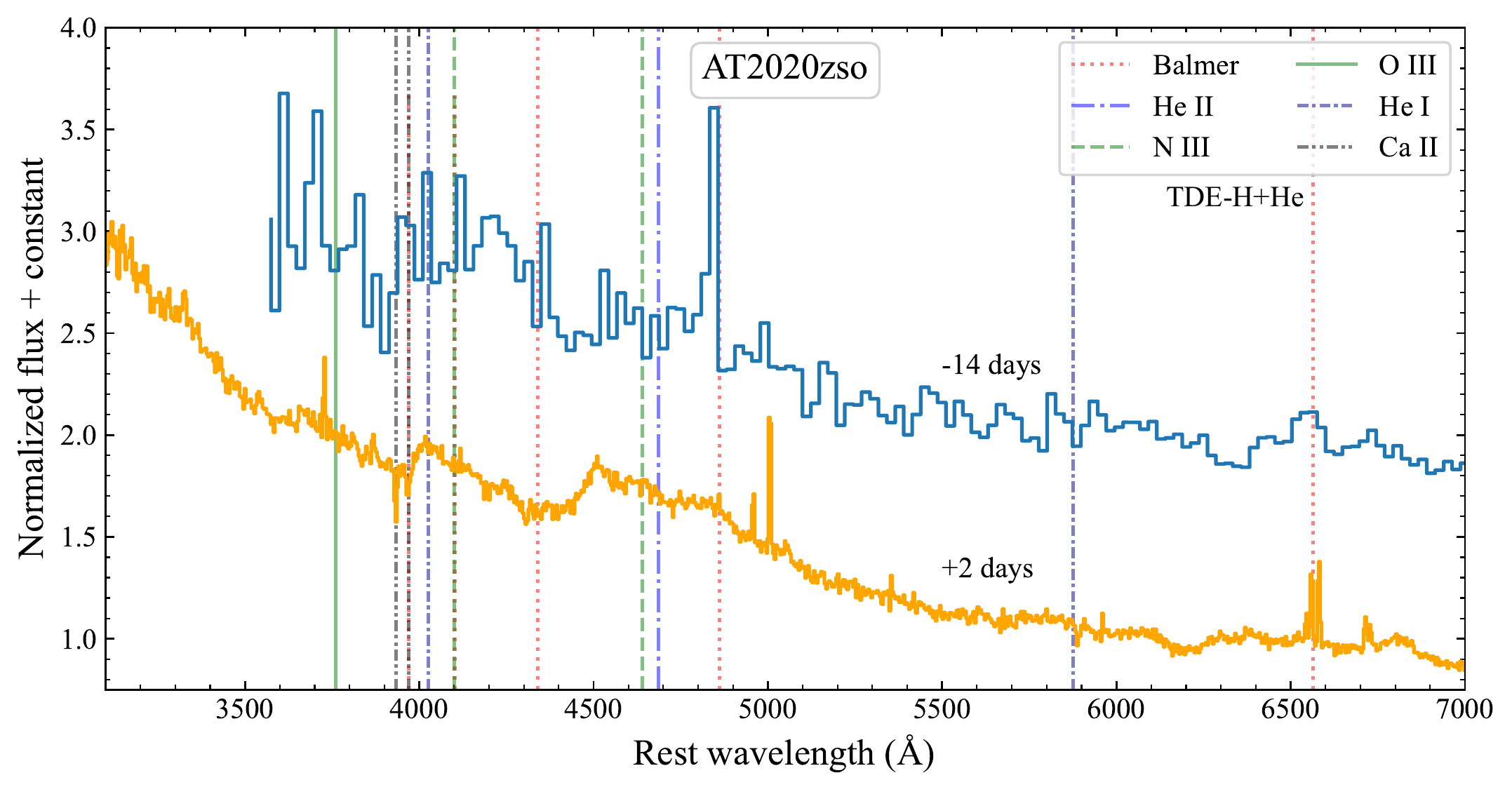}{0.45 \textwidth}{} 
            \fig{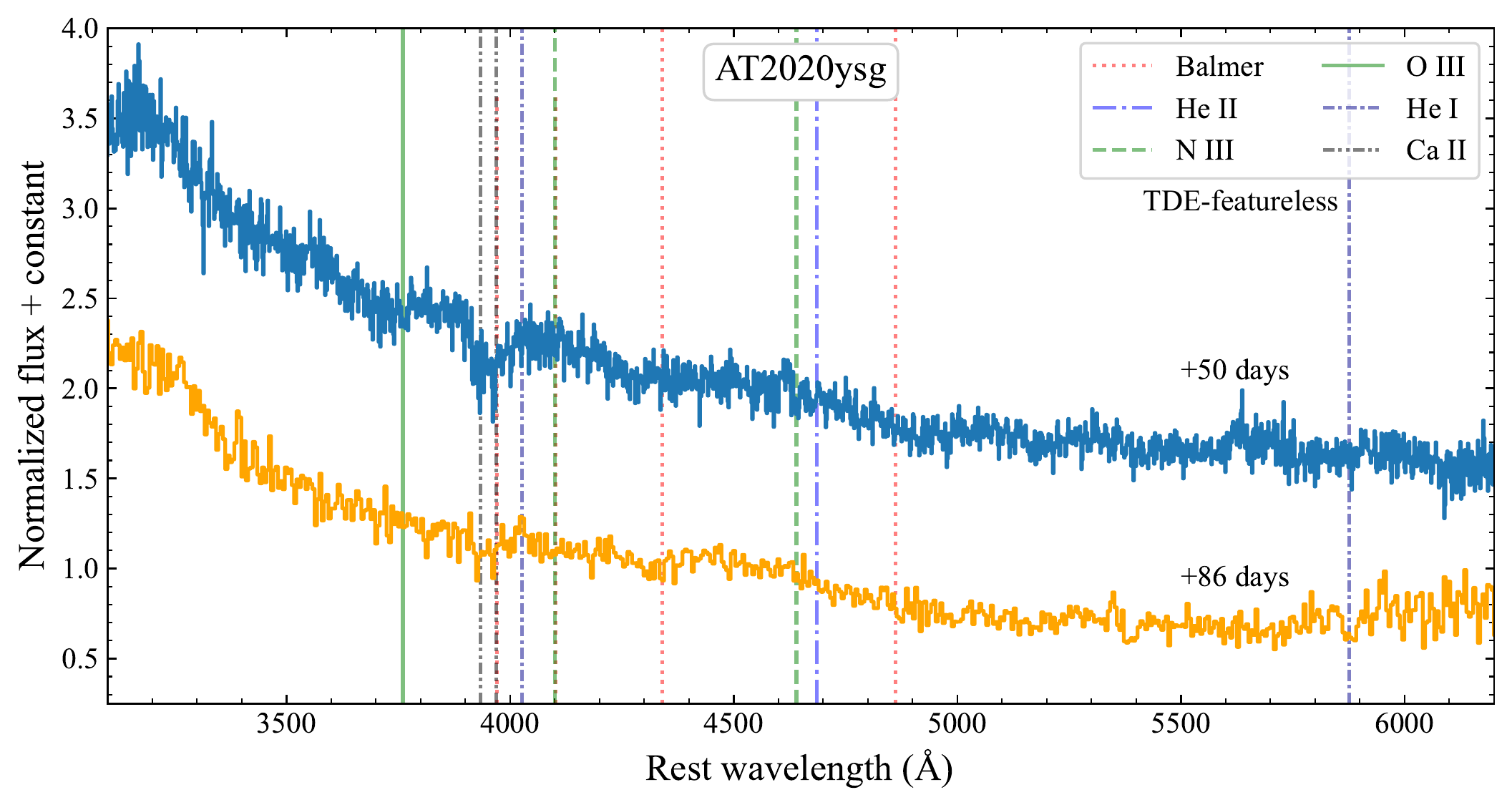}{0.45 \textwidth}{}
            \\[-30pt]}   

\caption{Optical spectra for the events in this sample. We provide an early and late time spectrum for each event when available and provide the approximate phase relative to peak that the spectrum was taken. We label common TDE emission lines and galaxy absorption lines. Spectra have not been host subtracted. Some spectra still contain telluric absorption lines, which have been labeled.}\label{fig:spec3}
\end{figure*}

\section{Light Curves \& Fits}

\begin{deluxetable*}{l r r r r r r r r r r r r}
\tablecaption{Light curve shape parameters.}
\tablehead{ \colhead{IAU Name} & \thead{$\sigma$ \\ $\log$ day} & \thead{$\tau$ \\ $\log$ day} & \thead{$T_{\rm peak}$ \\ $\log$ K} & \thead{$L_{\rm BB}$\\ $\log$ erg/s} & \thead{$t_0$ \\ $\log$ day} & \thead{$t_0$ ($p=5/3$) \\ $\log$ day} & \colhead{$p$} & \thead{$L_{\rm g}$ \\ $\log$ erg/s} & \thead{$dT/dt$ \\ $10^2$ K/day} & \thead{$t_{\rm peak}$ \\ MJD}}
\startdata
\input{tables/LCfittab}
\enddata
\label{tab:lcfitt}
\tablecomments{The light curve fitting parameters from the 3 different light curve models used.}
\end{deluxetable*}

\begin{deluxetable*}{l r r r r}
\tablecaption{Black hole and disrupted star masses.}
\tablehead{ \colhead{IAU Name} & \thead{$\log M_{\rm BH} / M_\odot$ \\ (TDEmass)} & \thead{$\log M_{\rm BH} / M_\odot$ \\ (MOSFiT)} & \thead{$M_\star / M_\odot$ \\ (TDEmass)} & \thead{$M_\star / M_\odot$ \\ (MOSFiT)}}
\startdata
\input{tables/MBH_MStar}
\enddata
\label{tab:bhmass}
\tablecomments{The black hole mass and the mass of the disrupted star from \texttt{TDEmass} and \texttt{MOSFiT}.}
\end{deluxetable*}

\begin{figure*}
\gridline{	\fig{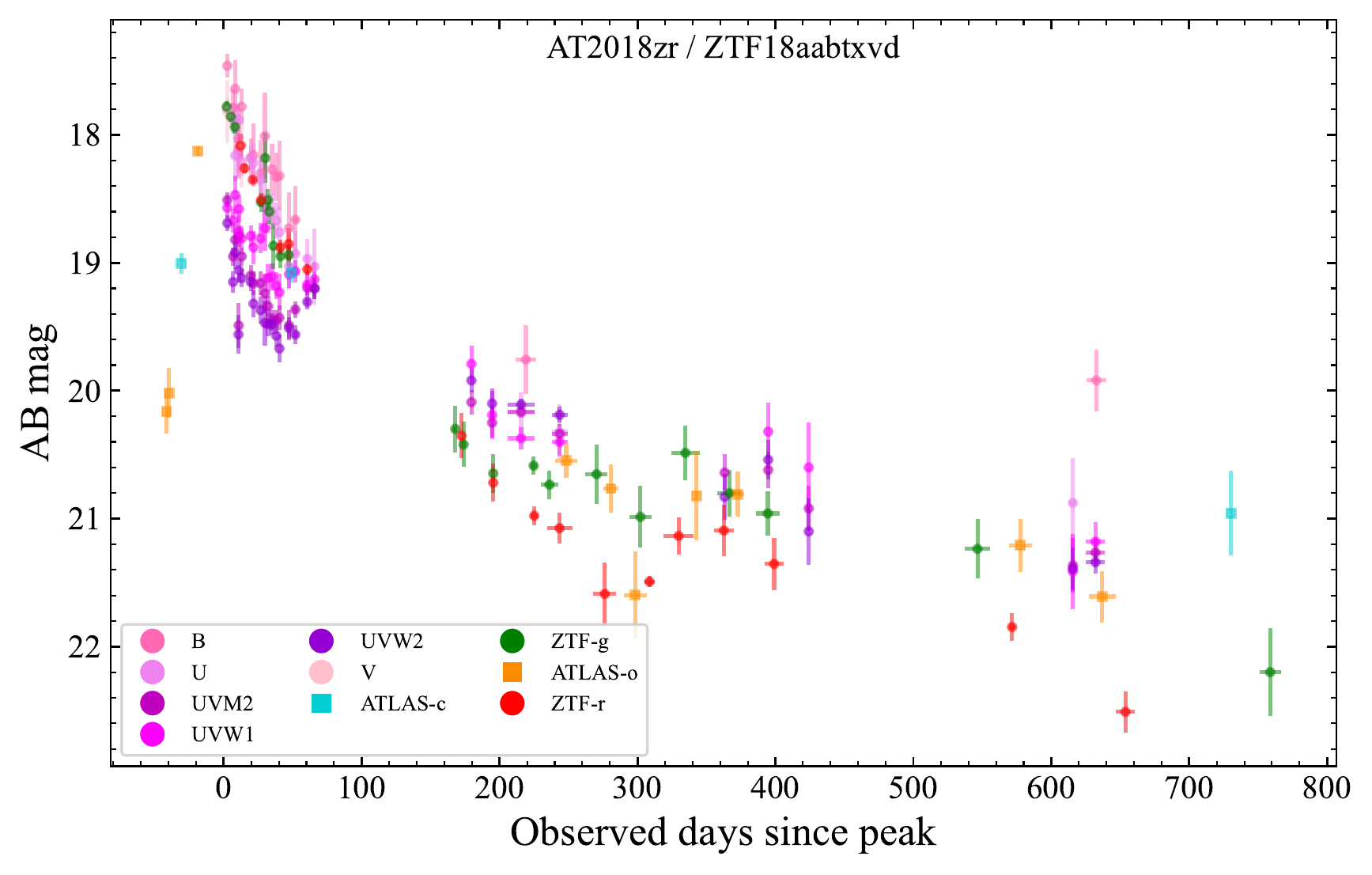}{0.33 \textwidth}{} 
			\fig{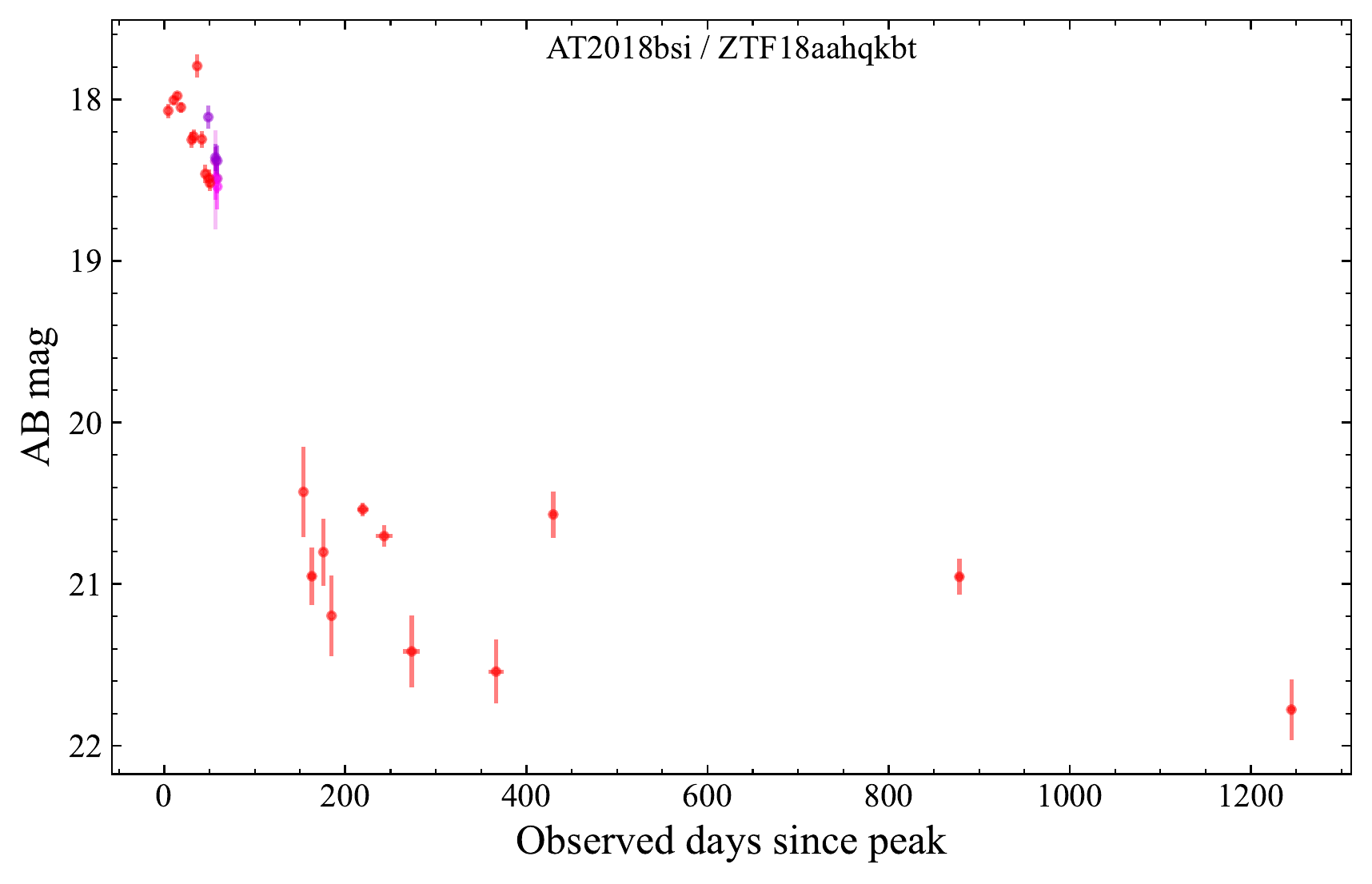}{0.33 \textwidth}{}
			\fig{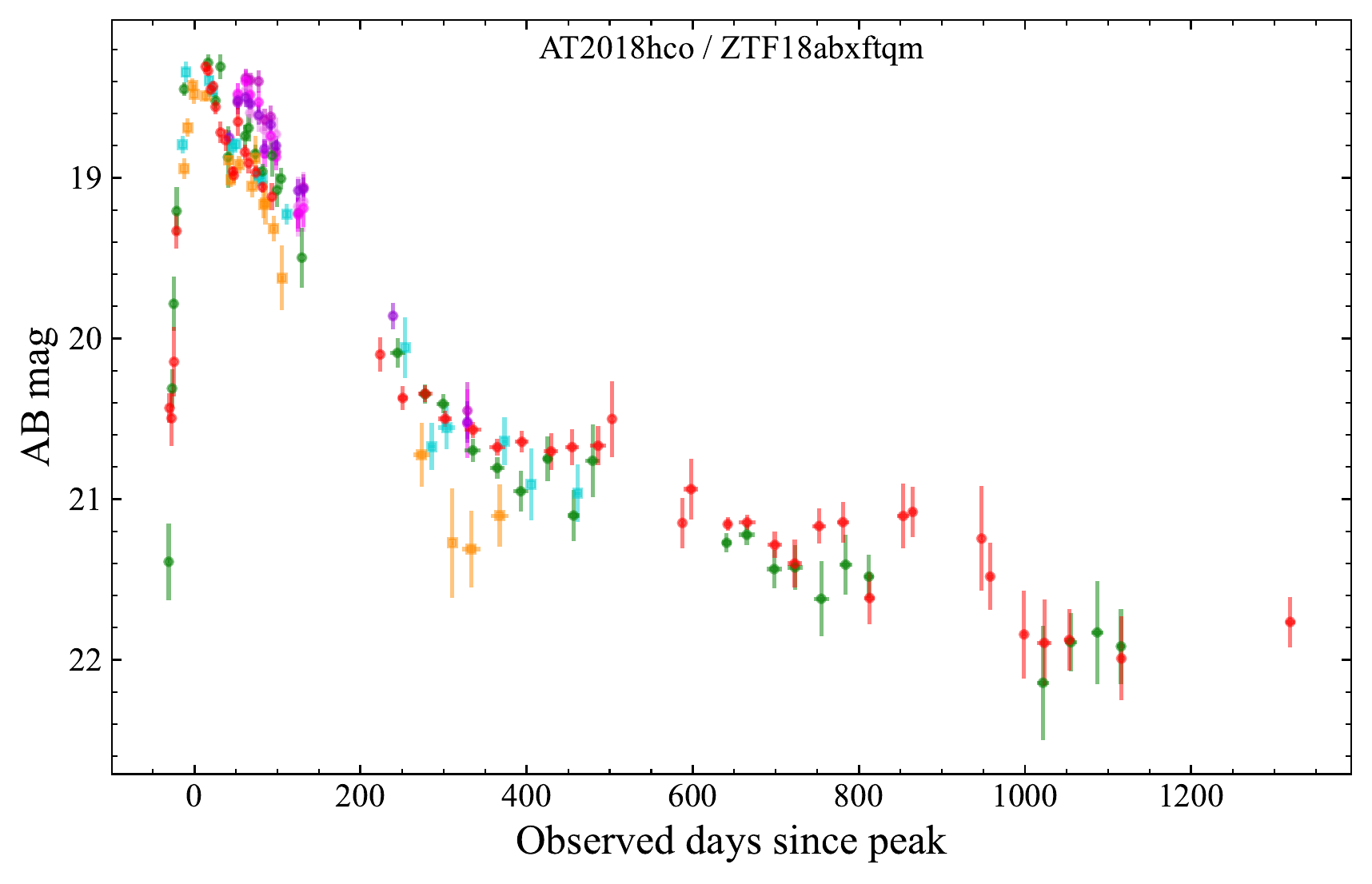}{0.33 \textwidth}{}
			\\[-20pt]}
			
\gridline{\fig{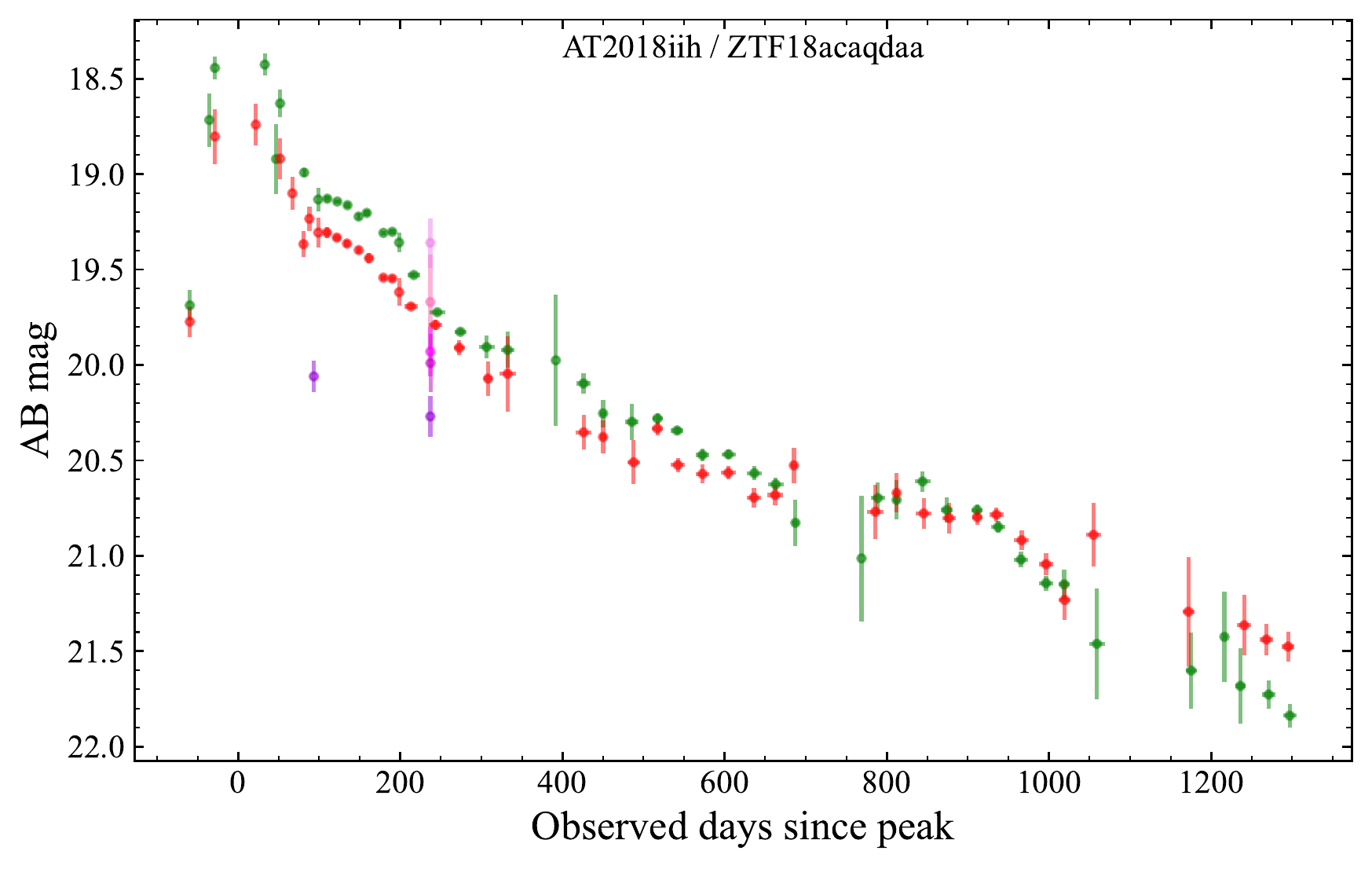}{0.33 \textwidth}{} 
\fig{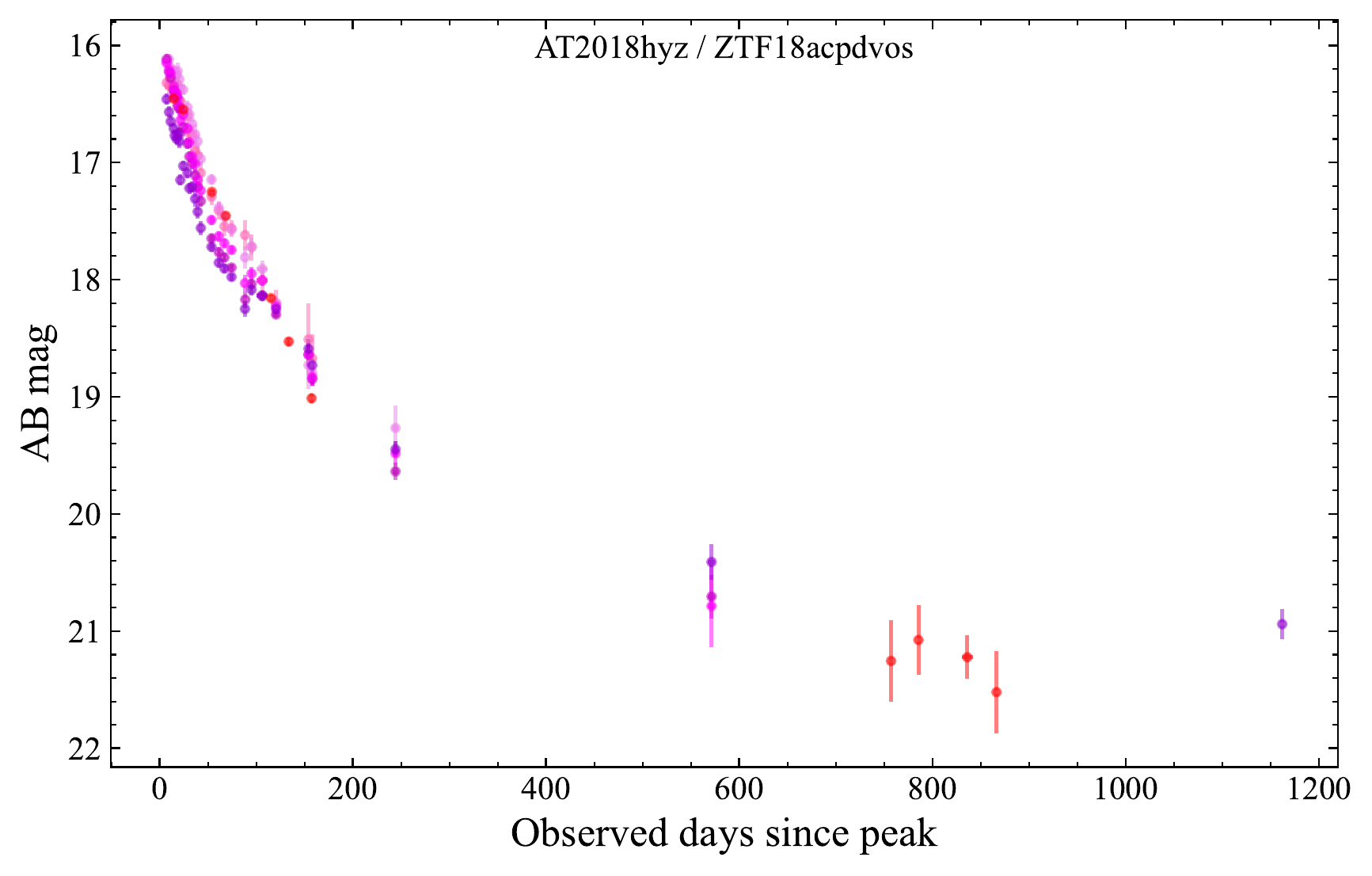}{0.33 \textwidth}{} 
            \fig{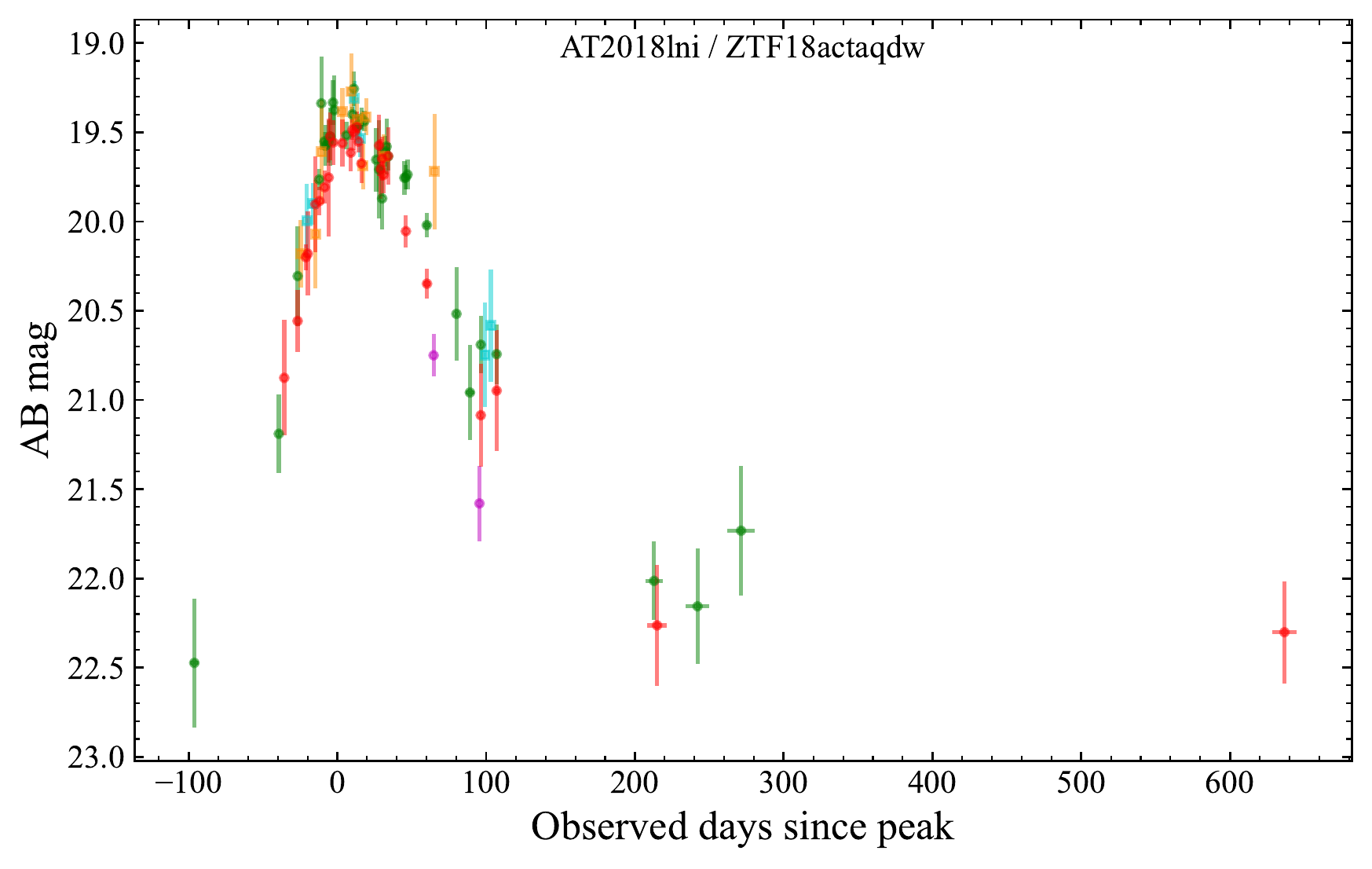}{0.33 \textwidth}{}
            \\[-20pt]}
     
\gridline{  \fig{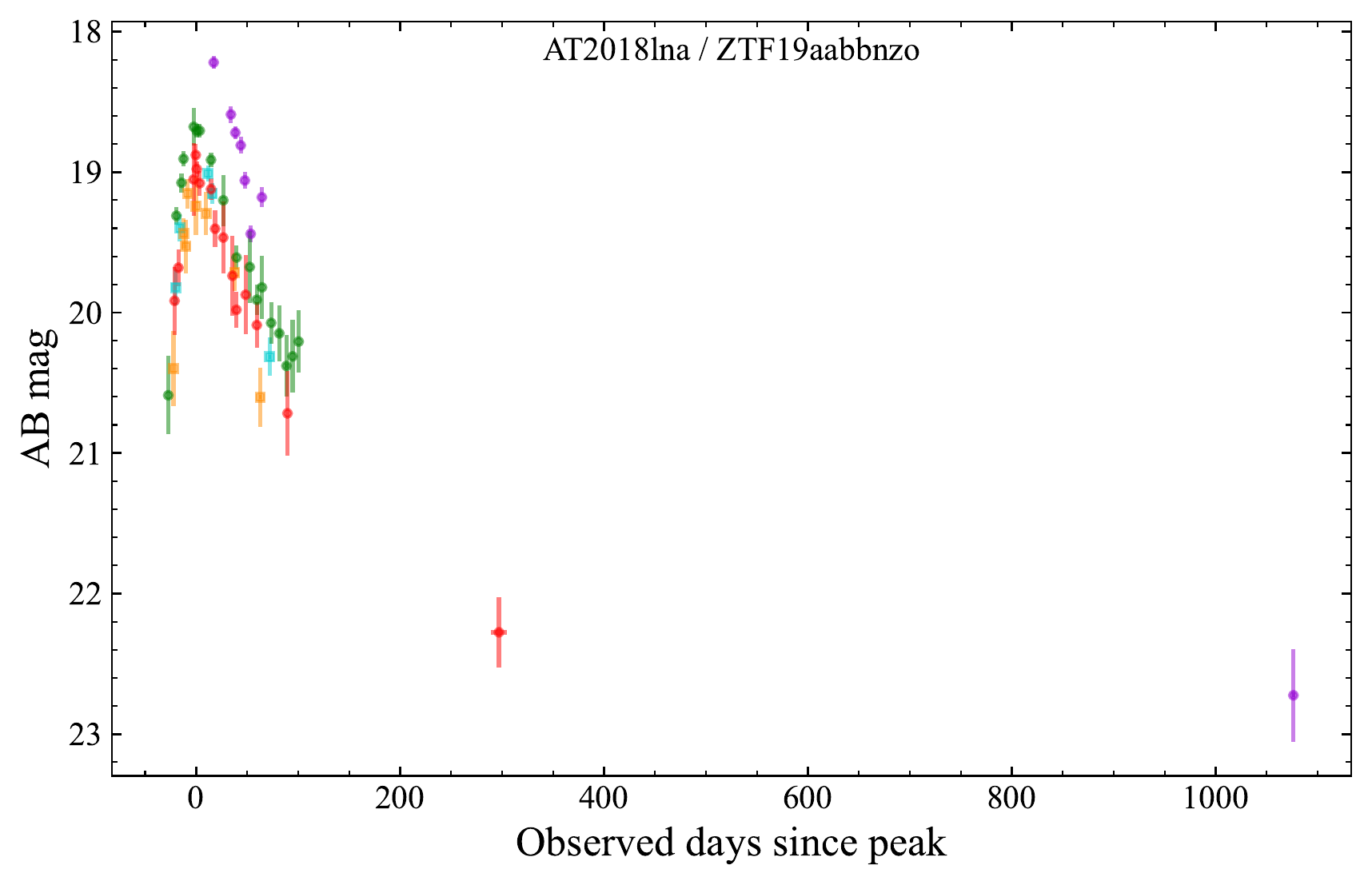}{0.33 \textwidth}{}
            \fig{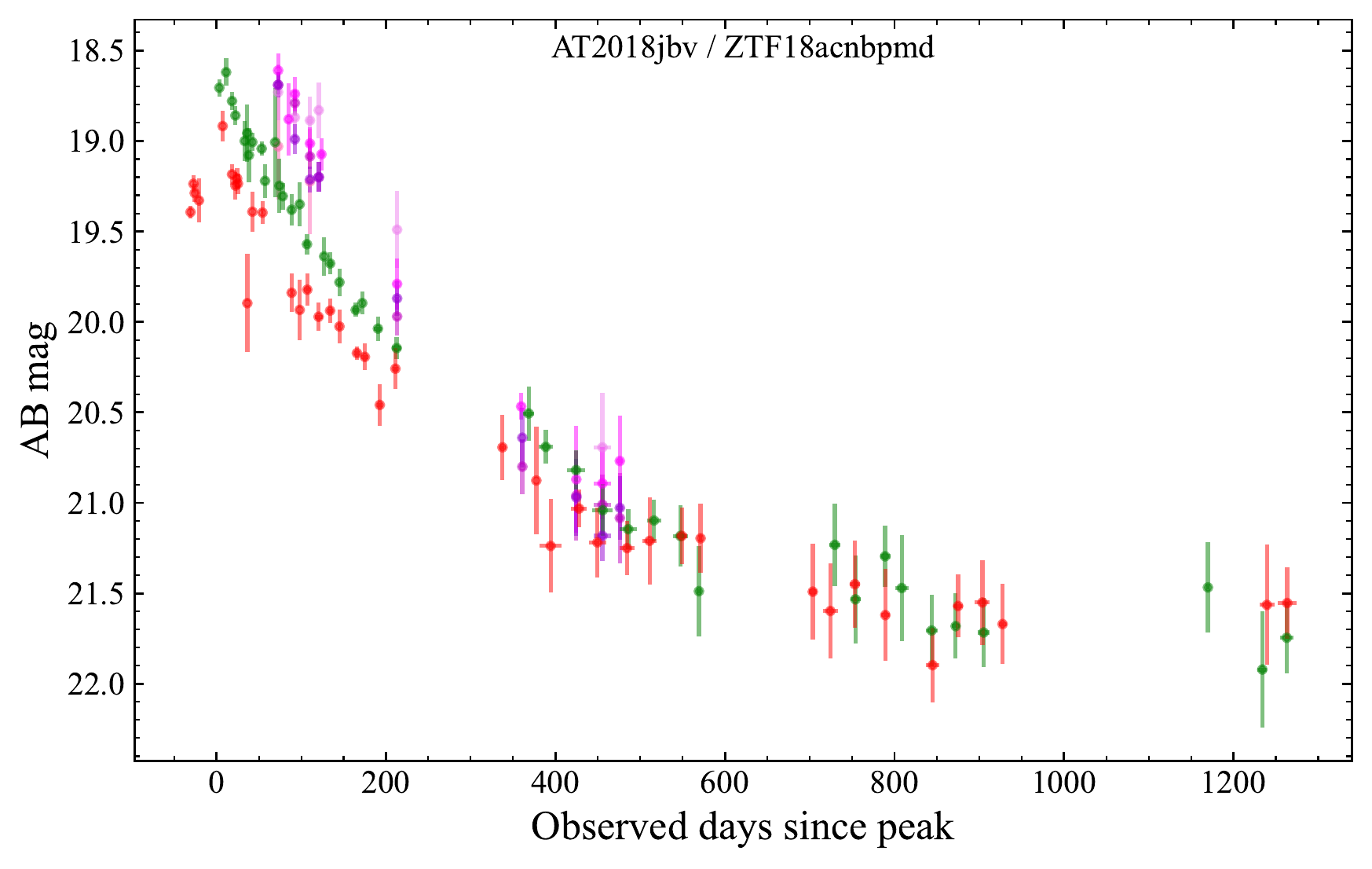}{0.33 \textwidth}{}
            \fig{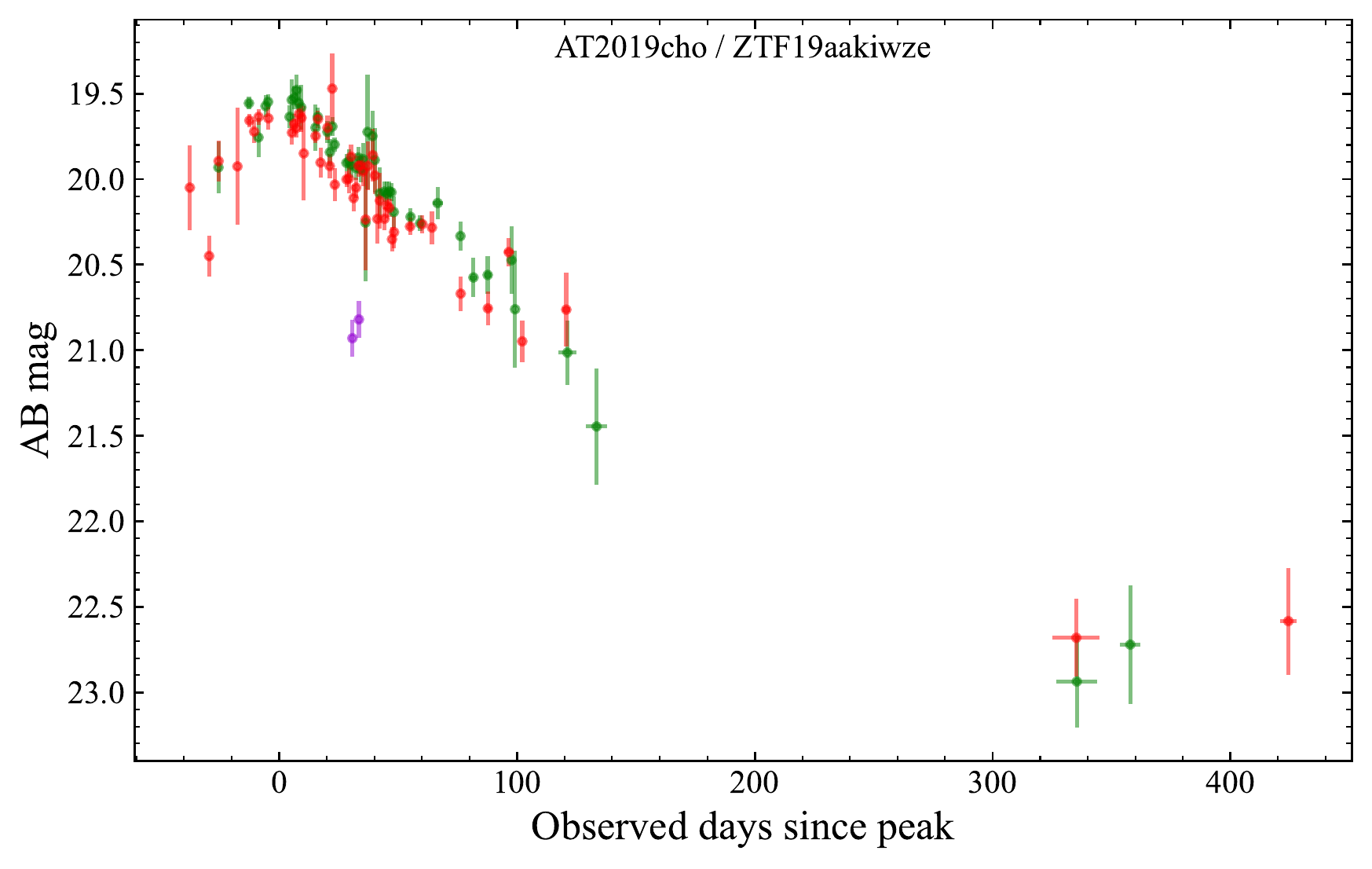}{0.33 \textwidth}{} 
             \\[-20pt]}

\gridline{\fig{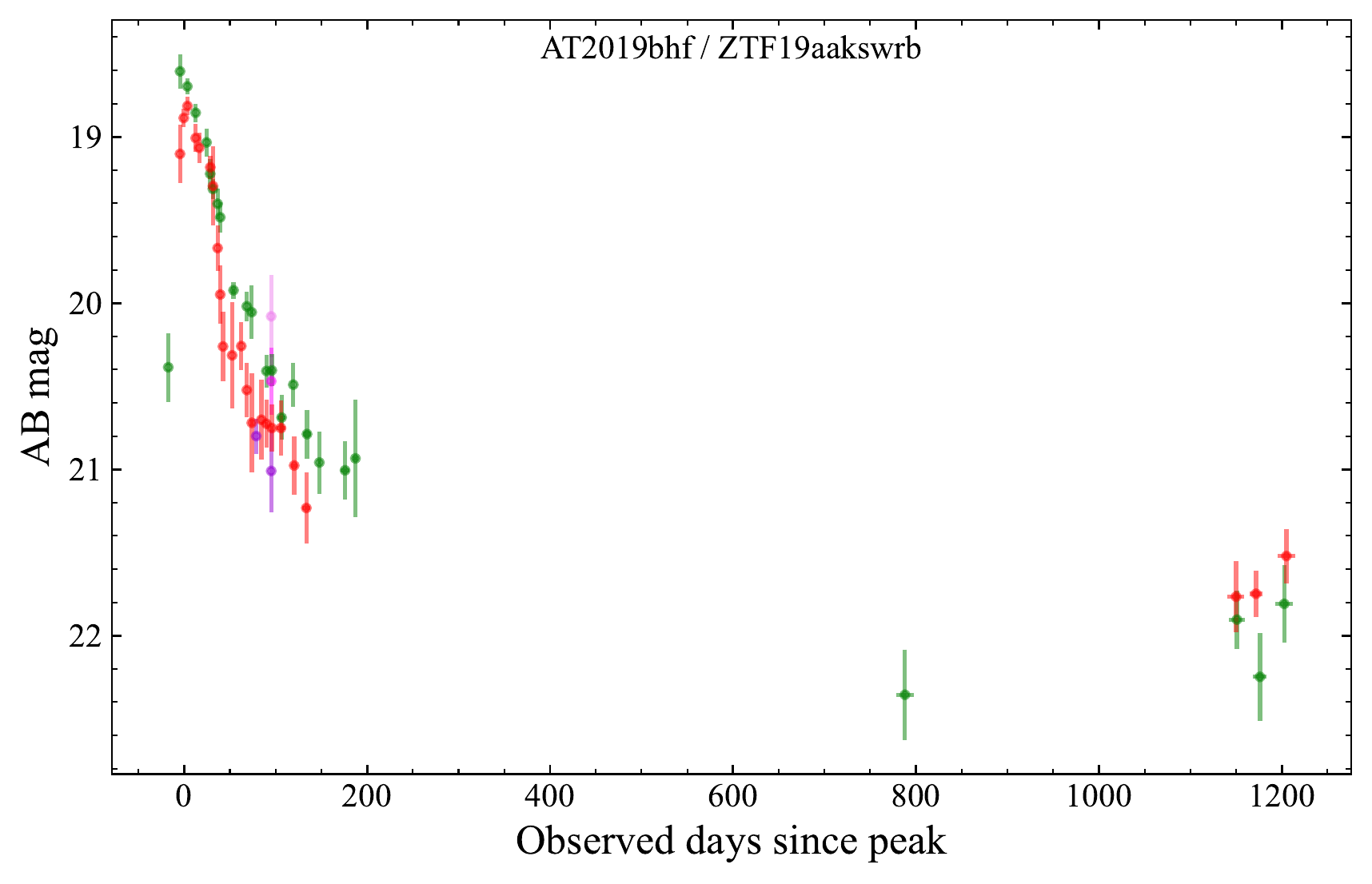}{0.33 \textwidth}{}
	\fig{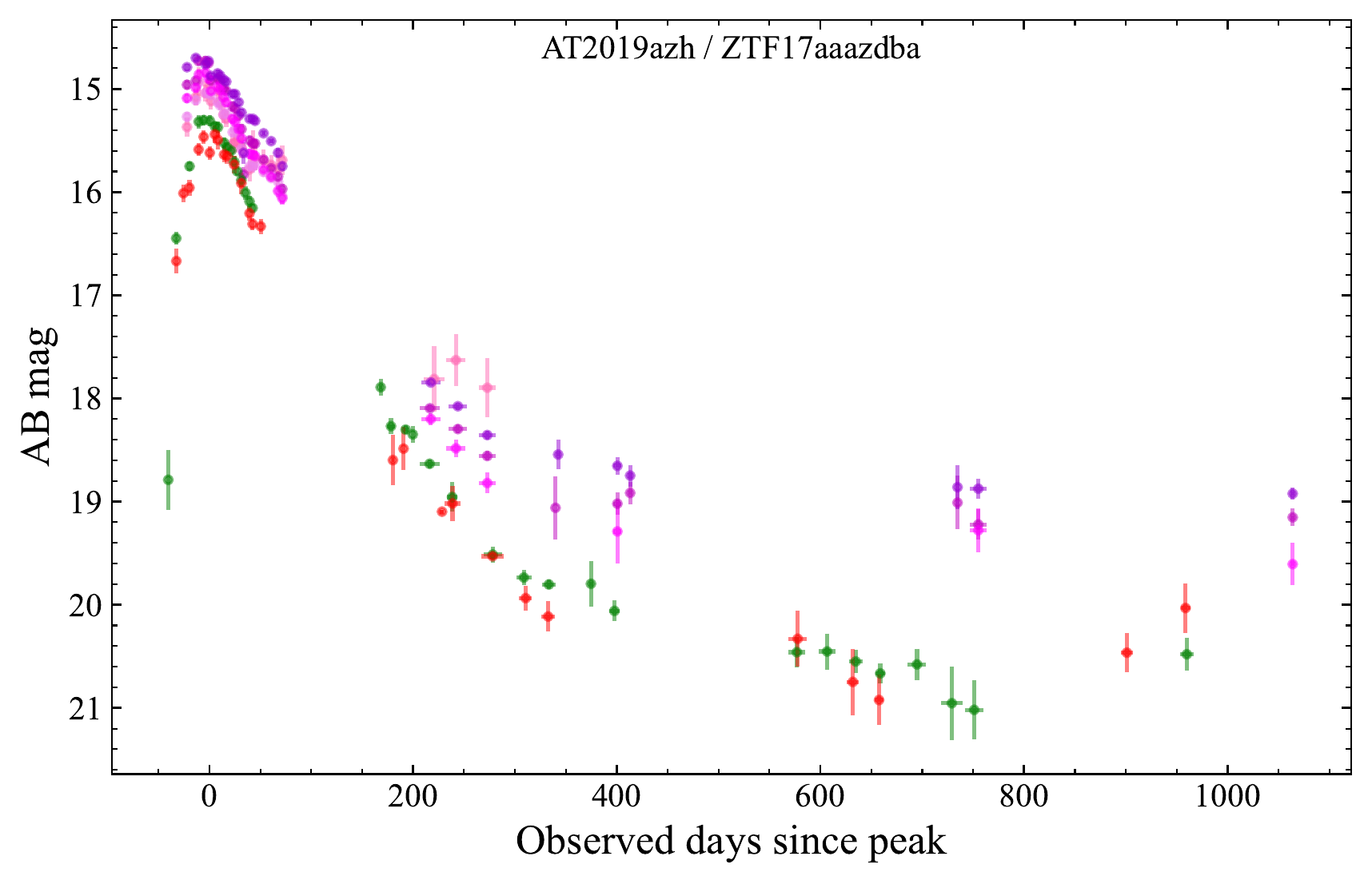}{0.33 \textwidth}{} 
            \fig{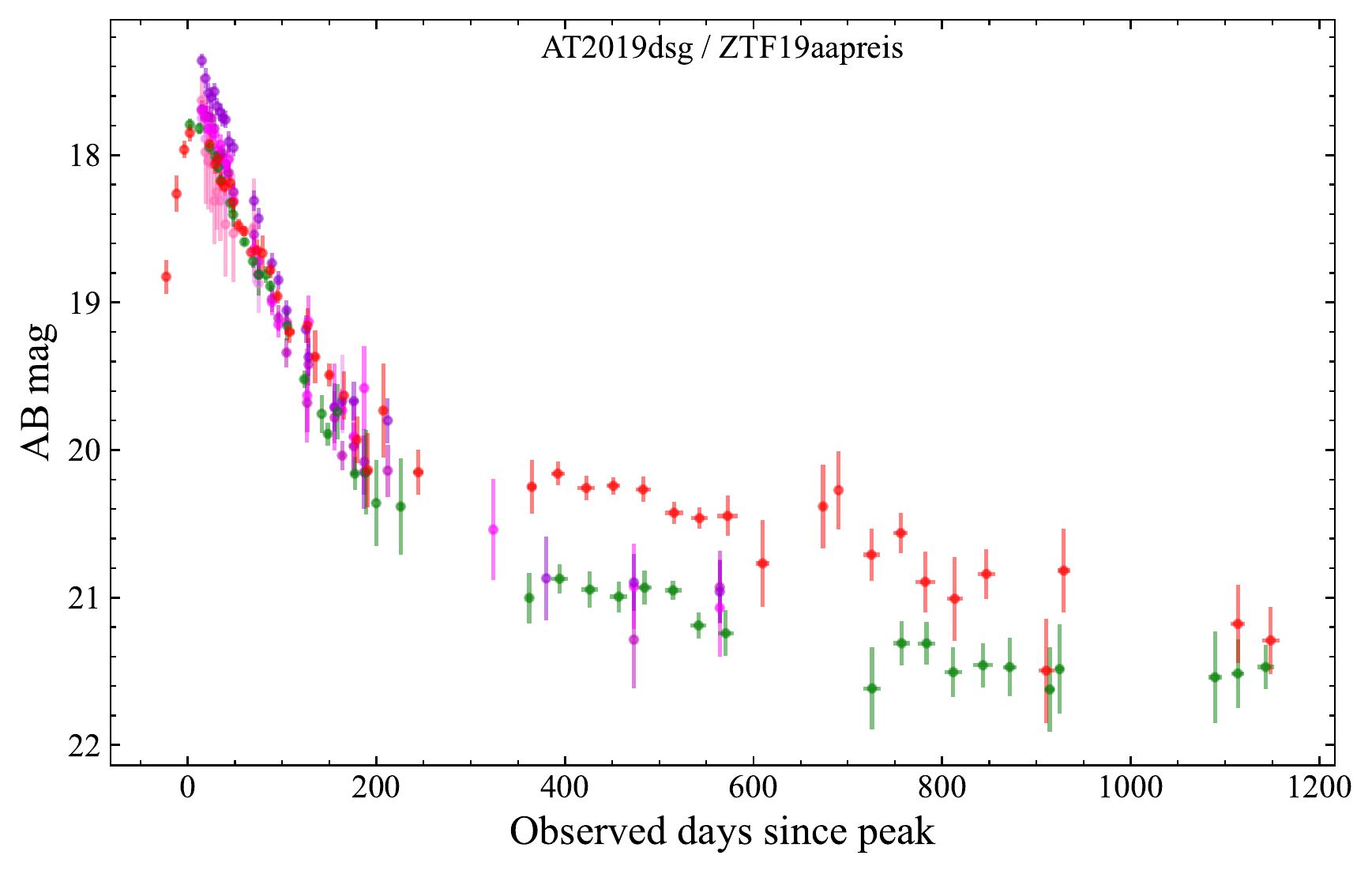}{0.33 \textwidth}{}
            \\[-20pt]} 
\gridline{  \fig{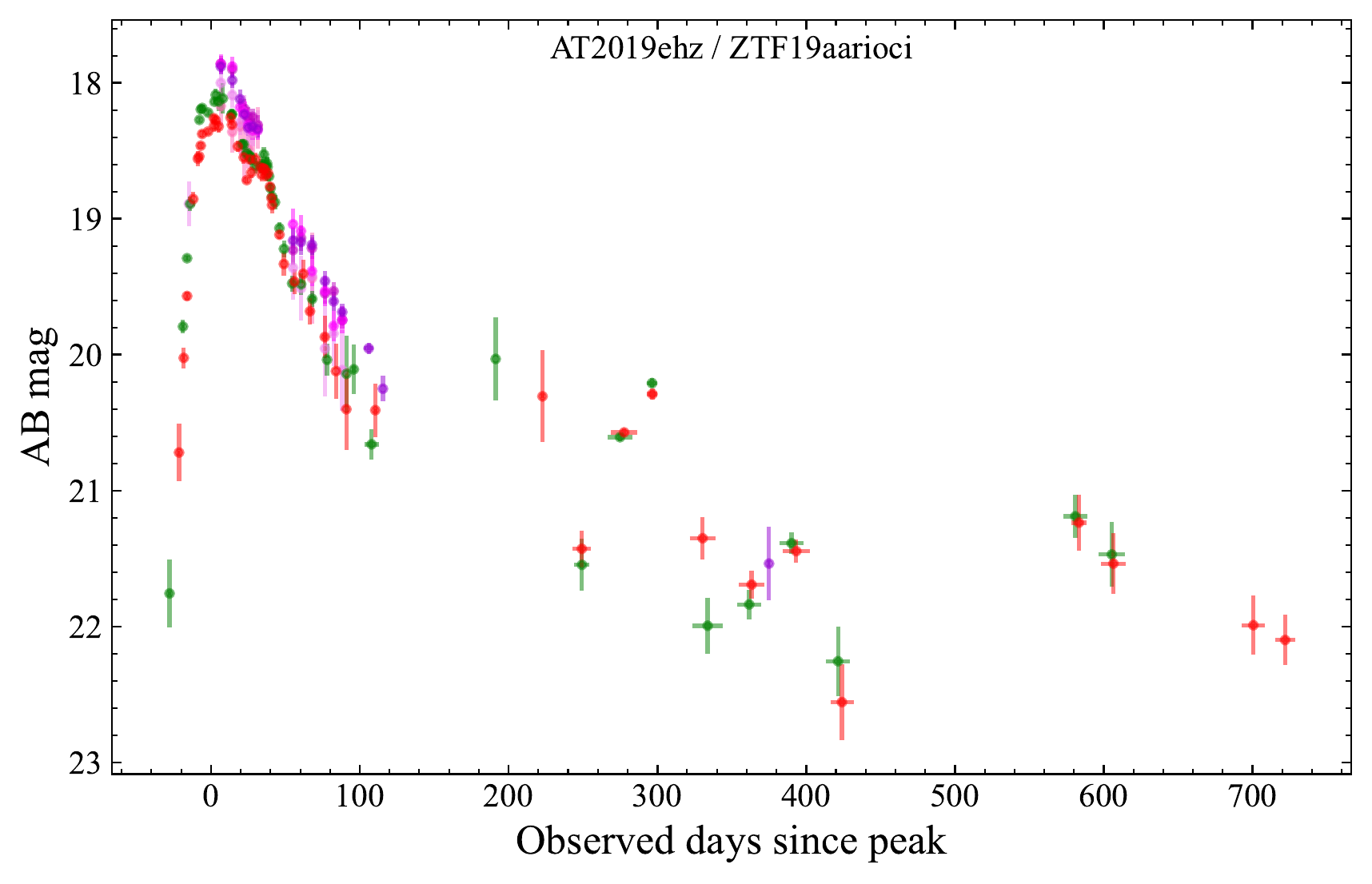}{0.33 \textwidth}{} 
            \fig{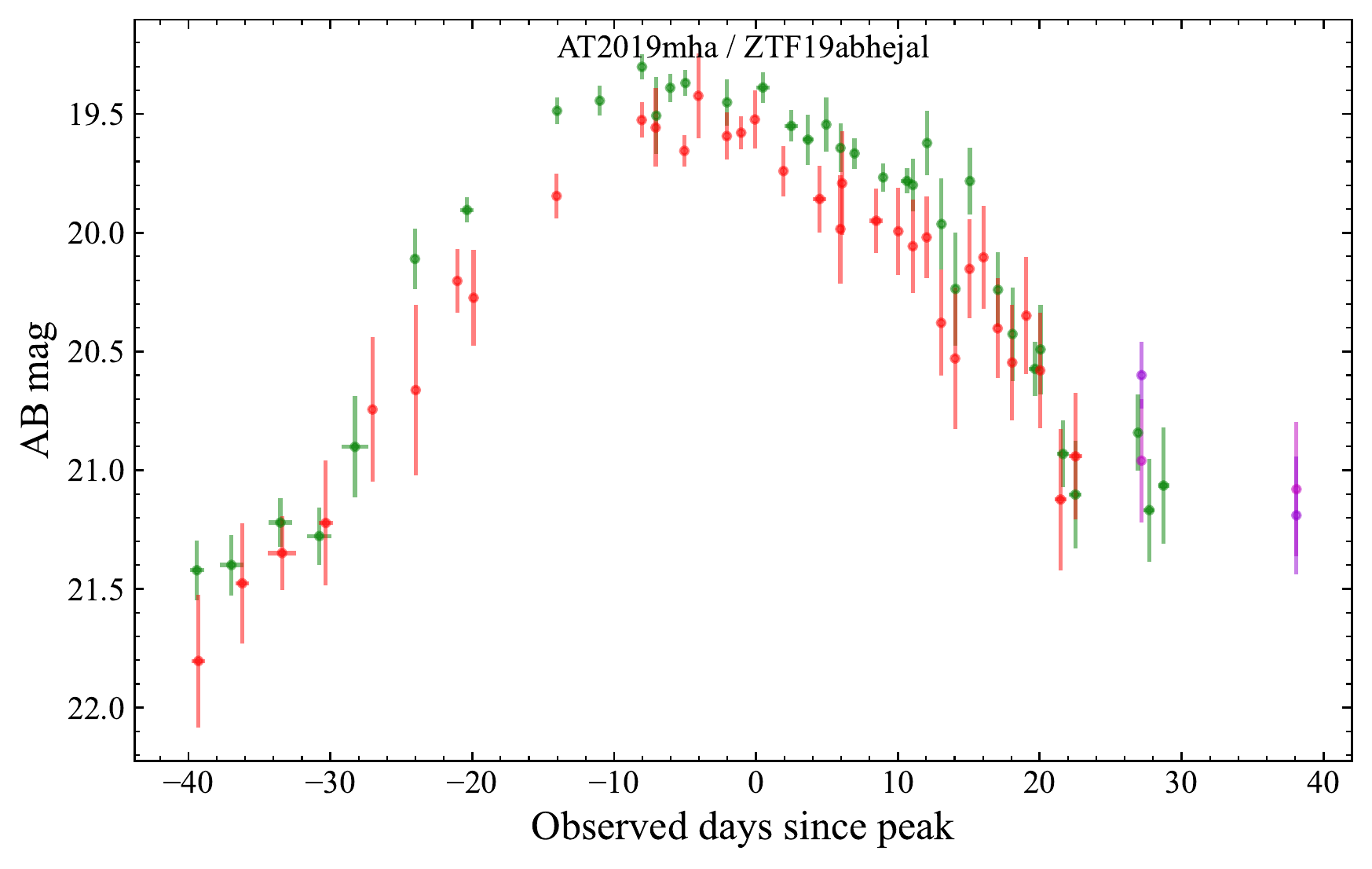}{0.33 \textwidth}{}
            \fig{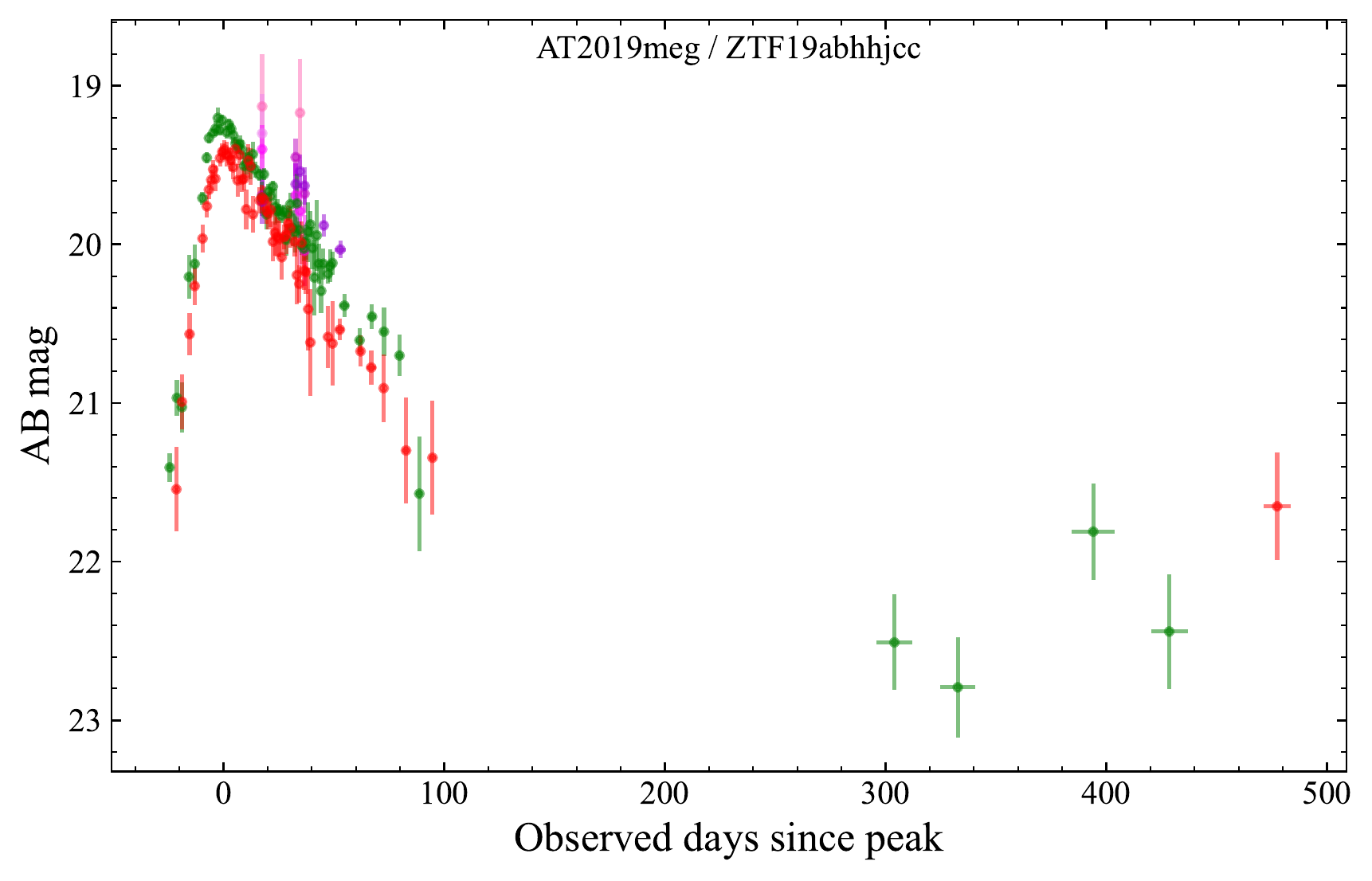}{0.33 \textwidth}{} 
            \\[-20pt]}

\caption{Optical/UV light curves from ZTF, {\it Swift}/UVOT, and ATLAS photometry. The light curves are 3-$\sigma$ detections binned based on time relative to peak, with observations $>$200 days post-peak binned by 30 days. The legend for the individual bands can be seen in the top left panel. Data for Figures 17--18 are available in the related files associated with Figure 17.}\label{fig:obslcs1}
\end{figure*}

\begin{figure*}
            
\gridline{\fig{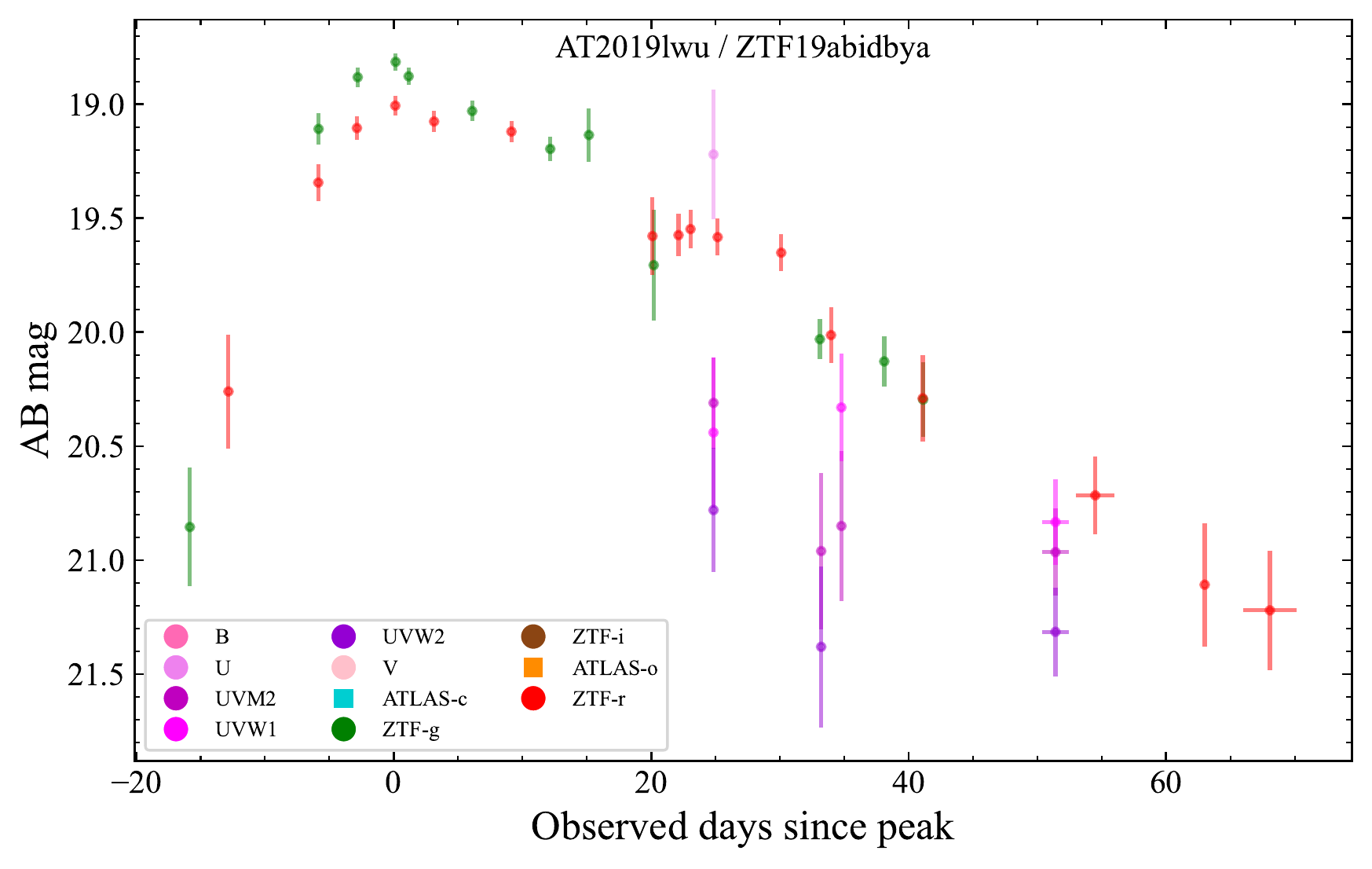}{0.33 \textwidth}{}
  \fig{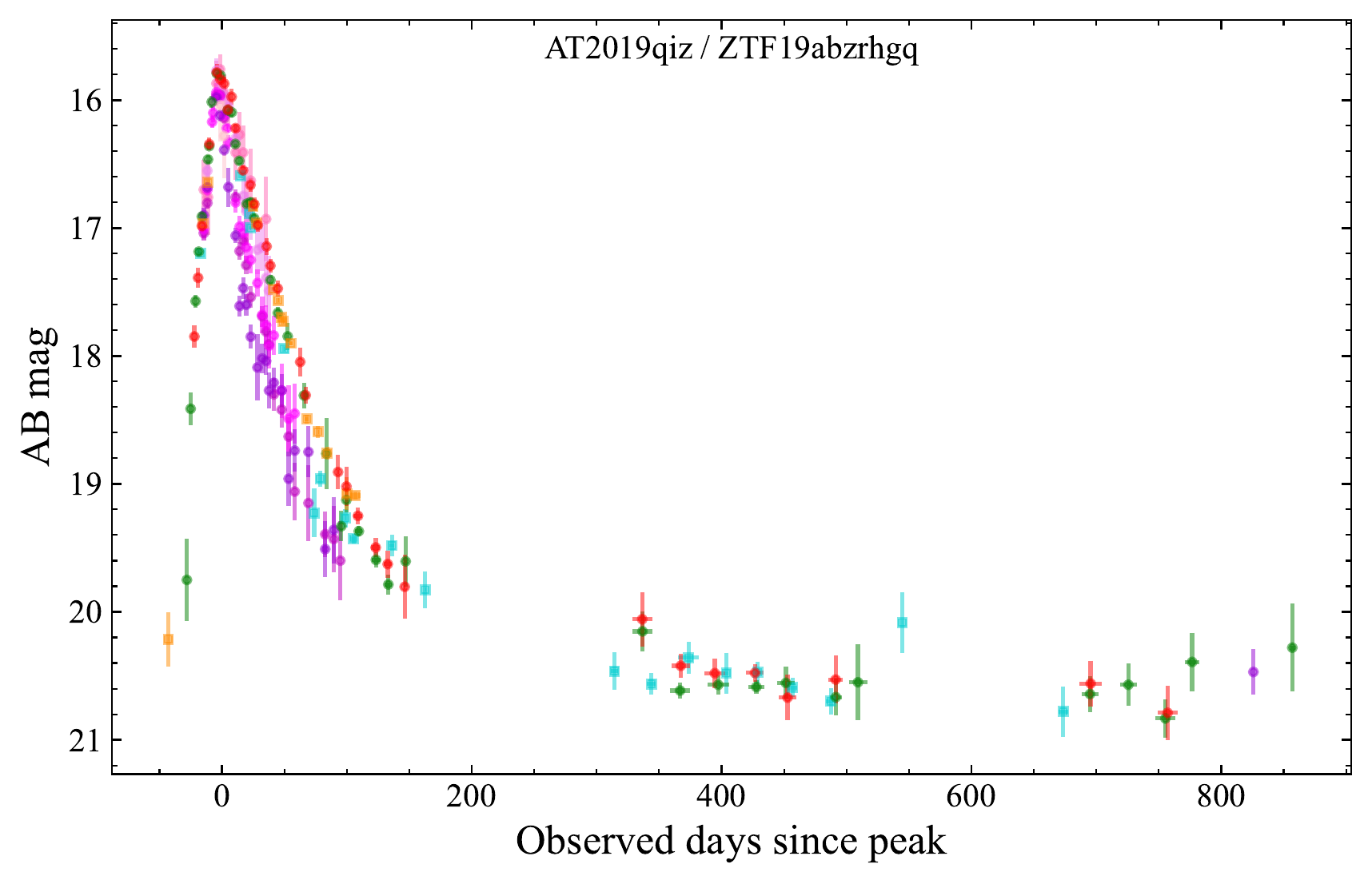}{0.33 \textwidth}{}
            \fig{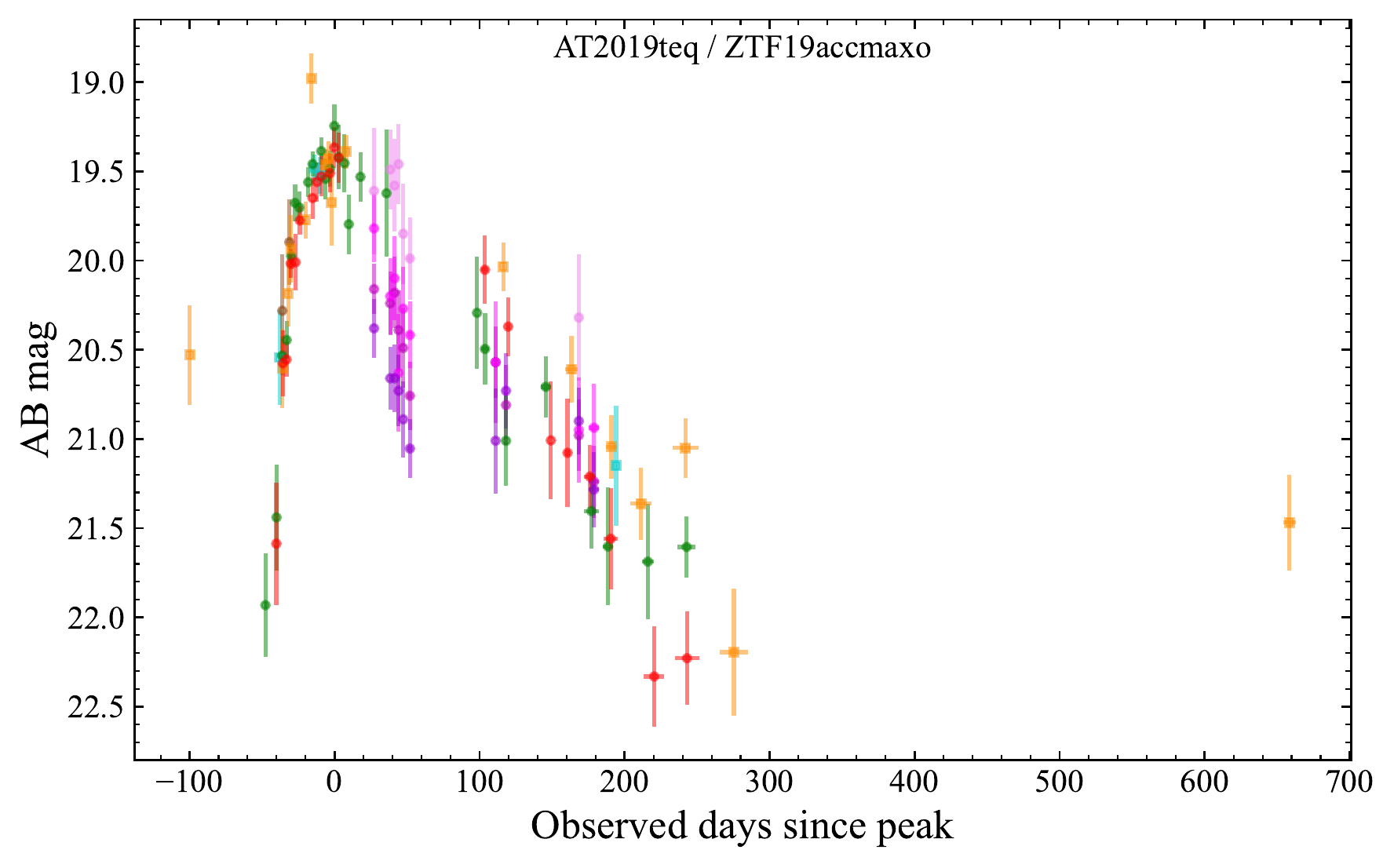}{0.33 \textwidth}{}
            \\[-20pt]}          

\gridline{  \fig{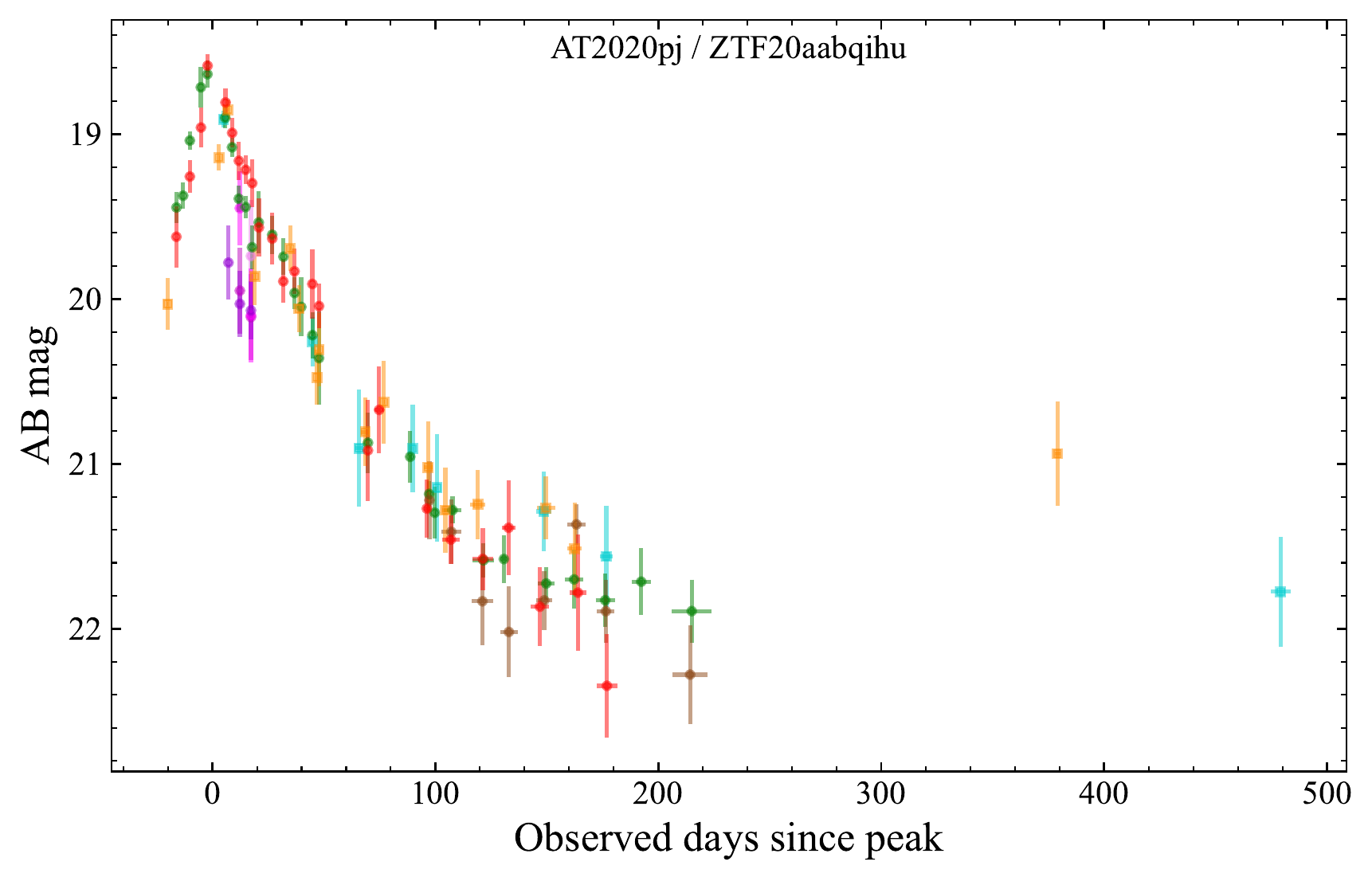}{0.33 \textwidth}{}
            \fig{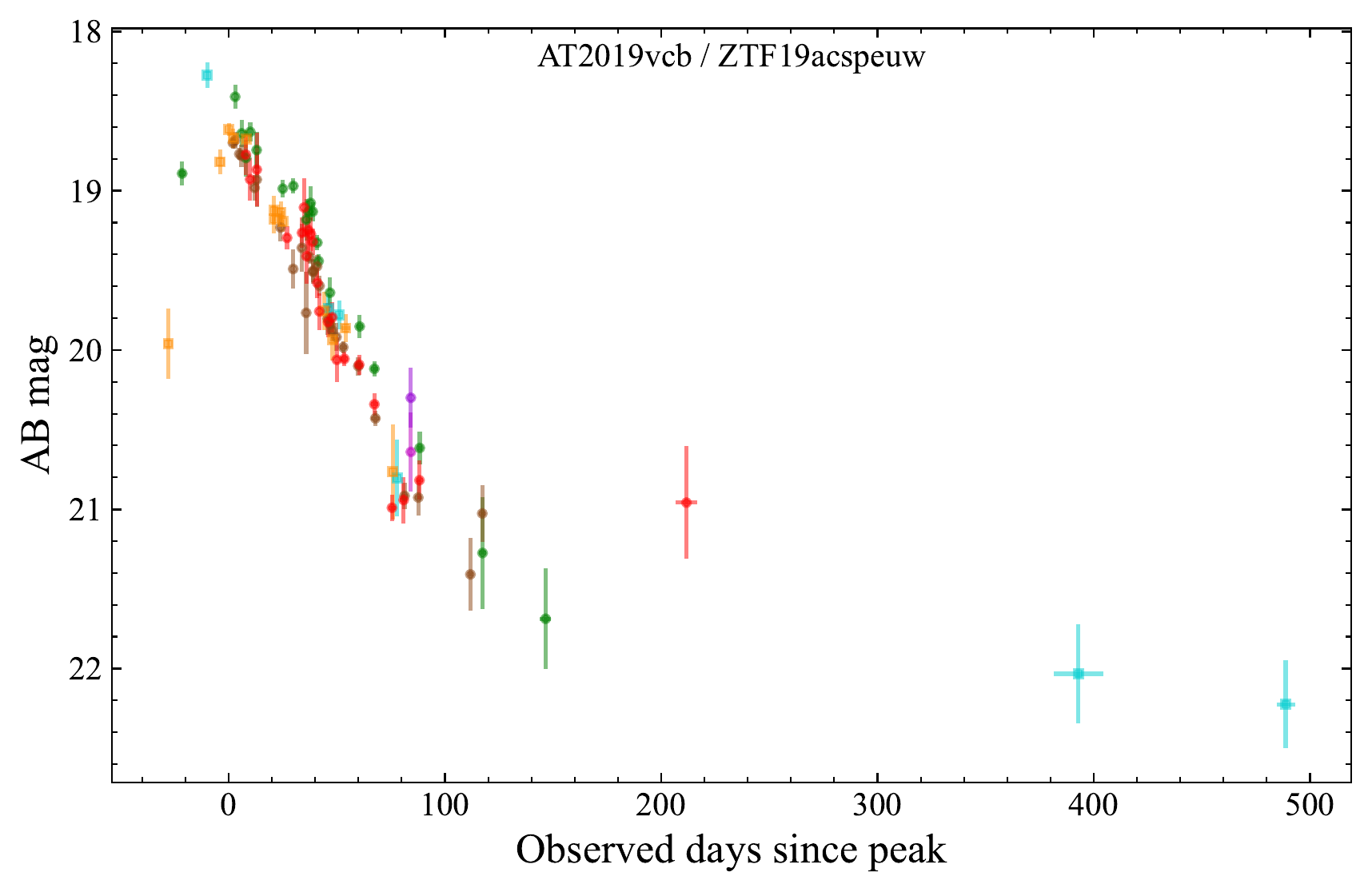}{0.33 \textwidth}{}
            \fig{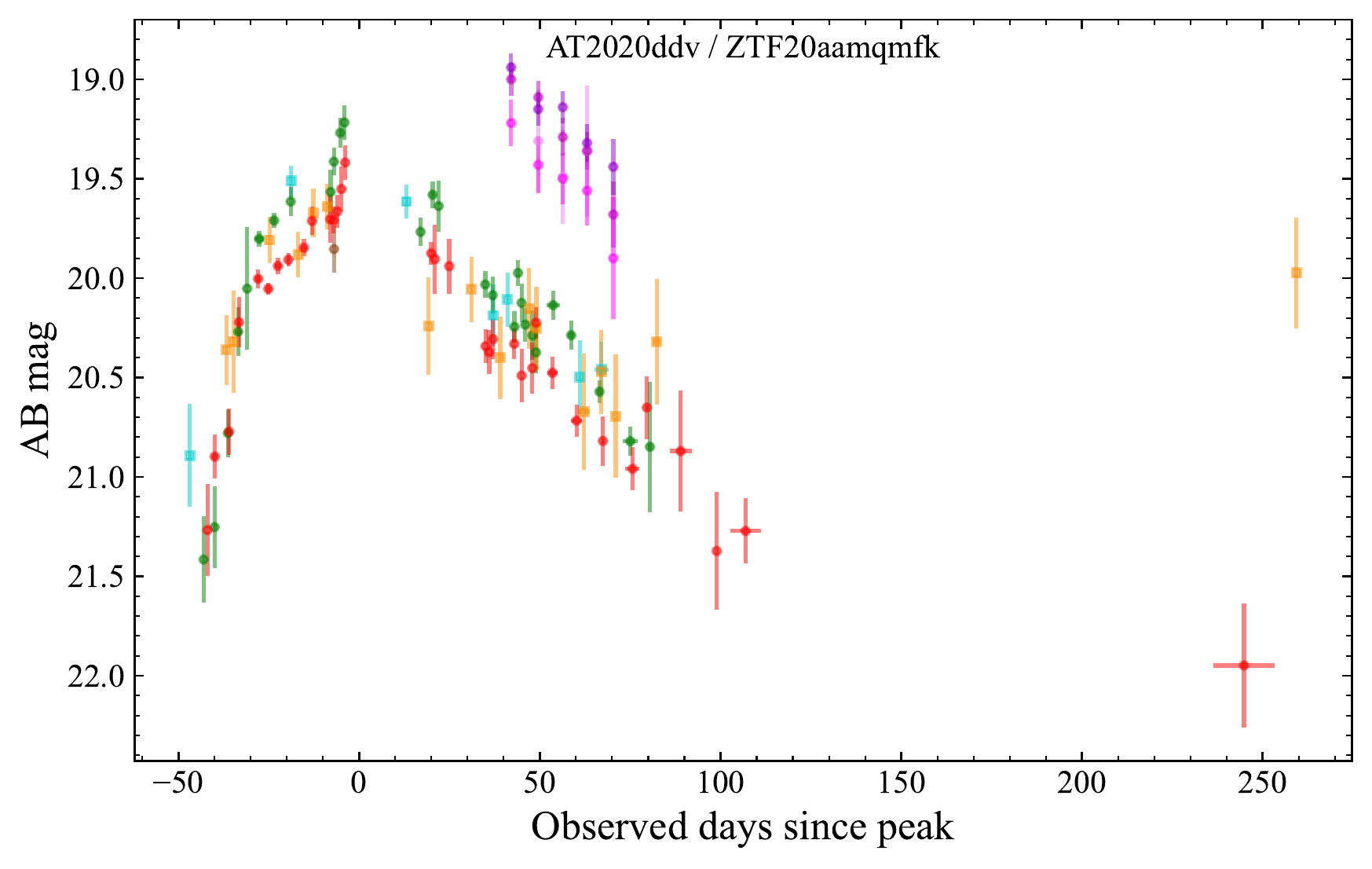}{0.33 \textwidth}{}
            \\[-20pt]}   

\gridline{	\fig{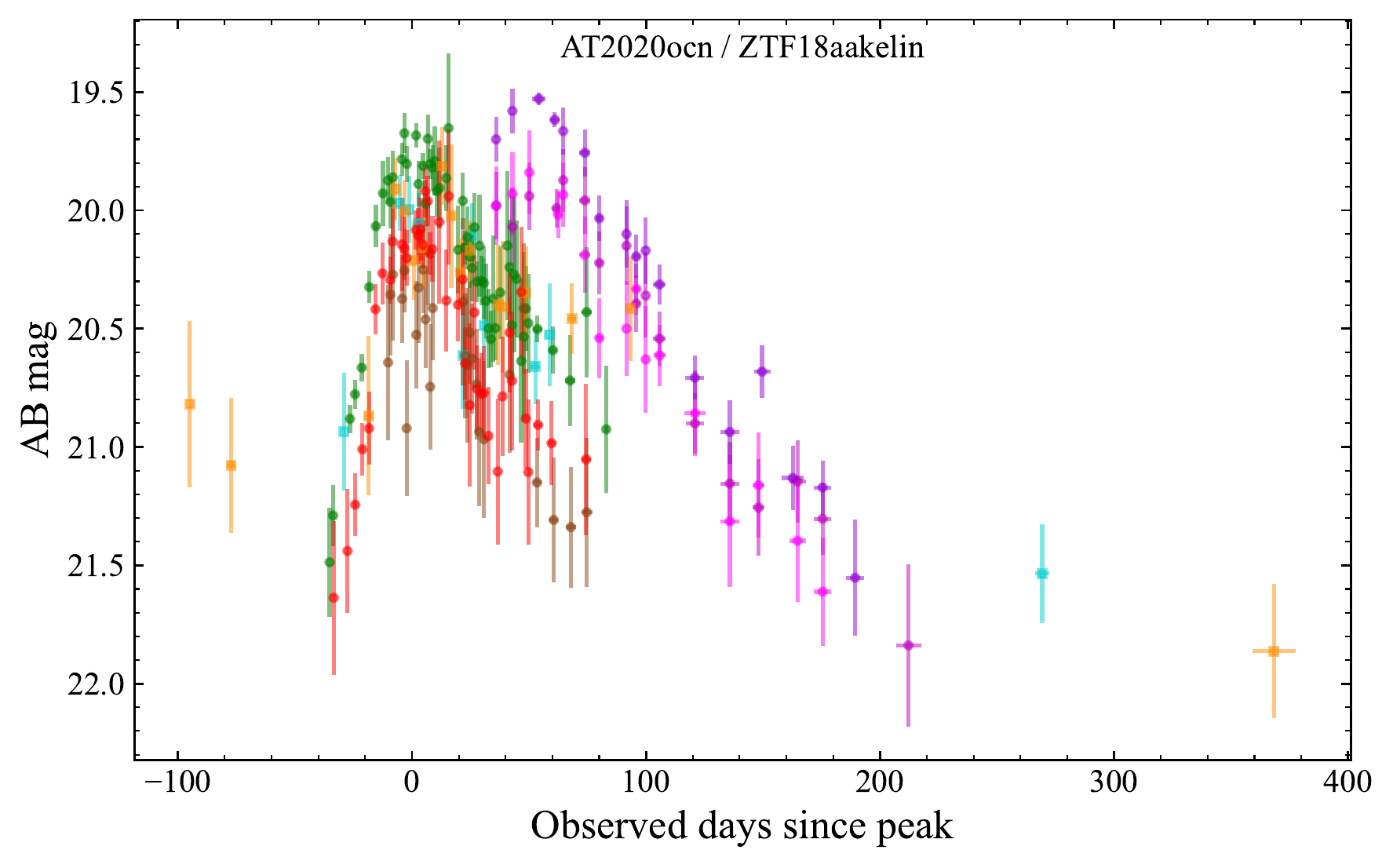}{0.33 \textwidth}{}
             \fig{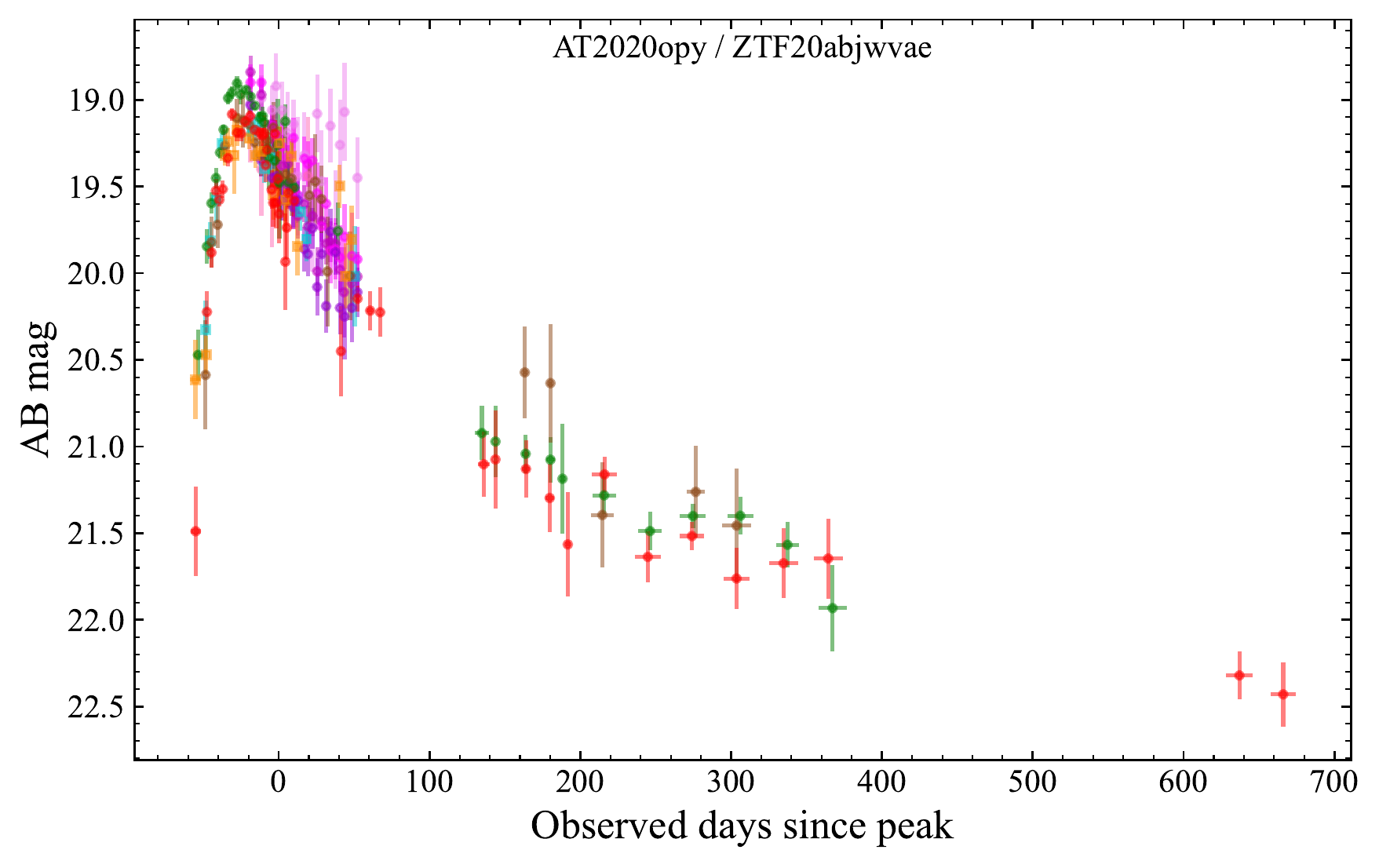}{0.33 \textwidth}{}
            \fig{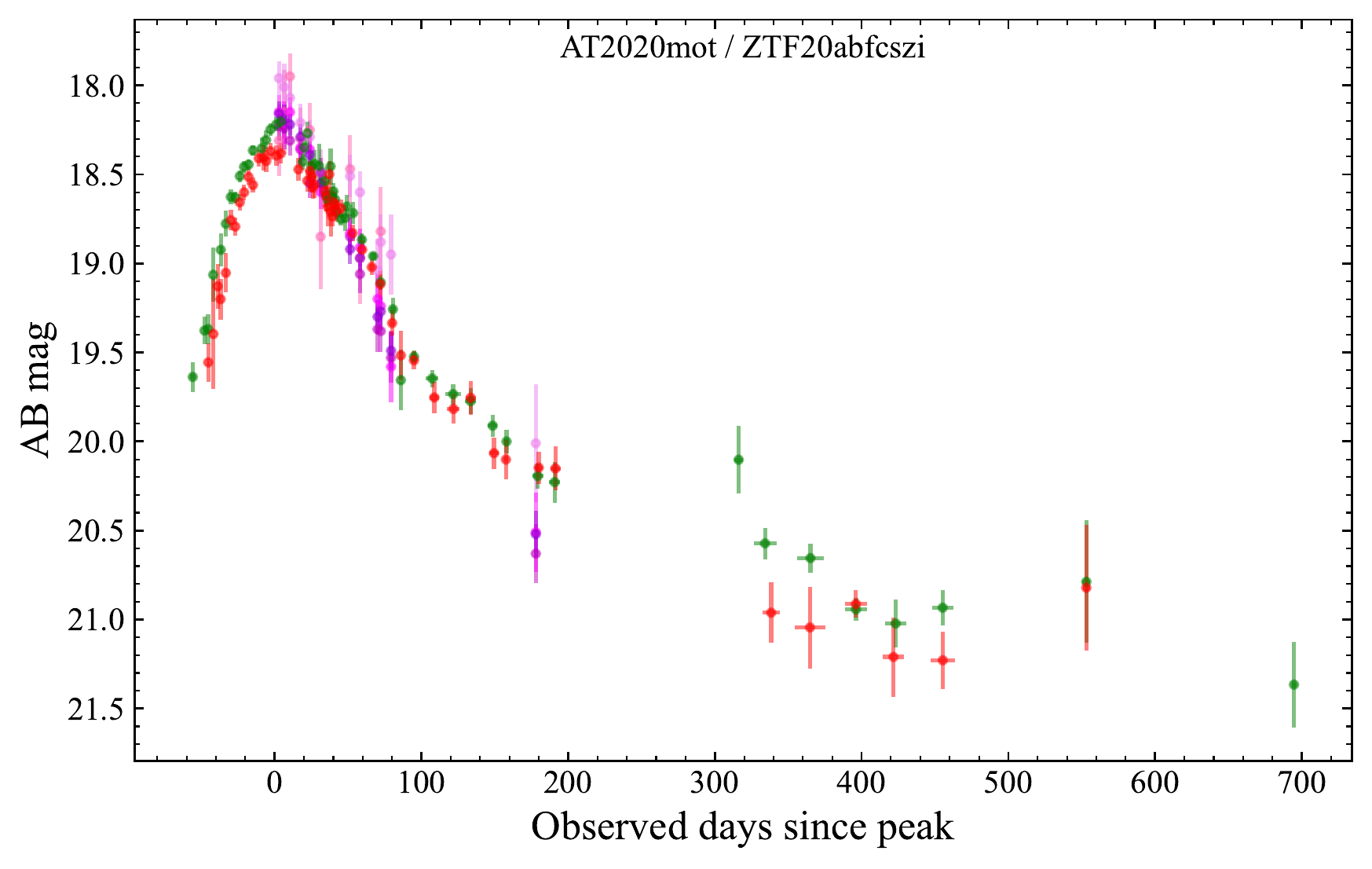}{0.33 \textwidth}{} 			\\[-20pt]}

\gridline{  \fig{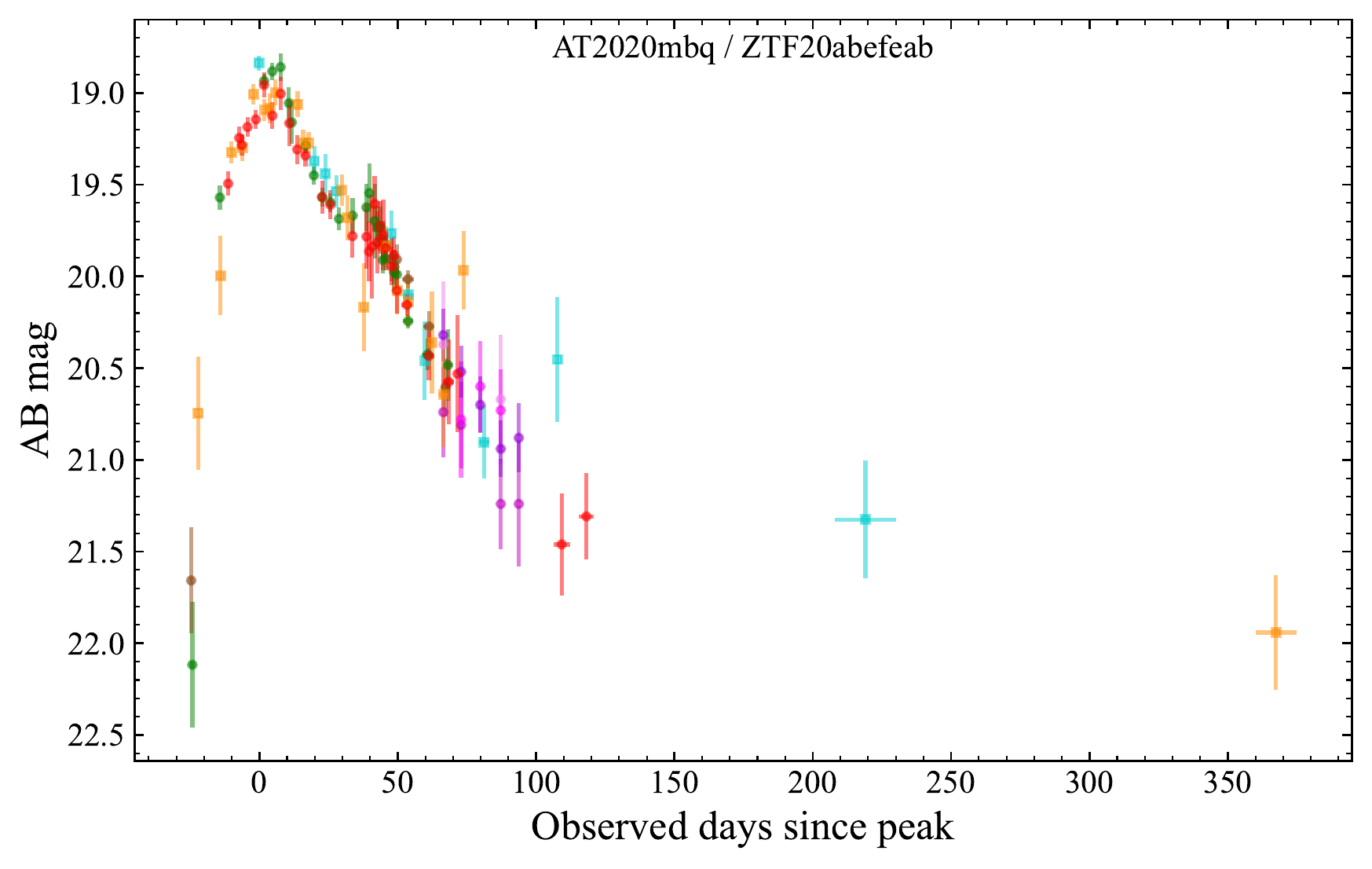}{0.33 \textwidth}{}
            \fig{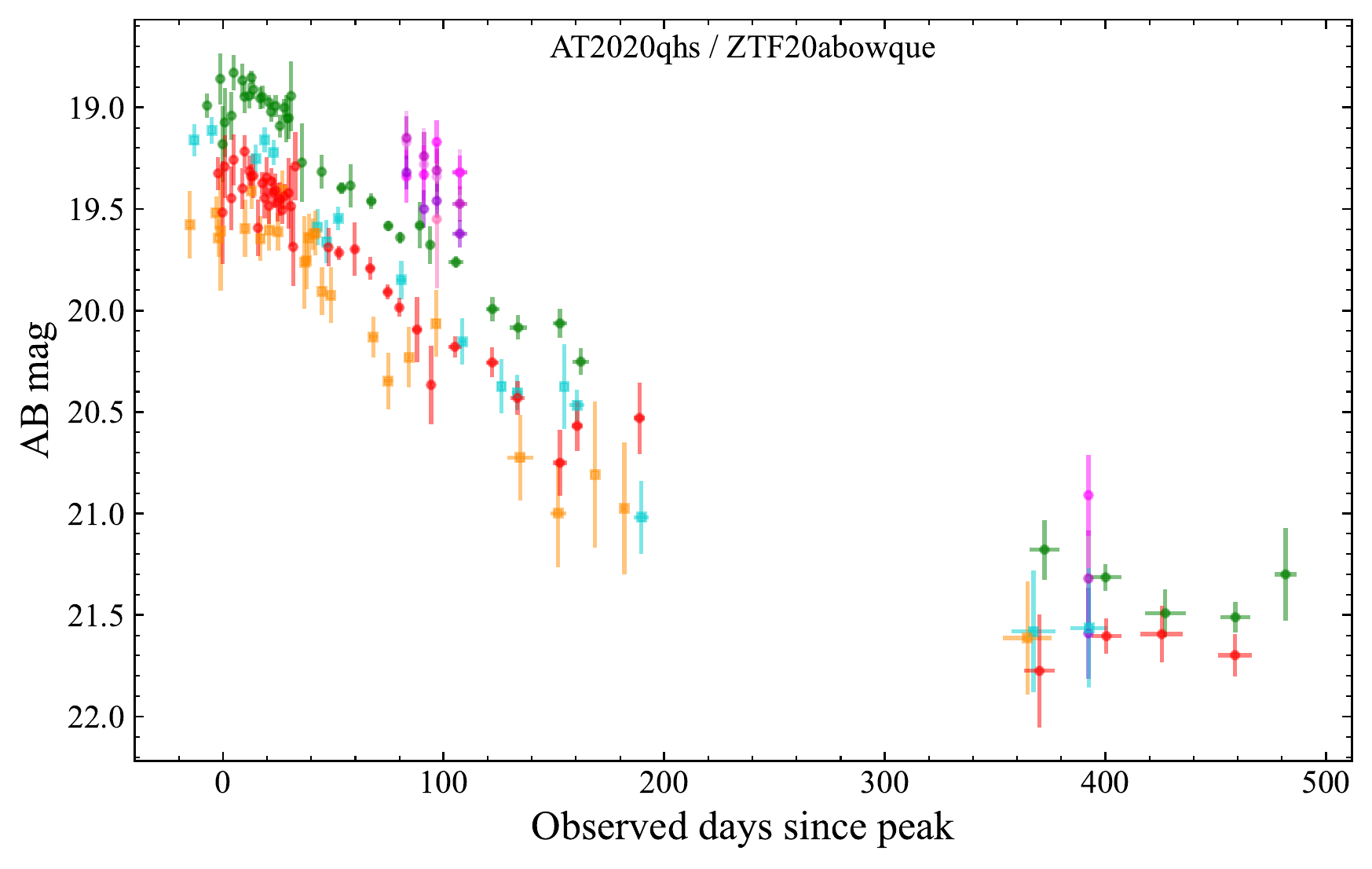}{0.33 \textwidth}{}
             \fig{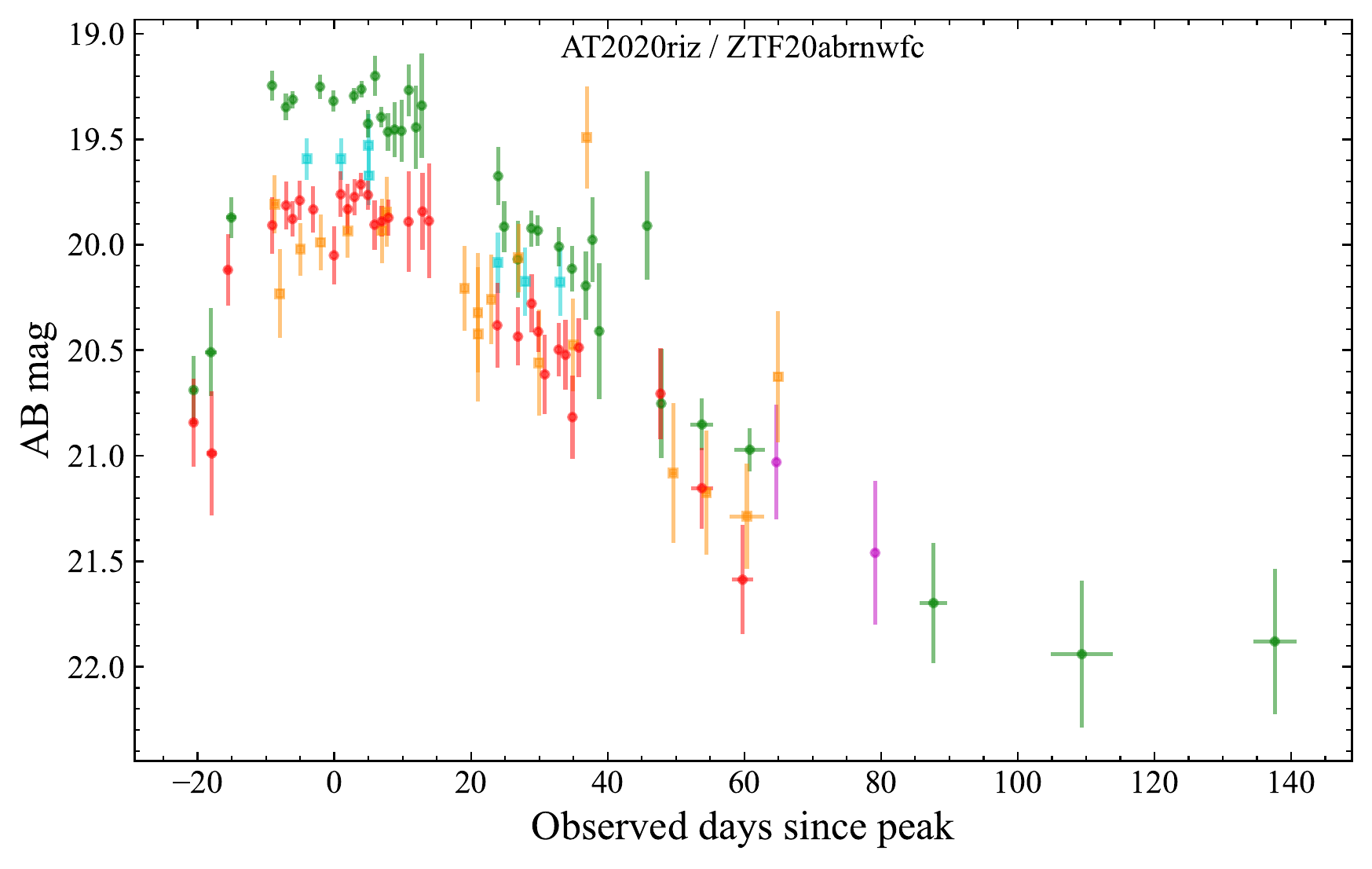}{0.33 \textwidth}{}
             \\[-20pt]}

\gridline{  \fig{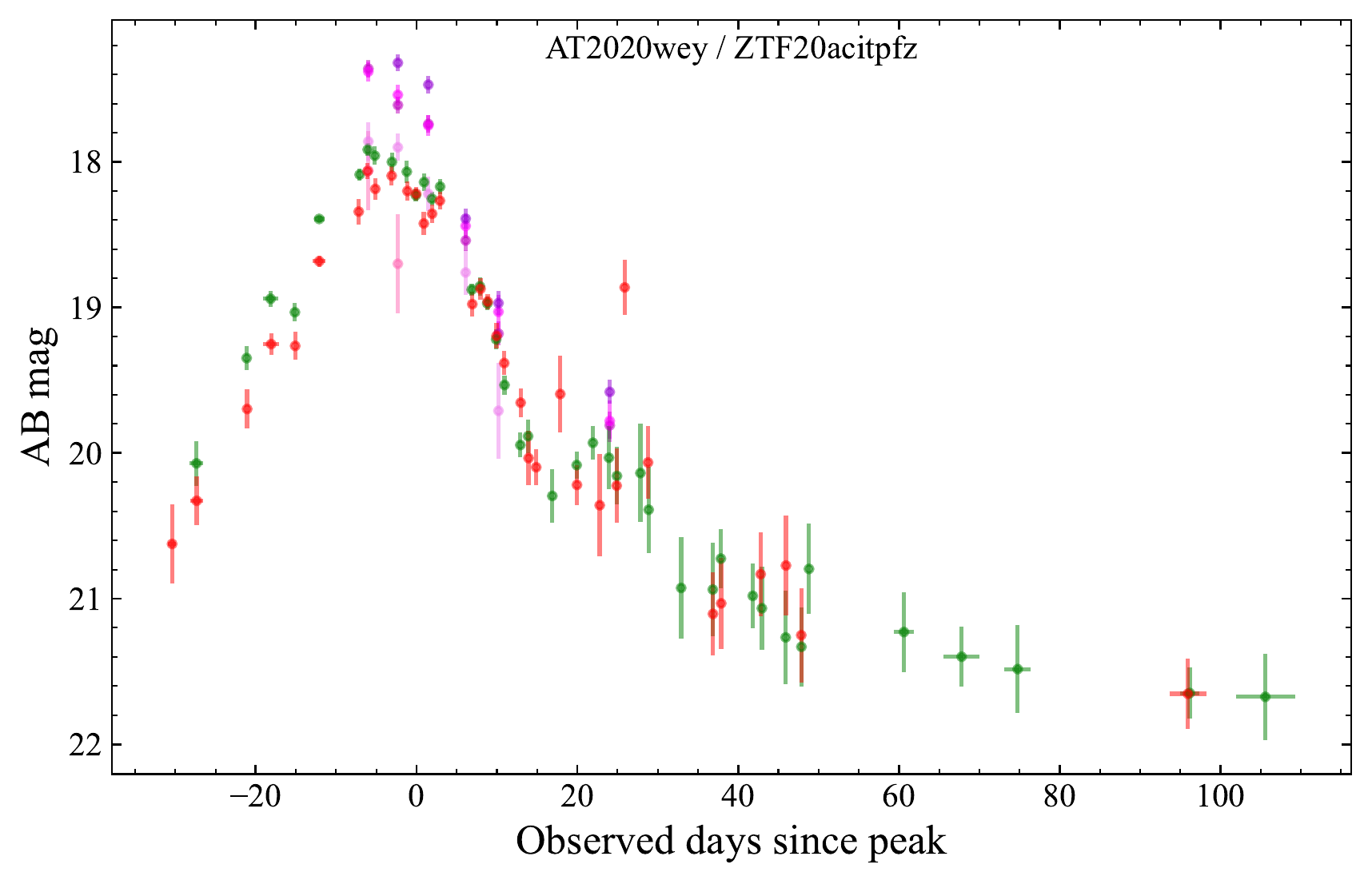}{0.33 \textwidth}{} 
            \fig{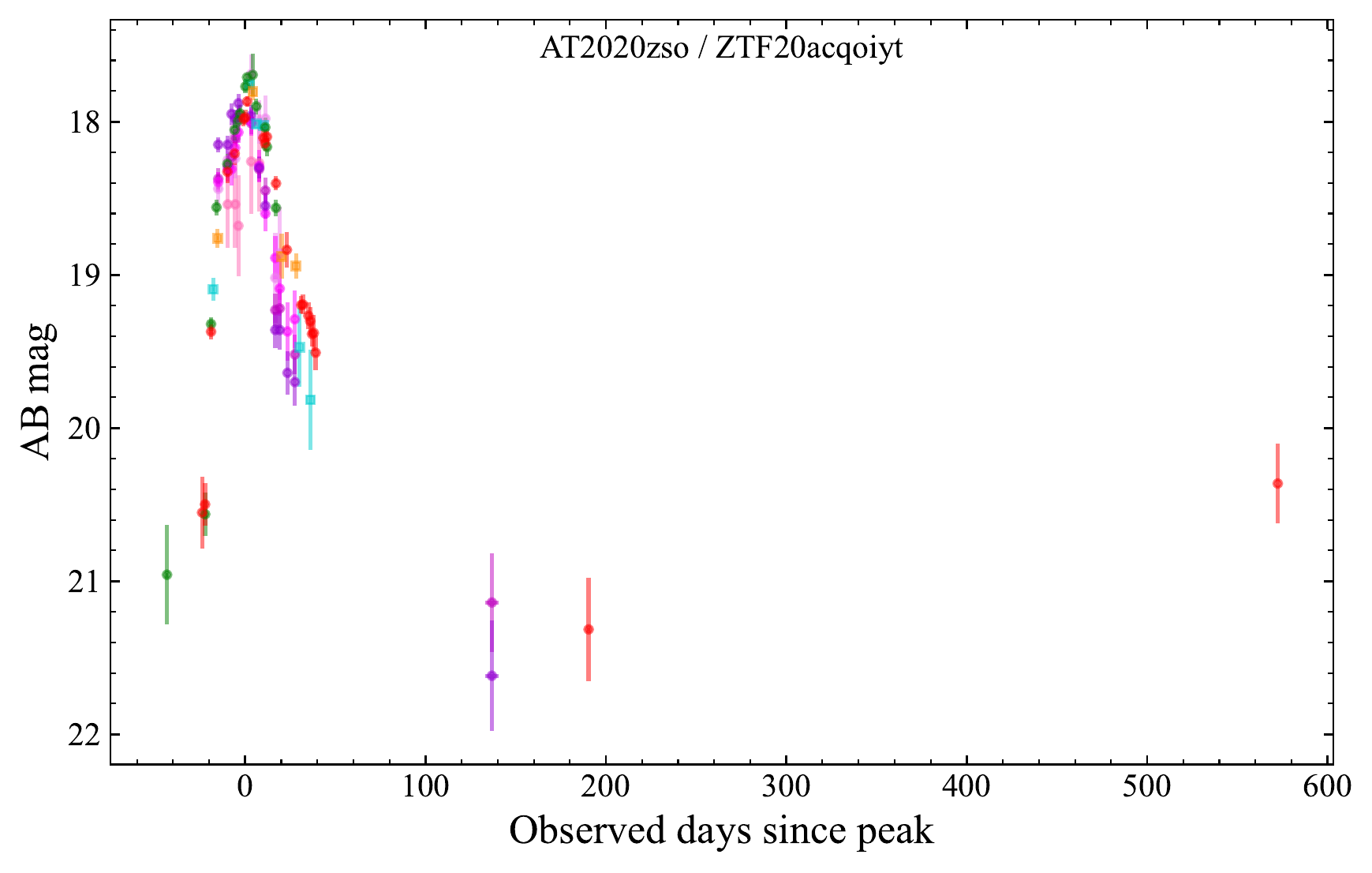}{0.33 \textwidth}{}
            \fig{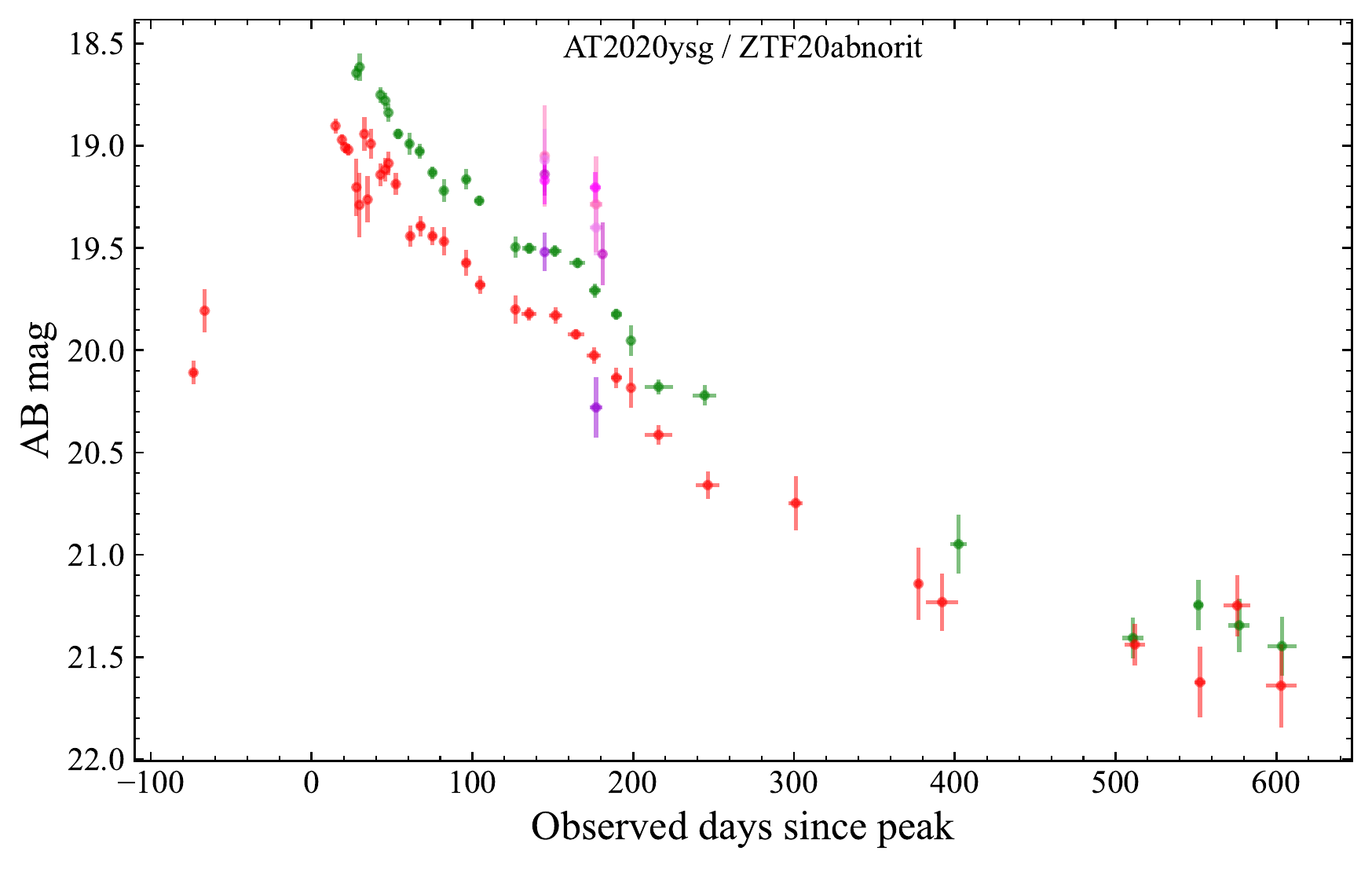}{0.33 \textwidth}{} 
            \\[-20pt]}   

\caption{Same as Figure \ref{fig:obslcs1}.}
\end{figure*}

\begin{figure*}
\gridline{	\fig{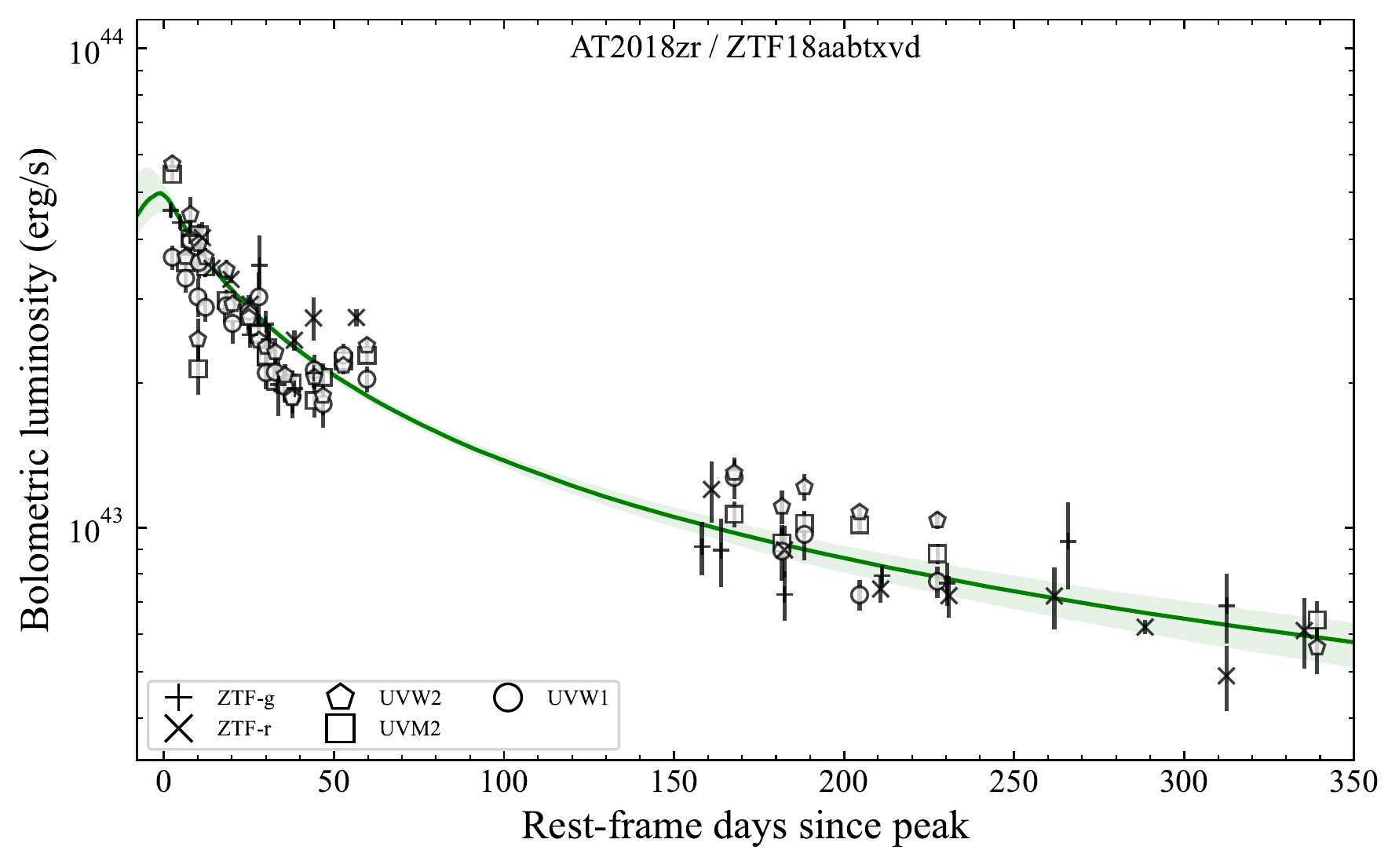}{0.33 \textwidth}{} 
			\fig{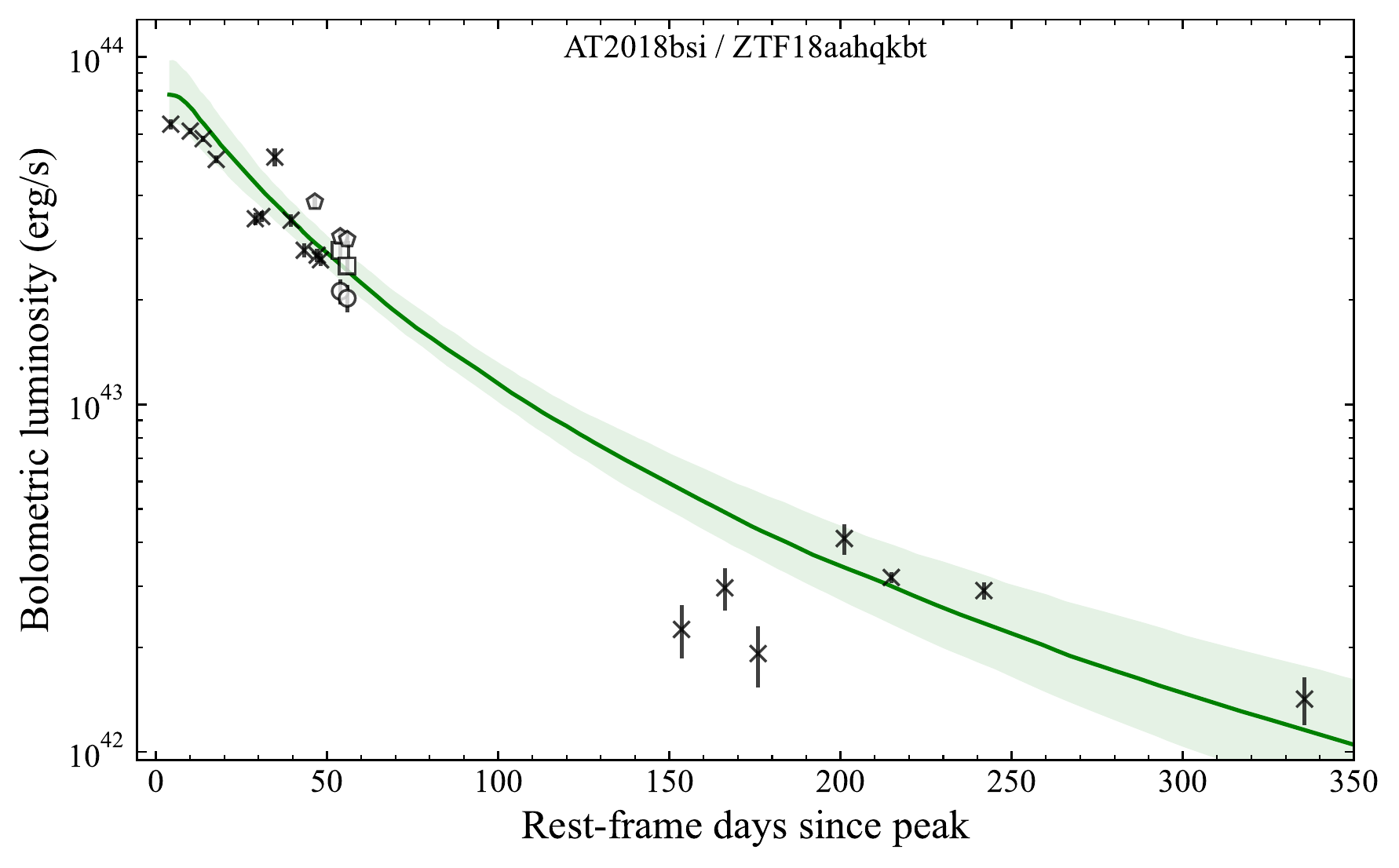}{0.33 \textwidth}{}
			\fig{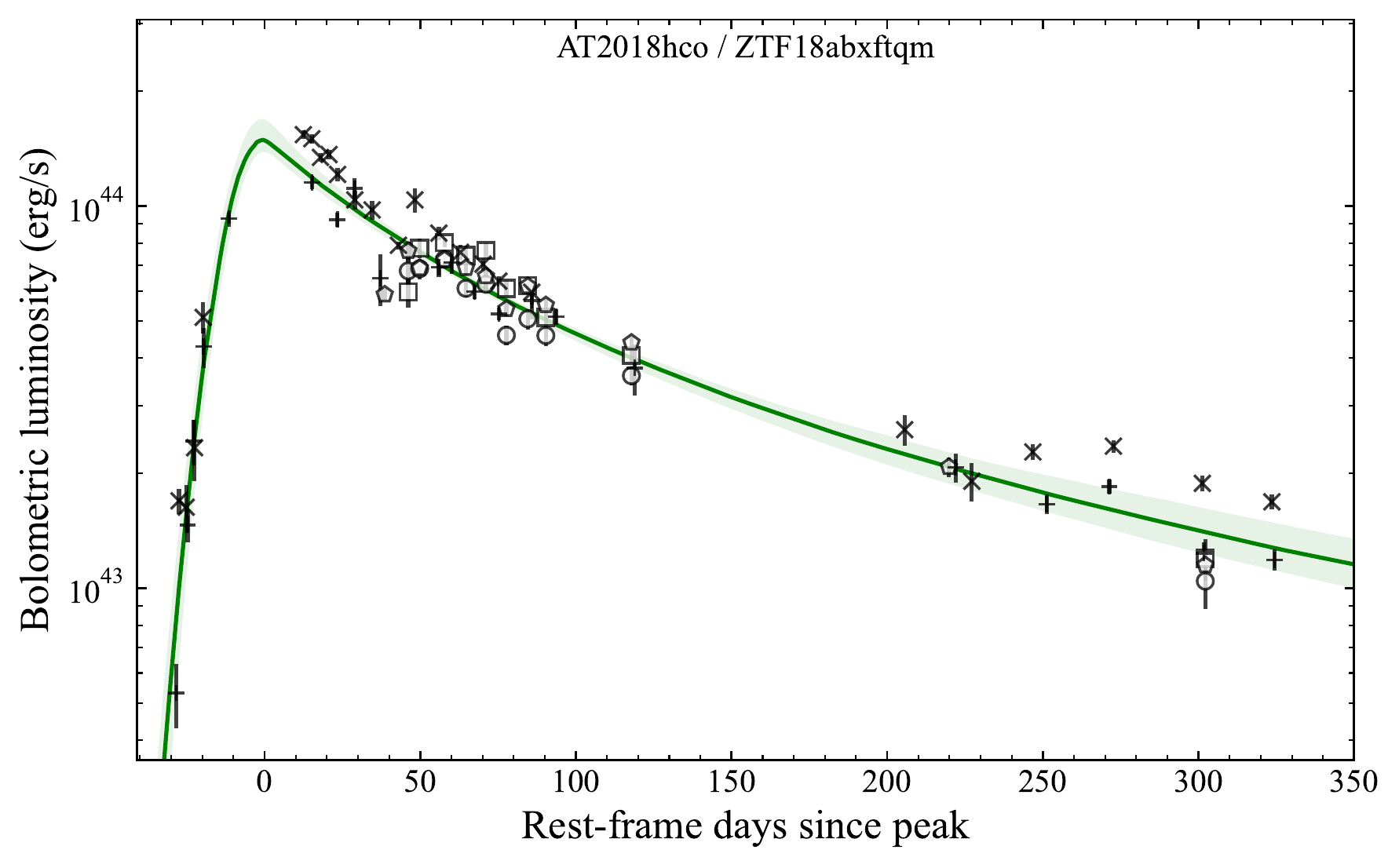}{0.33 \textwidth}{}
			\\[-20pt]}
			
\gridline{\fig{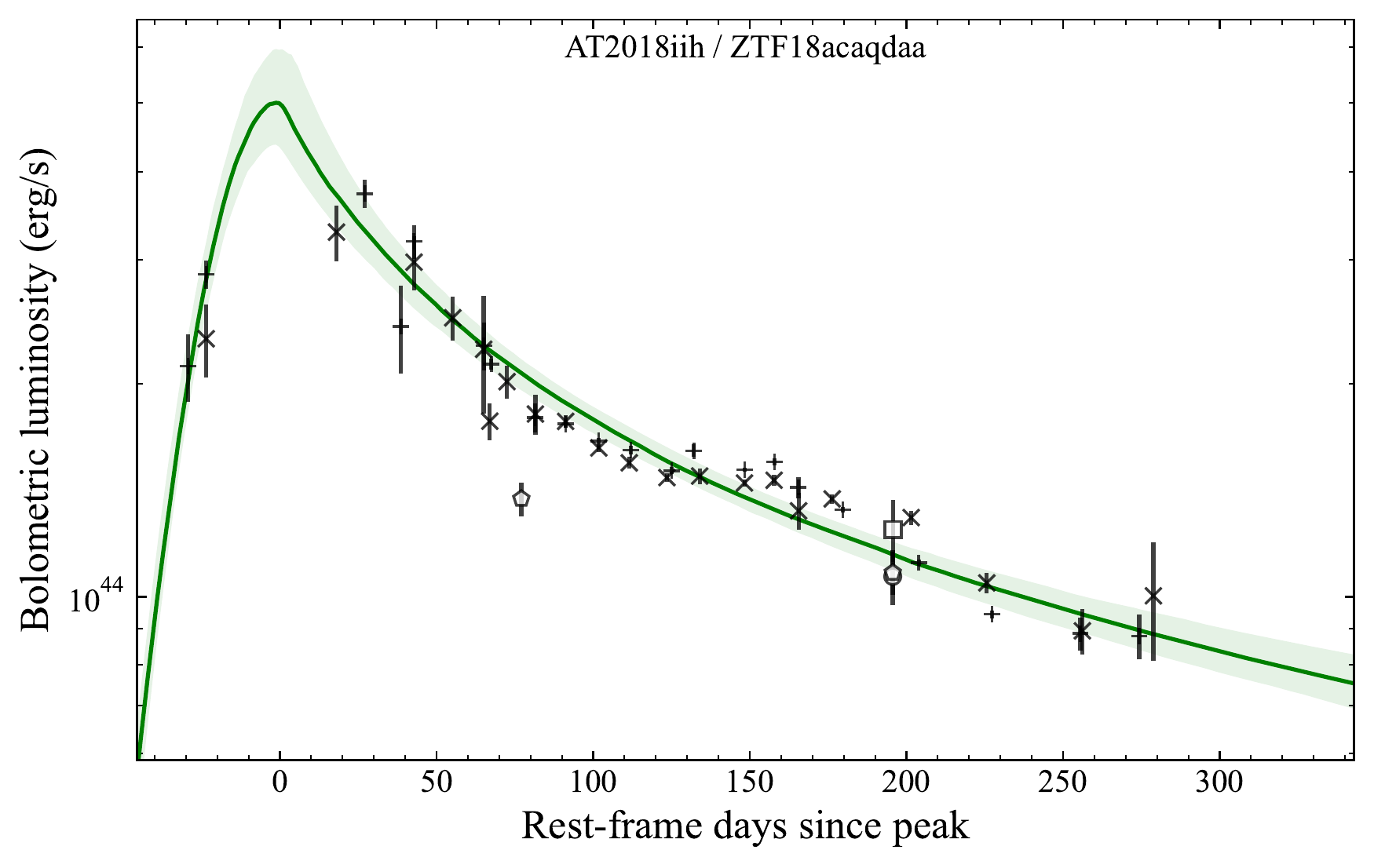}{0.33 \textwidth}{} 
\fig{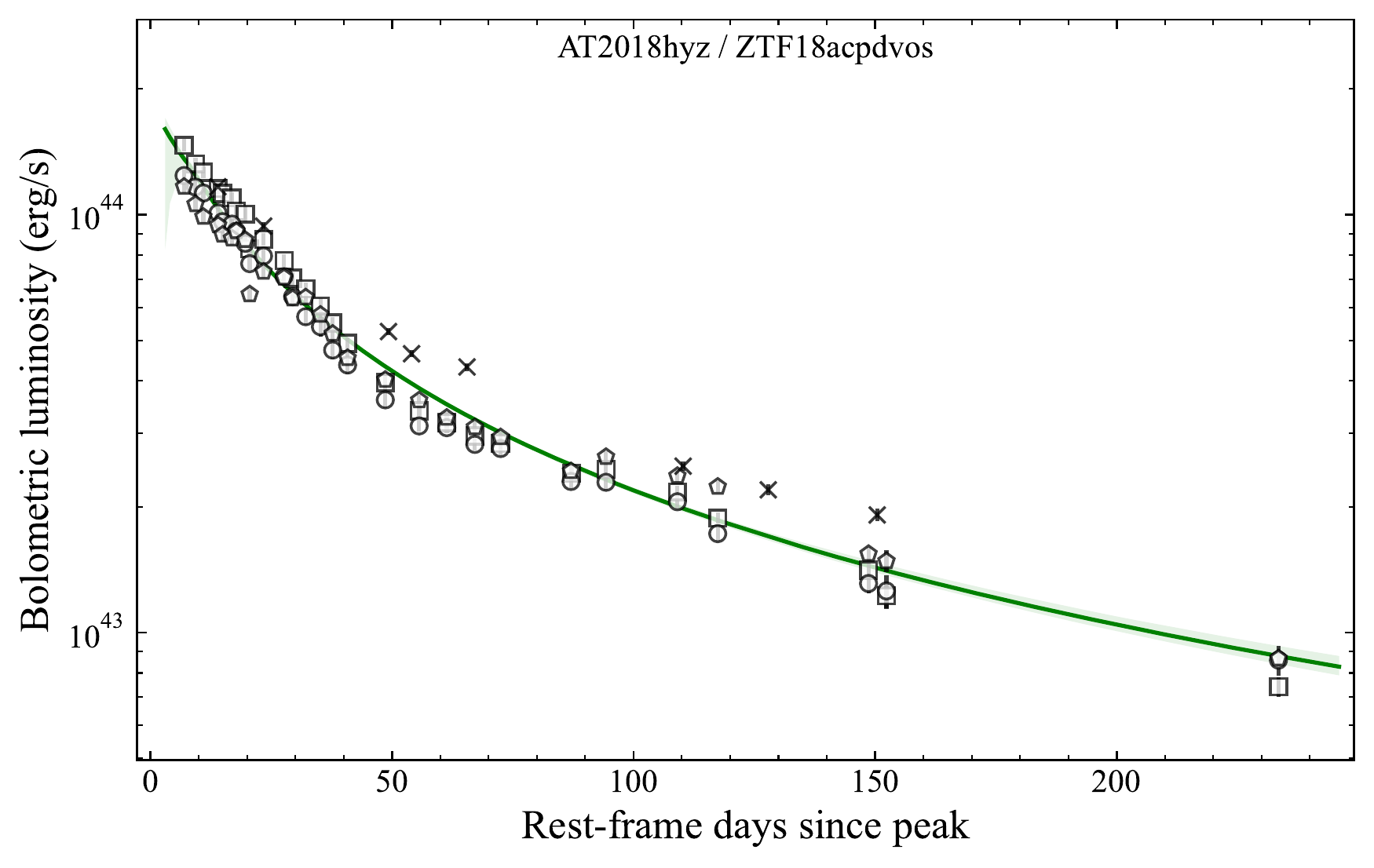}{0.33 \textwidth}{} 
            \fig{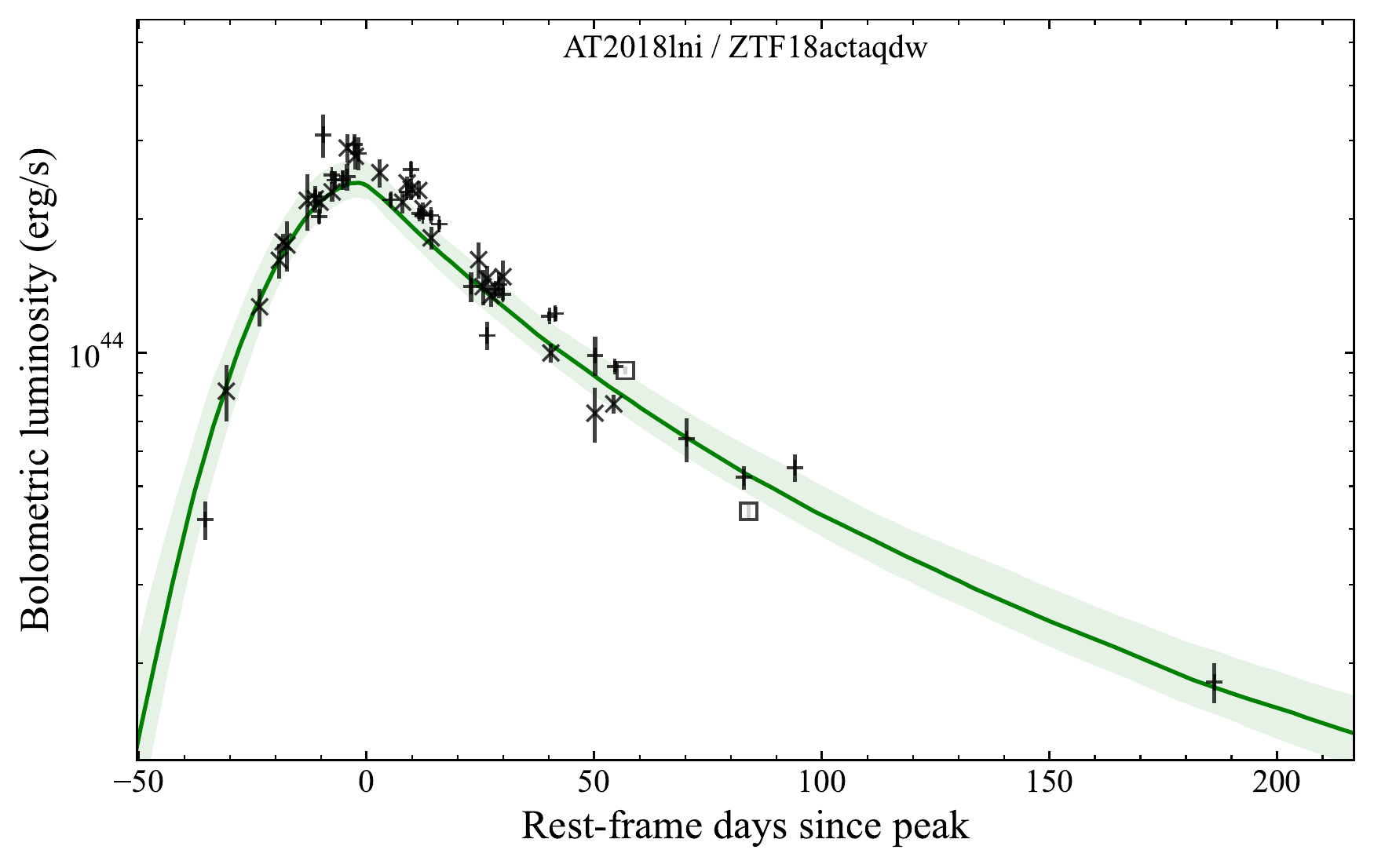}{0.33 \textwidth}{}
            \\[-20pt]}
     
\gridline{  \fig{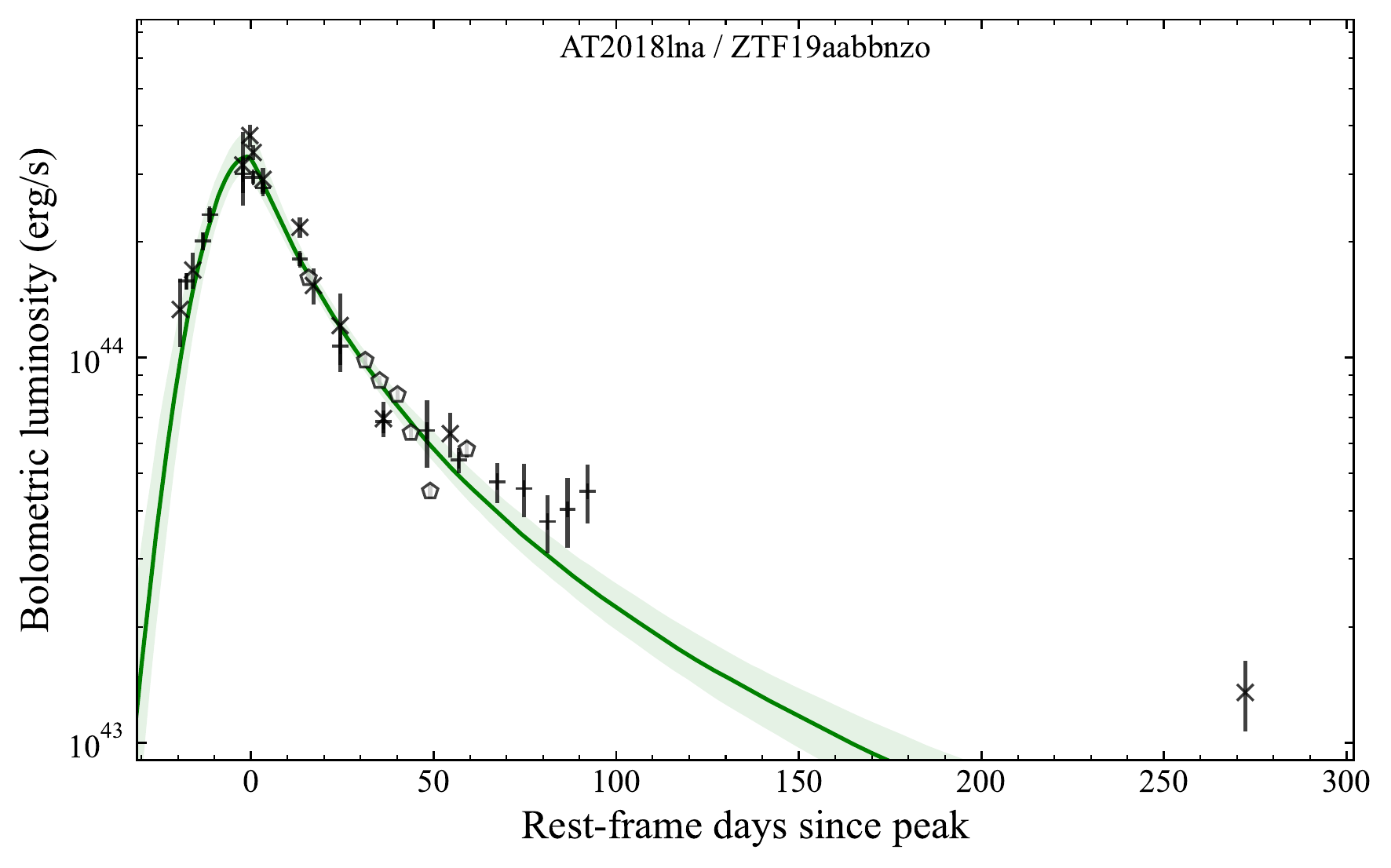}{0.33 \textwidth}{}
            \fig{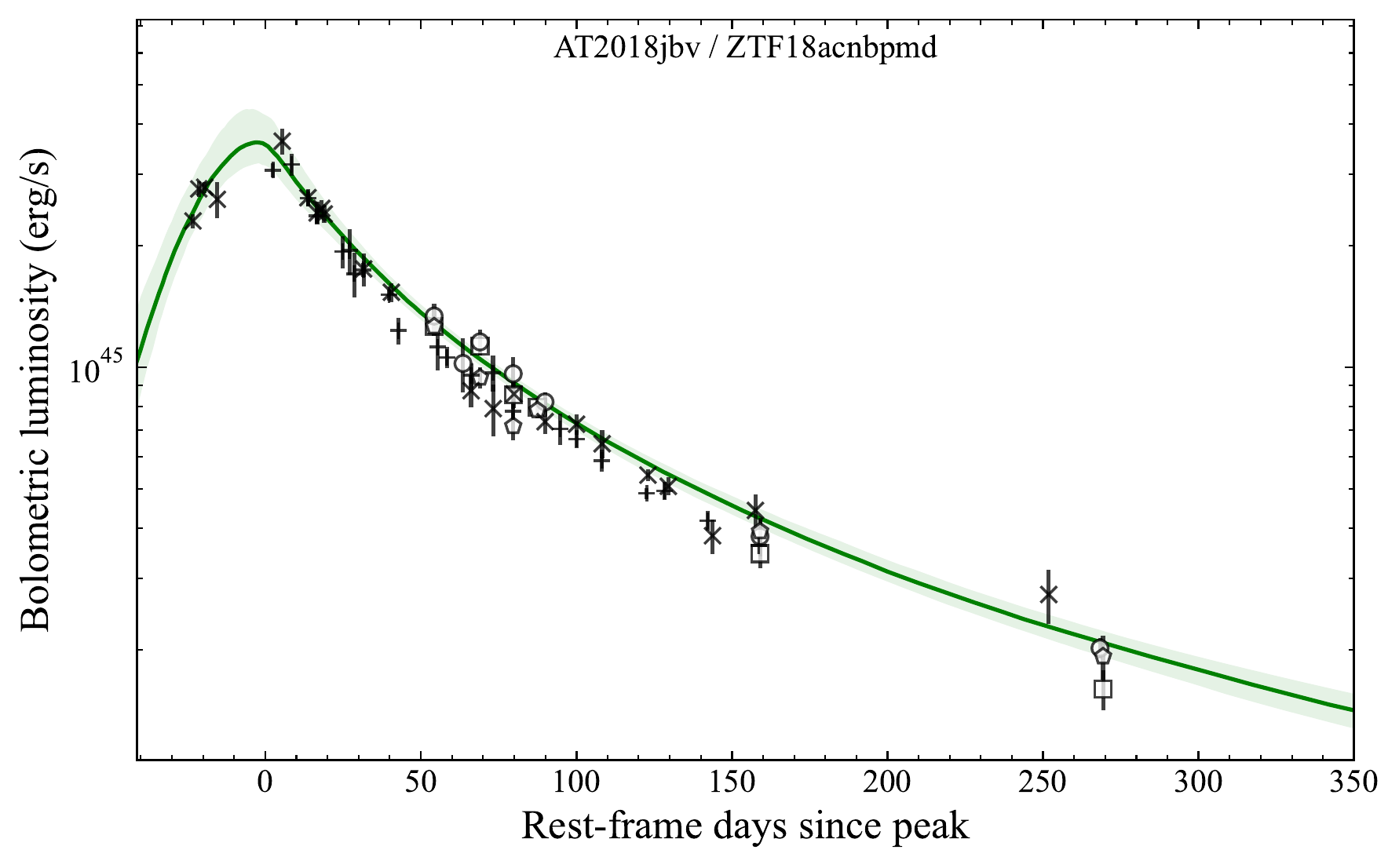}{0.33 \textwidth}{}
            \fig{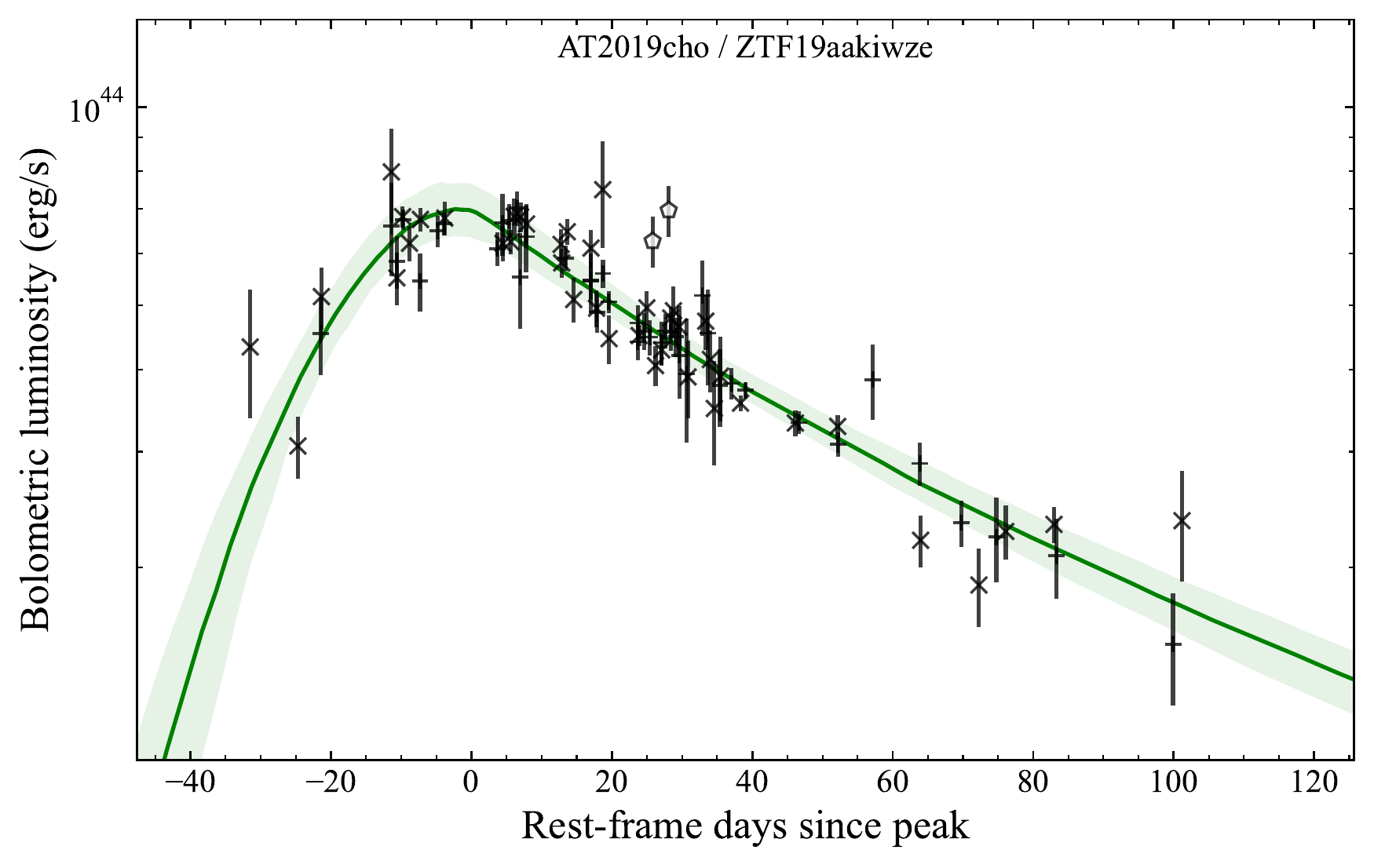}{0.33 \textwidth}{} 
             \\[-20pt]}

\gridline{\fig{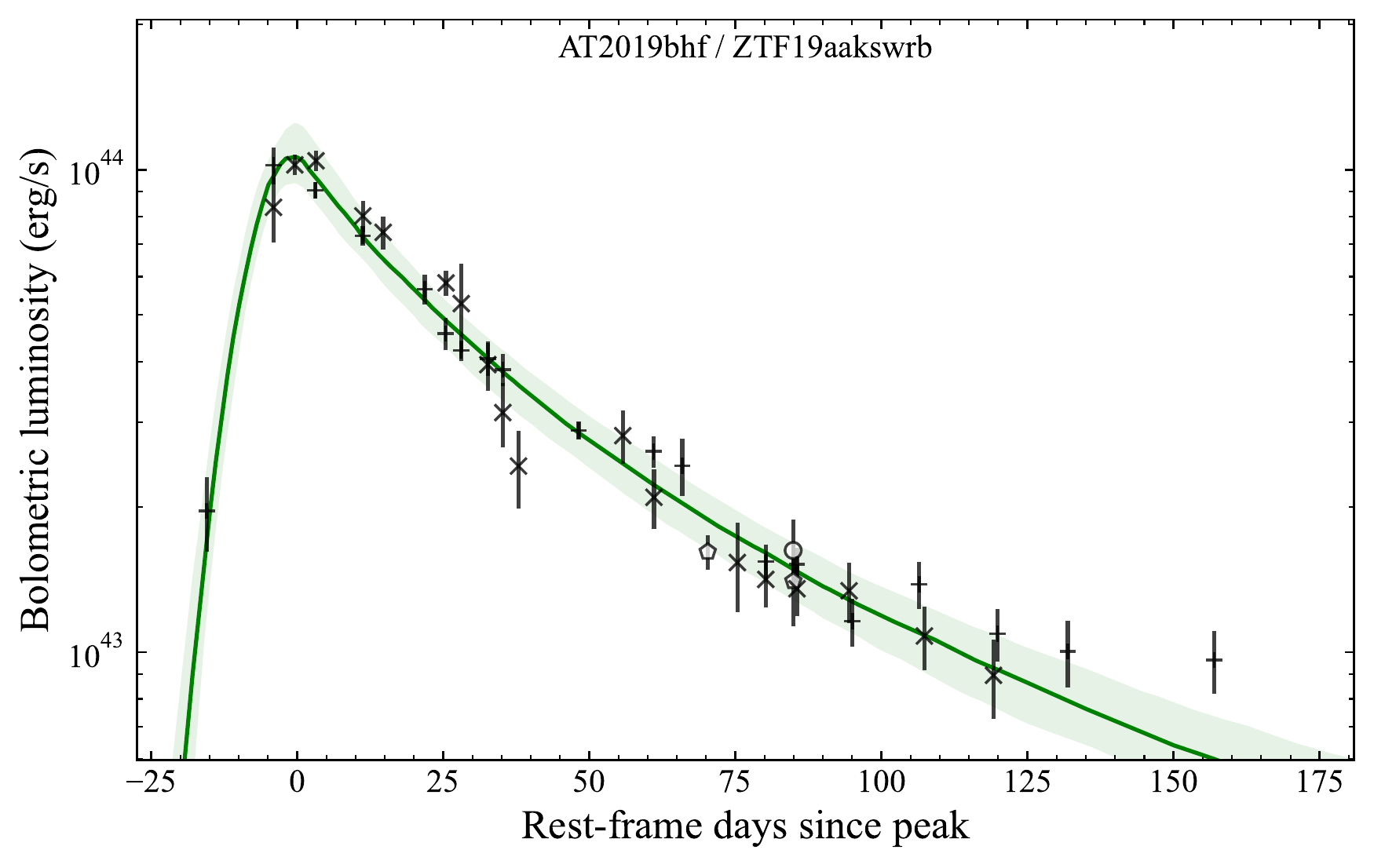}{0.33 \textwidth}{}
	\fig{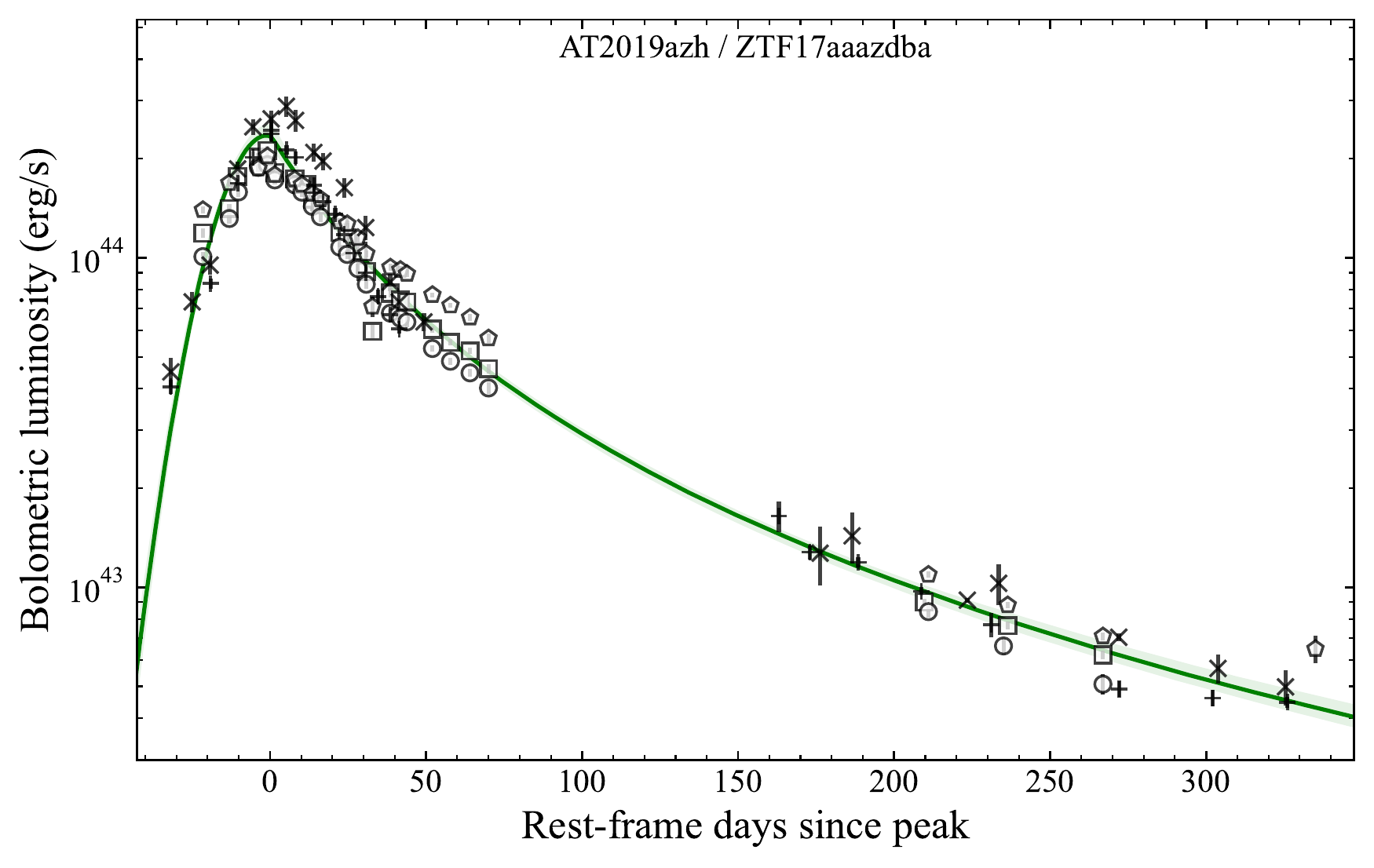}{0.33 \textwidth}{} 
            \fig{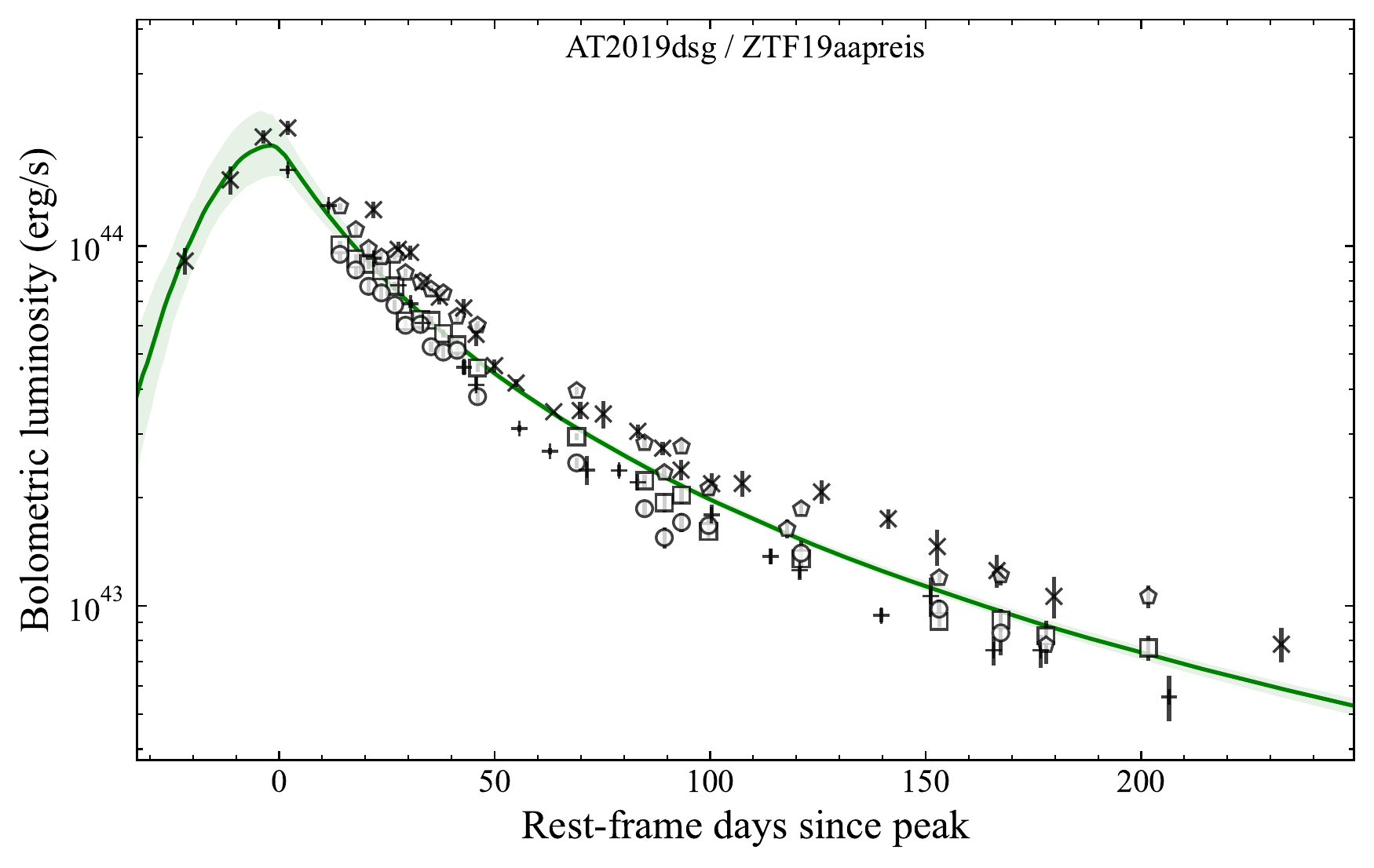}{0.33 \textwidth}{}
            \\[-20pt]} 
\gridline{  \fig{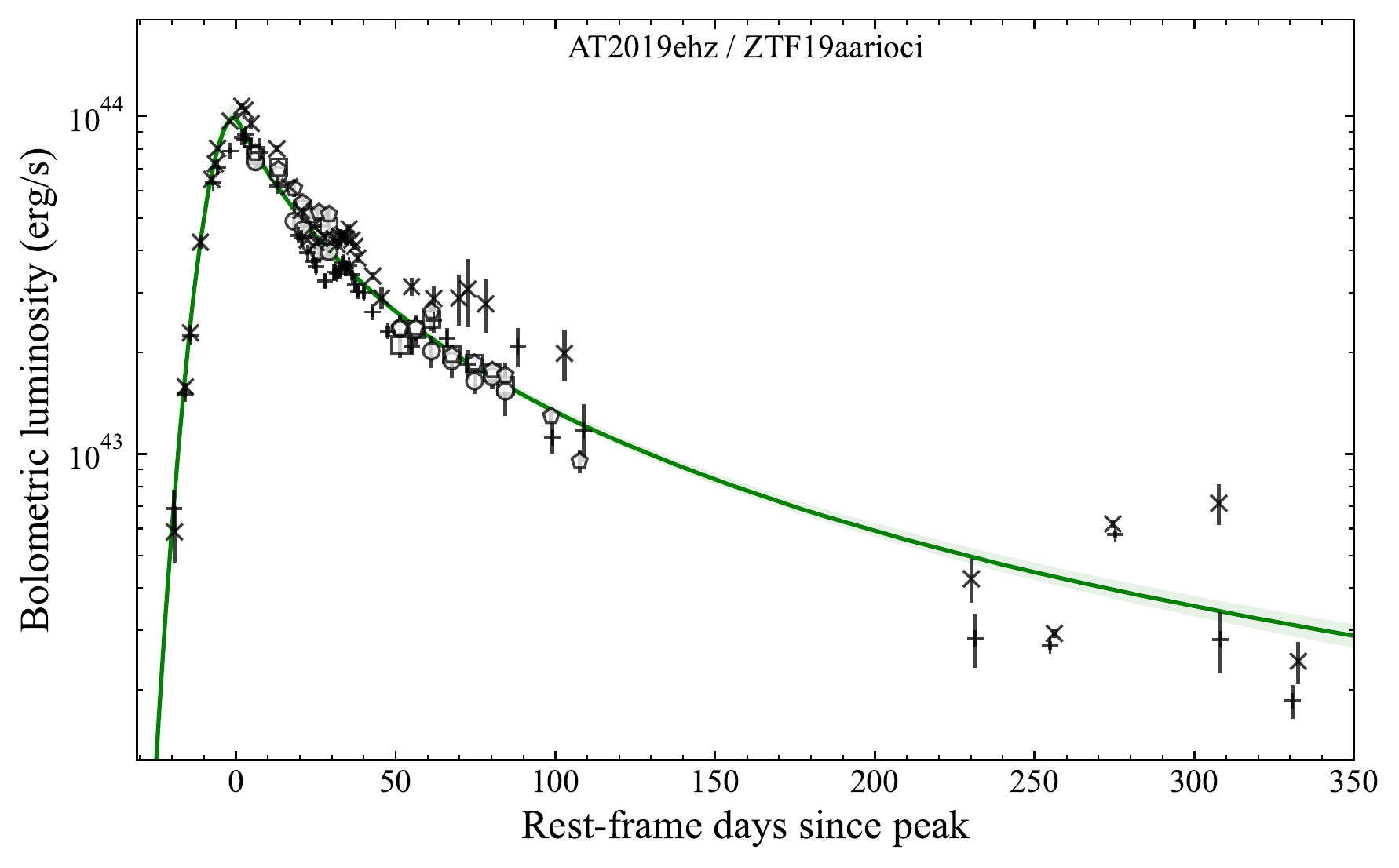}{0.33 \textwidth}{} 
            \fig{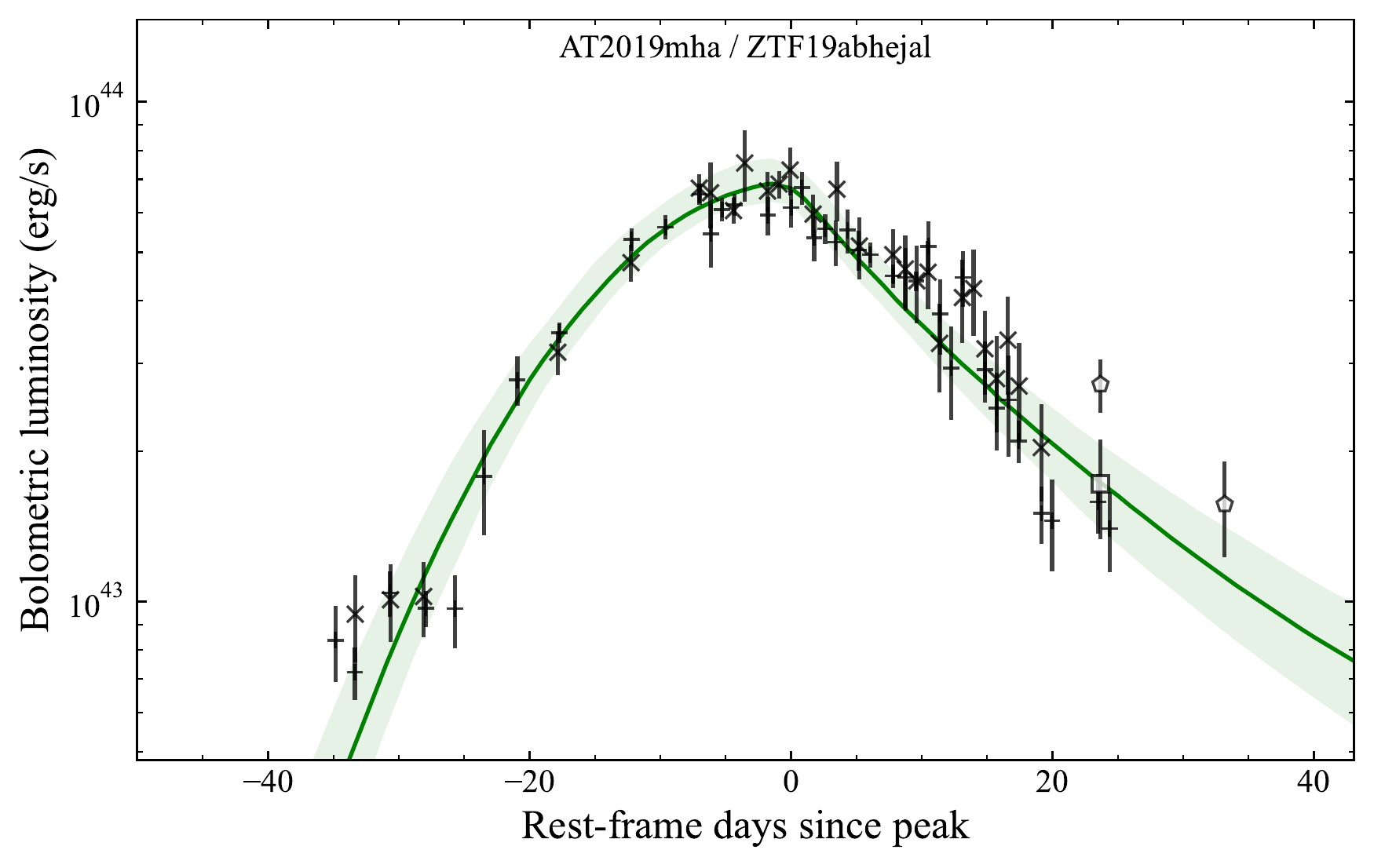}{0.33 \textwidth}{}
            \fig{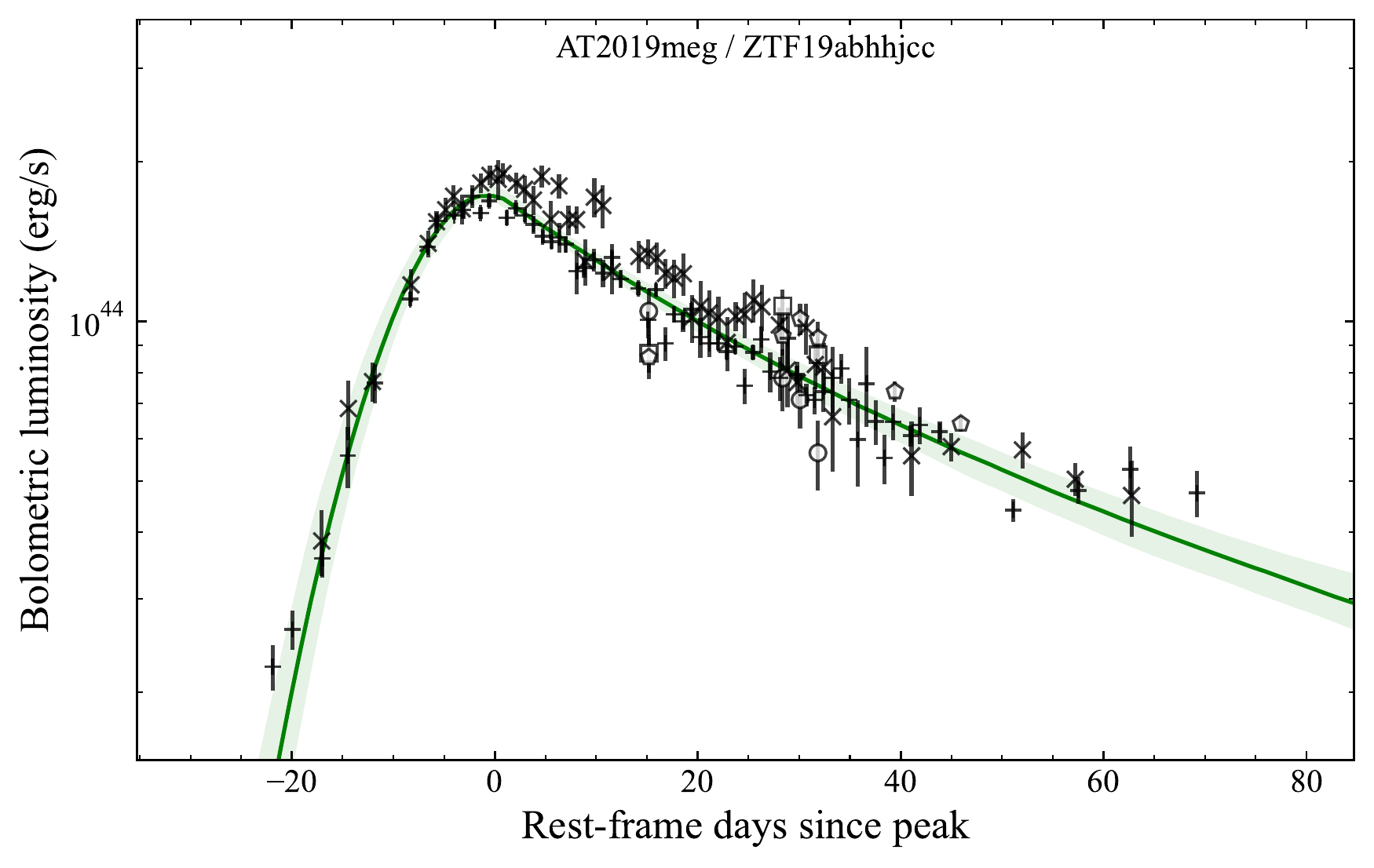}{0.33 \textwidth}{} 
            \\[-20pt]}

\caption{Gaussian rise and power-law decay fits with flexible temperature fitting, shown with the optical and UV 3-$\sigma$ detections binned as in Figure \ref{fig:obslcs1}. We also show the 1-$\sigma$ spread in uncertainty of the fit. The legend can be seen in the top left panel.}\label{fig:fitlcs1}
\end{figure*}

\begin{figure*}
            
\gridline{\fig{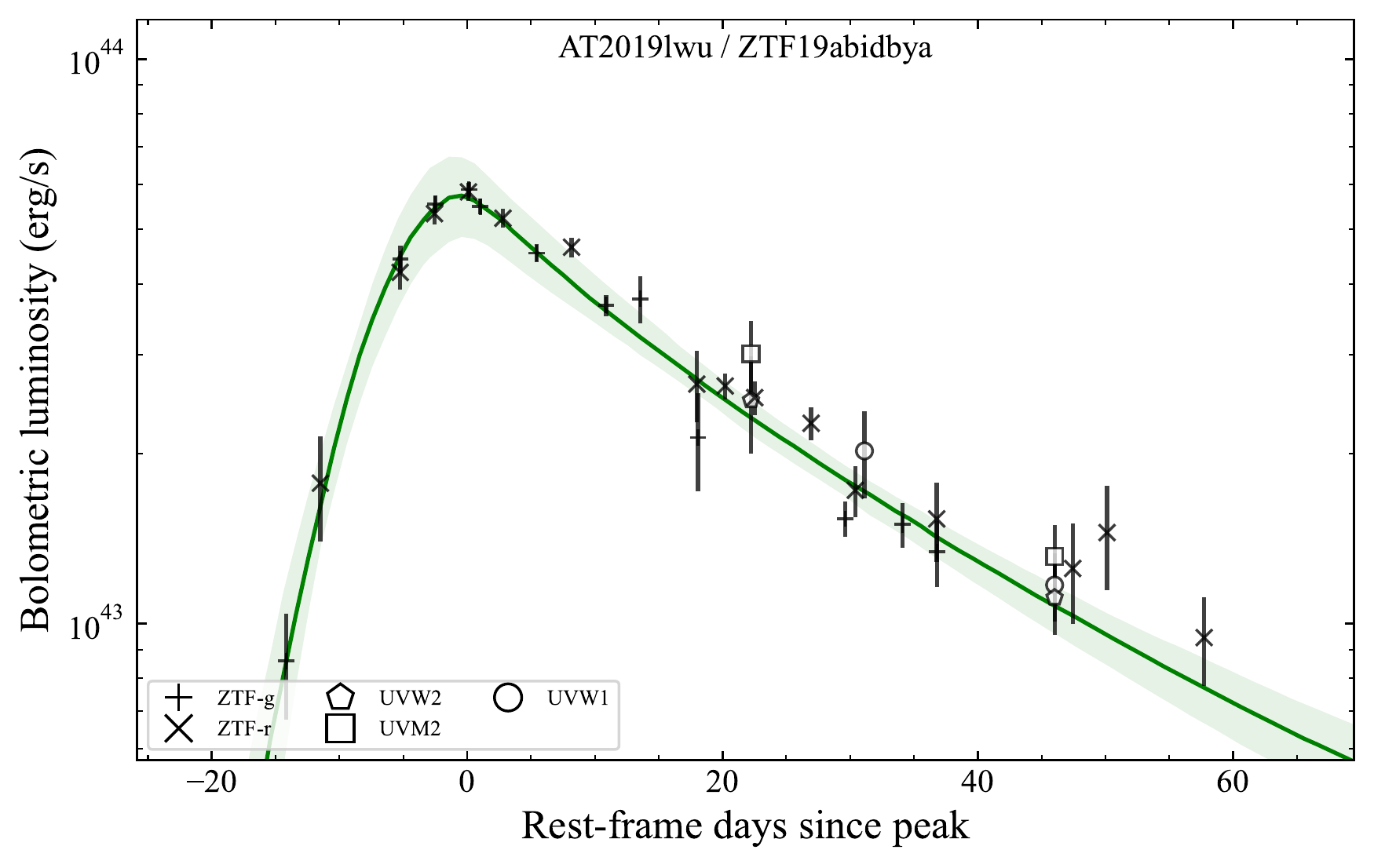}{0.33 \textwidth}{}
  \fig{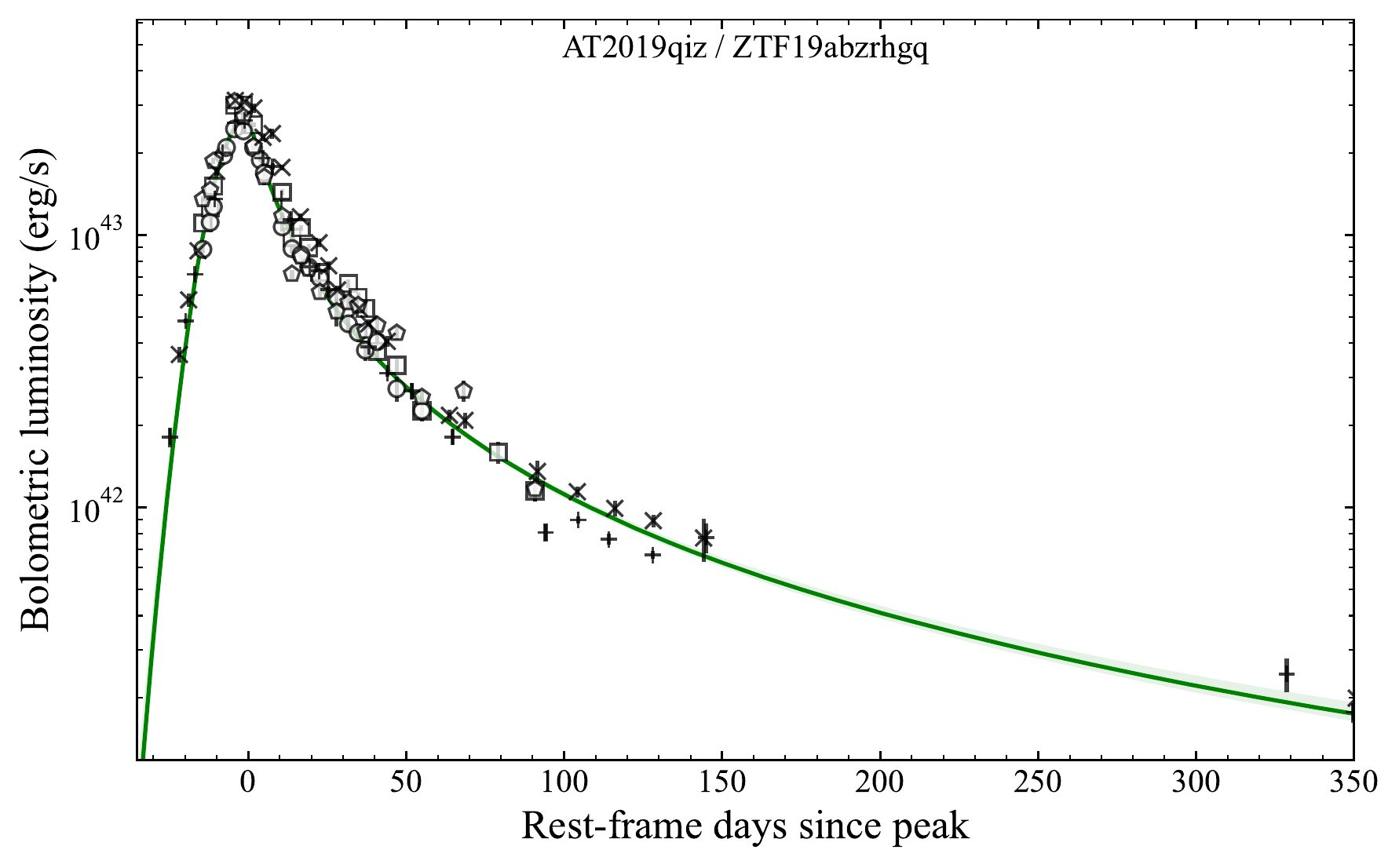}{0.33 \textwidth}{}
            \fig{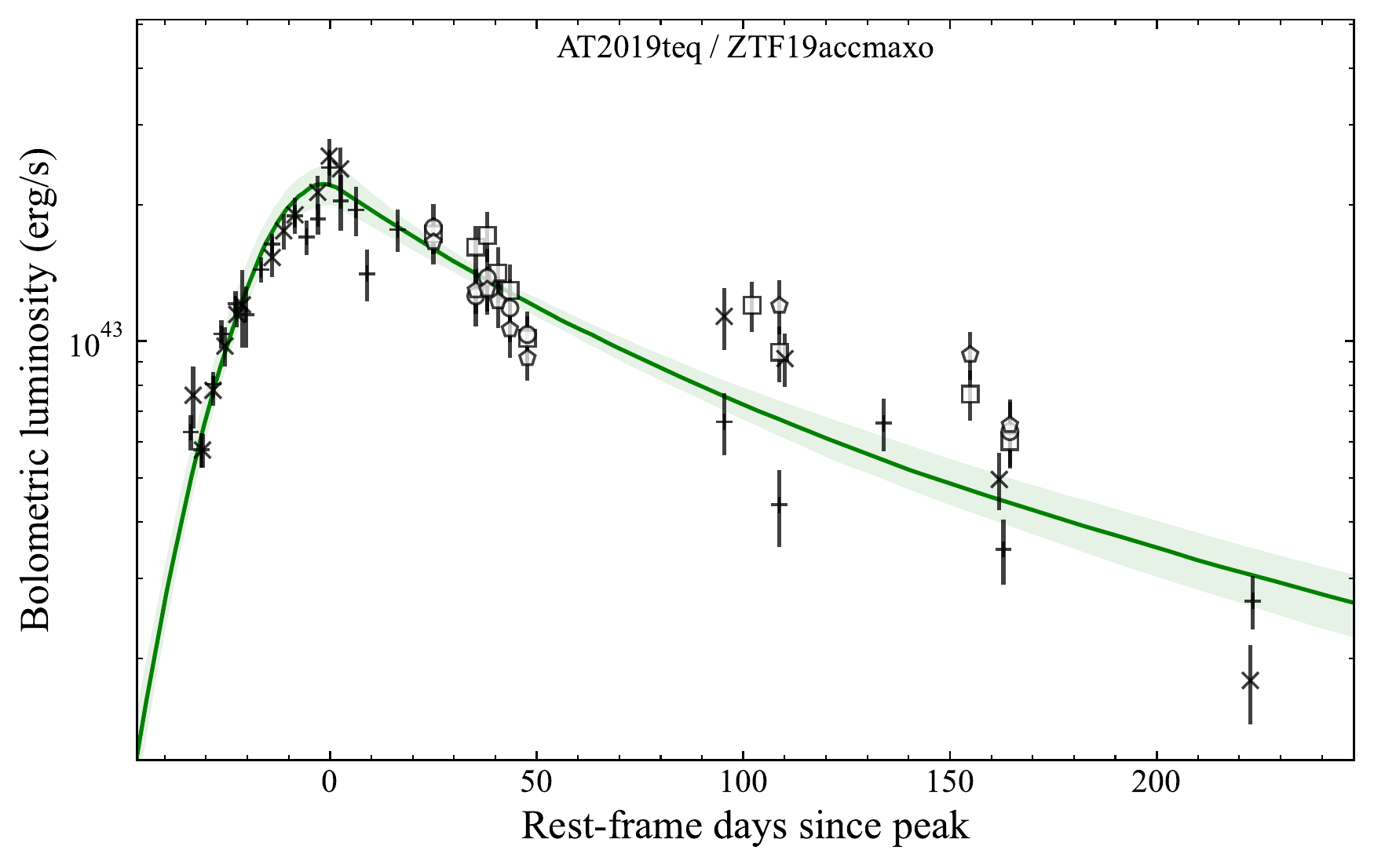}{0.33 \textwidth}{}
            \\[-20pt]}          

\gridline{  \fig{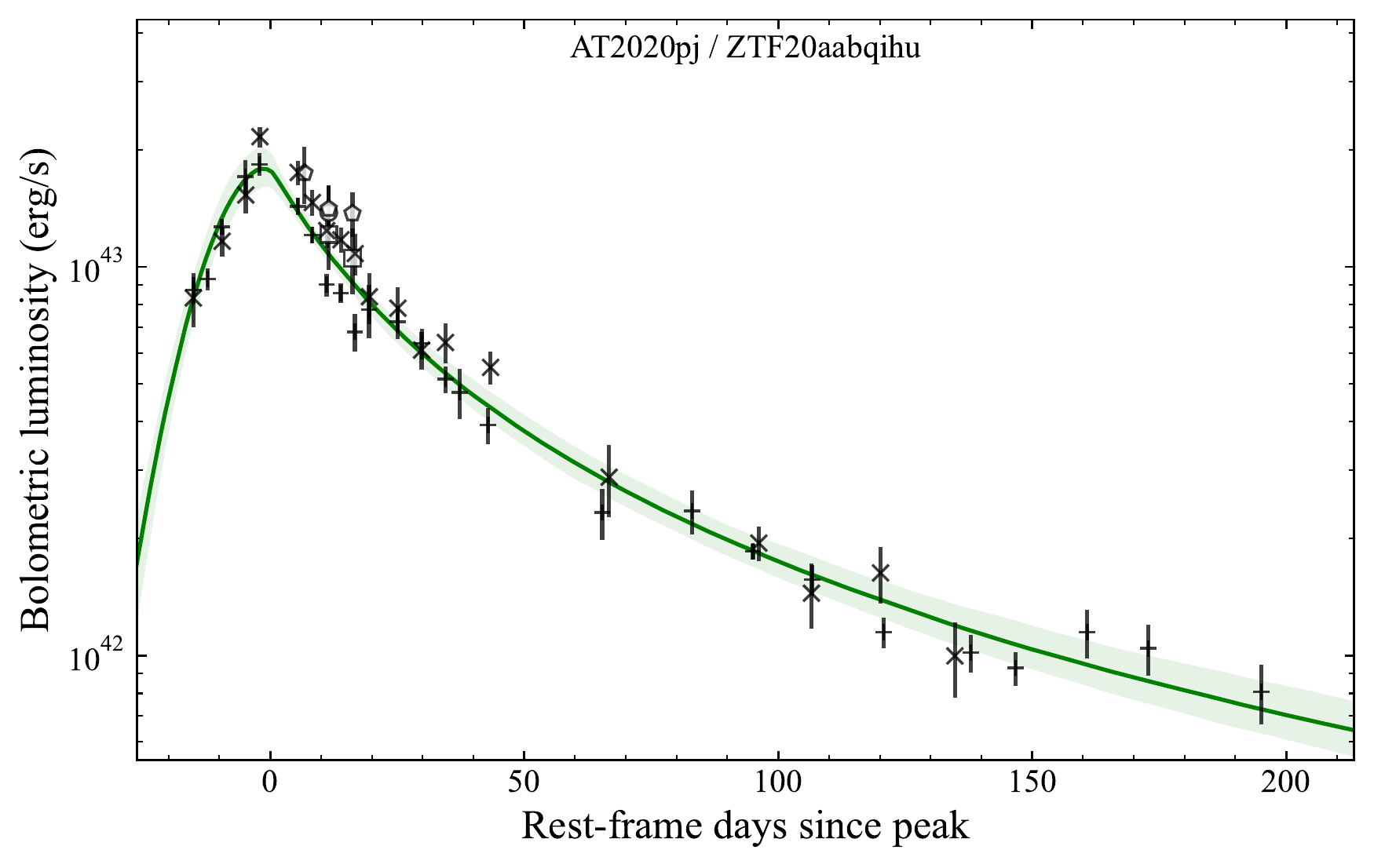}{0.33 \textwidth}{}
            \fig{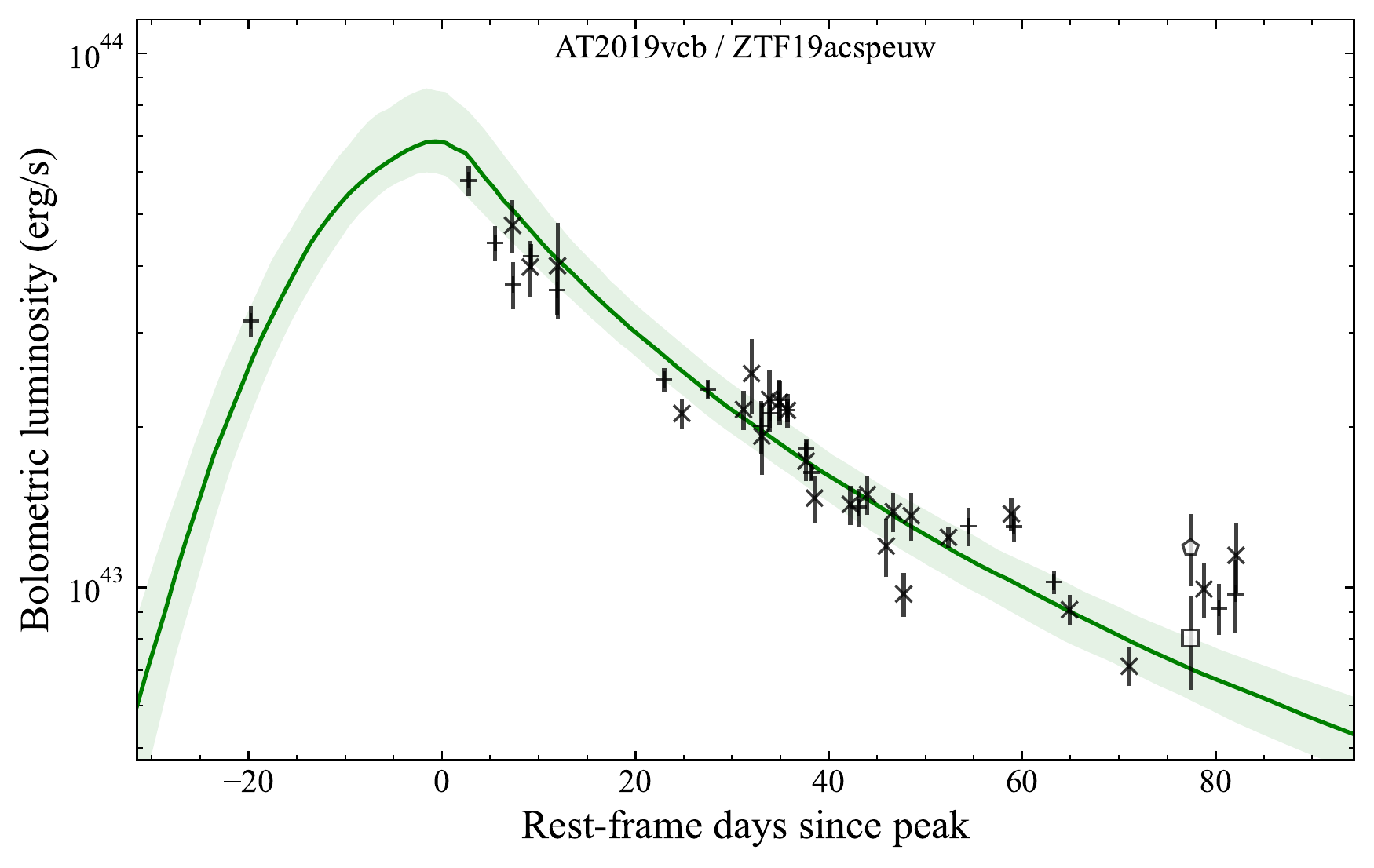}{0.33 \textwidth}{}
            \fig{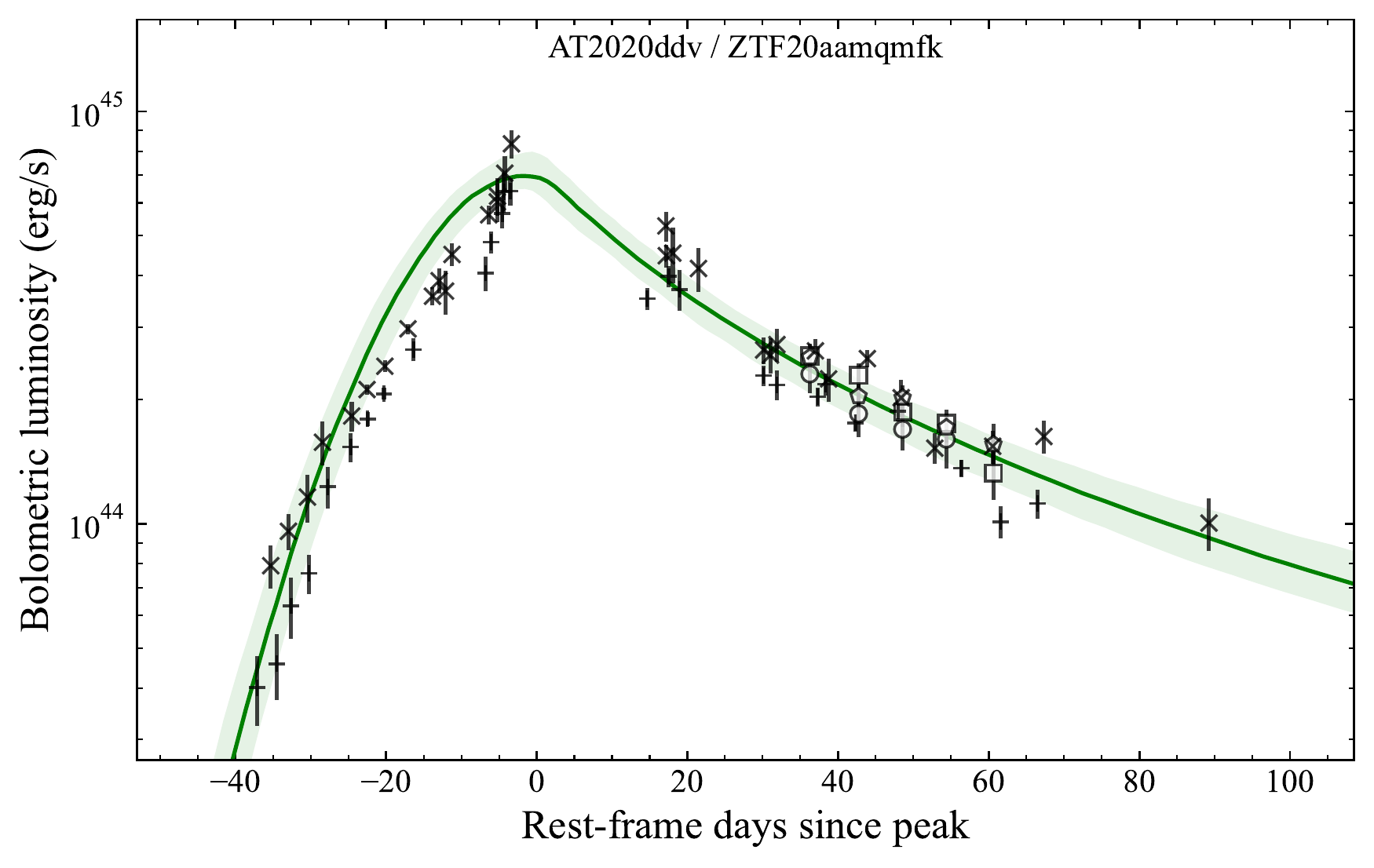}{0.33 \textwidth}{}
            \\[-20pt]}   

\gridline{	\fig{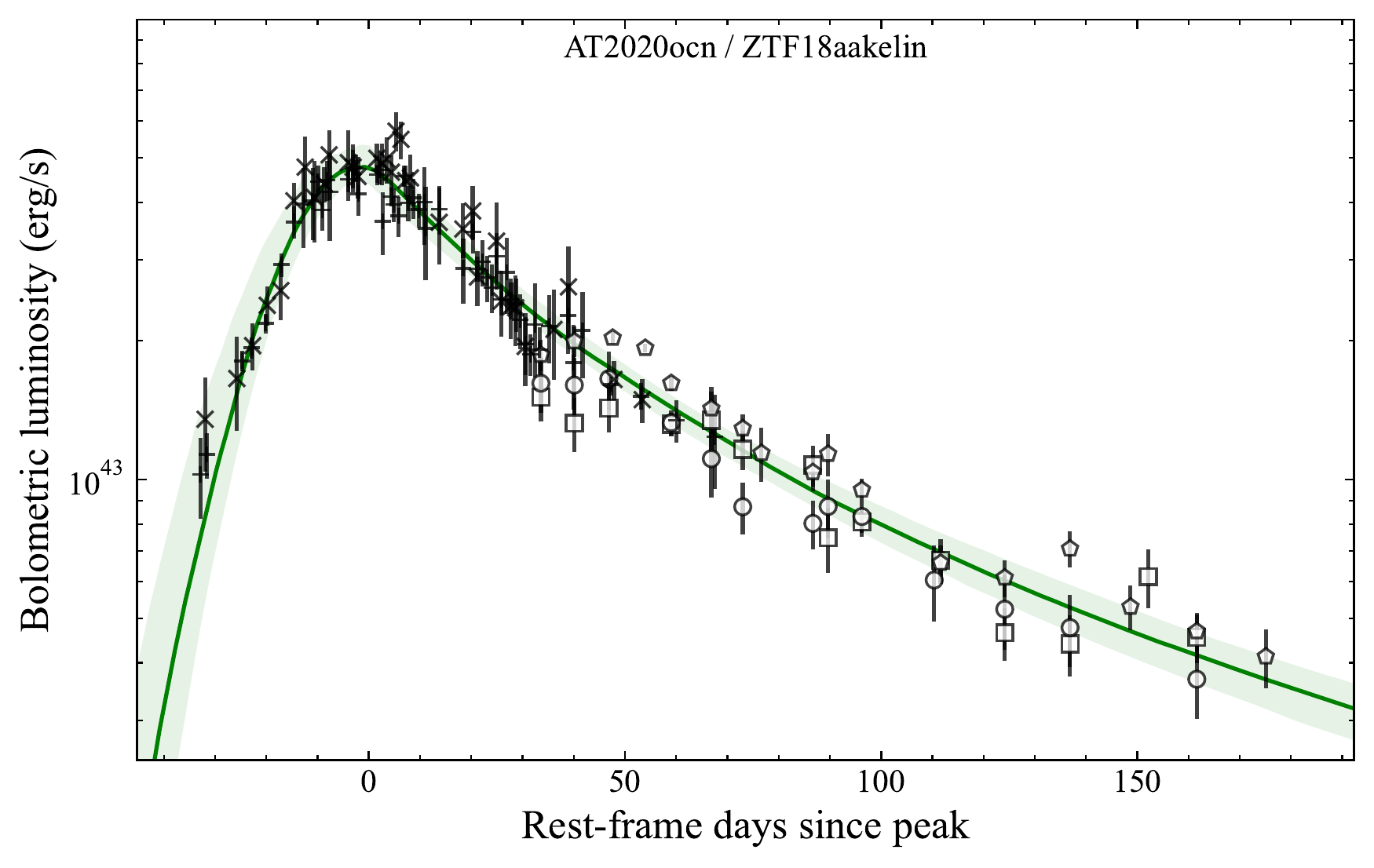}{0.33 \textwidth}{}
             \fig{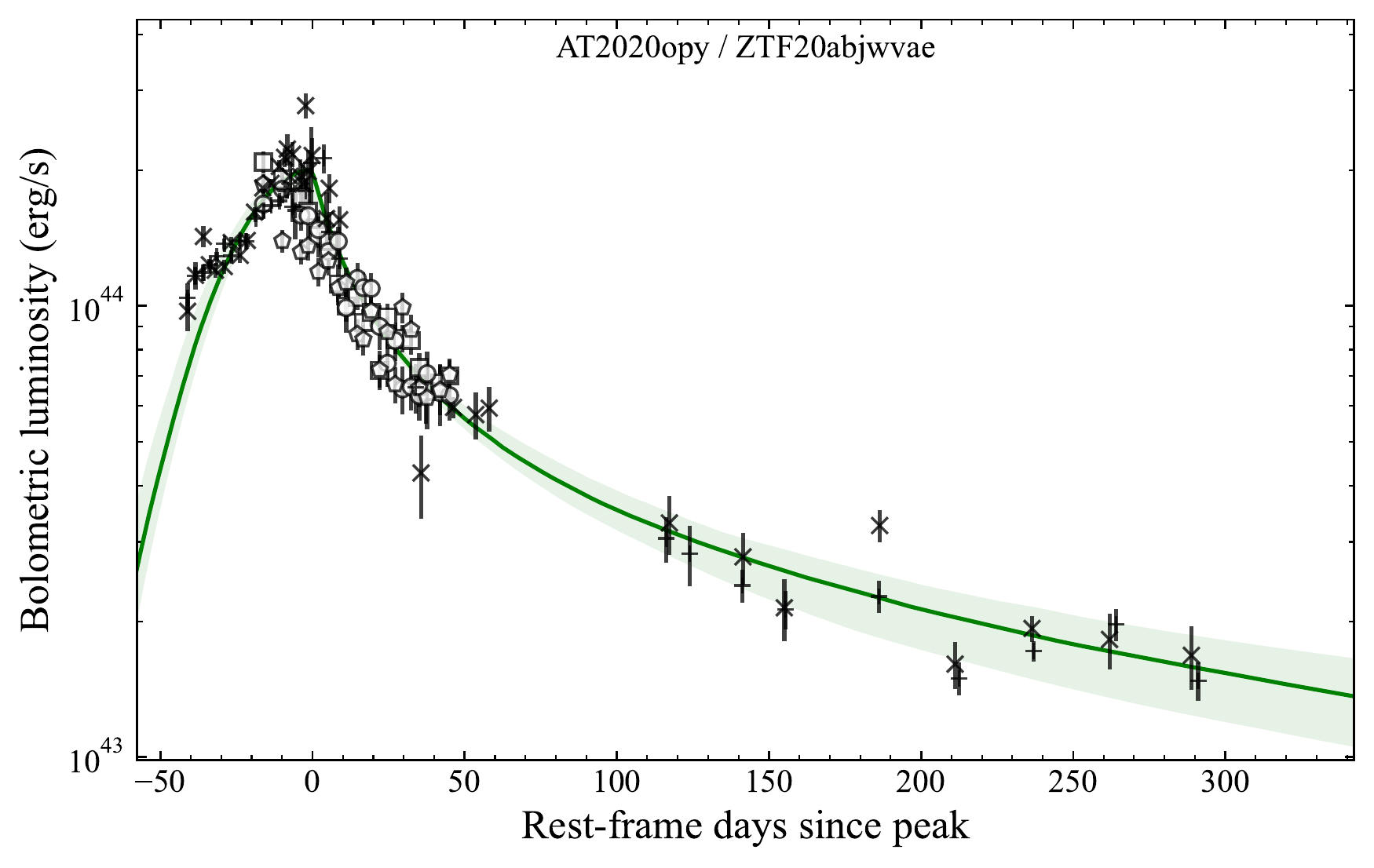}{0.33 \textwidth}{}
            \fig{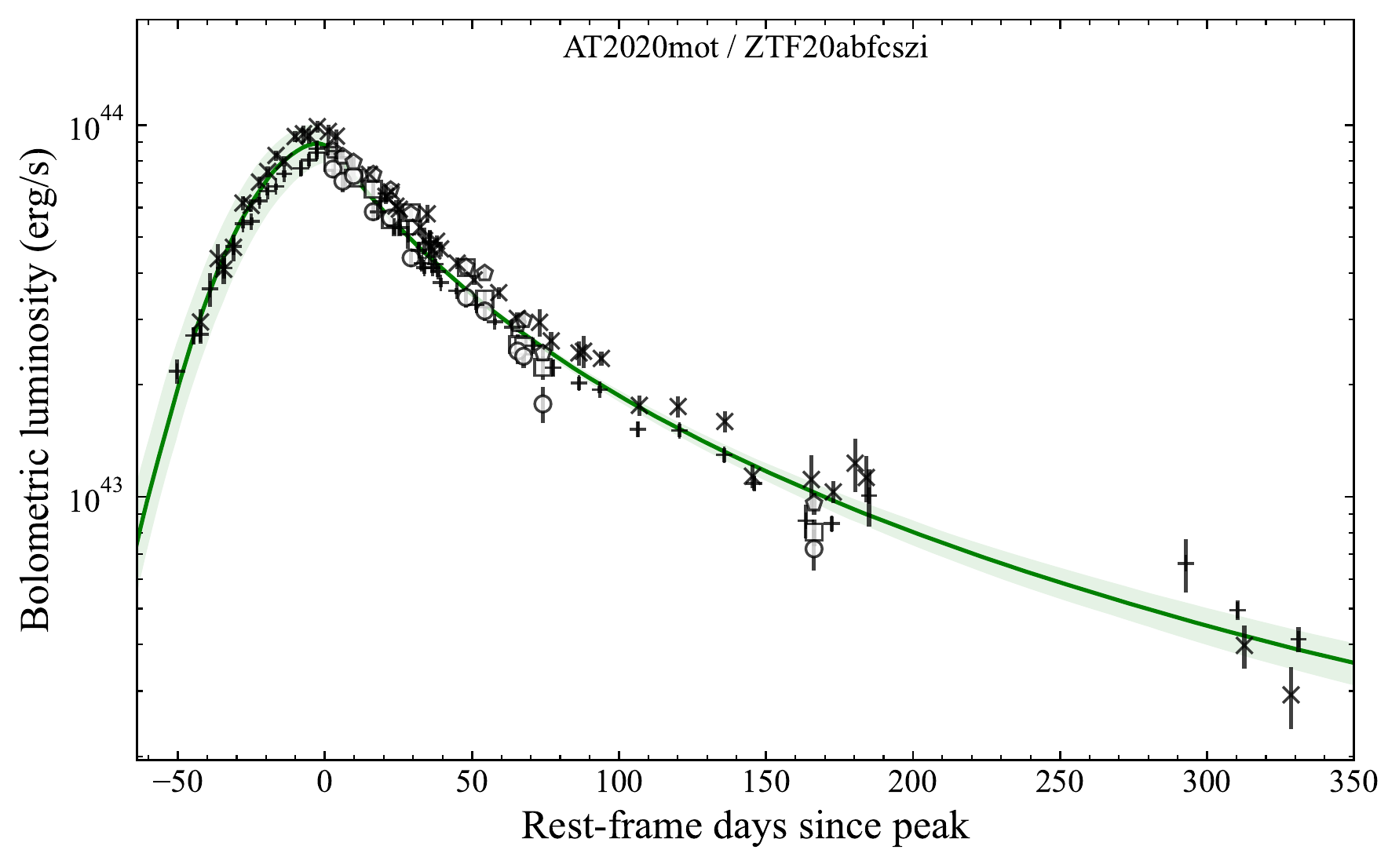}{0.33 \textwidth}{} 			\\[-20pt]}

\gridline{  \fig{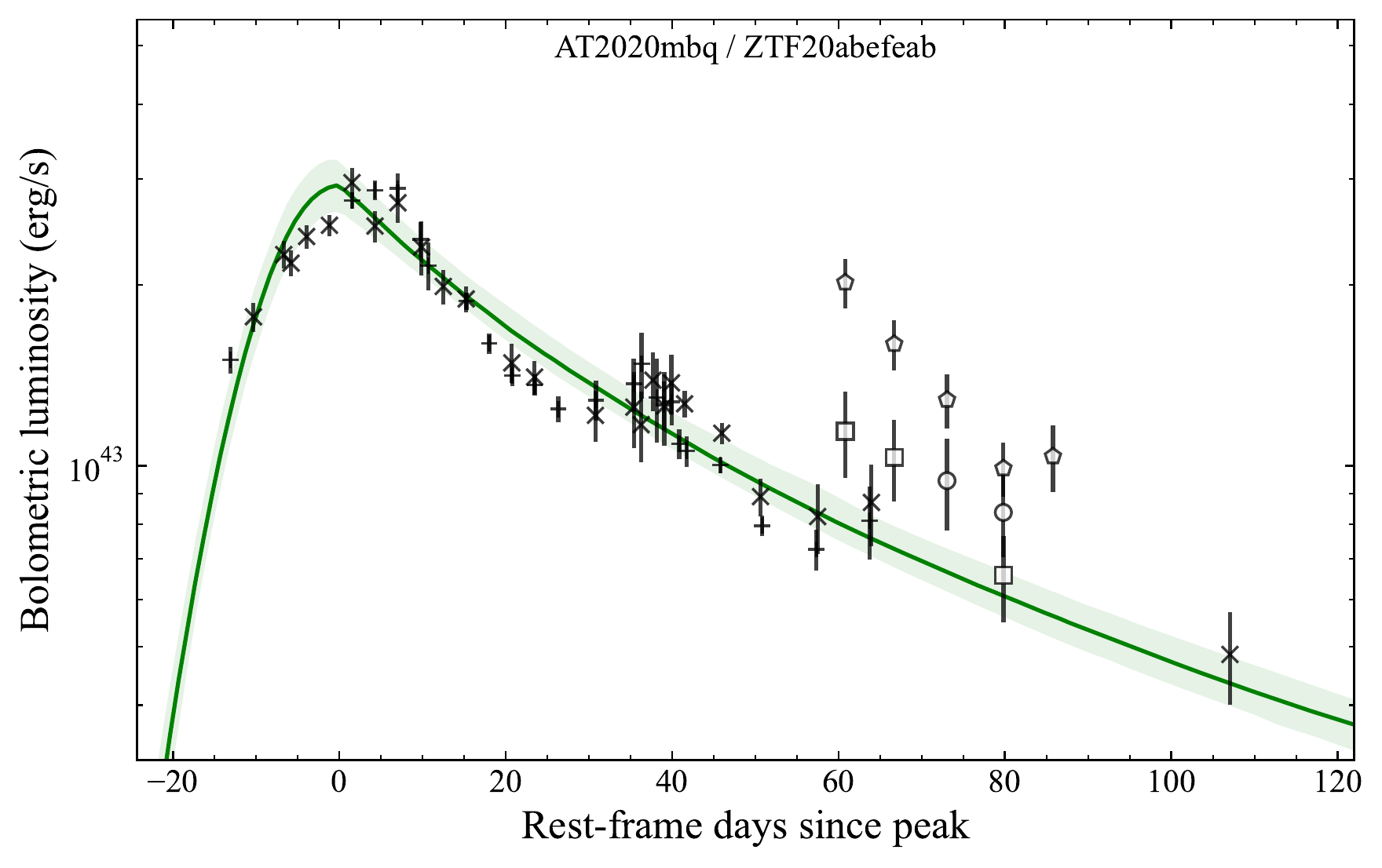}{0.33 \textwidth}{}
            \fig{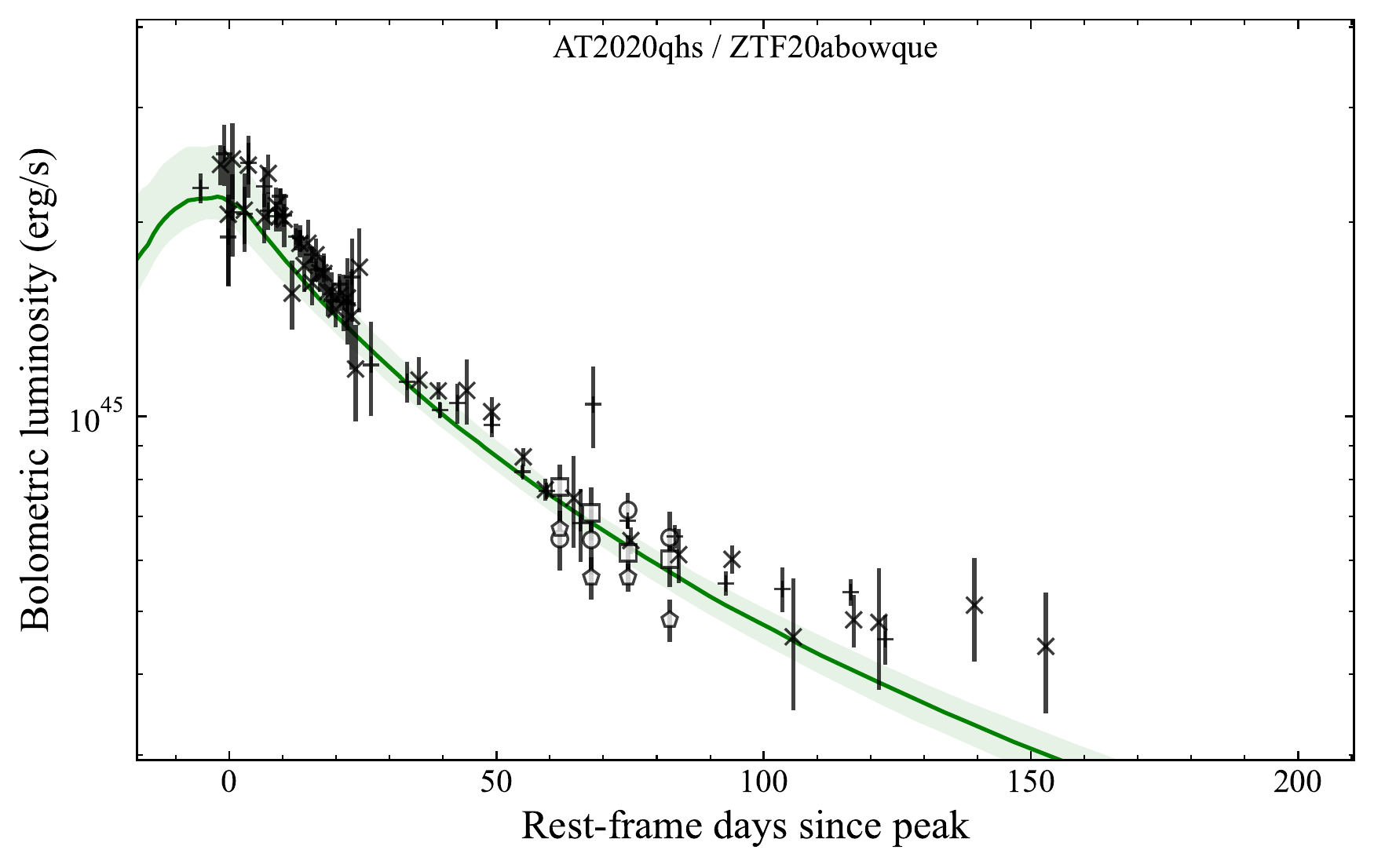}{0.33 \textwidth}{}
             \fig{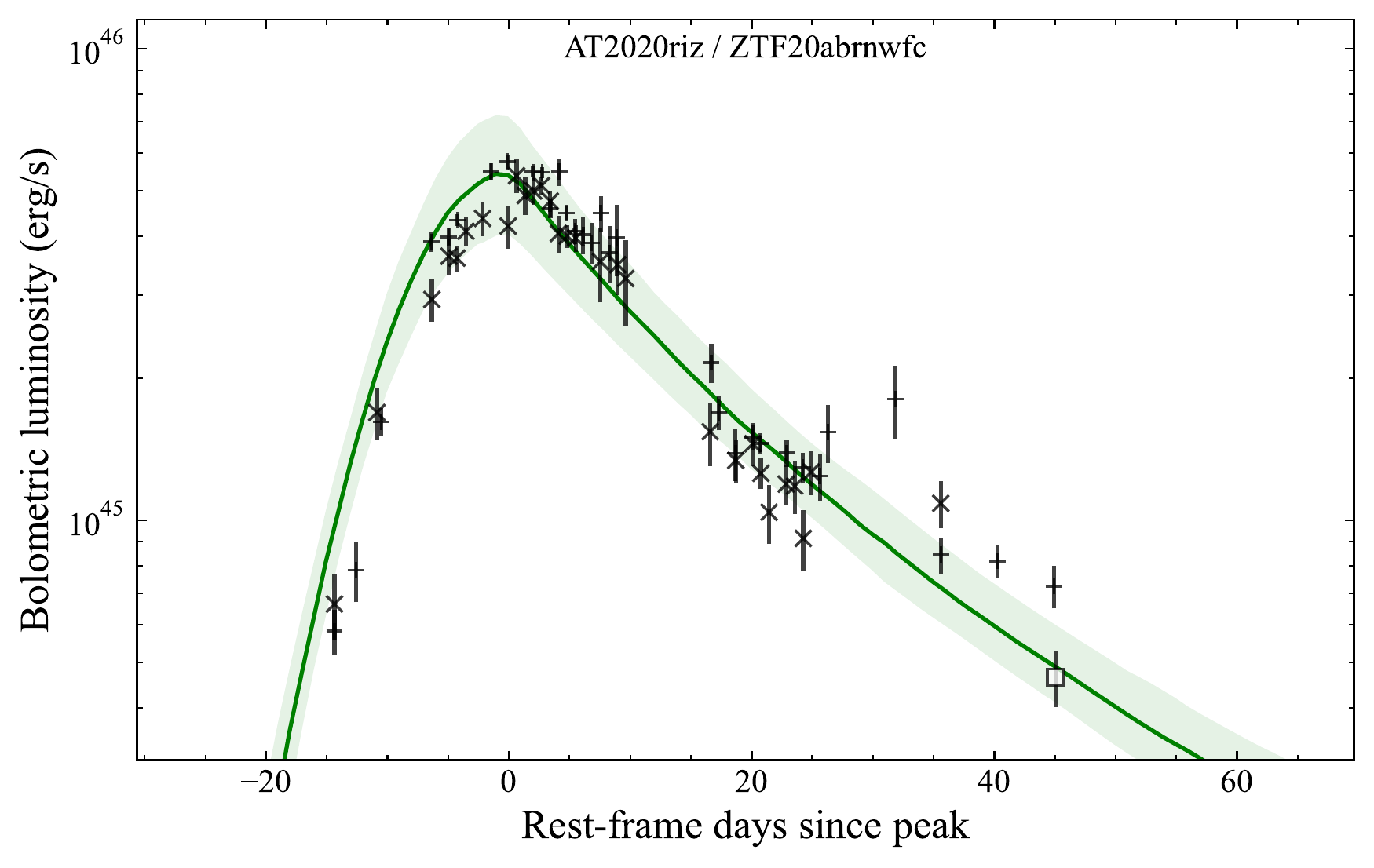}{0.33 \textwidth}{}
             \\[-20pt]}

\gridline{  \fig{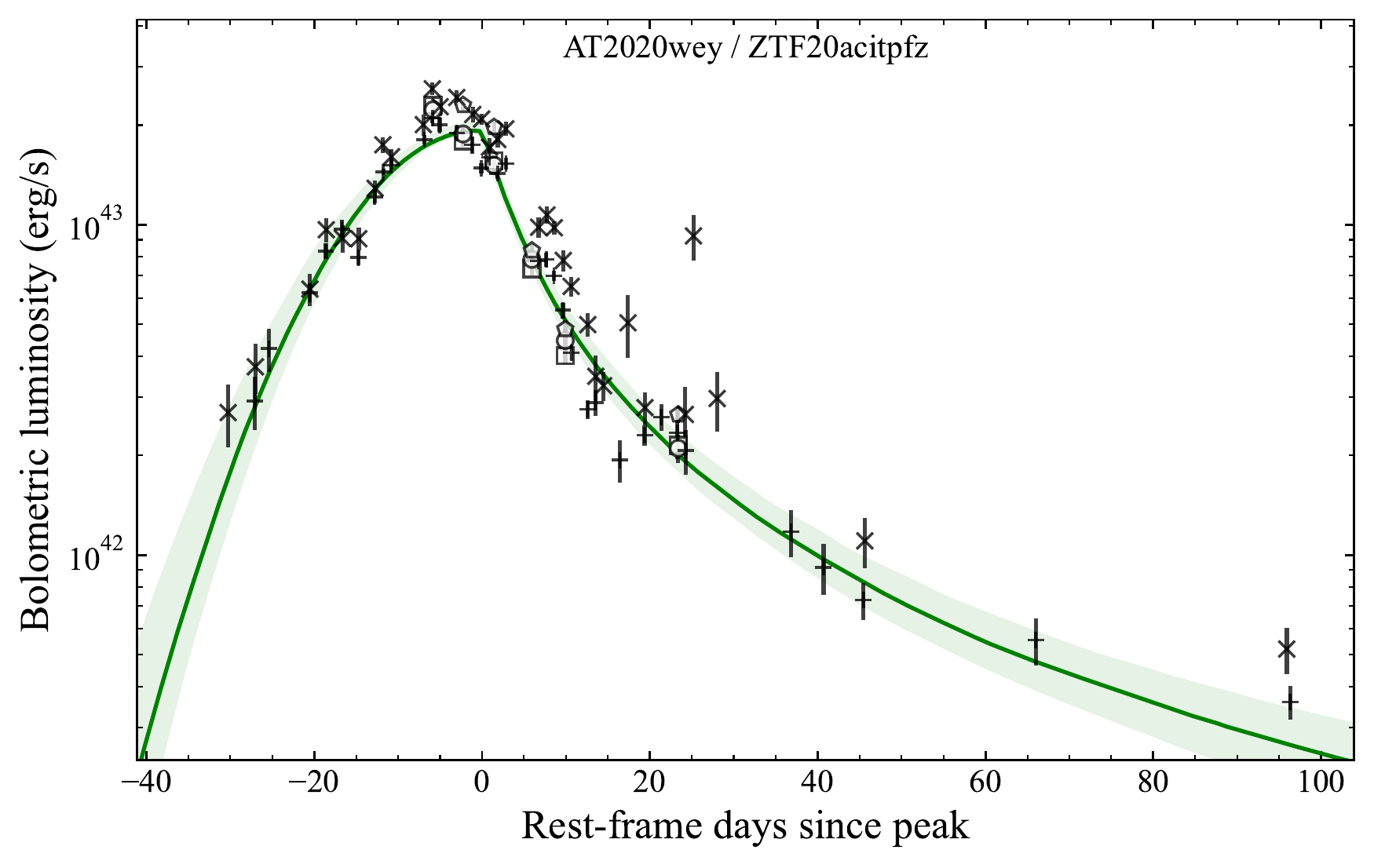}{0.33 \textwidth}{} 
            \fig{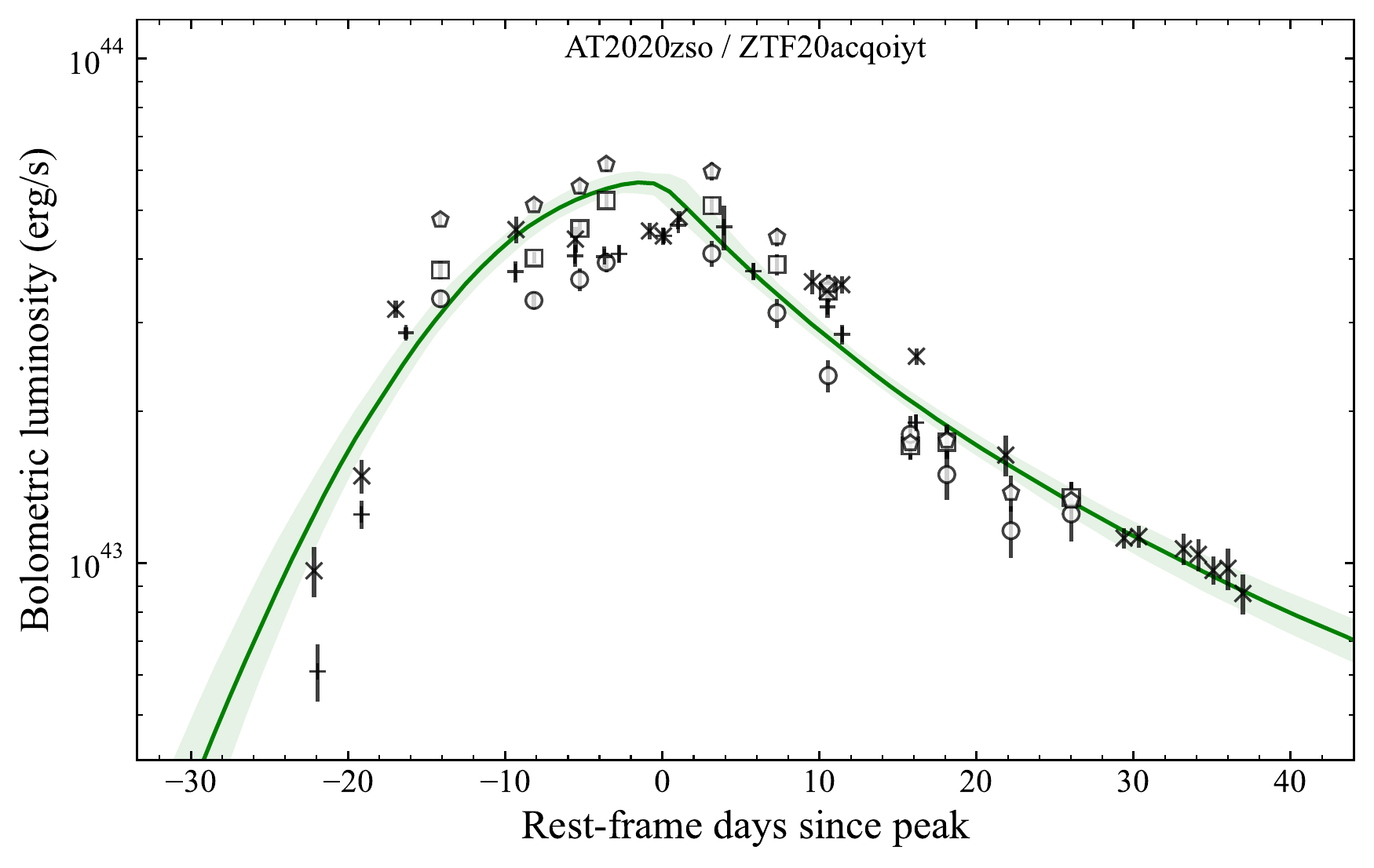}{0.33 \textwidth}{}
            \fig{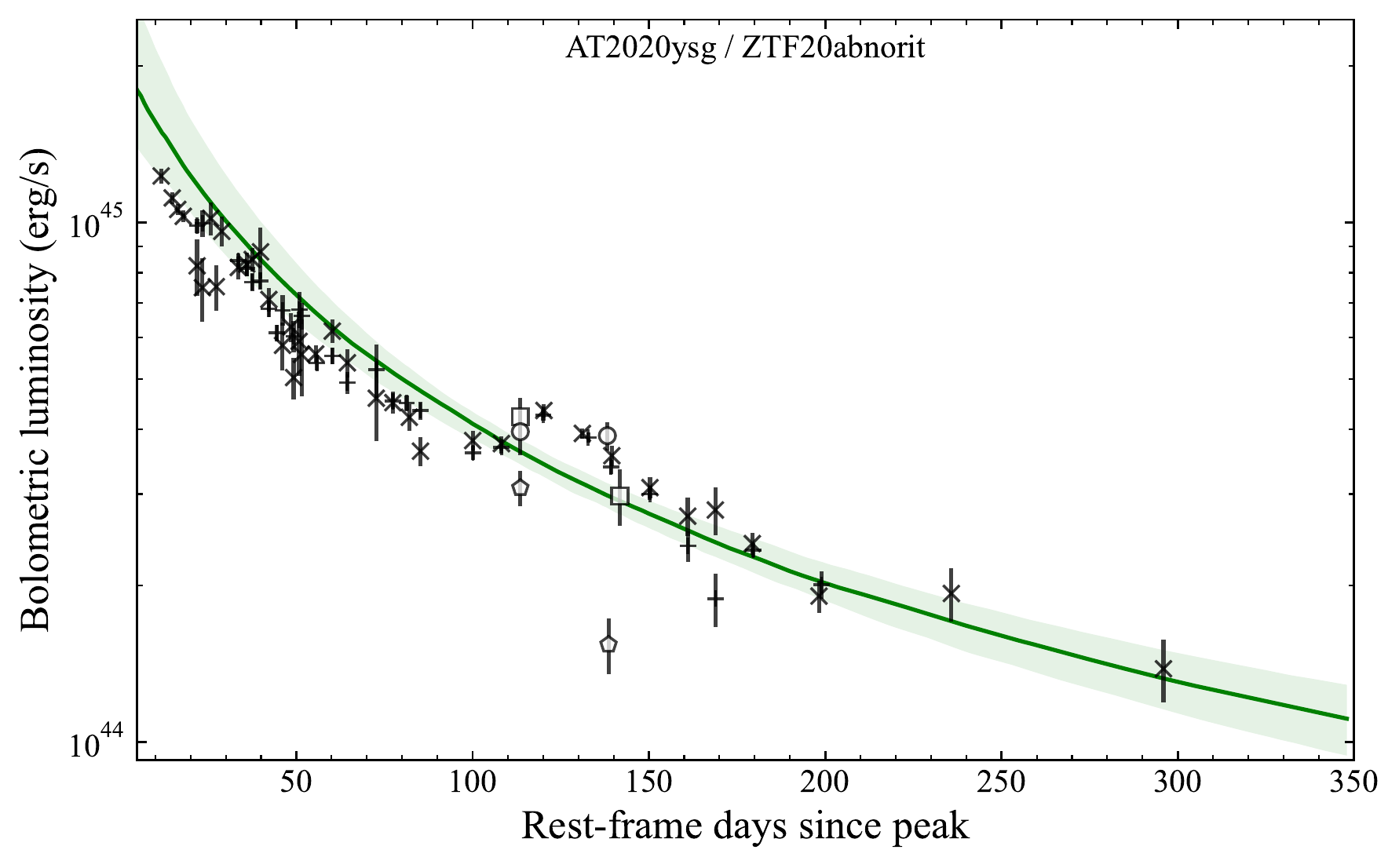}{0.33 \textwidth}{} 
            \\[-20pt]}   

\caption{Same as Figure \ref{fig:fitlcs1}.}
\end{figure*}

\begin{figure*}
\gridline{	\fig{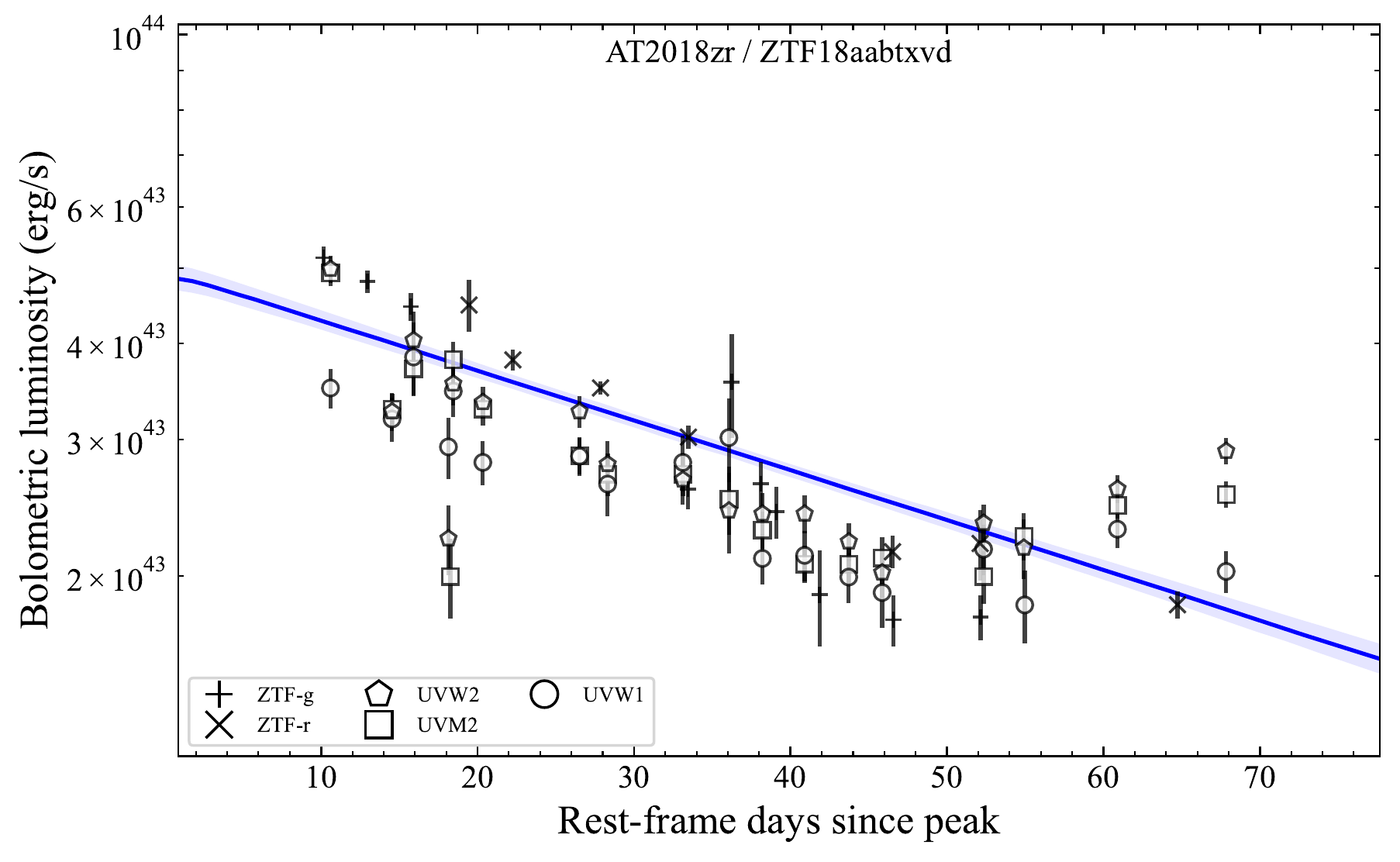}{0.33 \textwidth}{} 
			\fig{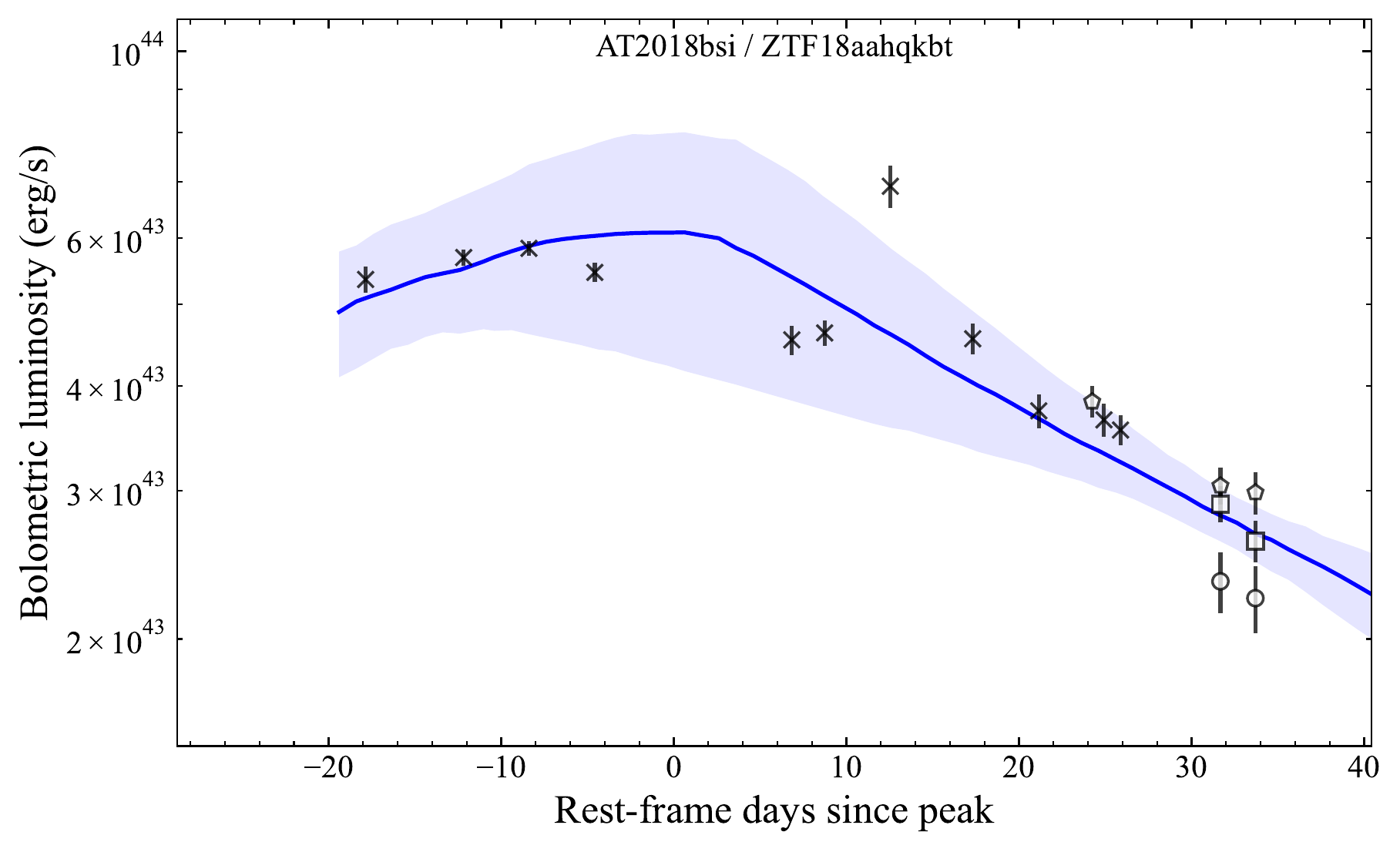}{0.33 \textwidth}{}
			\fig{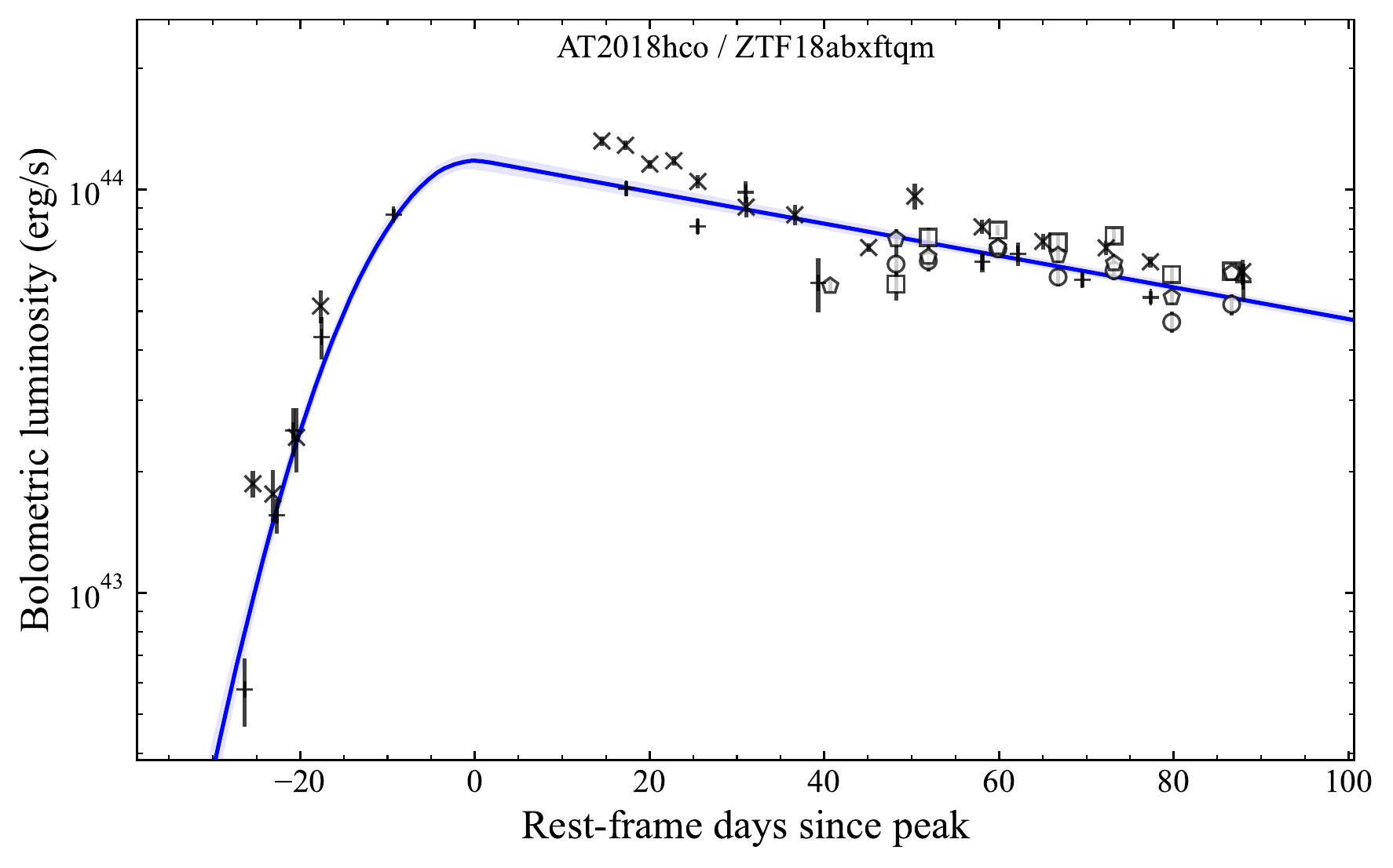}{0.33 \textwidth}{}
			\\[-20pt]}
			
\gridline{\fig{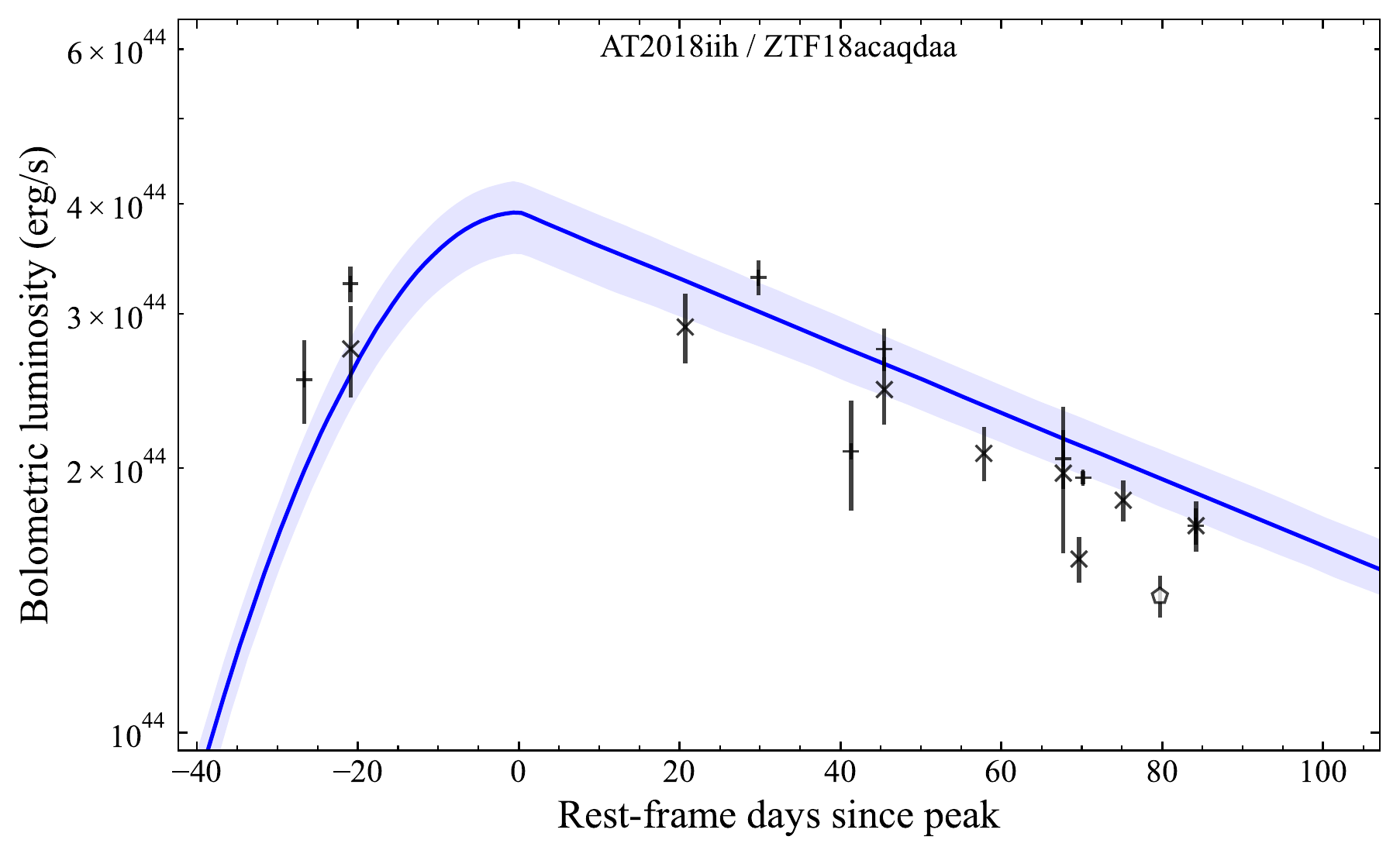}{0.33 \textwidth}{} 
\fig{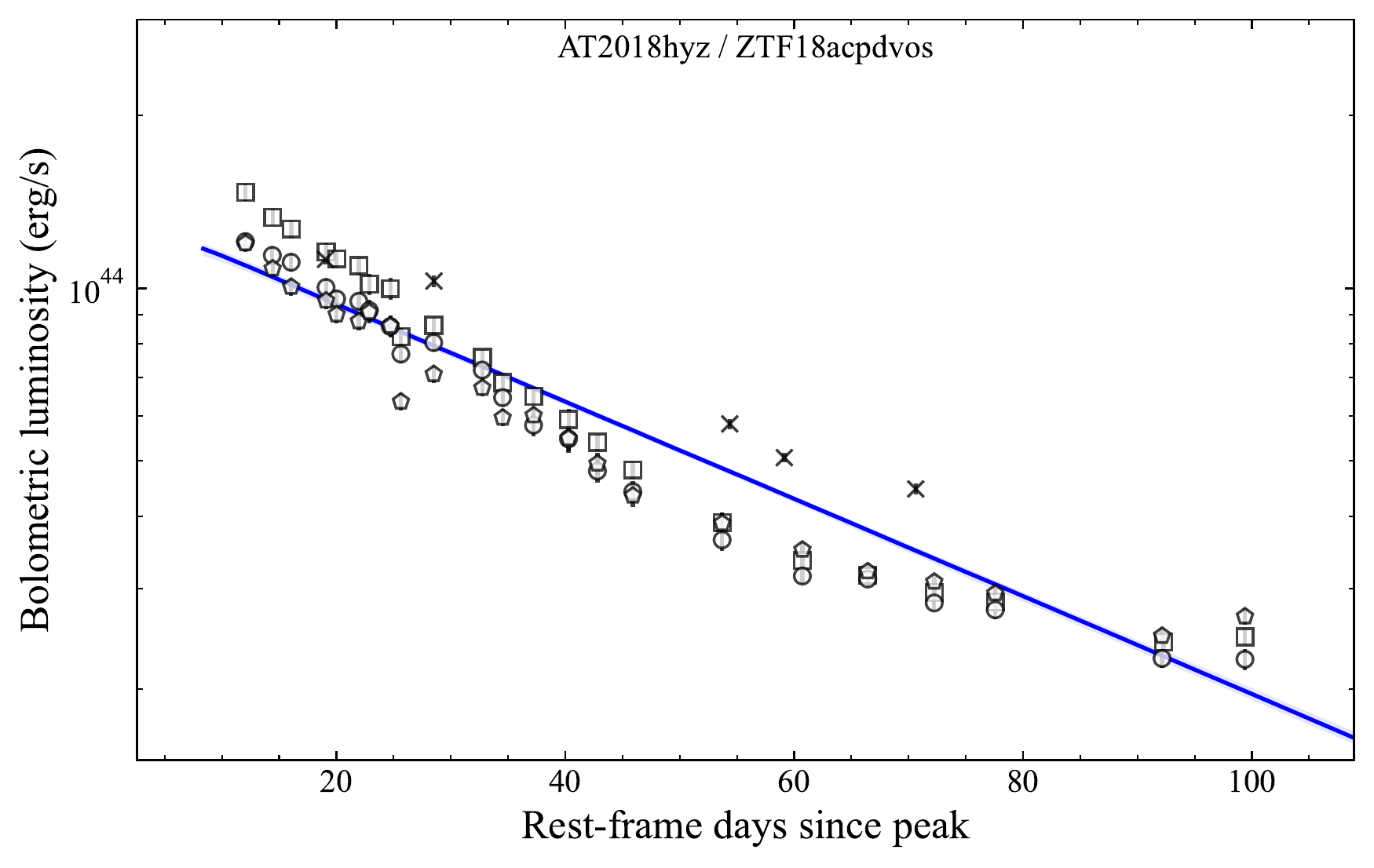}{0.33 \textwidth}{} 
            \fig{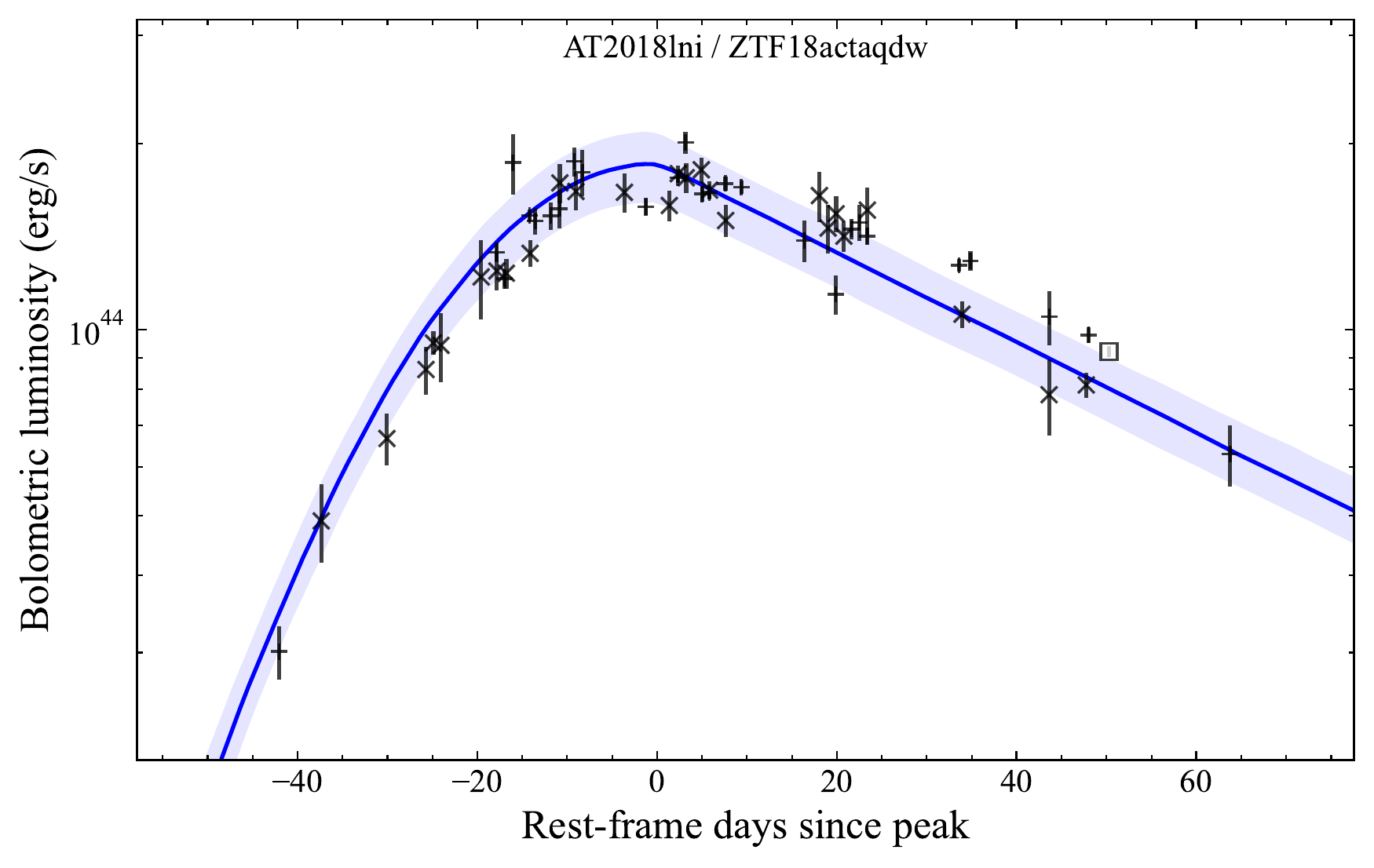}{0.33 \textwidth}{}
            \\[-20pt]}
     
\gridline{  \fig{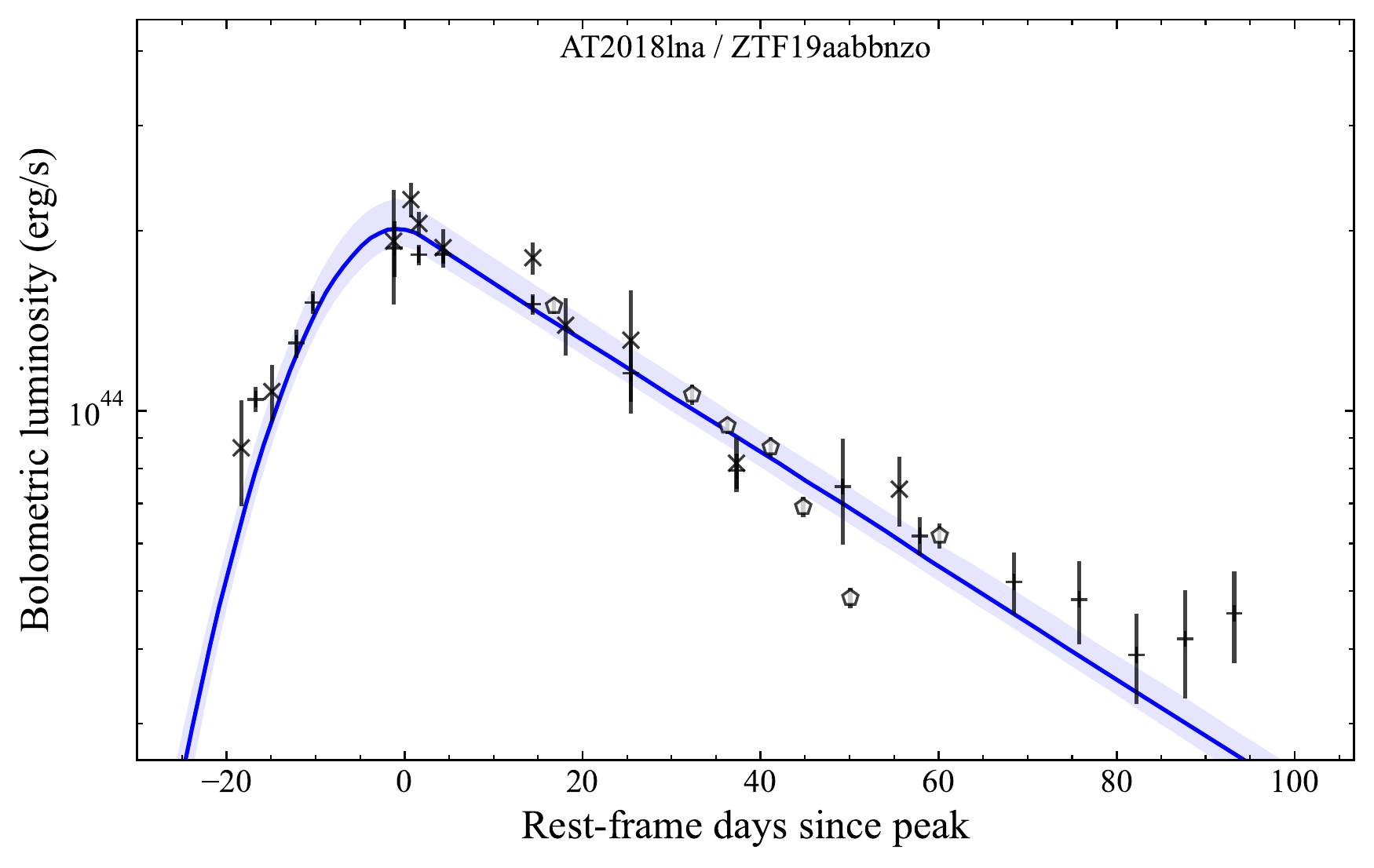}{0.33 \textwidth}{}
            \fig{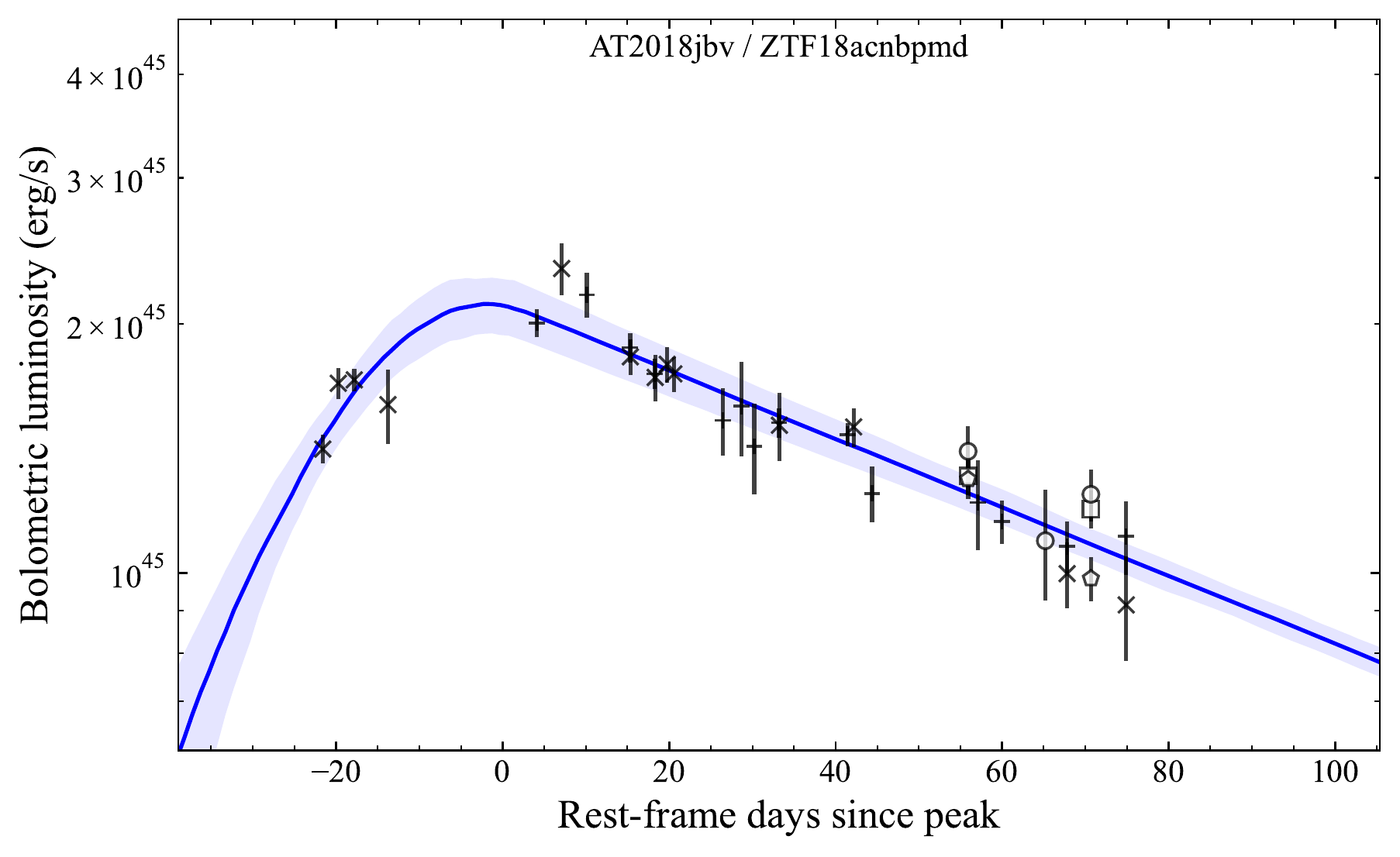}{0.33 \textwidth}{}
            \fig{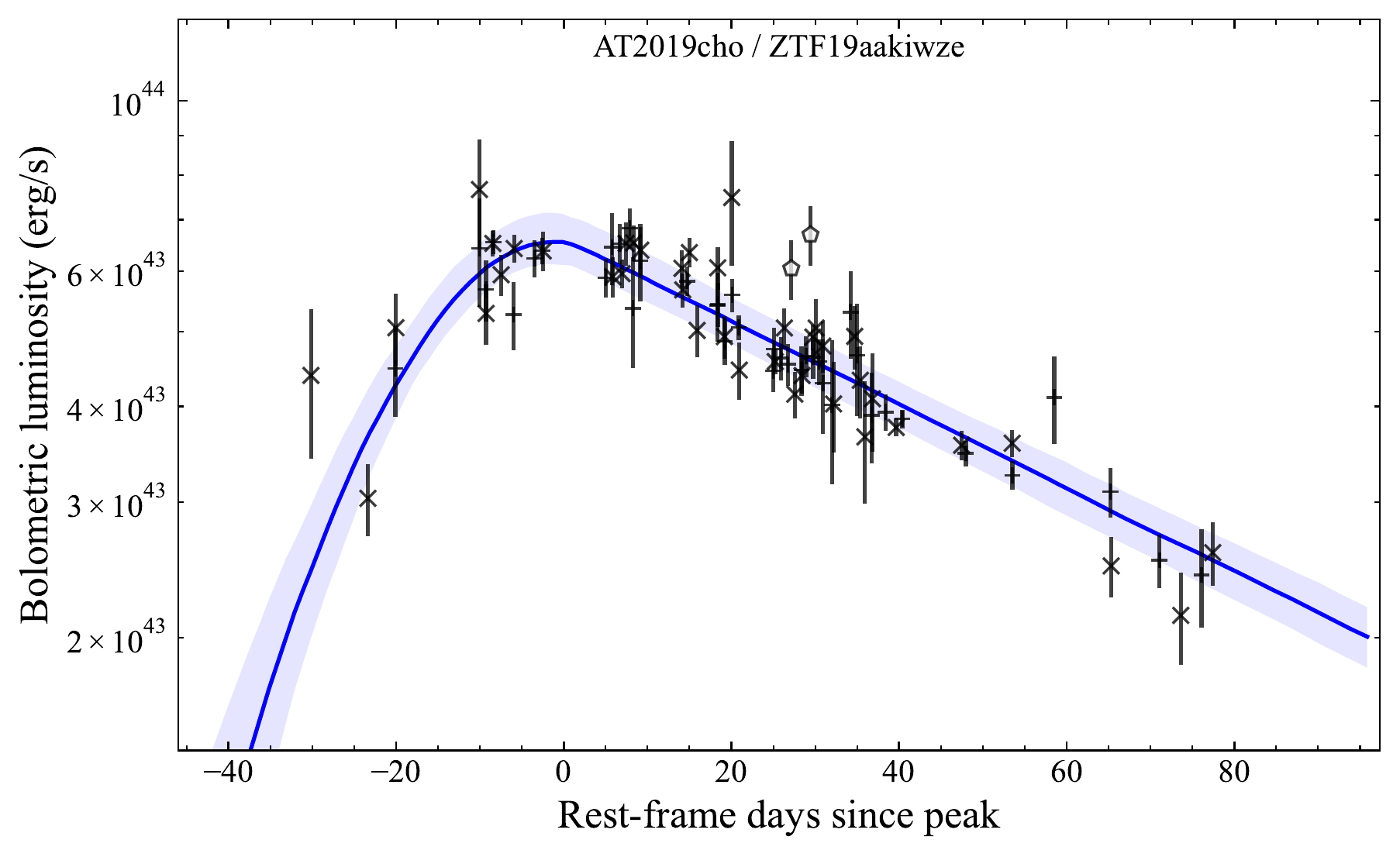}{0.33 \textwidth}{} 
             \\[-20pt]}

\gridline{\fig{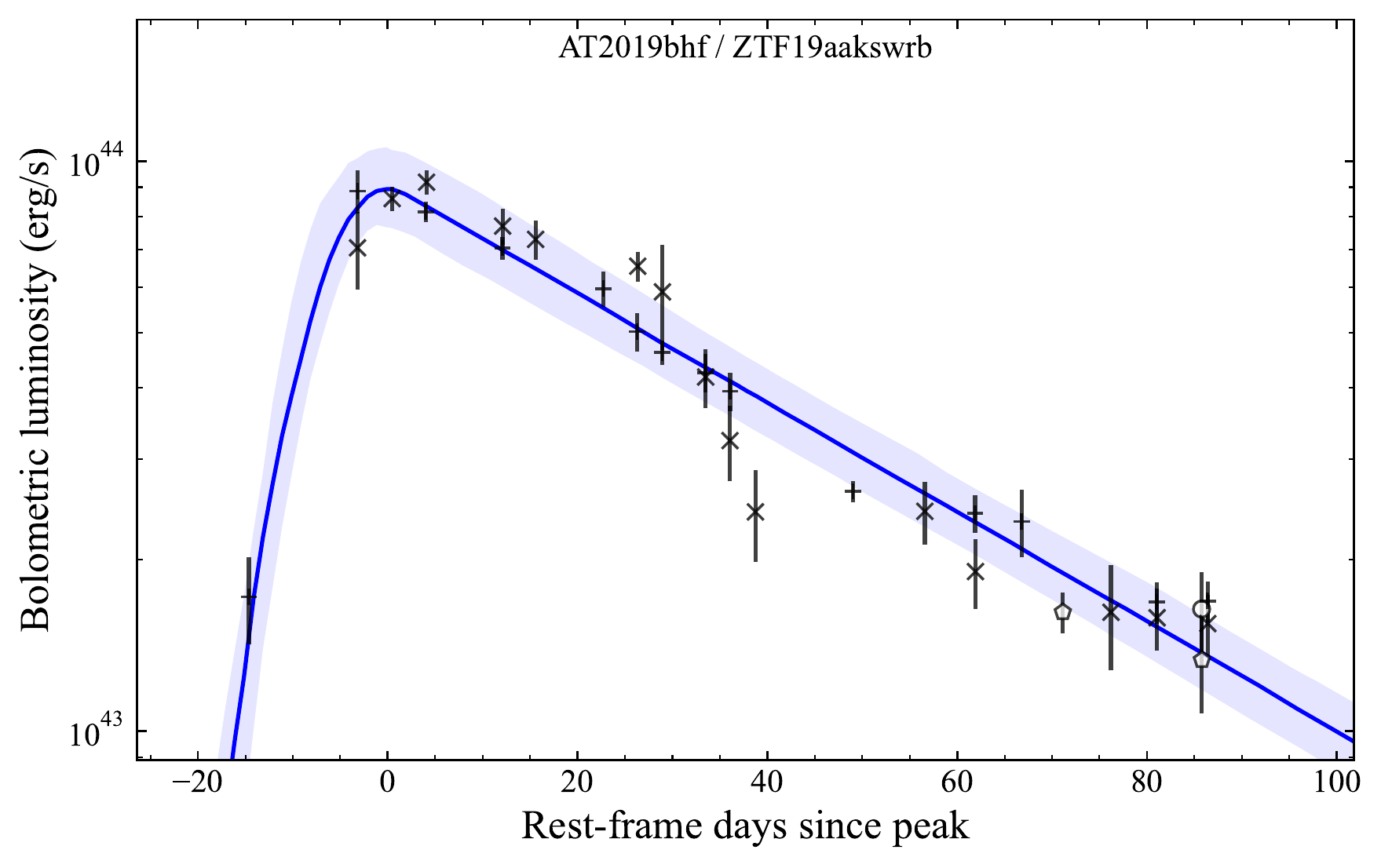}{0.33 \textwidth}{}
	\fig{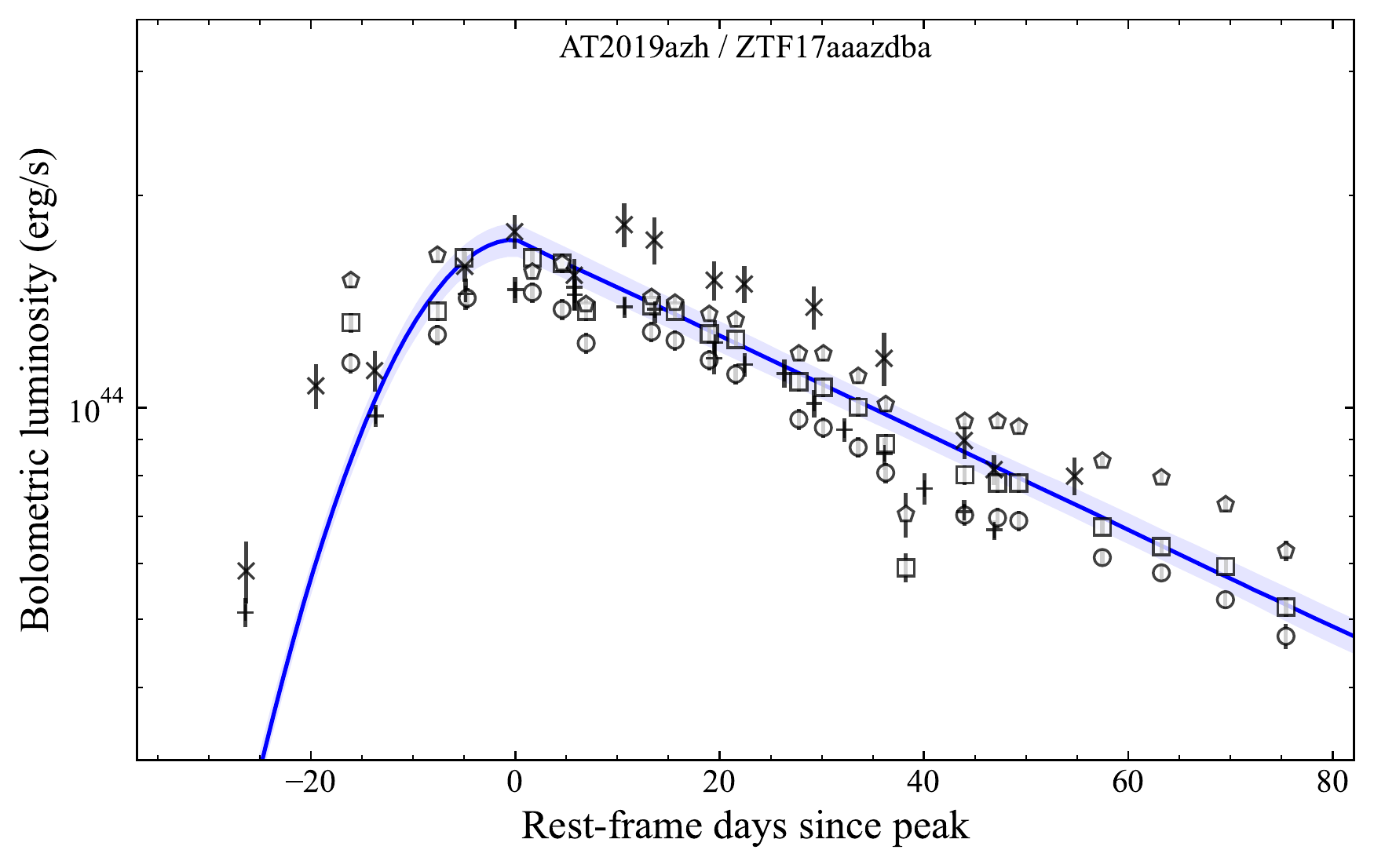}{0.33 \textwidth}{} 
            \fig{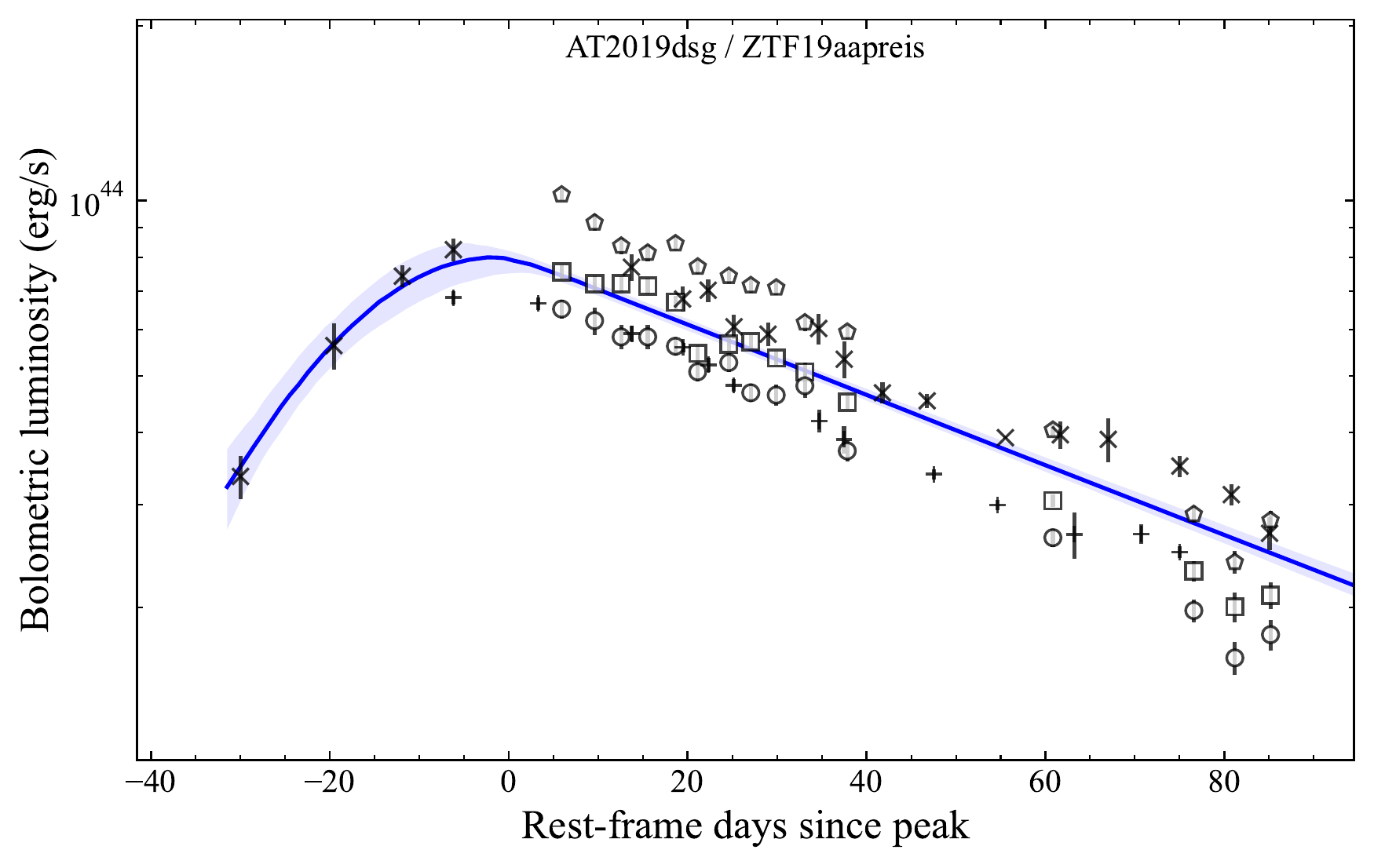}{0.33 \textwidth}{}
            \\[-20pt]} 
\gridline{  \fig{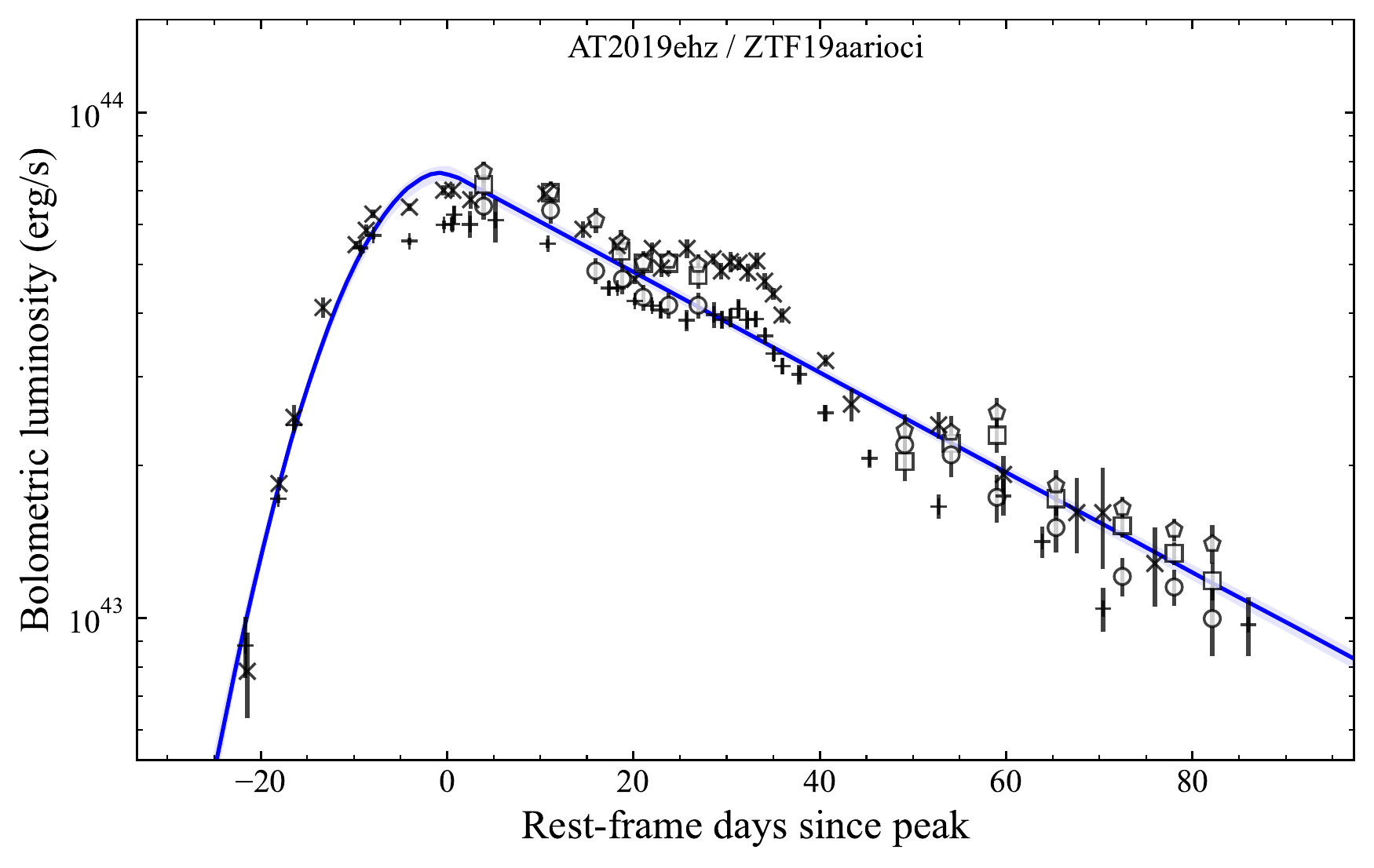}{0.33 \textwidth}{} 
            \fig{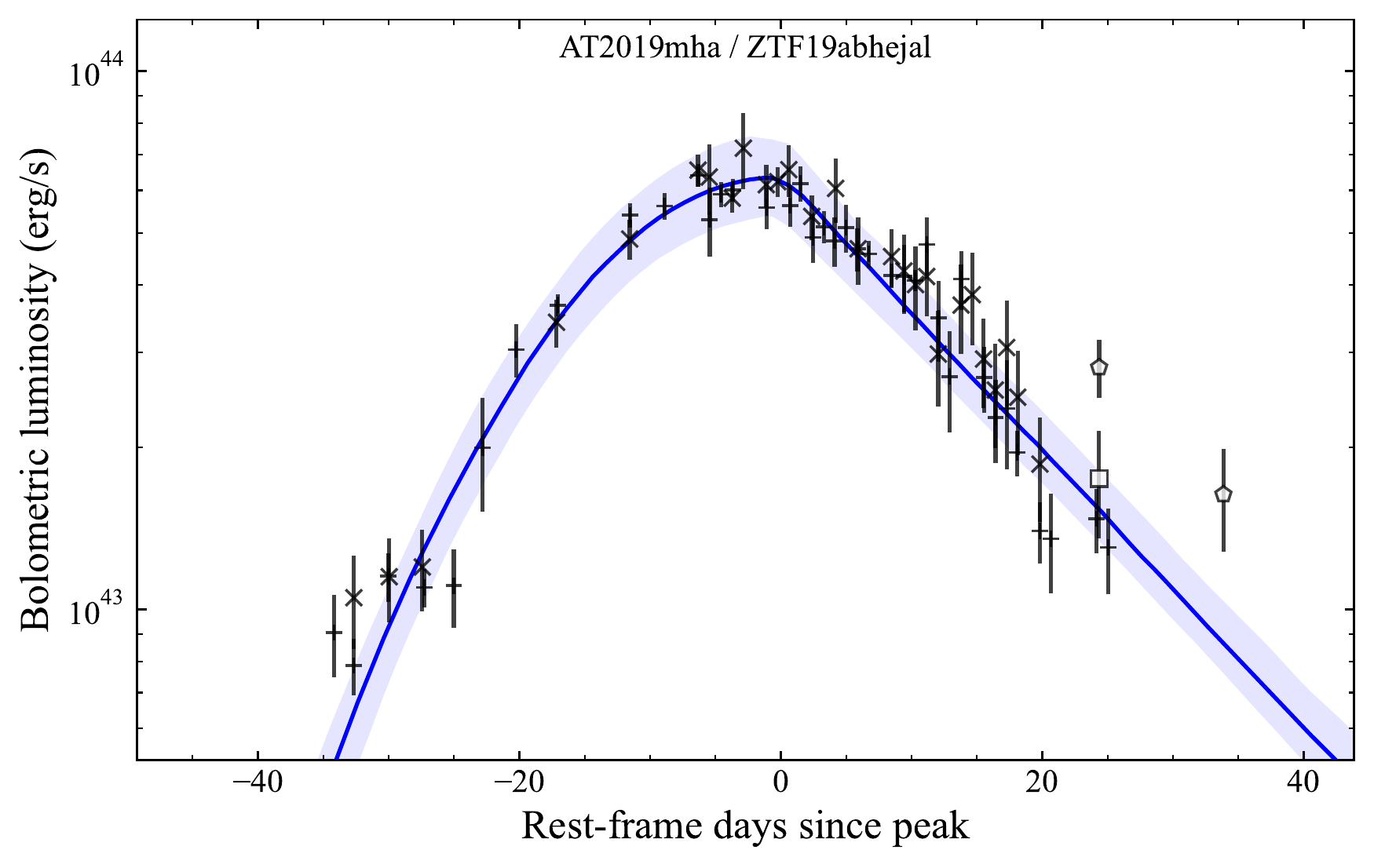}{0.33 \textwidth}{}
            \fig{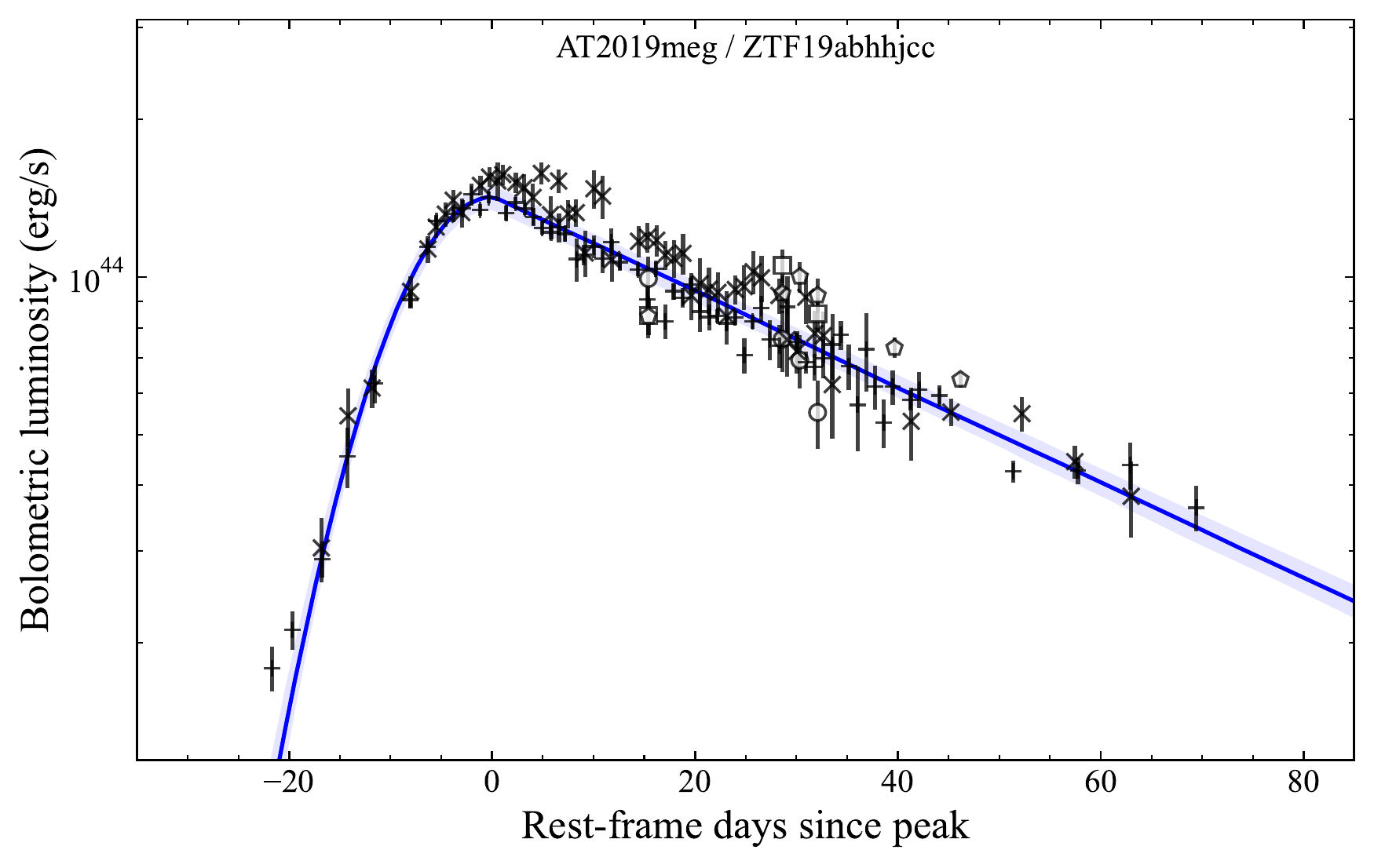}{0.33 \textwidth}{} 
            \\[-20pt]}

\caption{Gaussian rise and exponential decay fits with fixed temperature, shown with the optical and UV 3-$\sigma$ detections binned as in Figure \ref{fig:obslcs1}. We also show the 1-$\sigma$ spread in uncertainty of the fit. The legend can be seen in the top left panel.}\label{fig:expfitlcs1}
\end{figure*}

\begin{figure*}
            
\gridline{\fig{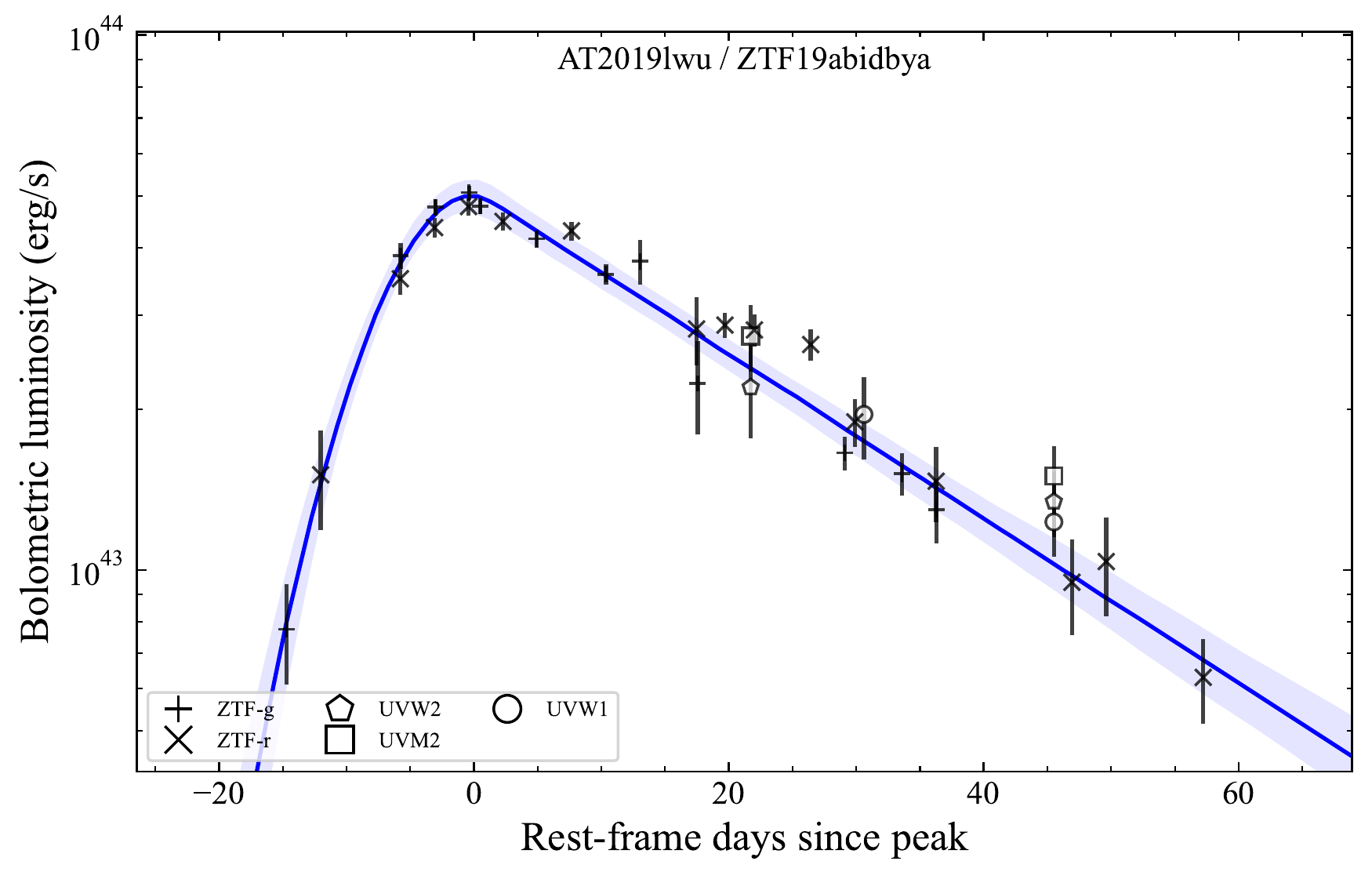}{0.33 \textwidth}{}
  \fig{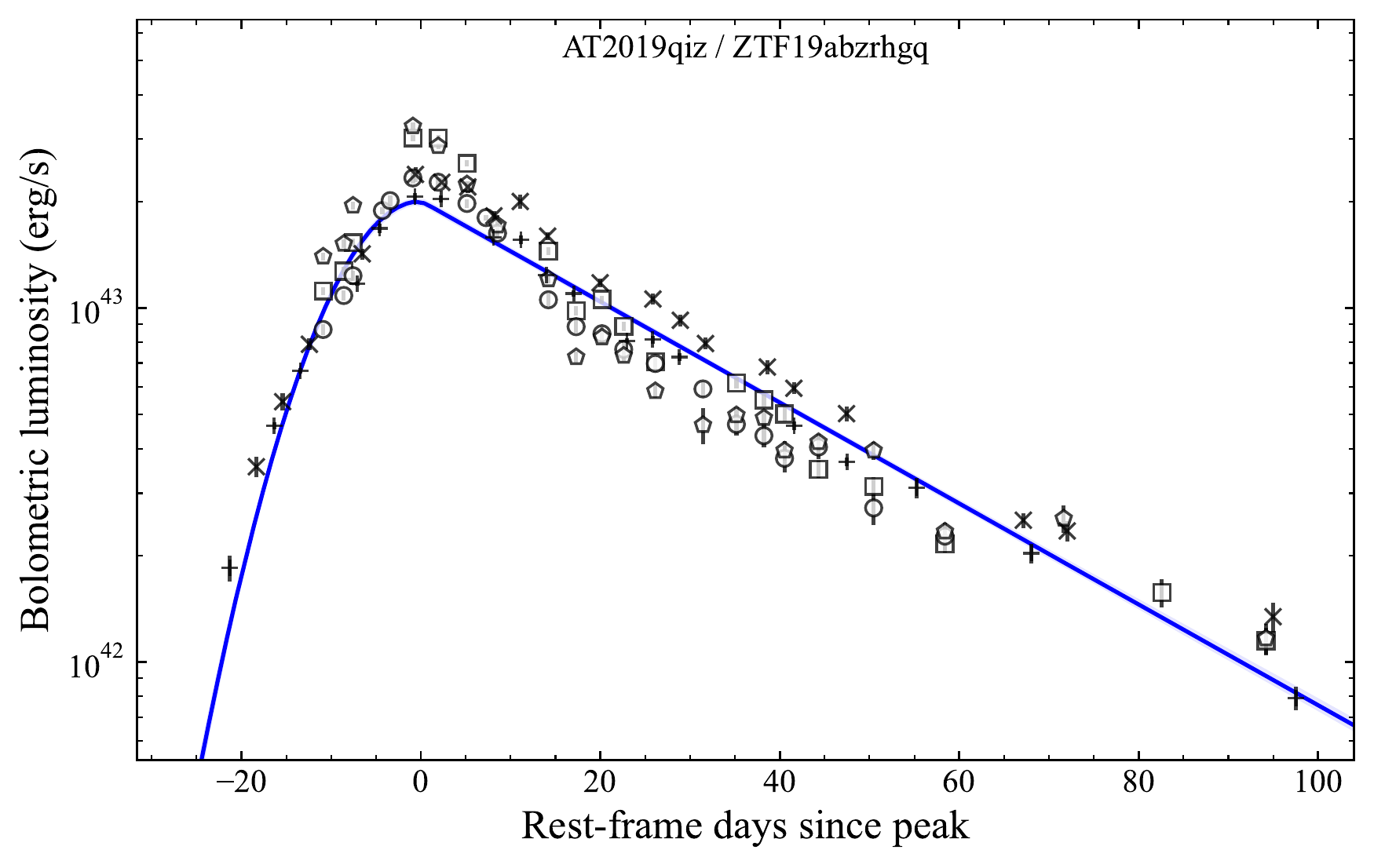}{0.33 \textwidth}{}
            \fig{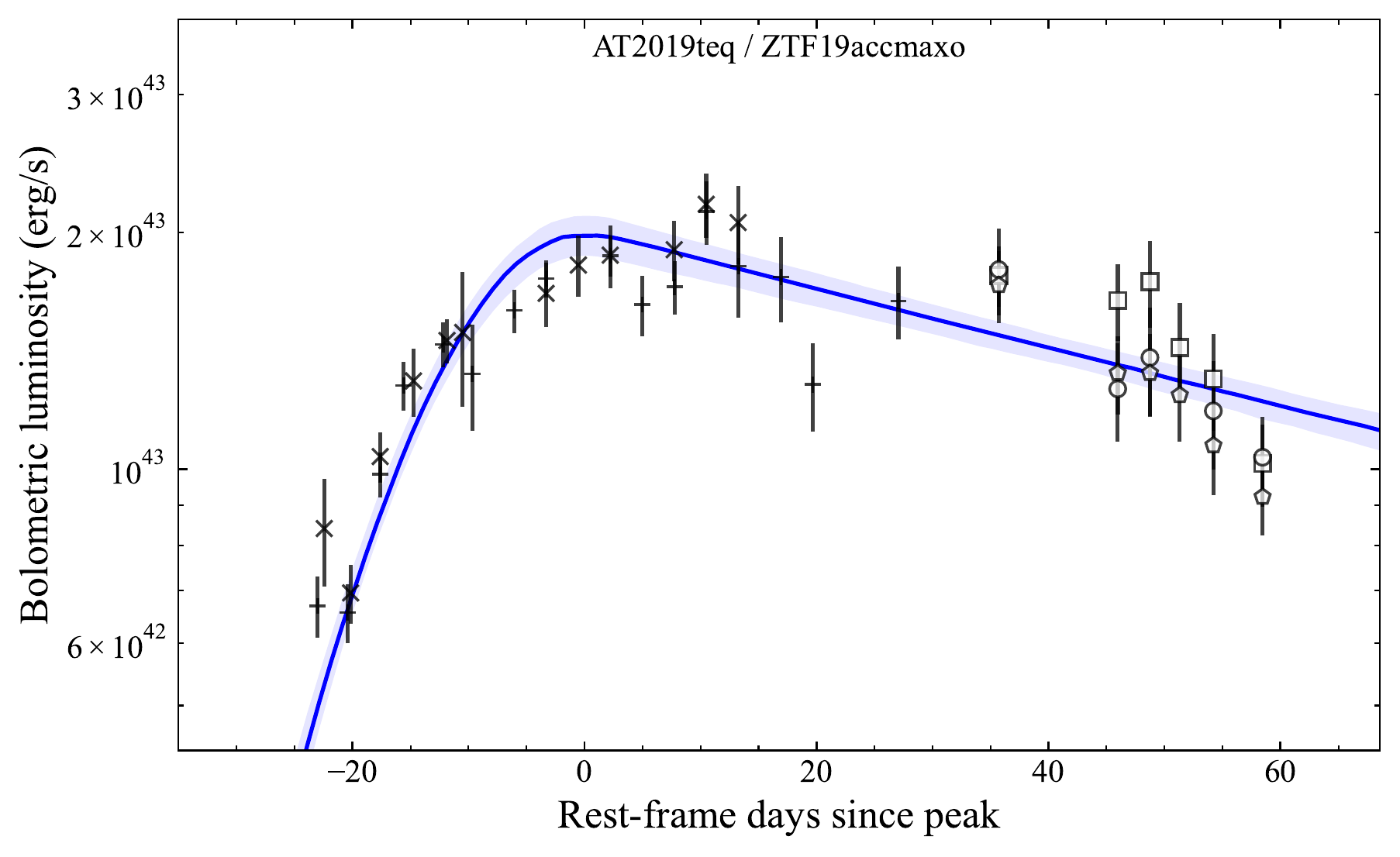}{0.33 \textwidth}{}
            \\[-20pt]}          

\gridline{  \fig{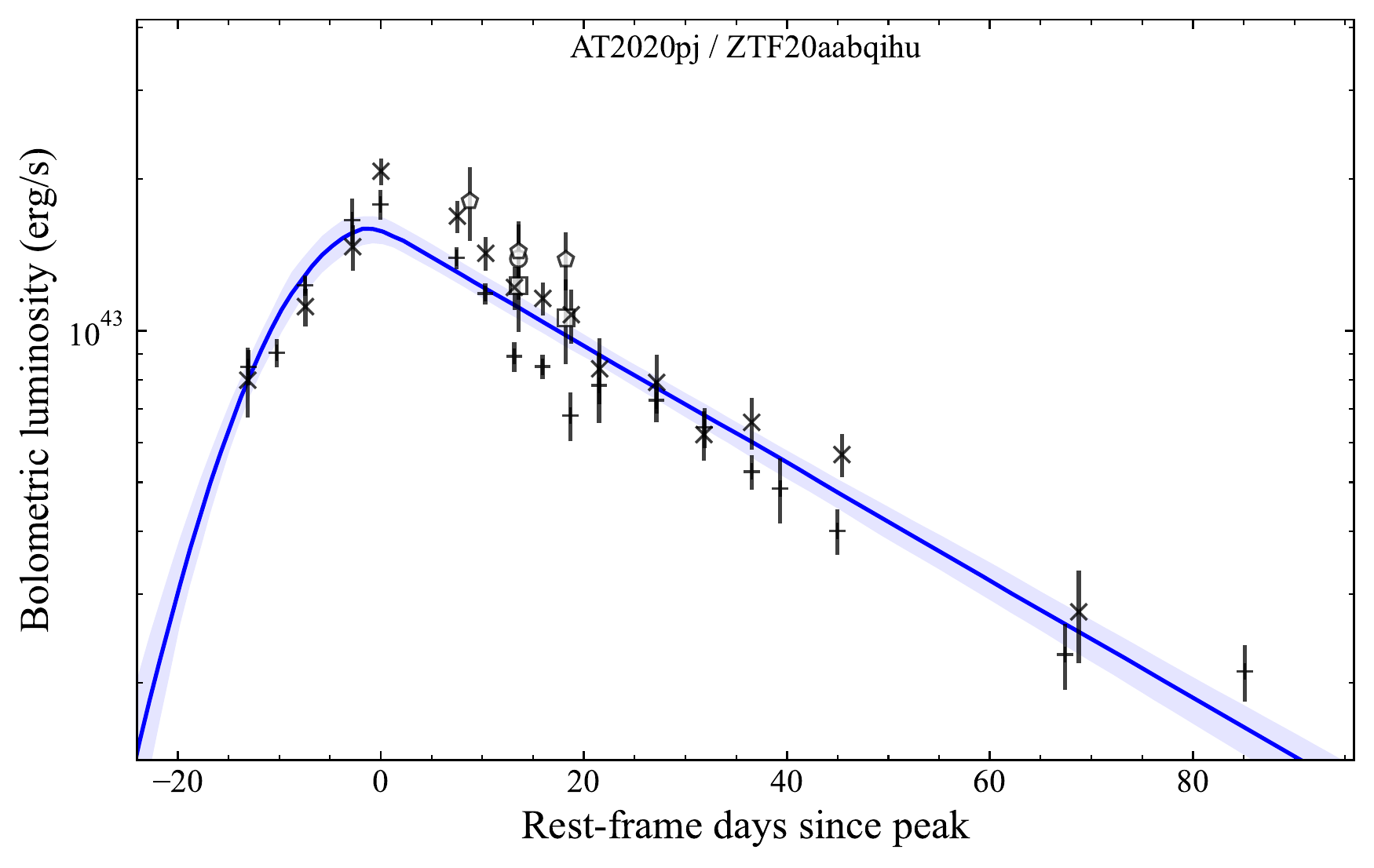}{0.33 \textwidth}{}
            \fig{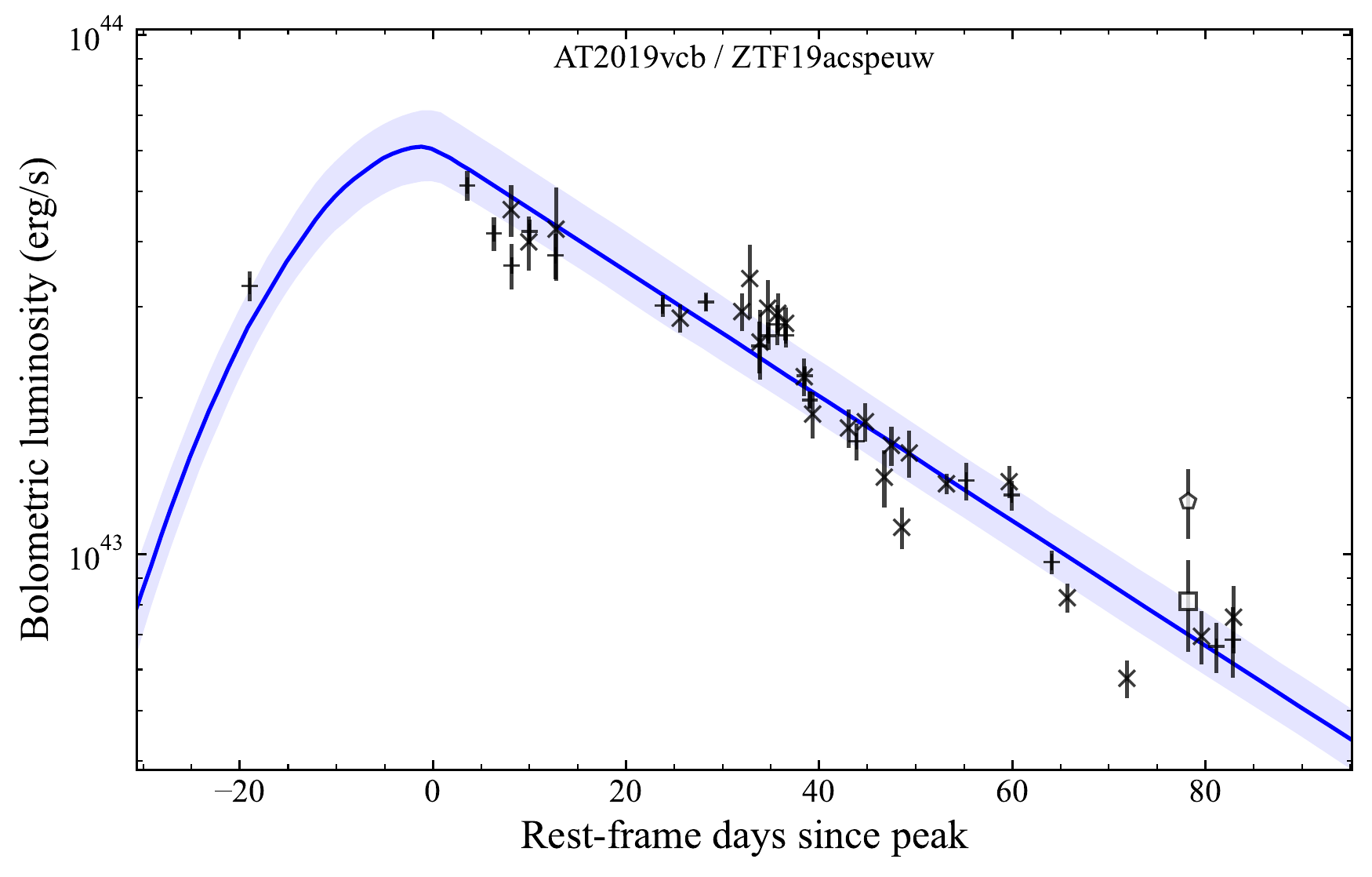}{0.33 \textwidth}{}
            \fig{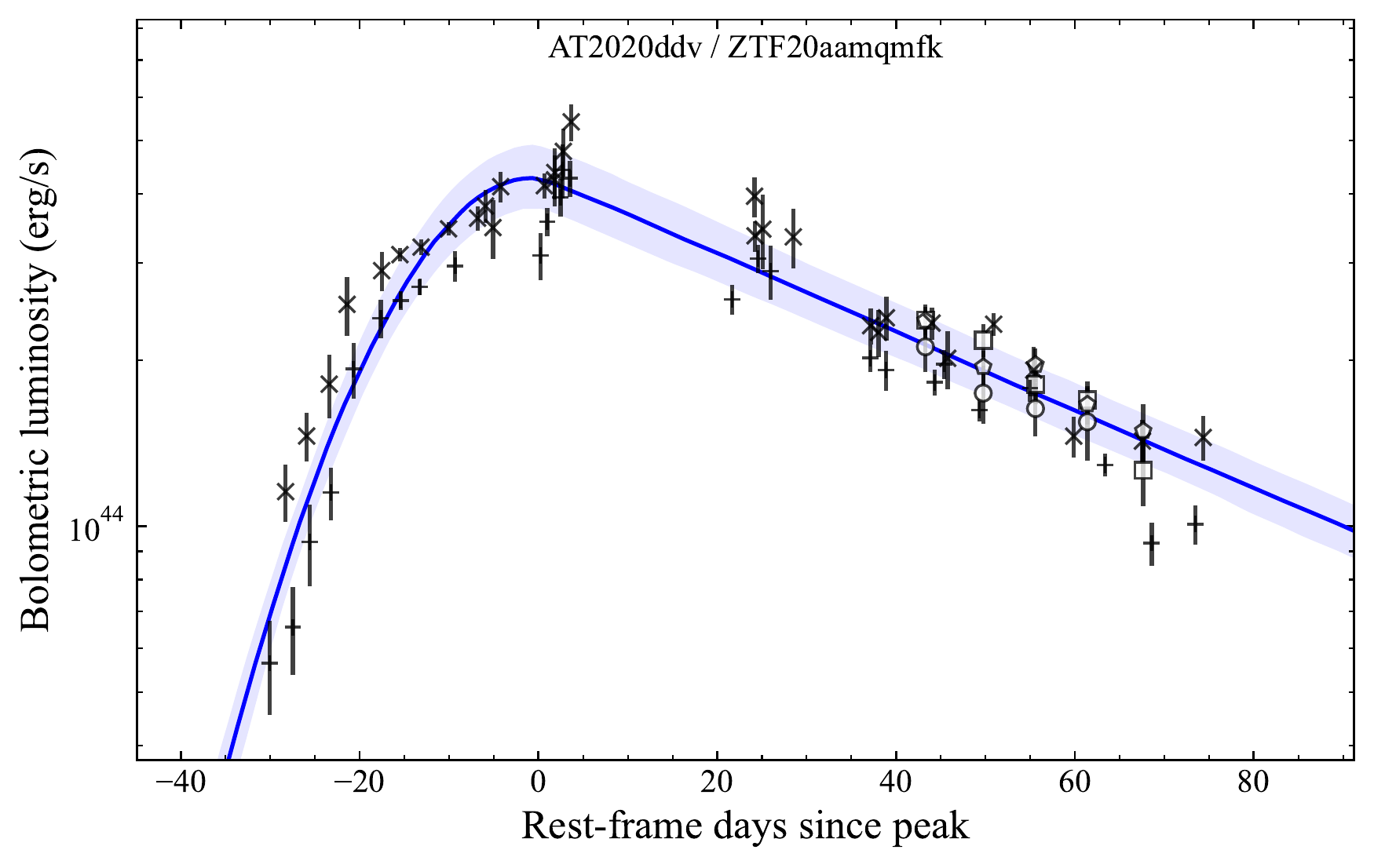}{0.33 \textwidth}{}
            \\[-20pt]}   

\gridline{	\fig{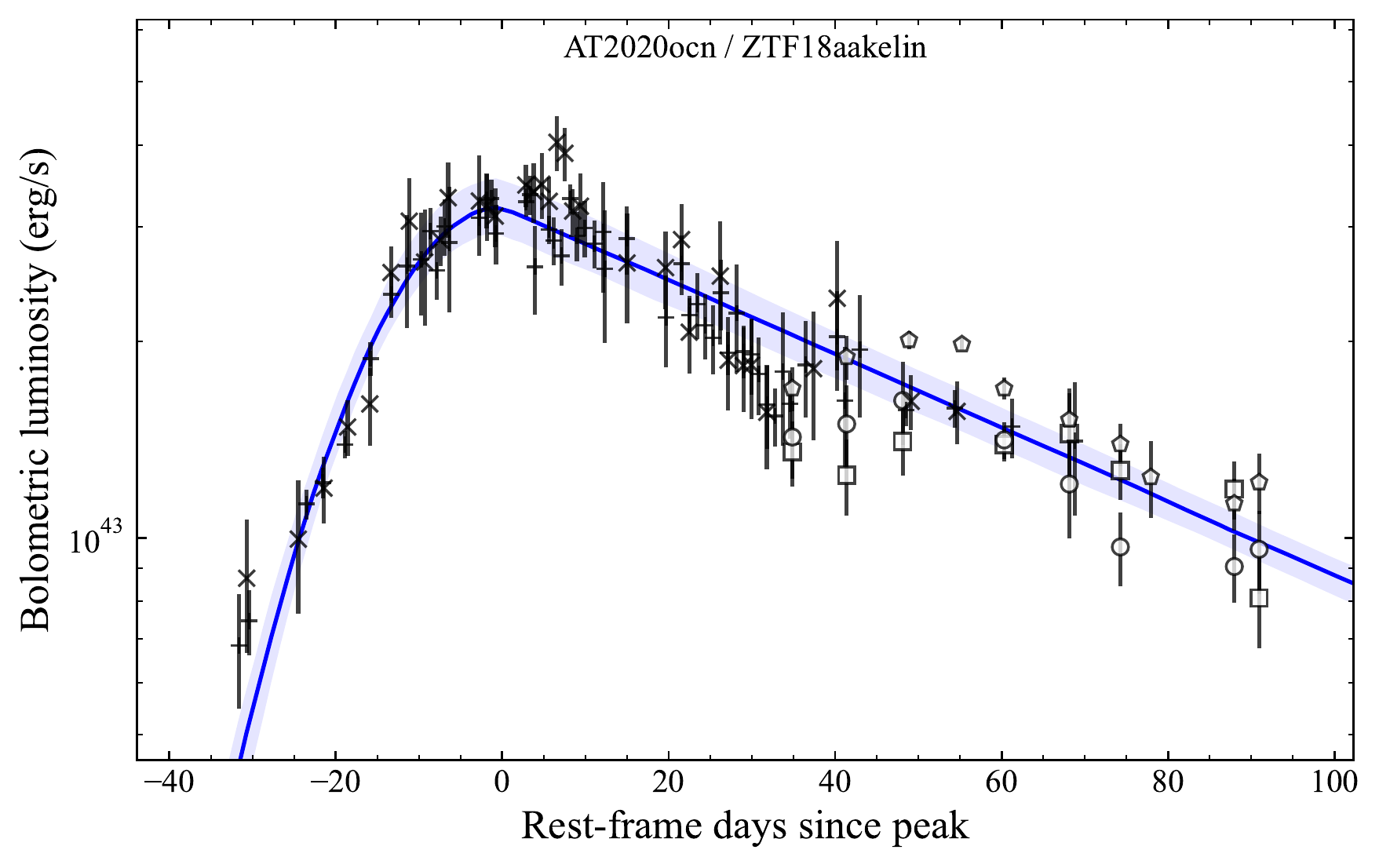}{0.33 \textwidth}{}
             \fig{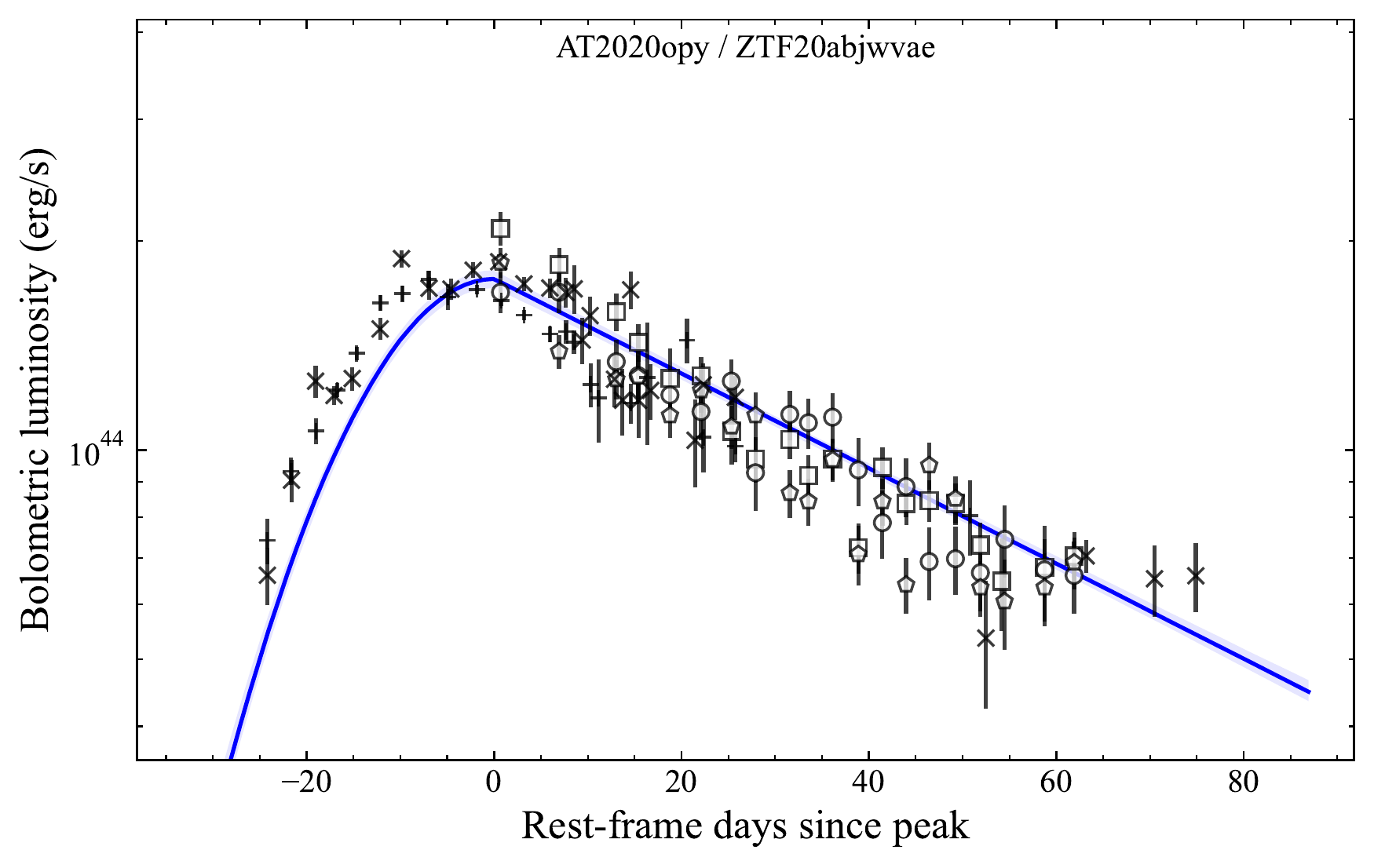}{0.33 \textwidth}{}
            \fig{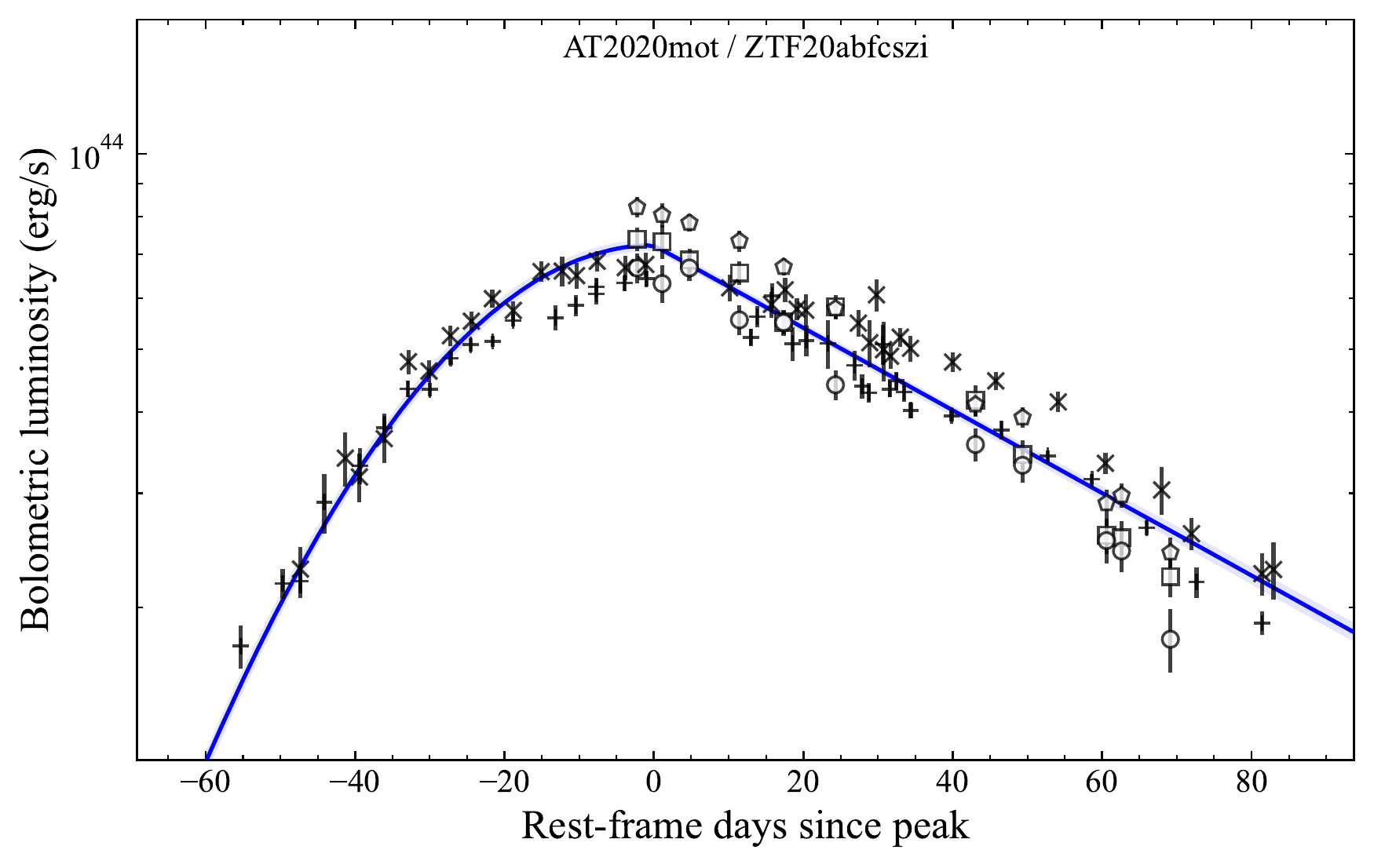}{0.33 \textwidth}{} 			\\[-20pt]}

\gridline{  \fig{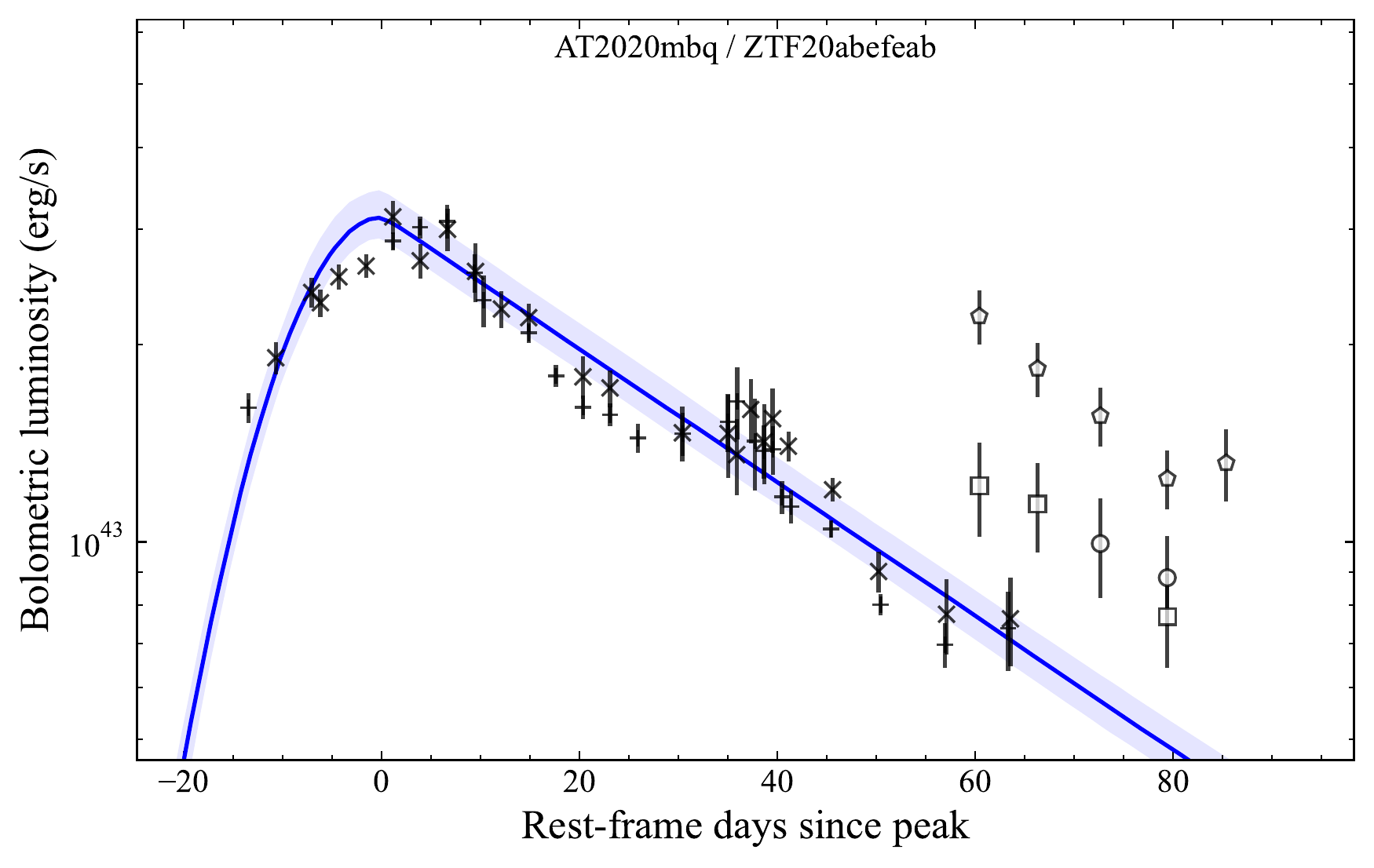}{0.33 \textwidth}{}
            \fig{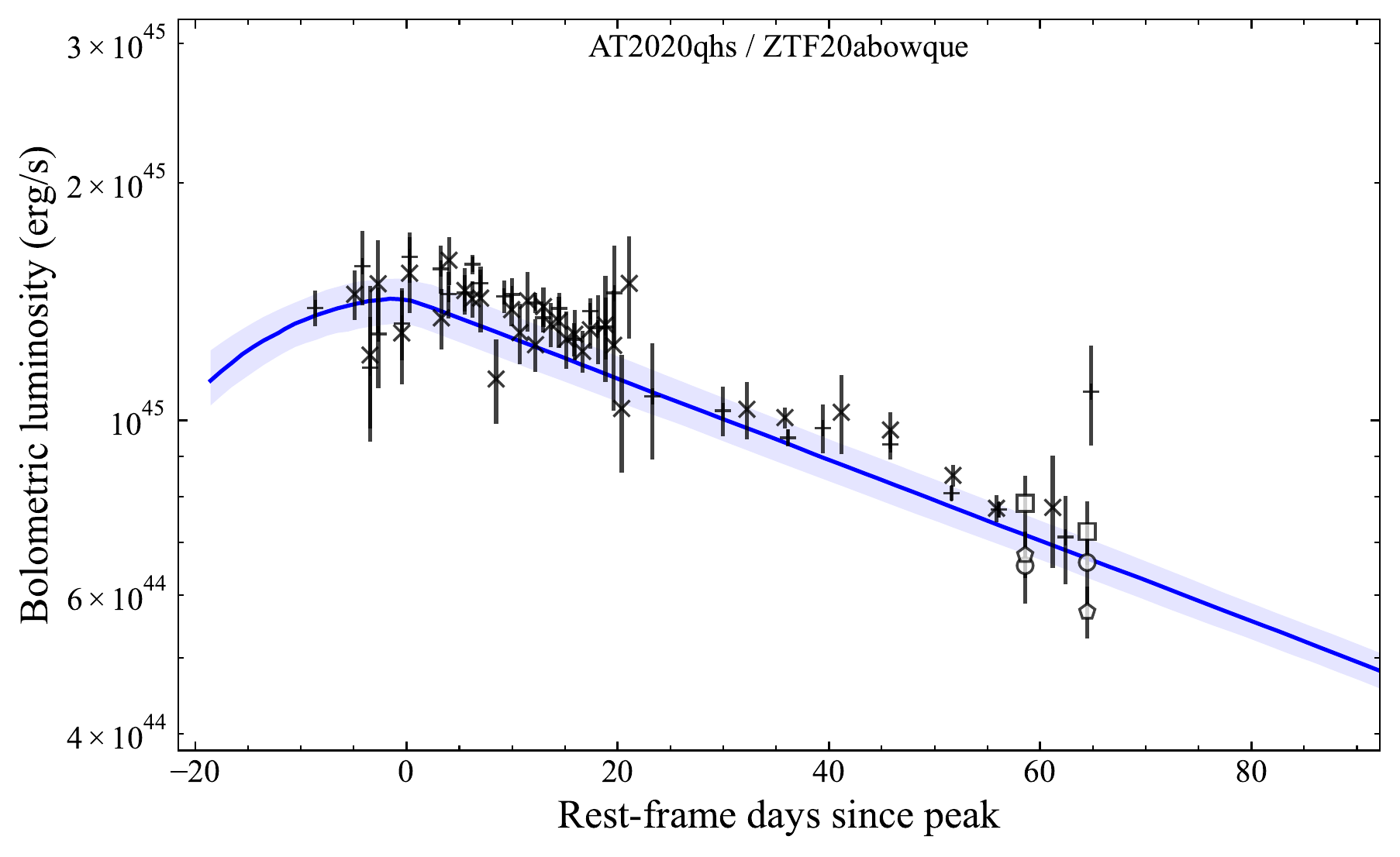}{0.33 \textwidth}{}
             \fig{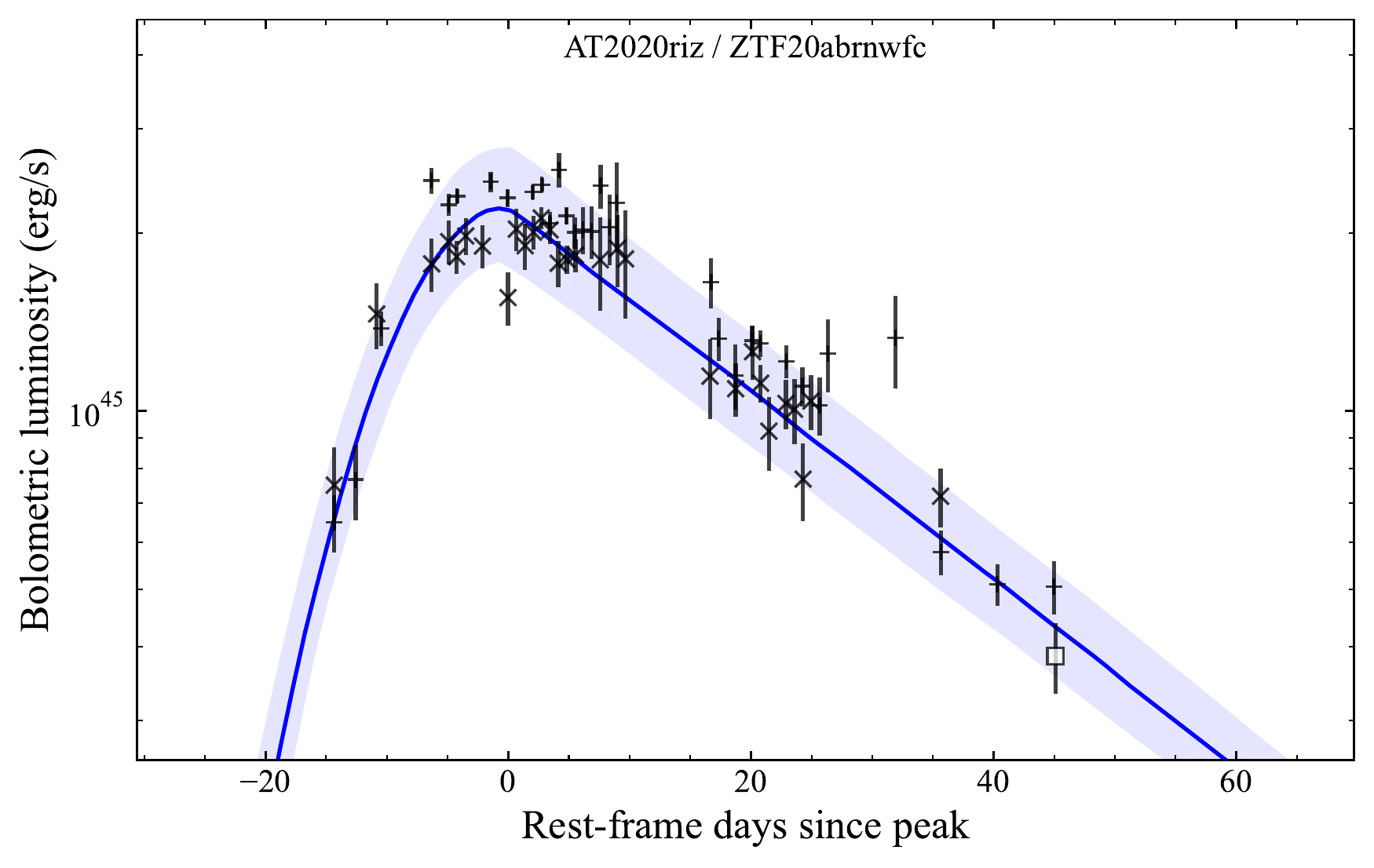}{0.33 \textwidth}{}
             \\[-20pt]}

\gridline{  \fig{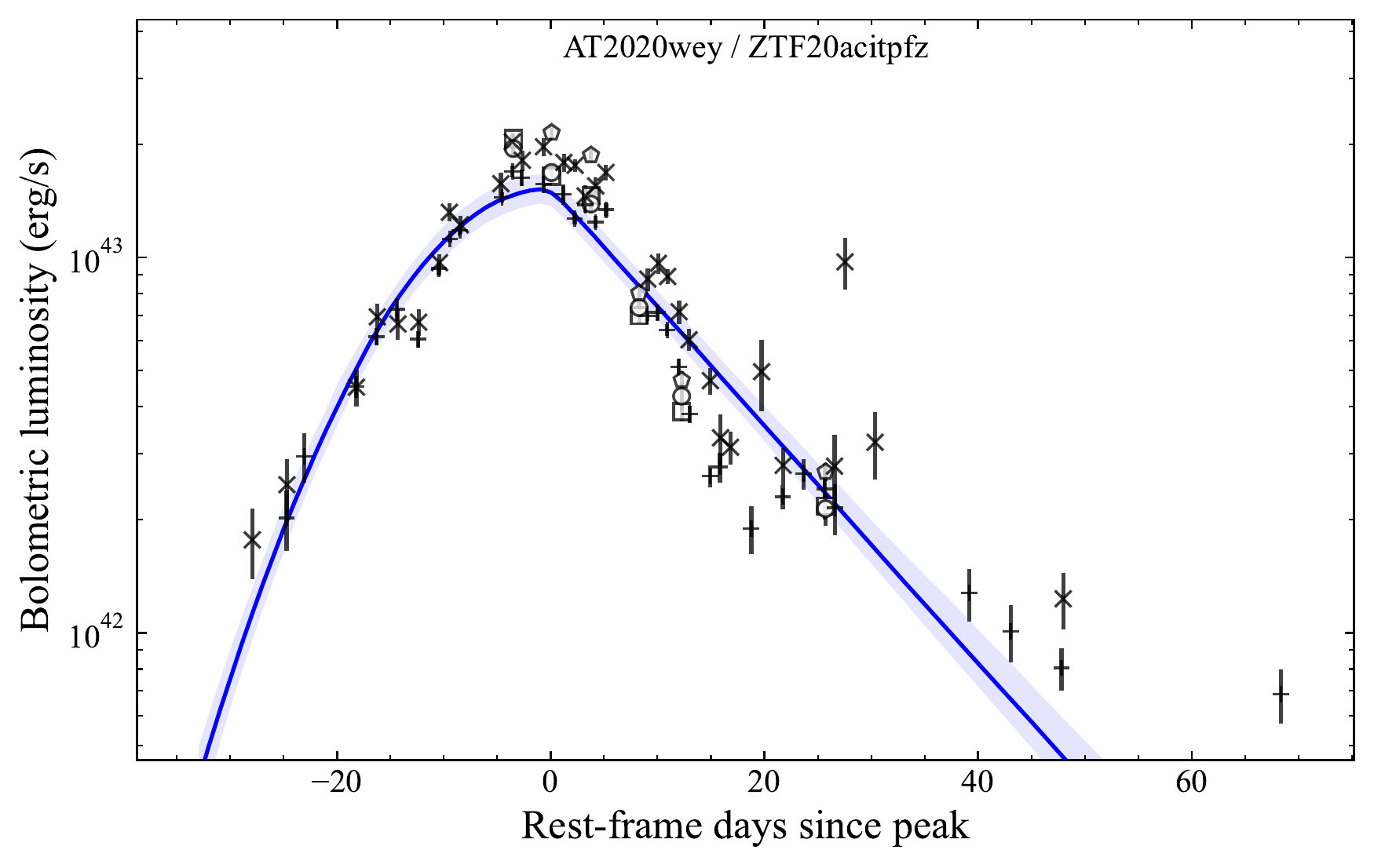}{0.33 \textwidth}{} 
            \fig{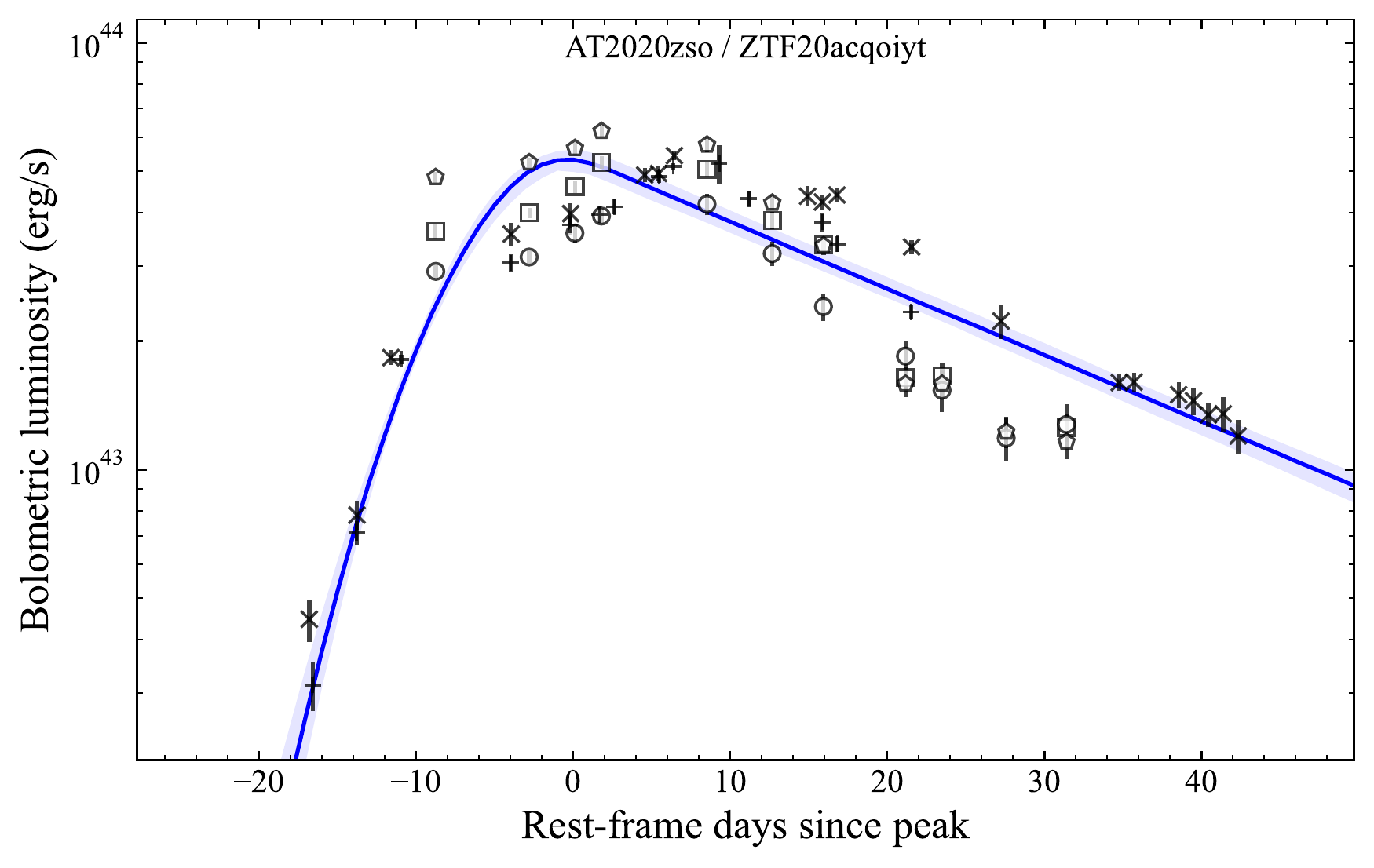}{0.33 \textwidth}{}
            \fig{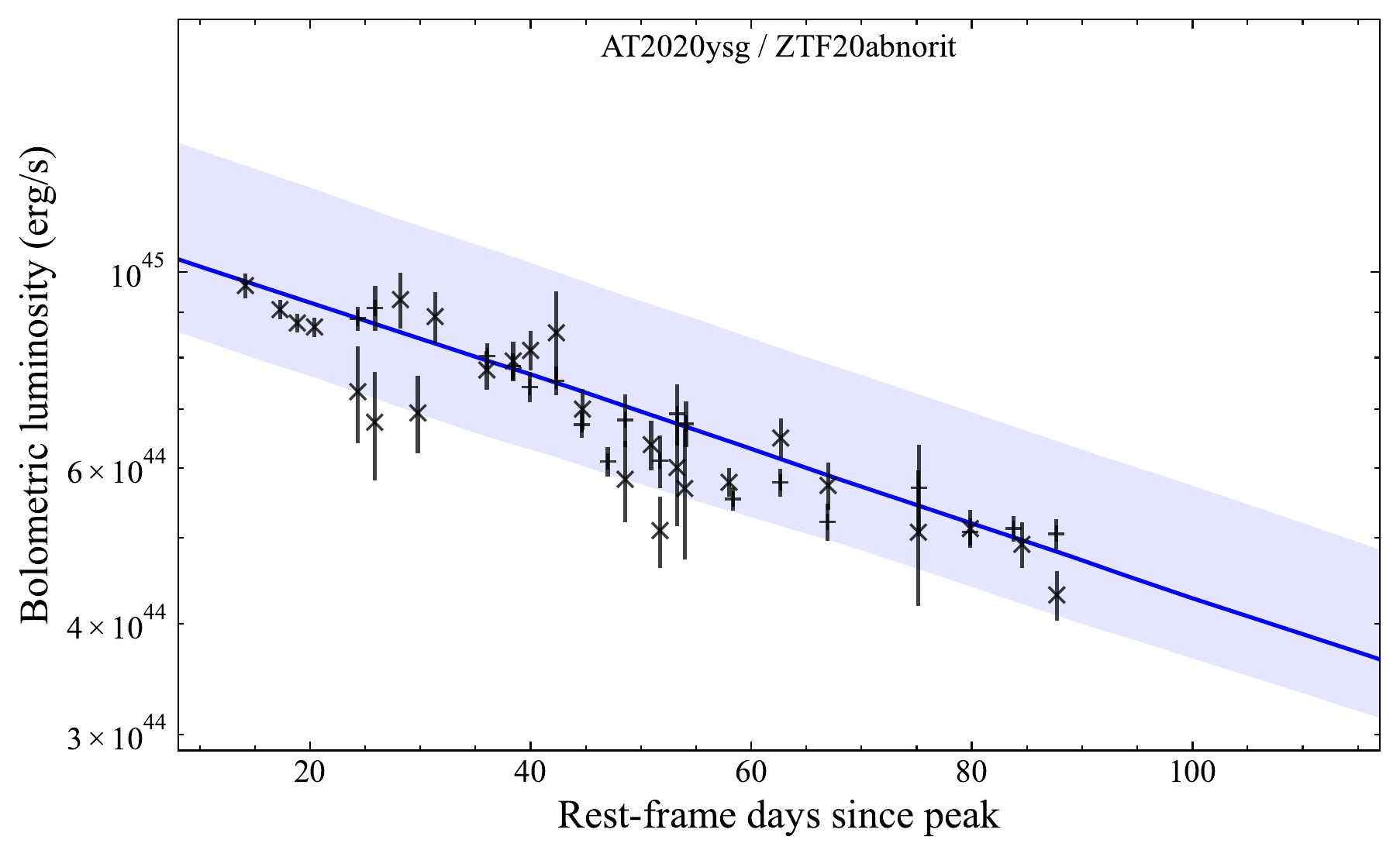}{0.33 \textwidth}{} 
            \\[-20pt]}   

\caption{Same as Figure \ref{fig:expfitlcs1}.}
\end{figure*}

\section{Results from Statistical Tests}
\movetabledown=1.0in
\begin{rotatetable*}
\centerwidetable
\begin{deluxetable*}{r | r r r r r r r r r }
\label{tab:kendalltau}
\tablecaption{Kendall's tau coefficient and associated $p$-value comparing light curve and host galaxy properties.}
\tablehead{ \colhead{} & \colhead{$u-r$} & \colhead{$M_{\rm gal}$} & \colhead{$p$} & \colhead{$t_0$ ($p=-5/3$)} & \colhead{$L_{\rm peak}$} & \colhead{$R_{\rm peak}$} & \colhead{$T_{\rm peak}$} & \colhead{$\sigma$} & \colhead{$\tau$} }
\startdata
\input{tables/lc_kendalltest_round2}
\enddata
\tablecomments{Listed are the Kendall's tau coefficient and associated $p$-value in parentheses for measured light curve and host galaxy properties. The comparisons which constitute statistically significant ($p < 0.05$) correlations are boldfaced. These correlations and the implications are discussed in Sections \ref{sec:results} and \ref{sec:discussion}.}
\end{deluxetable*}
\end{rotatetable*}

\begin{deluxetable*}{r | c c c c c c c}
\tablecaption{Anderson-Darling $p$-value comparing the four TDE spectral classes.}
\label{tab:spectest}
\tablehead{ \colhead{} & \colhead{H vs. H+He} & \colhead{H vs. He} & \colhead{H vs. Featureless} & \colhead{H+He vs. He} & \colhead{H+He vs. Featureless} & \colhead{He vs. Featureless} & \colhead{X-ray vs. non-X-ray}}
\startdata
\input{tables/class_test_round2}
\enddata
\tablecomments{Listed are the $p$-values from an Anderson-Darling test which tests the null hypothesis that the two samples are drawn from the same parent population. Cases where the null hypothesis can be rejected with $p < 0.05$ are boldfaced.}
\end{deluxetable*}

\end{document}

%% file: tables/host_tab.tex
AT2018zr & $10.01_{-0.14}^{+0.08}$ & $2.38_{-0.05}^{+0.06}$ & $0.06_{-0.04}^{+0.1}$ & $6.43_{-2.67}^{+1.87}$ & $0.24_{-0.11}^{+0.24}$ & $-0.09_{-0.11}^{+0.17}$ \\AT2018bsi & $10.62_{-0.07}^{+0.05}$ & $2.09_{-0.05}^{+0.03}$ & $0.76_{-0.25}^{+0.17}$ & $3.08_{-0.71}^{+0.63}$ & $0.8_{-0.19}^{+0.14}$ & $-0.3_{-0.4}^{+0.27}$ \\AT2018hco & $10.03_{-0.16}^{+0.12}$ & $1.85_{-0.05}^{+0.06}$ & $0.2_{-0.12}^{+0.13}$ & $7.44_{-3.04}^{+3.42}$ & $0.31_{-0.17}^{+0.32}$ & $-1.46_{-0.37}^{+0.44}$ \\AT2018iih & $10.81_{-0.14}^{+0.11}$ & $2.35_{-0.06}^{+0.08}$ & $0.38_{-0.24}^{+0.31}$ & $7.2_{-3.12}^{+3.39}$ & $0.27_{-0.12}^{+0.34}$ & $-0.51_{-0.44}^{+0.45}$ \\AT2018hyz & $9.96_{-0.16}^{+0.09}$ & $1.87_{-0.04}^{+0.05}$ & $0.23_{-0.14}^{+0.11}$ & $7.76_{-2.9}^{+2.61}$ & $0.27_{-0.14}^{+0.34}$ & $-1.48_{-0.38}^{+0.52}$ \\AT2018lni & $10.1_{-0.13}^{+0.1}$ & $1.99_{-0.07}^{+0.05}$ & $0.39_{-0.22}^{+0.2}$ & $7.93_{-2.86}^{+3.3}$ & $0.32_{-0.17}^{+0.33}$ & $-1.5_{-0.34}^{+0.44}$ \\AT2018lna & $9.56_{-0.14}^{+0.11}$ & $1.95_{-0.07}^{+0.05}$ & $0.19_{-0.14}^{+0.22}$ & $7.9_{-2.98}^{+3.16}$ & $0.3_{-0.16}^{+0.33}$ & $-1.23_{-0.45}^{+0.32}$ \\AT2018jbv & $10.34_{-0.18}^{+0.14}$ & $2.17_{-0.07}^{+0.07}$ & $0.44_{-0.31}^{+0.49}$ & $5.83_{-3.35}^{+4.09}$ & $0.43_{-0.27}^{+0.74}$ & $-0.7_{-0.55}^{+0.61}$ \\AT2019cho & $10.2_{-0.13}^{+0.14}$ & $2.12_{-0.07}^{+0.07}$ & $0.33_{-0.23}^{+0.31}$ & $5.95_{-2.38}^{+3.74}$ & $0.37_{-0.2}^{+0.39}$ & $-0.8_{-0.55}^{+0.53}$ \\AT2019bhf & $10.26_{-0.15}^{+0.16}$ & $2.09_{-0.06}^{+0.05}$ & $0.66_{-0.39}^{+0.22}$ & $3.58_{-1.46}^{+2.59}$ & $0.37_{-0.21}^{+0.39}$ & $-1.04_{-0.64}^{+0.56}$ \\AT2019azh & $9.74_{-0.05}^{+0.08}$ & $1.85_{-0.03}^{+0.04}$ & $0.24_{-0.17}^{+0.17}$ & $1.3_{-0.29}^{+0.88}$ & $0.15_{-0.03}^{+0.09}$ & $-0.27_{-0.39}^{+0.22}$ \\AT2019dsg & $10.55_{-0.12}^{+0.09}$ & $2.2_{-0.07}^{+0.05}$ & $0.31_{-0.19}^{+0.25}$ & $8.82_{-2.8}^{+2.42}$ & $0.34_{-0.19}^{+0.38}$ & $-0.82_{-0.38}^{+0.34}$ \\AT2019ehz & $9.81_{-0.12}^{+0.09}$ & $1.94_{-0.07}^{+0.06}$ & $0.27_{-0.18}^{+0.19}$ & $8.03_{-2.73}^{+3.02}$ & $0.31_{-0.16}^{+0.38}$ & $-1.33_{-0.41}^{+0.42}$ \\AT2019mha & $10.01_{-0.18}^{+0.14}$ & $1.99_{-0.07}^{+0.07}$ & $0.5_{-0.23}^{+0.23}$ & $3.24_{-1.61}^{+1.87}$ & $0.28_{-0.13}^{+0.41}$ & $-1.07_{-0.65}^{+0.79}$ \\AT2019meg & $9.64_{-0.08}^{+0.07}$ & $1.83_{-0.07}^{+0.05}$ & $0.71_{-0.27}^{+0.2}$ & $2.34_{-0.63}^{+0.82}$ & $0.7_{-0.2}^{+0.2}$ & $-0.61_{-0.53}^{+0.57}$ \\AT2019lwu & $9.99_{-0.15}^{+0.09}$ & $1.85_{-0.04}^{+0.06}$ & $0.15_{-0.11}^{+0.13}$ & $8.66_{-3.19}^{+2.55}$ & $0.26_{-0.12}^{+0.29}$ & $-1.43_{-0.39}^{+0.52}$ \\AT2019qiz & $10.01_{-0.12}^{+0.1}$ & $2.1_{-0.05}^{+0.04}$ & $0.36_{-0.24}^{+0.2}$ & $5.95_{-1.94}^{+3.17}$ & $0.3_{-0.15}^{+0.34}$ & $-0.97_{-0.71}^{+0.25}$ \\AT2019teq & $9.95_{-0.11}^{+0.07}$ & $1.86_{-0.04}^{+0.03}$ & $0.47_{-0.34}^{+0.33}$ & $3.34_{-1.27}^{+0.94}$ & $0.71_{-0.31}^{+0.21}$ & $-0.79_{-0.81}^{+0.39}$ \\AT2020pj & $10.07_{-0.13}^{+0.09}$ & $1.85_{-0.05}^{+0.06}$ & $0.74_{-0.47}^{+0.43}$ & $7.8_{-3.64}^{+3.21}$ & $2.74_{-1.94}^{+2.96}$ & $-1.44_{-0.38}^{+0.63}$ \\AT2019vcb & $9.49_{-0.06}^{+0.06}$ & $1.55_{-0.04}^{+0.03}$ & $0.67_{-0.22}^{+0.2}$ & $1.74_{-0.45}^{+0.54}$ & $0.69_{-0.19}^{+0.18}$ & $-1.01_{-0.37}^{+0.66}$ \\AT2020ddv & $10.3_{-0.16}^{+0.13}$ & $2.21_{-0.06}^{+0.06}$ & $0.58_{-0.35}^{+0.26}$ & $6.75_{-2.86}^{+3.79}$ & $0.31_{-0.15}^{+0.35}$ & $-1.03_{-0.66}^{+0.62}$ \\AT2020ocn & $10.28_{-0.17}^{+0.13}$ & $2.25_{-0.04}^{+0.06}$ & $0.76_{-0.44}^{+0.14}$ & $6.7_{-3.02}^{+4.09}$ & $0.27_{-0.14}^{+0.31}$ & $-1.26_{-0.55}^{+0.62}$ \\AT2020opy & $10.01_{-0.14}^{+0.13}$ & $1.78_{-0.06}^{+0.04}$ & $0.45_{-0.23}^{+0.21}$ & $2.25_{-0.97}^{+1.51}$ & $0.29_{-0.15}^{+0.38}$ & $-1.32_{-0.47}^{+0.52}$ \\AT2020mot & $10.42_{-0.11}^{+0.07}$ & $2.22_{-0.06}^{+0.07}$ & $0.13_{-0.09}^{+0.23}$ & $9.12_{-2.87}^{+2.32}$ & $0.29_{-0.17}^{+0.34}$ & $-0.65_{-0.35}^{+0.24}$ \\AT2020mbq & $9.64_{-0.15}^{+0.11}$ & $2.11_{-0.07}^{+0.07}$ & $0.42_{-0.26}^{+0.24}$ & $7.4_{-3.05}^{+3.22}$ & $0.29_{-0.15}^{+0.37}$ & $-1.11_{-0.59}^{+0.44}$ \\AT2020qhs & $11.23_{-0.07}^{+0.07}$ & $2.38_{-0.06}^{+0.05}$ & $0.09_{-0.06}^{+0.12}$ & $5.81_{-1.51}^{+1.96}$ & $0.45_{-0.3}^{+0.38}$ & $0.0_{-0.36}^{+0.05}$ \\AT2020riz & $11.1_{-0.13}^{+0.1}$ & $2.81_{-0.14}^{+0.13}$ & $0.82_{-0.22}^{+0.14}$ & $8.28_{-3.13}^{+2.75}$ & $0.36_{-0.21}^{+0.38}$ & $-0.19_{-0.39}^{+0.29}$ \\AT2020wey & $9.63_{-0.22}^{+0.18}$ & $2.11_{-0.07}^{+0.04}$ & $0.05_{-0.04}^{+0.06}$ & $5.63_{-3.42}^{+4.63}$ & $0.22_{-0.1}^{+0.26}$ & $-0.25_{-0.5}^{+0.37}$ \\AT2020zso & $10.05_{-0.12}^{+0.09}$ & $1.95_{-0.04}^{+0.04}$ & $0.47_{-0.26}^{+0.28}$ & $3.34_{-1.36}^{+1.37}$ & $0.59_{-0.3}^{+0.28}$ & $-0.96_{-0.69}^{+0.52}$ \\AT2020ysg & $10.72_{-0.12}^{+0.11}$ & $2.5_{-0.06}^{+0.07}$ & $0.45_{-0.23}^{+0.33}$ & $3.62_{-1.71}^{+2.27}$ & $0.17_{-0.05}^{+0.16}$ & $0.12_{-0.13}^{+0.05}$ \\

%% file: tables/LCfittab.tex
AT2018zr & $1.16_{-0.04}^{+0.04}$ & $1.83_{-0.03}^{+0.03}$ & $4.14_{-0.01}^{+0.01}$ & $43.71_{-0.04}^{+0.07}$ & $1.36_{-0.19}^{+0.15}$ & $2.04_{-0.03}^{+0.03}$ & $-0.78_{-0.12}^{+0.09}$ & $43.55_{-0.02}^{+0.02}$ & $0.56_{-0.04}^{+0.05}$ & $58202.06_{-4.24}^{+1.88}$ \\AT2018bsi & $1.40_{-0.35}^{+0.07}$ & $1.60_{-0.14}^{+0.26}$ & $4.30_{-0.03}^{+0.03}$ & $43.96_{-0.08}^{+0.14}$ & $1.97_{-0.29}^{+0.19}$ & $1.72_{-0.10}^{+0.09}$ & $-2.92_{-0.86}^{+0.93}$ & $43.51_{-0.05}^{+0.05}$ & $-0.76_{-0.15}^{+0.66}$ & $58212.63_{-5.33}^{+7.92}$ \\AT2018hco & $1.06_{-0.02}^{+0.02}$ & $2.04_{-0.02}^{+0.02}$ & $4.32_{-0.01}^{+0.01}$ & $44.18_{-0.03}^{+0.06}$ & $1.98_{-0.14}^{+0.11}$ & $2.06_{-0.02}^{+0.02}$ & $-1.68_{-0.26}^{+0.27}$ & $43.75_{-0.01}^{+0.01}$ & $-0.12_{-0.04}^{+0.05}$ & $58409.75_{-2.07}^{+2.12}$ \\AT2018iih & $1.36_{-0.01}^{+0.01}$ & $2.06_{-0.03}^{+0.03}$ & $4.22_{-0.02}^{+0.02}$ & $44.71_{-0.07}^{+0.08}$ & $1.62_{-0.19}^{+0.19}$ & $2.20_{-0.04}^{+0.04}$ & $-0.88_{-0.15}^{+0.14}$ & $44.39_{-0.02}^{+0.02}$ & $0.09_{-0.07}^{+0.04}$ & $58459.66_{-2.38}^{+2.14}$ \\AT2018hyz & $0.73_{-0.50}^{+0.55}$ & $1.71_{-0.01}^{+0.01}$ & $4.21_{-0.01}^{+0.01}$ & $44.30_{-0.16}^{+0.22}$ & $1.29_{-0.19}^{+0.10}$ & $1.54_{-0.09}^{+0.09}$ & $-1.20_{-0.09}^{+0.08}$ & $43.95_{-0.08}^{+0.07}$ & $0.13_{-0.04}^{+0.04}$ & $58424.96_{-6.63}^{+7.90}$ \\AT2018lni & $1.36_{-0.02}^{+0.02}$ & $1.78_{-0.02}^{+0.02}$ & $4.34_{-0.03}^{+0.03}$ & $44.39_{-0.04}^{+0.07}$ & $2.00_{-0.17}^{+0.17}$ & $1.96_{-0.06}^{+0.07}$ & $-2.51_{-0.71}^{+0.50}$ & $43.92_{-0.02}^{+0.02}$ & $-0.20_{-0.13}^{+0.17}$ & $58477.08_{-2.81}^{+2.16}$ \\AT2018lna & $1.08_{-0.04}^{+0.04}$ & $1.66_{-0.02}^{+0.02}$ & $4.50_{-0.01}^{+0.02}$ & $44.53_{-0.04}^{+0.07}$ & $1.62_{-0.16}^{+0.13}$ & $1.70_{-0.04}^{+0.05}$ & $-2.20_{-0.40}^{+0.38}$ & $43.67_{-0.01}^{+0.01}$ & $-0.36_{-0.16}^{+0.20}$ & $58507.58_{-2.49}^{+1.59}$ \\AT2018jbv & $1.40_{-0.10}^{+0.07}$ & $2.02_{-0.03}^{+0.03}$ & $4.50_{-0.01}^{+0.01}$ & $45.57_{-0.06}^{+0.09}$ & $1.82_{-0.15}^{+0.13}$ & $1.91_{-0.02}^{+0.02}$ & $-1.78_{-0.24}^{+0.22}$ & $44.69_{-0.02}^{+0.02}$ & $-0.46_{-0.05}^{+0.05}$ & $58469.19_{-2.40}^{+2.47}$ \\AT2019cho & $1.33_{-0.07}^{+0.06}$ & $1.89_{-0.03}^{+0.04}$ & $4.11_{-0.02}^{+0.02}$ & $43.85_{-0.04}^{+0.05}$ & $2.22_{-0.18}^{+0.15}$ & $2.32_{-0.14}^{+0.23}$ & $-3.00_{-0.72}^{+0.72}$ & $43.69_{-0.03}^{+0.03}$ & $0.47_{-0.31}^{+0.28}$ & $58552.00_{-2.32}^{+2.44}$ \\AT2019bhf & $0.87_{-0.22}^{+0.07}$ & $1.65_{-0.03}^{+0.03}$ & $4.23_{-0.04}^{+0.04}$ & $44.03_{-0.05}^{+0.08}$ & $1.76_{-0.25}^{+0.21}$ & $1.68_{-0.06}^{+0.06}$ & $-2.23_{-0.69}^{+0.61}$ & $43.76_{-0.04}^{+0.04}$ & $-0.09_{-0.21}^{+0.20}$ & $58543.90_{-1.87}^{+1.47}$ \\AT2019azh & $1.13_{-0.01}^{+0.01}$ & $1.80_{-0.02}^{+0.02}$ & $4.39_{-0.01}^{+0.01}$ & $44.37_{-0.02}^{+0.03}$ & $1.77_{-0.09}^{+0.08}$ & $1.77_{-0.02}^{+0.02}$ & $-2.10_{-0.18}^{+0.19}$ & $43.81_{-0.01}^{+0.01}$ & $0.09_{-0.03}^{+0.03}$ & $58566.78_{-1.75}^{+1.16}$ \\AT2019dsg & $1.37_{-0.08}^{+0.08}$ & $1.86_{-0.02}^{+0.02}$ & $4.33_{-0.01}^{+0.01}$ & $44.29_{-0.08}^{+0.10}$ & $1.61_{-0.13}^{+0.12}$ & $1.75_{-0.02}^{+0.02}$ & $-1.84_{-0.18}^{+0.17}$ & $43.57_{-0.02}^{+0.02}$ & $-0.23_{-0.04}^{+0.04}$ & $58605.38_{-2.38}^{+2.48}$ \\AT2019ehz & $1.03_{-0.02}^{+0.02}$ & $1.64_{-0.01}^{+0.01}$ & $4.27_{-0.01}^{+0.01}$ & $44.00_{-0.03}^{+0.04}$ & $1.51_{-0.10}^{+0.09}$ & $1.56_{-0.03}^{+0.02}$ & $-1.44_{-0.11}^{+0.11}$ & $43.63_{-0.01}^{+0.01}$ & $-0.39_{-0.02}^{+0.03}$ & $58614.21_{-0.62}^{+0.54}$ \\AT2019mha & $1.18_{-0.03}^{+0.03}$ & $1.23_{-0.03}^{+0.03}$ & $4.22_{-0.04}^{+0.05}$ & $43.84_{-0.04}^{+0.06}$ & $1.64_{-0.17}^{+0.16}$ & $1.38_{-0.07}^{+0.07}$ & $-3.26_{-1.03}^{+0.76}$ & $43.60_{-0.04}^{+0.04}$ & $0.95_{-0.81}^{+0.70}$ & $58705.23_{-1.01}^{+0.98}$ \\AT2019meg & $0.98_{-0.03}^{+0.02}$ & $1.68_{-0.02}^{+0.02}$ & $4.34_{-0.01}^{+0.01}$ & $44.24_{-0.01}^{+0.02}$ & $1.85_{-0.17}^{+0.17}$ & $1.77_{-0.04}^{+0.04}$ & $-2.28_{-0.73}^{+0.57}$ & $43.80_{-0.01}^{+0.01}$ & $-0.06_{-0.19}^{+0.21}$ & $58697.91_{-0.89}^{+0.73}$ \\AT2019lwu & $0.88_{-0.06}^{+0.05}$ & $1.45_{-0.03}^{+0.03}$ & $4.13_{-0.02}^{+0.02}$ & $43.76_{-0.07}^{+0.08}$ & $1.90_{-0.17}^{+0.13}$ & $1.60_{-0.07}^{+0.07}$ & $-3.80_{-0.84}^{+0.93}$ & $43.56_{-0.03}^{+0.03}$ & $0.47_{-0.36}^{+0.32}$ & $58694.28_{-1.04}^{+1.17}$ \\AT2019qiz & $0.96_{-0.01}^{+0.01}$ & $1.48_{-0.01}^{+0.01}$ & $4.18_{-0.01}^{+0.01}$ & $43.43_{-0.03}^{+0.03}$ & $1.20_{-0.05}^{+0.05}$ & $1.19_{-0.02}^{+0.02}$ & $-1.61_{-0.06}^{+0.07}$ & $43.13_{-0.01}^{+0.01}$ & $-0.21_{-0.02}^{+0.02}$ & $58767.61_{-0.61}^{+0.60}$ \\AT2019teq & $1.14_{-0.04}^{+0.04}$ & $2.08_{-0.06}^{+0.06}$ & $4.15_{-0.01}^{+0.01}$ & $43.35_{-0.04}^{+0.05}$ & $2.21_{-0.24}^{+0.18}$ & $2.37_{-0.08}^{+0.08}$ & $-2.32_{-0.65}^{+0.61}$ & $43.15_{-0.02}^{+0.02}$ & $0.29_{-0.09}^{+0.09}$ & $58794.31_{-3.54}^{+2.40}$ \\AT2020pj & $1.04_{-0.05}^{+0.05}$ & $1.57_{-0.02}^{+0.02}$ & $4.10_{-0.01}^{+0.01}$ & $43.26_{-0.05}^{+0.06}$ & $1.46_{-0.18}^{+0.19}$ & $1.50_{-0.04}^{+0.05}$ & $-1.56_{-0.35}^{+0.24}$ & $43.07_{-0.02}^{+0.03}$ & $-0.02_{-0.11}^{+0.17}$ & $58866.68_{-1.07}^{+1.10}$ \\AT2019vcb & $1.18_{-0.03}^{+0.04}$ & $1.56_{-0.01}^{+0.01}$ & $4.21_{-0.03}^{+0.03}$ & $43.85_{-0.06}^{+0.09}$ & $1.59_{-0.15}^{+0.14}$ & $1.83_{-0.09}^{+0.09}$ & $-2.14_{-0.44}^{+0.36}$ & $43.59_{-0.04}^{+0.04}$ & $0.63_{-0.20}^{+0.23}$ & $58825.55_{-1.53}^{+2.27}$ \\AT2020ddv & $1.19_{-0.03}^{+0.03}$ & $1.79_{-0.02}^{+0.02}$ & $4.56_{-0.02}^{+0.02}$ & $44.86_{-0.03}^{+0.06}$ & $1.60_{-0.19}^{+0.17}$ & $1.85_{-0.05}^{+0.06}$ & $-1.80_{-0.40}^{+0.40}$ & $43.86_{-0.01}^{+0.01}$ & $-0.44_{-0.17}^{+0.27}$ & $58919.26_{-2.36}^{+2.51}$ \\AT2020ocn & $1.20_{-0.04}^{+0.04}$ & $1.88_{-0.03}^{+0.03}$ & $4.47_{-0.02}^{+0.02}$ & $43.69_{-0.03}^{+0.05}$ & $1.90_{-0.13}^{+0.11}$ & $1.99_{-0.06}^{+0.07}$ & $-2.27_{-0.36}^{+0.36}$ & $42.94_{-0.01}^{+0.01}$ & $-0.01_{-0.19}^{+0.25}$ & $58989.56_{-3.16}^{+2.25}$ \\AT2020opy & $1.20_{-0.01}^{+0.01}$ & $1.80_{-0.02}^{+0.02}$ & $4.30_{-0.01}^{+0.01}$ & $44.30_{-0.00}^{+0.00}$ & $1.14_{-0.15}^{+0.17}$ & $1.83_{-0.02}^{+0.02}$ & $-0.83_{-0.17}^{+0.14}$ & $43.95_{-0.01}^{+0.01}$ & $-0.21_{-0.07}^{+0.07}$ & $59088.79_{-2.51}^{+0.62}$ \\AT2020mot & $1.50_{-0.00}^{+0.00}$ & $1.83_{-0.01}^{+0.01}$ & $4.27_{-0.00}^{+0.01}$ & $43.96_{-0.05}^{+0.05}$ & $1.87_{-0.09}^{+0.08}$ & $1.83_{-0.02}^{+0.02}$ & $-1.86_{-0.22}^{+0.21}$ & $43.60_{-0.01}^{+0.01}$ & $-0.22_{-0.03}^{+0.04}$ & $59072.20_{-2.86}^{+4.96}$ \\AT2020mbq & $1.01_{-0.02}^{+0.02}$ & $1.63_{-0.01}^{+0.02}$ & $4.12_{-0.02}^{+0.02}$ & $43.47_{-0.04}^{+0.05}$ & $1.80_{-0.16}^{+0.13}$ & $1.80_{-0.05}^{+0.05}$ & $-1.96_{-0.35}^{+0.35}$ & $43.36_{-0.03}^{+0.03}$ & $0.29_{-0.19}^{+0.18}$ & $59023.53_{-0.80}^{+0.81}$ \\AT2020qhs & $1.45_{-0.07}^{+0.04}$ & $1.93_{-0.02}^{+0.02}$ & $4.48_{-0.01}^{+0.01}$ & $45.36_{-0.04}^{+0.08}$ & $1.78_{-0.13}^{+0.15}$ & $1.92_{-0.02}^{+0.02}$ & $-1.65_{-0.33}^{+0.28}$ & $44.56_{-0.01}^{+0.01}$ & $-0.55_{-0.07}^{+0.08}$ & $59063.64_{-6.85}^{+2.52}$ \\AT2020riz & $0.97_{-0.04}^{+0.04}$ & $1.44_{-0.02}^{+0.02}$ & $4.52_{-0.04}^{+0.04}$ & $45.74_{-0.12}^{+0.13}$ & $1.69_{-0.16}^{+0.12}$ & $1.46_{-0.07}^{+0.07}$ & $-3.82_{-0.82}^{+0.89}$ & $44.68_{-0.01}^{+0.01}$ & $-1.36_{-0.38}^{+0.33}$ & $59082.56_{-0.68}^{+0.62}$ \\AT2020wey & $1.09_{-0.02}^{+0.02}$ & $1.14_{-0.02}^{+0.02}$ & $4.36_{-0.02}^{+0.02}$ & $43.29_{-0.01}^{+0.02}$ & $0.93_{-0.12}^{+0.13}$ & $0.93_{-0.02}^{+0.03}$ & $-1.71_{-0.31}^{+0.23}$ & $42.81_{-0.02}^{+0.02}$ & $-1.54_{-0.17}^{+0.31}$ & $59156.58_{-0.51}^{+0.43}$ \\AT2020zso & $0.84_{-0.05}^{+0.05}$ & $1.44_{-0.03}^{+0.03}$ & $4.23_{-0.01}^{+0.01}$ & $43.76_{-0.02}^{+0.02}$ & $1.41_{-0.12}^{+0.13}$ & $1.39_{-0.05}^{+0.05}$ & $-2.12_{-0.41}^{+0.28}$ & $43.53_{-0.02}^{+0.02}$ & $-1.75_{-0.16}^{+0.20}$ & $59188.04_{-1.35}^{+1.37}$ \\AT2020ysg & $1.49_{-0.01}^{+0.00}$ & $2.02_{-0.02}^{+0.02}$ & $4.41_{-0.04}^{+0.05}$ & $45.34_{-0.12}^{+0.16}$ & $1.53_{-0.19}^{+0.15}$ & $2.01_{-0.03}^{+0.03}$ & $-1.24_{-0.26}^{+0.22}$ & $44.59_{-0.03}^{+0.03}$ & $-0.24_{-0.07}^{+0.08}$ & $59122.64_{-2.20}^{+2.35}$ \\

%% file: tables/MBH_MStar.tex
AT2018zr&$ 6.97_{-0.05}^{+0.09}$&$ 6.53_{-0.17}^{+0.11}$&$ 1.20_{-0.06}^{+0.22}$&$ 3.52_{-1.60}^{+2.20}$\\AT2018bsi&$ 6.51_{-0.07}^{+0.10}$&$ 6.65_{-0.36}^{+0.52}$&$ 1.20_{-0.20}^{+1.30}$&$ 0.75_{-0.44}^{+1.69}$\\AT2018hco&$ 6.60_{-0.02}^{+0.02}$&$ 6.50_{-0.15}^{+0.36}$&$ 2.40_{-0.39}^{+0.75}$&$ 1.54_{-0.63}^{+0.84}$\\AT2018iih&$ 6.84_{-0.08}^{+0.08}$&$ 6.34_{-0.04}^{+0.05}$&$ 63.00_{-31.00}^{+59220.00}$&$ 3.98_{-0.88}^{+1.51}$\\AT2018hyz&$ 6.77_{-0.02}^{+0.02}$&$ 6.58_{-0.06}^{+0.05}$&$ 4.60_{-1.80}^{+13.00}$&$ 0.99_{-0.06}^{+0.05}$\\AT2018lni&$ 6.45_{-0.02}^{+0.01}$&$ 6.60_{-0.17}^{+0.12}$&$ 5.10_{-1.10}^{+2.20}$&$ 0.79_{-0.23}^{+0.31}$\\AT2018lna&$ 6.11_{-0.01}^{+0.01}$&$ 6.83_{-0.13}^{+0.19}$&$ 5.50_{-1.40}^{+3.00}$&$ 2.85_{-1.76}^{+2.82}$\\AT2018jbv& -- &$ 7.55_{-0.14}^{+0.10}$& -- &$ 4.61_{-1.63}^{+2.18}$\\AT2019cho&$ 7.00_{-0.13}^{+0.12}$&$ 6.41_{-0.10}^{+0.10}$&$ 1.50_{-0.25}^{+0.63}$&$ 2.77_{-0.92}^{+1.96}$\\AT2019bhf&$ 6.77_{-0.07}^{+0.03}$&$ 6.80_{-0.21}^{+0.19}$&$ 1.80_{-0.40}^{+0.94}$&$ 0.83_{-0.49}^{+0.86}$\\AT2019azh&$ 6.34_{-0.00}^{+0.01}$&$ 7.43_{-0.31}^{+0.11}$&$ 3.60_{-0.39}^{+0.70}$&$ 3.59_{-0.95}^{+2.55}$\\AT2019dsg&$ 6.30_{-0.03}^{+0.01}$&$ 6.86_{-0.08}^{+0.09}$&$ 2.10_{-0.67}^{+1.60}$&$ 8.71_{-6.46}^{+3.04}$\\AT2019ehz&$ 6.51_{-0.03}^{+0.03}$&$ 6.78_{-0.07}^{+0.08}$&$ 1.20_{-0.08}^{+0.14}$&$ 9.81_{-3.41}^{+2.89}$\\AT2019mha&$ 6.68_{-0.08}^{+0.08}$&$ 6.64_{-0.12}^{+0.12}$&$ 1.10_{-0.11}^{+0.20}$&$ 5.09_{-1.89}^{+2.51}$\\AT2019meg&$ 6.54_{-0.01}^{+0.00}$&$ 6.68_{-0.14}^{+0.47}$&$ 3.10_{-0.20}^{+0.34}$&$ 0.96_{-0.54}^{+1.48}$\\AT2019lwu&$ 6.79_{-0.15}^{+0.20}$&$ 6.37_{-0.20}^{+0.21}$&$ 1.10_{-0.17}^{+0.35}$&$ 1.10_{-0.22}^{+2.17}$\\AT2019qiz&$ 6.20_{-0.02}^{+0.02}$&$ 6.31_{-0.29}^{+1.06}$&$ 0.64_{-0.03}^{+0.03}$&$ 3.00_{-0.76}^{+0.57}$\\AT2019teq&$ 6.30_{-0.04}^{+0.05}$&$ 6.05_{-0.40}^{+0.37}$&$ 0.62_{-0.06}^{+0.06}$&$ 0.49_{-0.20}^{+0.34}$\\AT2020pj&$ 6.46_{-0.05}^{+0.05}$&$ 7.98_{-0.03}^{+0.02}$&$ 0.59_{-0.05}^{+0.05}$&$ 10.37_{-4.77}^{+3.78}$\\AT2019vcb&$ 6.81_{-0.12}^{+0.14}$&$ 7.92_{-0.04}^{+0.04}$&$ 1.20_{-0.17}^{+0.52}$&$ 12.21_{-3.33}^{+2.36}$\\AT2020ddv&$ 5.96_{-0.01}^{+0.00}$&$ 7.93_{-0.13}^{+0.05}$&$ 22.00_{-4.50}^{+9.90}$&$ 13.08_{-5.34}^{+3.88}$\\AT2020ocn&$ 5.65_{-0.00}^{+0.01}$&$ 7.06_{-0.35}^{+0.16}$&$ 0.68_{-0.03}^{+0.04}$&$ 18.12_{-6.03}^{+6.53}$\\AT2020opy&$ 6.46_{-0.00}^{+0.00}$&$ 6.85_{-0.15}^{+0.16}$&$ 3.50_{-0.02}^{+0.09}$&$ 2.69_{-0.82}^{+1.64}$\\AT2020mot&$ 6.51_{-0.04}^{+0.06}$&$ 6.67_{-0.19}^{+0.17}$&$ 1.10_{-0.10}^{+0.17}$&$ 1.01_{-0.12}^{+1.50}$\\AT2020mbq&$ 6.67_{-0.08}^{+0.09}$&$ 6.82_{-0.30}^{+0.33}$&$ 0.79_{-0.05}^{+0.07}$&$ 2.39_{-1.20}^{+2.25}$\\AT2020qhs& -- &$ 7.22_{-0.07}^{+0.07}$& -- &$ 1.01_{-0.28}^{+0.42}$\\AT2020riz& -- &$ 7.37_{-0.10}^{+0.14}$& -- &$ 4.99_{-1.51}^{+2.89}$\\AT2020wey&$ 5.63_{-0.00}^{+0.00}$&$ 7.36_{-0.03}^{+0.04}$&$ 0.48_{-0.01}^{+0.02}$&$ 4.34_{-1.53}^{+1.96}$\\AT2020zso&$ 6.72_{-0.03}^{+0.03}$&$ 6.25_{-0.16}^{+0.79}$&$ 1.00_{-0.04}^{+0.04}$&$ 0.97_{-0.71}^{+0.06}$\\AT2020ysg& -- &$ 7.02_{-0.08}^{+0.08}$& -- &$ 1.26_{-0.40}^{+0.87}$\\

%% file: tables/lc_kendalltest_round2.tex
$u-r$&& \textbf{0.462  ($<$0.001)} & 0.085  (0.524) & 0.149  (0.256) & 0.154  (0.241) & 0.099  (0.457) & 0.209  (0.109) & \textbf{0.26  (0.045)} & 0.159  (0.227) \\ $M_{\rm  gal}$ & \textbf{0.462  ($<$0.001)} && 0.007  (0.972) & 0.154  (0.241) & \textbf{0.361  (0.005)} & 0.195  (0.135) & \textbf{0.278  (0.031)} & \textbf{0.301  (0.019)} & \textbf{0.31  (0.016)} \\$p$& 0.085  (0.524) & 0.007  (0.972) && 0.136  (0.304) & 0.067  (0.62) & 0.067  (0.62) & -0.044  (0.75) & 0.172  (0.188) & 0.228  (0.08) \\$t_0$ ($p=-5/3$)& 0.149  (0.256) & 0.154  (0.241) & 0.136  (0.304) && 0.177  (0.177) & 0.214  (0.101) & 0.048  (0.724) & \textbf{0.421  ($<$0.001)} & \textbf{0.706  ($<$0.001)} \\ $L_{\rm  peak}$ & 0.154  (0.241) & \textbf{0.361  (0.005)} & 0.067  (0.62) & 0.177  (0.177) && \textbf{0.32  (0.013)} & \textbf{0.605  ($<$0.001)} & 0.168  (0.201) & \textbf{0.278  (0.031)} \\ $R_{\rm  peak}$ & 0.099  (0.457) & 0.195  (0.135) & 0.067  (0.62) & 0.214  (0.101) & \textbf{0.32  (0.013)} && -0.076  (0.571) & 0.094  (0.479) & 0.186  (0.155) \\ $T_{\rm  peak}$ & 0.209  (0.109) & \textbf{0.278  (0.031)} & -0.044  (0.75) & 0.048  (0.724) & \textbf{0.605  ($<$0.001)} & -0.076  (0.571) && 0.131  (0.321) & 0.122  (0.357) \\ $\sigma$ & \textbf{0.26  (0.045)} & \textbf{0.301  (0.019)} & 0.172  (0.188) & \textbf{0.421  ($<$0.001)} & 0.168  (0.201) & 0.094  (0.479) & 0.131  (0.321) && \textbf{0.439  ($<$0.001)} \\ $\tau$ & 0.159  (0.227) & \textbf{0.31  (0.016)} & 0.228  (0.08) & \textbf{0.706  ($<$0.001)} & \textbf{0.278  (0.031)} & 0.186  (0.155) & 0.122  (0.357) & \textbf{0.439  ($<$0.001)} &\\

%% file: tables/class_test_round2.tex
$\tau$ &$>0.25$&0.146&$>0.25$&0.118&0.082&$>0.25$&0.062\\ $\sigma$ &0.082&0.064&0.086&$>0.25$&0.114&0.215&$>0.25$\\ $T_{\rm  peak}$ &$>0.25$&0.105& \textbf{0.005} &0.102& \textbf{0.002} &$>0.25$&$>0.25$\\ $R_{\rm  peak}$ &0.157&0.127& \textbf{0.005} &0.134& \textbf{0.001} &0.223&$>0.25$\\ $L_{\rm  peak}$ &$>0.25$&0.105& \textbf{0.005} &0.054& \textbf{0.001} & \textbf{0.017} & \textbf{0.049} \\ $M_{\rm  gal}$ &0.079& \textbf{0.009} & \textbf{0.005} &0.057& \textbf{0.002} &0.163&$>0.25$\\$u-r$&$>0.25$&0.098& \textbf{0.025} & \textbf{0.003} & \textbf{0.001} &0.215&$>0.25$\\$t_0$ ($p=-5/3$)&$>0.25$&0.219&$>0.25$& \textbf{0.046} &$>0.25$&$>0.25$&0.098\\$p$&$>0.25$&$>0.25$&$>0.25$&$>0.25$&$>0.25$&$>0.25$&$>0.25$\\$L_g$&$>0.25$&0.127& \textbf{0.005} &$>0.25$& \textbf{0.001} & \textbf{0.017} & \textbf{0.045} \\